\documentclass[aps,prb,showpacs,twocolumn,superscriptaddress]{revtex4-1}
\usepackage{amsmath}
\usepackage{amssymb,amscd,bm,dsfont,wasysym,mathrsfs,latexsym,psfrag,accents,mathtools}
\usepackage{amsthm}
\usepackage{graphicx}
\usepackage{lipsum}
\usepackage{multirow}
\usepackage{dcolumn}
\usepackage{float}
\usepackage{color}
\usepackage{xcolor}
\usepackage{braket} 
\usepackage{comment}
\usepackage{hyperref}
\usepackage{bbold}
\usepackage{xcolor} 

\newcolumntype{L}[1]{>{\raggedright\arraybackslash}p{#1}}
\newcolumntype{C}[1]{>{\centering\arraybackslash}p{#1}}
\newcolumntype{R}[1]{>{\raggedleft\arraybackslash}p{#1}}

			\newcommand{\e}[1]{\begin{align}{#1}\end{align}}	
			\newcommand{\es}[1]{\begin{align*}{#1}\end{align*}}	
			\newcommand{\m}[1]{\begin{multline}{#1}\end{multline}}	
		
		\newcommand{\f}[2]{\frac{#1}{#2}}
		\newcommand{\tf}[2]{\tfrac{#1}{#2}}
		
		\newcommand{\p}[2]{\frac{\partial #1}{\partial #2}}

		\newcommand{\la}[1]{\label{#1}}

		\newcommand{\q}[1]{Eq.\ (\ref{#1})}
		\newcommand{\qq}[2]{Eqs.\ (\ref{#1}-\ref{#2})}
		\newcommand{\s}[1]{Sec.\ \ref{#1}}
		\newcommand{\fig}[1]{Fig.\ \ref{#1}}		
		\newcommand{\tab}[1]{Table\ \ref{#1}}	
		\newcommand{\app}[1]{App.\ \ref{#1}}
		\newcommand{\ocite}[1]{Ref.\ \onlinecite{#1}}

		\newcommand{\eq}{=&\;}

		\newcommand{\limit}[1]{\substack{\text{lim} #1}\as}

		\newcommand{\R}{\mathbb{R}}
		
		\newcommand{\C}{\mathbb{C}}

	\newcommand{\eikr}{e^{i\bk \cdot \br}}

\newcommand{\nabk}{\nabla_{\boldsymbol{k}}}

\newcommand{\var}{\varepsilon}

\newcommand\as{\;\;\;\;}
\newcommand\ass{\;\;\;\;\;\;\;\;}

\newcommand{\ba}{\boldsymbol{a}}
\newcommand{\bb}{\boldsymbol{b}}

\newcommand{\be}{\boldsymbol{e}}
\newcommand{\bff}{\boldsymbol{f}}

\newcommand{\bk}{\boldsymbol{k}}

\newcommand{\bp}{\boldsymbol{p}}
\newcommand{\bq}{\boldsymbol{q}}
\newcommand{\br}{\boldsymbol{r}}
\newcommand{\bs}{\boldsymbol{s}}
\newcommand{\bt}{\boldsymbol{t}}

\newcommand{\bv}{\boldsymbol{v}}
\newcommand{\bx}{\boldsymbol{x}}

\newcommand{\bE}{\boldsymbol{E}}
\newcommand{\bA}{\boldsymbol{A}}

\newcommand{\bF}{\boldsymbol{F}}
\newcommand{\bG}{\boldsymbol{G}}
\newcommand{\bK}{\boldsymbol{K}}

\newcommand{\bR}{\boldsymbol{R}}
\newcommand{\bS}{\boldsymbol{S}}
\newcommand{\bV}{\boldsymbol{V}}

\newcommand{\bsigma}{\boldsymbol{\sigma}}

\newcommand{\bvarpi}{\boldsymbol{\varpi}}

\newcommand{\frakP}{\mathfrak{P}}
\newcommand{\frakQ}{\mathfrak{Q}}

\newcommand{\W}{{\cal W}}

\newcommand{\ins}[1]{\;\;\;\;\text{#1}\;\;\;\;}

\newcommand{\calc}{{\cal C}}

\newcommand{\calf}{{\cal F}}

\newcommand{\calh}{{\cal H}}
\newcommand{\cali}{{\cal I}}

\newcommand{\calp}{{\cal P}}
\newcommand{\calq}{{\cal Q}}

\newcommand{\calt}{{\cal T}}

\newcommand{\calv}{{\cal V}}

\newcommand{\noi}[1]{\noindent (#1)}

\newcommand{\mo}{\text{-}1}

\newcommand{\braketaa}[2]{\big\langle #1 \big| #2 \big\rangle}
\newcommand{\ketbra}[2]{\big| #1 \big\rangle \big\langle #2 \big| }
\newcommand{\braopket}[3]{\big\langle #1 \big| #2 \big| #3 \big\rangle}

\newcommand{\lin}{\notag \\}

\newcommand{\ab}{\alpha\beta}

\newcommand{\low}{L$\ddot{\text{o}}$wdin\;}

\newcommand{\bpm}{\begin{pmatrix}}
\newcommand{\epm}{\end{pmatrix}}

\newcommand{\dg}[1]{#1^{\scriptstyle{\dagger}}}

\newcommand{\sma}[1]{\scriptscriptstyle{#1}}

\newcommand{\Z}{\mathbb{Z}}

\newcommand{\nocontentsline}[3]{}
\newcommand{\tocless}[2]{\bgroup\let\addcontentsline=\nocontentsline#1{#2}\egroup}

\newtheorem*{theorem*}{Theorem} 
\newtheorem{definition}{Definition} 
\newtheorem*{definition*}{Definition} 

\begin{document}

\title{Topological Bloch oscillations} 
\author{J. H\"oller}
\author{A. Alexandradinata}
\affiliation{Department of Physics, Yale University, P.O. Box 208120, New Haven, CT 06520-8120, USA}

\begin{abstract}
Bloch oscillations originate from the translational symmetry of crystals. These oscillations occur with a fundamental period that a semiclassical wavepacket takes to traverse a Brillouin-zone loop. We introduce a new type of Bloch oscillations whose periodicity is an integer ($\mu{>}1$) multiple of the fundamental period. The period multiplier $\mu$ is a topological invariant protected by the space groups of crystals, which include more than just translational symmetries. For example, $\mu$ divides $n$ for crystals with an $n$-fold rotational or screw symmetry; with a reflection, inversion or glide symmetry, $\mu$ equals two. We identify the commonality underlying all period-multiplied oscillations: the multi-band Berry-Zak phases, which encode the holonomy of adiabatic transport of Bloch functions in quasimomentum space, differ pairwise by integer multiples of $2\pi/\mu$. For a  class of multi-band subspaces whose projected-position operators commute, period multiplication has a complementary explanation through the real space distribution of Wannier functions. This complementarity follows from a  one-to-one correspondence between Berry-Zak phases and the centers of Wannier functions.  A Wannier description of period multiplication does not always exist, as we exemplify with band subspaces with either a nonzero Chern number or $\Z_2$ Kane-Mele topological order. In the former case, we present general constraints between Berry-Zak phases and Chern numbers, as well as introduce a recipe to construct nontrivial Chern bands -- by splitting elementary band representations. To help identify band subspaces with $\mu{>}1$, a general theorem is presented that outputs Zak phases that are symmetry-protected to integer multiples of $2\pi/n$, given the point-group symmetry representation of any gapped band subspace. A cold-atomic experiment that has observed period-multiplied Bloch oscillations is discussed, and directions are provided for future experiments. 
\end{abstract}

\date{\today}

\maketitle

{\tableofcontents \par}

\newpage
\section{Introduction}\la{sec:intro}

The highly-anticipated discovery of Bloch oscillations in superlattices\cite{Voisin1988,Mendez1993} was a crowning achievement of early solid-state physics.\cite{Wannier1937,Wannier1962a} Bloch oscillations essentially rely on the periodicity of crystal quasimomentum ($\bk$), as well as the existence of an energy gap -- both are basic features of a quantum theory of solids. From a semiclassical perspective, Bloch oscillations originate from the dynamics of a wavepacket formed from a single band. The fundamental period ($T_B$) of this oscillation is the time taken by a wavepacket in traversing a loop across the Brillouin torus; by the acceleration theorem,\cite{Nenciu1980} $T_B{=}\hbar |\bG|/|\bF|$, with $\bG$ the {smallest reciprocal vector  parallel} to a time-independent driving force $\bF$. Fundamental Bloch oscillations may equivalently be understood as coherent Bragg reflection originating from the discrete translational symmetry of the lattice.\cite{Ashcroft} \\

However, translational symmetry does not exhaust the manifold symmetries of crystals, which are classified by 230 space groups in three spatial dimensions.\cite{Hiller1986}
Occurring ubiquitously in crystals are the symmetries of rotations, reflections, inversions, screw rotations and glide reflections. All such elements of a space group $G$ (which are  not purely lattice translations) are nontrivial elements of the point group $\calp$ of $G$.  For crystals with a nontrivial point group, we will demonstrate how  Bloch oscillations arise with an integer $(\mu {\in} \mathbb{N})$ multiple of the fundamental period $T_B$.    \\


This period multiplication occurs only in the dynamics of multi-band wavepackets, i.e., wavepackets that are linear combinations\cite{Culcer2005} of ($N{>}1$) independent Bloch waves at each wavevector. To clarify, $\mu$ characterizes a finite set of  bands (numbering $N$) that are separated from all other bands by energy gaps -- above and below -- at each wavevector ($\bk$) in the Brillouin torus.\footnote{We do not assume that these energy gaps are direct.} One characteristic feature of multi-band dynamics is that the expectation value $\braket{O}(t)$ of an observable $O$ evolves quasiperiodically. That is to say, all frequencies of the Fourier peaks are generated additively by  $N$ frequencies; the generating frequencies are generically incommensurate if the point group is trivial. Generally, the smallest generating frequency that is expressible as ${2\pi}/({\mu T_{B}})$, with $\mu$ a positive integer, defines the period multiplier $\mu$ for \textit{continuous-time Bloch oscillations}. \\

Translational symmetry guarantees that $2\pi/T_{B}$ is always a generating frequency, so $\mu$ is minimally one. A nontrivial point-group element $g$ may result in  $\mu{>}1$, for certain field orientations relative to a crystallographic axis associated to $g$. Generally, $\mu$ divides the order ($n$) of $g$, where $n$ is defined as the smallest integer such that the repeated transformation $g^n$ is a translation by an  integer ($p$) multiple of a primitive lattice vector ($\ba$). If $\mu{>}1$, commensuration with $2\pi/T_B$ leads to fewer-than-$N$ {independent} generating frequencies. \\


$\mu{=}1$ Bloch oscillations describe an alternating current in the presence of a static electric field, owing to the periodicity of the band velocity ($dE/d\bk$) in $\bk$.\cite{Ashcroft} In contrast, $\mu{>}1$ is not generally explainable by the energy-momentum dispersion ($E(\bk)$), but is intrinsically a geometric property of the band wavefunctions -- precisely, $\mu$ can be formulated equivalently as quantized differences in the Berry-Zak phases\cite{Berry1984,Zak1989,WilczekFrank1984} of wavefunctions. From this perspective, $\mu$ may broadly be applied to band systems of any particle statistics -- bosonic or fermionic. Applying $\mu$ to bands of bosonic particles turns out to be particularly fruitful -- some of the most realistic experiments to observe period-multiplied Bloch oscillations involve {bosonic} cold atoms in optical lattices, as have been performed by T. Li et al. in \ocite{Li2016} and will be discussed in greater detail.\\

$\mu$  is insensitive to slight variations of experimental parameters that preserve the space group, as well as \textit{both} energy gaps (above and below). That is to say, $\mu$ is a space-group-protected topological invariant for  a fixed number of bands.\footnote{In the language of bundle theory,\cite{Alexandradinata2018,Read2017,Shiozaki2017} two vector bundles having the same  rank and symmetry -- but different $\mu$ -- cannot be isomorphic.}\\

From a broader perspective, not all topological invariants in band systems have been linked to definitive signatures in transport experiments.\cite{Konig2007,Vayrynen2013} By `transport', we mean to induce dynamics in $\bk$-space  by applying electromagnetic fields (or generalized forces in cold-atomic systems). A celebrated example is the Thouless-Kohmoto-Nightingale-den Nijs invariant\cite{Thouless1982} which is linked to a quantization of the Hall conductance; this invariant is protected by the symmetry of charge conservation, and has become the paradigm for a topological phase of matter.  Recently, a notion of space-group-protected topological invariants has been formulated (by one of us) for \textit{magnetotransport},\cite{Alexandradinata2018a} which is measurable in the phase offset of quantum oscillations.\cite{Alexandradinata2018b} In contrast, the present work represents the first formulation of a space-group-protected topological invariant ($\mu$) in \textit{electric} transport.

\section{Summary and organization}\la{sumo}

\begin{figure}
\centering
\includegraphics[width=8.5cm]{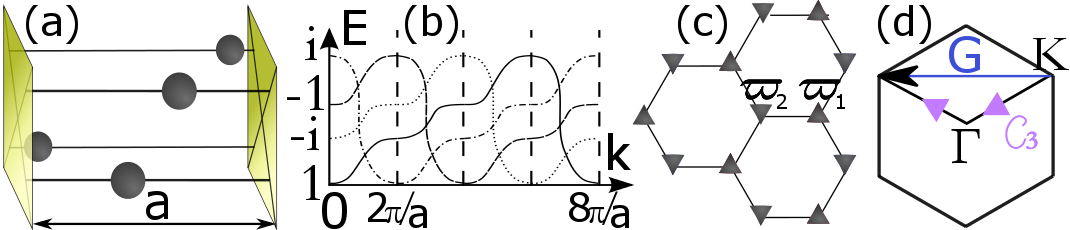}
\caption{(a) Screw-symmetric chain with lattice period $a$ and  four sites per unit cell. (b) The corresponding energy bands ($E(\bk)$) are four-fold connected. Each energy band can be labeled by one of four symmetry eigenvalues,\cite{Michel1999} as indicated by the different characters of the lines. (c) The unit cell of a honeycomb lattice contains two sites at coordinates $\bvarpi_1$ and $\bvarpi_2$. (d) Brillouin zone corresponding to the honeycomb; the loop  $\calc_3$ (violet) intersects the $C_{3,z}$-invariant wavevectors $\Gamma$ and $K$. \la{fig:helix} }
\end{figure}

Let us summarize our key results and the organization of this work. We consider two classes of Bloch oscillations, which are geometrically distinguished by the direction of the field relative to certain crystallographic axes. To summarize this geometric distinction, we associate the first class of Bloch oscillations  to a $\parallel$-field, and  the second to a $\perp$-field; often we will employ $\perp$ vs. $\parallel$ as subscripts on various quantities (e.g. $\mu$) to remind the reader of the different contexts. \\

The first class of Bloch oscillations occurs  only for space groups with a nonsymmorphic element ($g$), as exemplified by screw rotations and glide reflections (in short, screws and glides). A nonsymmorphic element involves a translation by a rational fraction of a primitive lattice vector: $p\ba/n$; $\ba$ lies along the screw axis (for $g$ a screw), or lies in the glide plane (for $g$ a glide). We may always choose a representative for $g$ such that  $0{<}p{<}n$.\cite{MelvinLax1974,Hiller1986} \fig{fig:helix}(a) exemplifies a screw-symmetric lattice with $n{=}4$ and $p{=}1$. Period-multiplied Bloch oscillations occur when we align  the field \emph{parallel} to the reciprocal vector ($\bG$) dual to $\ba$.   In all of our case studies, $\bG$ is also parallel to $\ba$.\\

The second class of Bloch oscillations is applicable to any space group with a nontrivial point group $\calp$. Precisely, we mean that the quotient group $\calp{=}G/\calt$, of the space group $G$ with its translational subgroup $\calt$, is not a trivial group. Included are space groups with nonsymmorphic elements (glides and screws, as discussed above), as well as space groups with symmorphic  elements (e.g.\ rotations and reflections). For symmorphic $g$, a spatial origin exists for which $g$ involves no fractional translations ($p{=}0$). In the second class of Bloch oscillations, we align the field \emph{perpendicular} to the screw/rotational axis (for $g$ a screw/rotation) or to the glide/reflection plane (for $g$ a glide/reflection).\\

Throughout this work, we will employ the same symbol $g$ to denote a space-group  element that is not purely translational, i.e., $g$ represents a nontrivial  element in $\calp$. In most contexts, $g$ refers to a single symmetry (symmorphic or nonsymmorphic) which should be deducible from the context.  For any space group $G$, $\calp{=}G/\calt$  is isomorphic to a group comprising isometries of a lattice that fix a point in space.\cite{MelvinLax1974} Point groups are sometimes defined as point-fixing isometries.  However, for this work,  a `point group'  ($\calp$) should be understood as the quotient group $G/\calt$ unless specified otherwise. For nonsymmorphic space groups, it is possible that elements in $\calp$ (such as a screw or glide) do \textit{not} preserve a spatial point.   \\


With a $\parallel$-field, Bloch oscillations occur with period multiplier
\e{ \mu_{\parallel} = \tf{n}{\text{gcd}(p,n)} \la{munon} }
with gcd ${=}$ greatest common divisor; $\mu_{\parallel}{=}4$ for the lattice of \fig{fig:helix}(a).  There are two complementary perspectives -- from real and quasimomentum  spaces -- to understand how period multiplication occurs. \\

\noi{A} In the real-space perspective,  bands are represented by exponentially-localized Wannier functions with well-defined average positions $\bvarpi$ (henceforth referred to as Wannier centers). The  nonsymmorphic symmetry $g$ results in Wannier functions coming in $\mu_{\parallel}$-plets per unit cell, and their corresponding Wannier centers are mapped by symmetry as $\bG {\cdot} \bvarpi {\rightarrow} \bG {\cdot} \bvarpi  {+} 2\pi/\mu_{\parallel}$. In other words, the translational period between Wannier centers  is $1/\mu_{\parallel}$ times the Bravais lattice period, hence the Bloch-oscillatory period is multiplied by $\mu_{\parallel}$, as further elaborated in \s{sec:BOparWF}. \\   

\noi{B} In the quasimomentum space ($\bk$-space) perspective, the nonsymmorphic symmetry representations of energy bands are permuted as $\bk$ is advanced by the primitive reciprocal vector $\bG$.\cite{Michel1999} The failure of the representation of $g$ to be single-valued -- known as monodromy -- results  in bands being connected as a graph, as illustrated in \fig{fig:helix}(b). This monodromy is also imprinted on the adiabatic transport of Bloch functions within each connected component -- specifically on the geometric component of the adiabatic transport, which encodes the multi-band, non-Abelian holonomy\cite{WilczekFrank1984} of noncontractible Brillouin-zone loops.\cite{Zak1989,Alexandradinata2014c} The eigenvalues of the holonomy matrix, as defined for transport over a single fundamental period $T_B$, are known as the Zak phase factors ($e^{i\phi}$). One of our key results is that, owing to the monodromy of symmetry representations, the set of $\phi$  has the translational property: $\phi {\rightarrow} \phi{+}2\pi/\mu_{\parallel}$. Only after $\mu_{\parallel}$ fundamental periods do all pairwise Zak phase differences  equal an integer multiple of $2\pi$. This provides a complementary explanation for  period multiplication that we  further develop in \s{parbloch}.     \\

\noindent The similar translational properties in (A{-}B) are not accidental, but reflect a one-to-one correspondence between Wannier centers and Zak phases: $\bG {\cdot} \bvarpi{=}\phi$ for a \textit{multi-band} subspace. Such a correspondence is already known for a single band;\cite{Zak1989} in subsequent work, Kingsmith and Vanderbilt have shown that the \textit{sum} of all Zak phases is related to  the \textit{total}  polarization  of a multi-band subspace.\cite{King-Smith1993} Our contribution -- for nonsymmorphic space groups -- is to relate \textit{individual} Zak phases to the polarization of individual Wannier functions. Such correspondences will henceforth be referred to as \textit{multi-band Zak-Wannier relations}.\\



Let us next consider Bloch-oscillatory phenomena under a $\perp$-field, with the corresponding period-multiplier $\mu_{\perp}$. In two-dimensional crystals, $\mu_{\perp}{>}1$ occurs only for \textit{certain} multi-band subspaces, which can be characterized from two complementary perspectives.\\





\noi{A'} In the real-space perspective, we are interested in multi-band subspaces whose Wannier functions are so strongly localized (in 2D space) as to resemble point charges. Such band subspaces are not generic for the following reason:  while the bare position operators ($x$ and $y$) commute according to basic quantum-mechanical principles, position operators that are projected to a multi-band subspace (by the operator $P$) do not generally commute, no matter the space-time symmetry of $P$. However, by combining a symmetry condition with a condition on the tunneling strength between spatially-separated Wannier functions, we have identified a class of multi-band subspaces for which the projected-position operators $PxP$ and $PyP$ commute to exponential accuracy. Such subspaces will be referred to as `strong elementary band representations', and they are exhaustively identified by us for all 2D space groups  in \s{perp}. Our nomenclature builds upon the theory of elementary band representations (EBRs),\cite{Zak1981,Evarestov1984,Bacry1993,Bradlyn2017,Cano2017} which are representations of a space group on locally-symmetric Wannier functions that  cannot be split into smaller representations on locally-symmetric Wannier functions. By `locally-symmetric', we mean that for any spatial position $\br$, all Wannier functions centered on $\br$ form a representation of the site-stabilizer of $\br$ (i.e., the subgroup of the space group that preserves $\br$).\footnote{More details on EBRs are provided in \s{perp}} The description `strong' alludes to the just-mentioned commutation, {which implies that Wannier functions are strongly localized.} \\

Given that the Wannier centers ($\bvarpi$) are uniquely defined as simultaneous eigenvalues of both projected position operators,  we further particularize (in \s{perp}) to a subclass of strong EBRs for which a nearest-neighbor pair of Wannier centers ($\bvarpi_1,\bvarpi_2$) satisfies $|\bG {\cdot} (\bvarpi_1 {-} \bvarpi_2)|{=}2\pi/\mu_{\perp}$ for $\mu_{\perp}{>}1$ and $\bG$ a primitive reciprocal vector.  $2\pi/\mu_{\perp}$ may be viewed as the dynamical phase difference acquired by two Wannier functions over one fundamental period, in the presence of a field $\bF$ that integrates to $\bG{=}\int_0^{T_B}\bF dt/\hbar$.  Only after $\mu_{\perp}$ fundamental periods do pairwise phase differences acquired by Wannier functions return to an integer multiple of $2\pi$. For illustration, we consider a two-band subspace which comprises two $s$-orbitals centered on a honeycomb lattice, as illustrated in \fig{fig:helix}(c). Since each honeycomb vertex is invariant under three-fold rotation (modulo lattice translations), $|\bG {\cdot} (\bvarpi_1 {-} \bvarpi_2)|{=}2\pi/3$, leads to period-tripled Bloch oscillations.  \\

\noi{B'} In the $\bk$-space perspective, $[PxP,PyP]{=}0$ is equivalent\cite{Marzari1997} to the condition that the multi-band, non-Abelian Berry curvature vanishes at each $\bk$. The $\bk$-space analog of the dynamical phase condition for Wannier functions is that  the geometric Zak phases $\phi$ (associated to $g$-symmetric $\bk$-space loops, e.g.\ $\mathcal{C}_3$ illustrated in \fig{fig:helix}(d)) differ pairwise by integer multiples  $2\pi/\mu_{\perp}$; in particular, one pair of Zak phases must differ exactly by $2\pi/\mu_{\perp}$. One other key finding of this work is a theorem in \s{Zak} that aids us in identifying band subspaces whose Zak phases satisfy the just-mentioned conditions. For any $g$-symmetric band subspace, this theorem inputs the symmetry representations of $g$ at $g$-invariant $\bk$-points, and outputs the Zak phases which are symmetry-protected to the $n$'th roots of unity. By a `symmetry-protected' quantity, we mean a quantity (the Zak phase, in the present context) that is invariant under deformations of the Hamiltonian that preserve the symmetry $g$ as well as both energy gaps (above and below the $N$-band subspace). To recapitulate, the theorem is a means to systematically calculate symmetry-protected Zak phases, given the representation of Bloch functions at high-symmetry wavevectors; these representations can easily be determined from tight-binding models.\cite{Alexandradinata2014c} Differences in Zak phases not only determine the period of Bloch oscillations (as elaborated in \s{perp}), but they are in principle also measurable by generalized Ramsey interferometry in cold-atom experiments.\cite{Abanin2013,Atala2013a} A limited form of the present  theorem already exists for $g$ being a spatial-inversion symmetry;\cite{Alexandradinata2014c} this work provides the generalization to any space-group symmetry, including rotations, reflections, screws and glides.  \\

\noindent In \s{ZakEBR}, we propose another Zak-Wannier relation which underlies the similarities between (A') and (B'). We remark that Zak-Wannier relations have previously been formulated for Wannier functions which are maximally localized\cite{Marzari1997} in one direction,\cite{Zak1989,King-Smith1993,Alexandradinata2014c} but do not necessarily have any symmetry.  In comparison, \s{ZakEBR} describes a novel \textit{symmetry-based Zak-Wannier relation} for locally-symmetric Wannier functions that are localized in two independent directions, but such localization need not be maximal. A general formulation of symmetry-based multi-band Zak-Wannier relations (with applications going beyond strong EBRs) is presented in \s{composite}.  \\


Finally, we discuss the possibility of period multiplication (in a $\perp$-field) for band subspaces that have \textit{no} symmetry-based Zak-Wannier relation. These are topologically nontrivial band subspaces with symmetry-protected Zak phases differing by $2\pi/\mu_{\perp}$, but with no locally-$g$-symmetric representation on Wannier functions:\\

\noi{C'} Time-reversal-\textit{asymmetric} band subspaces -- in Wigner-Dyson symmetry\cite{Dyson1962,Kitaev2009a,Schnyder2009}  class A -- are classified by  a Chern number; band subspaces with a nonzero Chern number (in short, Chern bands) do not admit a representation by Wannier functions.\cite{Brouder2007} Our case study of Chern bands in \s{classA} demonstrates that quantum fluctuations -- with respect to the noncommuting projected position operators -- can produce a classically-forbidden symmetry-protected Zak phase (cf.\ \qq{c4constraint}{c6constraint}). This suggests the possibility of a classically-forbidden period multiplier $\mu_{\perp}$, for which we have constructed a model as proof of principle (in \s{sec:proofofprinciple}).  In constructing this model, we have utilized a novel recipe to generally obtain Chern bands  by splitting a multi-band EBR into multiple fewer-band subspaces. Our recipe provides a direct roadmap toward model realizations and concrete materializations of Chern bands.\\
 
\noi{D'} Time-reversal-symmetric band subspaces -- in Wigner-Dyson symmetry class\cite{Dyson1962,Kitaev2009a,Schnyder2009} AII -- are characterized by a $\Z_2$ Kane-Mele invariant.\cite{Kane2005b} The paradigmatic example of $\Z_2$ topological order is the Kane-Mele honeycomb model\cite{Kane2005} which realizes Bloch oscillations with multiplier $\mu_{\perp}{=}3$, as we demonstrate in \s{classAII}. \\

\noindent While the Chern and Kane-Mele invariants were first studied in the context of band insulators, these same invariants may more generally be used to characterize band wavefunctions -- independent of the statistics of the particles that fill these bands. Our proposal for Bloch oscillations does not apply to band insulators,\cite{Ashcroft} but may in principle apply to band metals. Realistically, Bloch oscillations are measurable in cold-atomic experiments: \s{cold} describes an existing experimental setup\cite{Li2016} that simulates \fig{fig:helix}(c) with bosonic $^{87}$Rb atoms in an optical honeycomb lattice. We further explain how reported measurements\cite{Li2016} may be interpreted as period-tripled Bloch oscillations, and provide directions for future cold-atomic experiments. The experimental feasibility of period-multiplied Bloch oscillations is discussed more generally in \s{exp}, where we address the validity of the adiabatic assumption, as well as the effect of finite relaxation time.\\


We conclude in \s{sec:discussin} by summarizing the unifying themes of this work, with additional elaboration on Zak-Wannier relations and the role of hybrid Wannier functions.\cite{Maryam2014} An outlook is provided for future investigations.

\section{Bloch oscillations in a $\parallel$-field}\la{sec:parallelfield}

\subsection{Motivational example on a screw-symmetric chain}

\begin{figure}[ht]
\centering
\includegraphics[width=8cm]{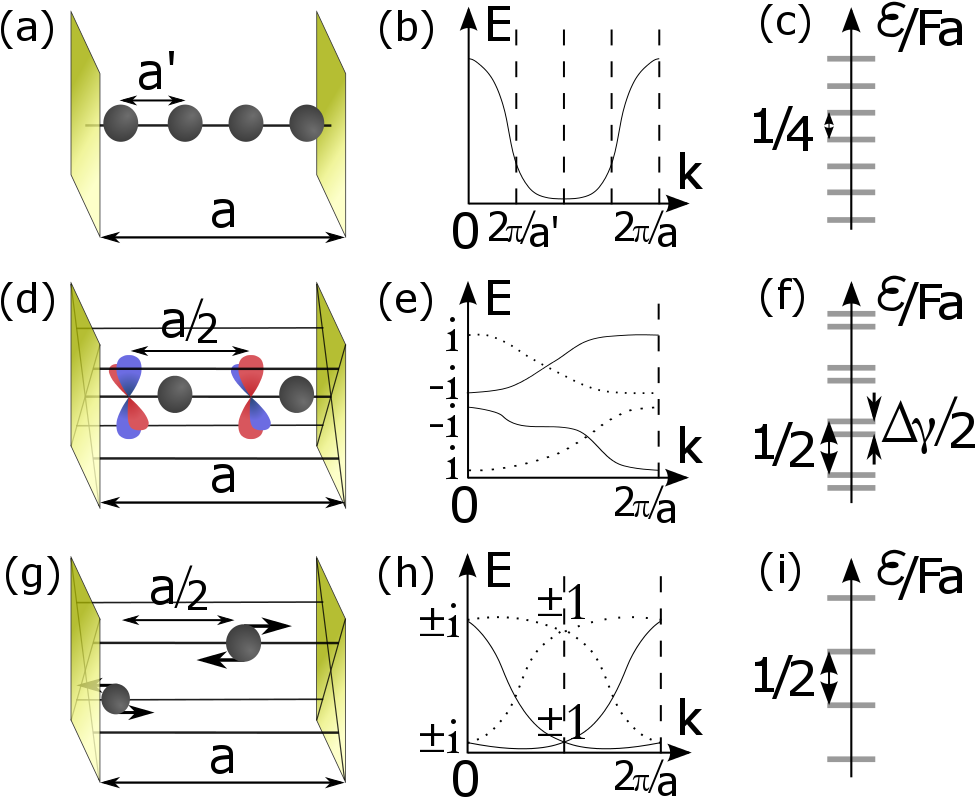}
\caption{Three 1D lattices with the following symmetries:  (a-c) trivial point group and a primitive translation period $a'{=}a/4$,  (d-f)  $g_{4,2}$-symmetry, and (g-i) $g_{2,1}$-symmetry. (a,d,g) illustrate the real-space distribution of Wannier functions (grey spheres indicate $s$-orbitals, red-blue dumbbells $p_x{+}i p_y$-orbitals), (b,e,h) the corresponding energy bands ($E(\bk)$) at zero field ($F{=}0$), (c,f,i)  the ladder-like spectra $(\varepsilon)$ of $P(H_0{-}Fz)P$ at nonzero field, in units of $Fa$. The Wannier functions in (g) are Kramers degenerate, as indicated by two arrows per sphere; each energy level in (i) is Kramers degenerate. \la{fig:bands} }
\end{figure}

We begin with a Gedankenexperiment to motivate period multiplication in a $\parallel$-field ($F$) that is aligned parallel to the fractional translation of a nonsymmorphic symmetry. Consider a one-dimensional chain with primitive lattice period $a'$, as illustrated in \fig{fig:bands}(a). This chain is described by a translation-invariant, single-particle Hamiltonian $H_0$; for simplicity we assume only translational symmetry, so the eigenvalues of $H_0$ are nondegenerate at each $k$, as exemplified in \fig{fig:bands}(b). A field applied parallel to the chain adds to the Hamiltonian a term ${-}Fz$, with $z$ the position operator along the chain. Adiabatic evolution of a wavepacket in a nondegenerate band results in Bloch oscillations with the fundamental oscillation period $2\pi \hbar/(Fa')$. \\

Suppose we deform the straight lattice into the helix of \fig{fig:helix}(a) with new lattice period $a{=}4a'$. The standard argument for Bloch oscillations (protected by translational symmetry) would predict a quarter reduction of the oscillation period to $T_B{=}2\pi\hbar/(Fa)$. However, we propose that the oscillation period persists at $2\pi \hbar/(Fa'){=}4T_B$, owing to a new kind of Bloch oscillation that is protected by a nonsymmorphic symmetry of the deformed lattice, as we explain in the next section.

\subsection{Bloch oscillations with $\mu_{\parallel}{>}1$ from the perspective of Wannier functions} \la{sec:BOparWF}

The persistence of the oscillation period in the Gedankenexperiment may be understood from the following symmetry analysis. Any symmetry ($g$) of a space group ($G$) is composed of a component $\check{g}$ that leaves the spatial origin invariant, as well as a translation by $\bt$; notationally, $g{=}(\check{g}|\bt)$. A nonsymmorphic symmetry is a symmetry that involves a translation by a rational fraction of a primitive lattice vector; such a symmetry may be represented in a particular coordinate system as $g_{n,p} = (\check g_n| p a \be_z/n)$ with $0{<} p{<} n$.\cite{MelvinLax1974,Hiller1986} This is exemplified in \fig{fig:helix}(a) by a screw, which is the composition of a four-fold rotation about the crystal-axis with a translation by $a/4$ along it; notationally, $g_{n,p}{=}(C_{4,z}|a \be_z/4)$ with $n{=}4$ and $p{=}1$, where $C_{n,j}$ denotes an $n$-fold rotation about the $j$th coordinate axis (with unit vector $\be_j$). \\

Let us consider a low-energy, $N$-band subspace that is energetically separated from other bands, and comprises $N$ Bloch functions $\{\psi_{j,k}\}_{j=1}^N$ at each $k$.  In the adiabatic approximation,\cite{Nenciu1991} field-induced dynamics within this low-energy subspace  is described by the effective Hamiltonian 
\e{ \calh=P(H_0-Fz)P, \;\;\text{with}\;\;P=\sum_k\sum_{j=1}^N\ket{\psi_{j,k}}\bra{\psi_{j,k}}\la{effham}}
a projector to the $N$-band subspace, where $\sum_k$ is a short-hand for a sum over the Brillouin zone. The eigenstates of $\calh$ are Wannier functions which are exponentially-localized in $z$, and long-lived in the sense of a resonance;\cite{Nenciu1991} the corresponding spectrum is well-known to have the structure of a Wannier-Stark ladder\cite{Wannier1937,Wannier1962a} owing to translational symmetry, i.e., for any eigenvalue $\var$, there exists another at $\var{+}Fa$ due to the invariance of $P$ and $H_0$ under $z{\rightarrow} z{-}a$. \\

From this perspective, it is quite natural that $g_{n,p}$, which involves a fractional translation, generates a ladder with a fractional spacing: 
\e{\var \rightarrow \var{+}Fap/n.\la{fractionalladder}} 
In the $g_{4,1}$-symmetric example  of \fig{fig:helix}(a), we therefore find that adjacent levels are separated by $Fa/4$, as illustrated in \fig{fig:bands}{(c)}.\\ 

Let us then consider the quantum expectation value of an observable $\braket{O}(t)$, for a generic state that linearly combines the eigenfunctions of $\calh$. The generically-allowable frequencies in $\braket{O}(t)$ are, up to a proportionality constant of $\hbar$, equal to pairwise differences in the eigenvalues of $\calh$. In particular, the smallest energy difference between adjacent (i.e., nearest-neighbor) levels in the spectrum of $\calh$ determines the smallest frequency in $\braket{O}(t)$; in the example of \fig{fig:helix}(a), the nearest-neighbor spacing of $Fa/4$ determines that $2\pi/(4T_B)$ is the smallest allowable frequency, leading to period-quadrupled, continuous-time Bloch oscillations (cf. \s{sec:intro}).  \\

Generally, for any $g_{n,p}$-symmetric lattice, we would like to determine the period multiplier  for continuous-time Bloch oscillations, as defined in \s{sec:intro}.  By combining \q{fractionalladder} and Bezout's identity,\cite{Bezout} we obtain that the nearest-neighbor spacing within one ladder is simply $Fa/\mu_{\parallel}$, with $\mu_{\parallel}$ defined in \q{munon} and identified with the period multiplier.\footnote{To further motivate the form of the period multiplier, we may consider that the same $g_{4,1}{=}(C_{4,z}|a \be_z/4)$-symmetric lattice in \fig{fig:helix}(a) is also invariant under $g_{4,1}^{-1} (0|a\be_z){=}(C^{\mo}_{4,z}|3a \be_z/4)$, which is an order four  symmetry with $p{=}3$. However, $\mu_{\parallel}{=}4$ independent of whether $(n,p){=}(4,1)$ or ${=}(4,3)$.} \\




Thus far, we have only discussed examples with the minimal number of Wannier functions allowable by $g_{n,p}$-symmetry alone -- namely, $\mu_{\parallel}$ Wannier functions per unit cell. In general, 
 the total number ($N$) of Wannier functions per unit cell may exceed $\mu_{\parallel}$; we may further show that $N $ is always an integer multiple ($J$) of  $\mu_{\parallel}$.\footnote[43]{\label{note:Supp} See Supplementary material, which includes \ocite{Nenciu1987,Fiorenza2015,Slater1954,Lodwin1950,Winkler2003,MelvinLax1974,Fang2013a,Evarestov,Michel1999,Alexandradinata2014c,Bacry1988,Bacry1988a,Bradlyn2017,Cano2017,Michel2001,MichelIV,Michel2000,Evarestov1984,WilczekFrank1984,Berry1984,Zak1989,Nenciu1980,Kato1950,Budich2013,Messiah,Nenciu2008,Avron1977,Nenciu1991,Avron1979} } In the absence of any other symmetry beyond $g_{n,p}$, there would then be $J$ non-degenerate ladders in the spectrum of $\calh$.
A simple example of $N{=}4,\mu_{\parallel}{=}2,J{=}2$ is illustrated in \fig{fig:bands}(d), with two $s$- and two $d$-orbitals in the primitive unit cell of a $g_{4,2}$-symmetric lattice; then, the Fourier peaks of  $\braket{O}(t)$ lie at frequencies generated additively by $2\pi/(\mu_{\parallel} T_B)$ and the frequency offset ($\Delta \gamma/(\mu_{\parallel} T_B)$) between the two ladders, as illustrated in \fig{fig:bands}(f); these two generators are generically incommensurate.\\


In contrast, for half-integer-spin systems that respect time-reversal and glide ($g_{2,1}$) symmetries, the Kramers degeneracy of each rung (as illustrated in \fig{fig:bands}(g{-}i)) ensures there is only one generating frequency ($2\pi/(\mu_{\parallel} T_B)$ with $\mu_{\parallel}{=}2$) in the Fourier spectrum. The above two examples with $J{>}1$ demonstrate that: (a) a nonsymmorphic symmetry $g_{n,p}$ guarantees that $2\pi/(\mu_{\parallel} T_B)$ is always a generating frequency, but (b) the presence of other independent generating frequencies depends on symmetries other than $g_{n,p}$.  A general symmetry-based criterion to determine the degeneracy of Wannier-Stark ladders, for arbitrary $J$, is presented in \app{deg}.


\subsection{Bloch oscillations with $\mu_{\parallel}{>}1$ from the perspective of Bloch functions}\la{parbloch}
One subtlety in the above proof is that the position operator only has a well-defined action on spatially-localized states (precisely, functions in the domain of $z$).\cite{Blount1962} However, a natural basis to describe band subspaces are spatially-extended Bloch functions $\psi_{j,k}$, owing to their forming a representation of translational symmetry. We are therefore motivated to derive period multiplication from a complementary perspective with Bloch functions as initial states. \\

Let us then consider the time evolution of Bloch functions that are initially restricted to a single wavevector $k_0$; let $\{\psi_{j,k_0}\}_{j=1}^N$ span the low-energy subspace at wavevector $k_0$. In the adiabatic limit, the time-evolution propagator is  the unitary generated by $\calh$ over one fundamental Bloch period; this propagator can be expressed in the basis of $\{\psi_{j,k_0}\}_{j=1}^N$  as the time-ordered exponential (denoted by $\overline{\mathrm{exp}}$)
\e{ U(T_B) =  \overline{\mathrm{exp}}\big[i{\int}_{0}^{T_B} \Big(\; F \, A \big(k(t)\big) - \mathfrak{E}\big(k(t)\big)\;\Big) \mathrm dt/\hbar \big], \la{adevo} }
with the non-Abelian Berry connection\cite{Berry1984,WilczekFrank1984}
\e{ A_{j,j'}(k) = \braket{ u_{j,k}| i \partial_k| u_{j',k} }_{\text{cell}}, \la{nonAB} }
and the energy matrix
\e{ \mathfrak{E}_{j,j'}(k) = \braket{ u_{j,k}| e^{-i k x} H_0 e^{i k x} | u_{j',k} }_{\text{cell}}; \la{ematrix}}
 $u_{j,k}$ is the cell-periodic component of $\psi_{j,k}$, and $\braket{f|g}_{\text{cell}}$ denotes the inner product over one unit cell.\\

Subsequently, we will exploit symmetry to deduce spectral constraints on the  propagator over one fundamental period (cf.\ \q{adevo}). Therefore, it is useful to have a definition of period multiplication for the stroboscopic time-evolution of the quantum expectation value $\{\braket{O}(t{=}jT_B)\}_{j{\in}\Z_{\ge 0}}$ of a translation-invariant observable $O$; this definition would be different but closely analogous to our previous definition for continuous-time evolution (cf.\ \s{sec:intro} and \s{sec:BOparWF}).  \\

\begin{definition} \la{define:stroboscopic}
\normalfont Fourier transform $\{\braket{O}(t{=}jT_B)\}_{j{\in}\Z_{\ge 0}}$, and identify the Fourier peak lying at the smallest rational multiple of the fundamental frequency $2\pi/T_B$. The rational multiplier is of the form $1/{\mu}$, with $\mu$ a positive integer that is defined as the period multiplier for \textit{stroboscopic Bloch oscillations}.\\
\end{definition}




Defining $\{e^{i\varphi_j}\}_{j{=}1}^N$ as the eigenvalues of $U(T_B)$, pairwise differences in the adiabatic phase ($\varphi_j{-}\varphi_{j'}$) manifest as Fourier peaks in the stroboscopic time evolution. The goal of this section is to prove that the spectrum has a ladder-like structure: $\varphi {\rightarrow} \varphi{+}2\pi/\mu_{\parallel}$. That is, for every $\varphi_j$, there exists a $\varphi_{j'}{=}\varphi_j{+}2\pi/\mu_{\parallel}$ mod $2\pi$. For a state initialized as a Bloch function, stroboscopic Bloch oscillations therefore occur with multiplier $\mu_{\parallel}$. This complements our previous result for the adiabatic time-evolution of a spatially-localized initial state, for which continuous Bloch oscillations occur with multiplier $\mu_{\parallel}$. \\

What is the origin of this ladder structure? A clue to its origin arises from the following observation: the ladder structure exists even in the purely-geometric component of the adiabatic propagator, but not in the purely-dynamical component. To elaborate, let us define $\W$ as equal to $U(T_B)$ with a zero energy matrix; $\W$ is an $N$-by-$N$ matrix representation  of holonomy, and is expressible as a path-ordered exponential (also denoted $\overline{\text{exp}}$) of the non-Abelian Berry connection:\cite{Berry1984,WilczekFrank1984}
\e{ \W = \overline{\mathrm{exp}} \big[ {i \int_0^{G} A (k) dk } \big], \la{Wilsonloopst} }
with $G{=}\int_0^{T_B}Fdt/\hbar$ a primitive reciprocal vector. Defining $\{e^{i\phi_j}\}_{j{=}1}^N$ as the eigenvalues of $\W$, then the Zak phases $\phi_j$ satisfy the same translational property: $\phi {\rightarrow} \phi{+}2\pi/\mu_{\parallel}$. \\

Let us explain this ladder for the screw-symmetric ($g_{4,1}$) four-band subspace  illustrated in \fig{fig:helix}(a{-}b); $N{=}\mu_{\parallel}{=}4$. Since $(g_{4,1})^4$ is a lattice translation that is represented on Bloch functions as $\mathrm{e}^{-i k a}$, the eigenvalues of $g_{4,1}$ fall into four $k$-dependent branches: $\{i^j e^{-ika/4}\}_{j=1}^4$. When $k$ is advanced by a primitive reciprocal period $G$, the representation indexed by $j$ is permuted to $j{-}1$ in the reduced-zone scheme;  only after four  periods does the representation recur: $j{-}4{\equiv}j$, implying that bands are connected minimally in sets of four, as illustrated in \fig{fig:helix}(b). This symmetry permutation during adiabatic transport is imprinted on $\W$, which encodes how the wavefunction evolves as a function of $k$; note that the connection (cf.\ \q{nonAB}) involves derivatives of the cell-periodic function with respect to $k$. Precisely, $\W$ is equivalent (modulo a band-independent phase) to a $\mu_{\parallel}$-cycle, i.e., $\W$ cyclically permutes all $\mu_{\parallel}$ basis vectors. The ladder structure arises because such a permutation matrix has eigenvalues $\{e^{2\pi i j/\mu_{\parallel}}\}_{j=1}^{\mu_{\parallel}}$.\\

It is instructive to compare the Zak phase ladder to the Wannier-Stark ladder described in \s{sec:BOparWF}. A comparison can be made by utilizing the known   spectral equivalence between the projected position operator $PzP$ and  ${-}i$log$\W$:\cite{Alexandradinata2014c} namely, for any eigenvalue $\varpi_j$ of $PzP$, there exists a Zak phase such that $\phi_j{=}G\varpi_j$ mod $2\pi$. This equivalence shows that the ladder structures of $\phi$ and $PzP$ are complementary -- the latter is a simple consequence of Wannier functions being related by a nonsymmorphic symmetry (cf.\ \s{sec:BOparWF}). The above spectral equivalence is the first appearance of a multi-band Zak-Wannier relation, which is a one-to-one correspondence between Zak phases (defined modulo $2\pi$) and Wannier centers (defined modulo lattice translations). \\ 

The ladder structure of $\phi$ persists when dynamical contributions are included, i.e., the adiabatic phase $\varphi$ (inclusive of geometric and dynamical contributions) also has a ladder structure, as proven in \app{app:perm}. Let us first consider the simplest cases with the minimal number of bands allowable by $g_{n,p}$ and time-reversal symmetries: $N{=}\mu_{\parallel}$ bands for integer-spin representations (as exemplified by the four-fold-screw-symmetric chain in \fig{fig:helix}(a{-}b)), or  $N{=}2\mu_{\parallel}$ bands for half-integer-spin representations (exemplified by a glide-symmetric chain with the hourglass dispersion of \fig{fig:bands}(c)). In these cases,  $\{\varphi_j\}_{j=1}^N$ and $\{\phi_j\}_{j=1}^N$ differ only by a constant offset (modulo $2\pi$). The general structure of the eigenvalues and -vectors of $U(T_B)$ for possibly more than one connected component, is detailed in \app{app:perm}. \\

We remark that a ladder-like structure (with $\mu_{\parallel}=2$) for the Zak phase $\phi$ has been previously noted in a glide-symmetric generalization of the Su-Schrieffer-Heeger model,\cite{Zhang2016} however it was not appreciated therein that the complete adiabatic phase $\varphi$ would have the same ladder structure. The proposed realization of this `two-leg' Su-Schrieffer-Heeger model by ultracold atoms\cite{Zhang2016} offers an experimental test of period-doubled Bloch oscillations.  

\subsection{Generalization to three dimensions}

The above discussion is simply generalized to higher-dimensional Bravais lattices with a nonsymmorphic symmetry $g_{n,p}{=}(\check g_n|p\ba/n)$ (we remind the reader of the definitions of $g_{n,p}, \check g_n, p, n, \ba$ in \tab{symbols}). Stroboscopic Bloch oscillations with $\mu_{\parallel} {>} 1$ occur if: (i) the force is directed parallel to a primitive reciprocal vector $\bG$ dual to $\ba$, and (ii) the initial state is a linear combination of Bloch functions with $g_{n,p}$-invariant wavevectors, i.e., wavevectors which satisfy $\check g_{n} \bk_0 {=}\bk_0$ modulo reciprocal vectors. For screw (resp.\ glide) symmetry, the $g_{n,p}$-invariant wavevectors form a line (resp.\ a plane) in the 3D Brillouin-zone (BZ).

\section{Bloch oscillations in a $\perp$-field}\la{sec:perpfield}

\subsection{Motivational example on a honeycomb lattice}\la{honey}

Let us begin with the conceptually-simplest example of period-multiplied Bloch oscillations in a $\perp$-field; this example has been realized experimentally with ultracold atoms in optical lattices, as we elaborate in \s{cold}.  In real space, the band subspace is represented by spatially-localized Wannier functions (s-orbitals, for simplicity) centered on the honeycomb lattice, as illustrated in \fig{fig:helix}(c). Denoting the coordinates of two nearest-neighbor Wannier centers as $\bvarpi_1$ and $\bvarpi_2$, all other Wannier centers are obtained by lattice translations. The spatial distribution of all Wannier centers is invariant under the space group $G{=}p6m$, whose point group is generated by a six-fold rotation  and a reflection. \\

We align a field $\bF$ perpendicular to the rotational axis (hence the name $\perp$-field) and parallel to a primitive reciprocal vector $\bG$; we assume that the field induces transport that is adiabatic with respect to the above-mentioned band subspace.\footnote{A justification of the adiabatic approximation is provided in \s{adiabatic}.} Let us demonstrate that this band subspace realizes three-fold period multiplication. Field-induced dynamics of Wannier functions can heuristically be modelled by the dynamics of point charges on the honeycomb lattice; we will justify this heuristic model in the next section. The field couples to a point charge with coordinate $\bvarpi$ as a scalar potential ${-}\bF{\cdot} \bvarpi$, and the dynamical phase acquired by this charge over one fundamental Bloch period is ${-}\int_0^{T_B}\bF{\cdot}\bvarpi dt {=}{-}\bG{\cdot} \bvarpi$. In particular, due to $\bvarpi_1$ and $\bvarpi_2$ being related by a six-fold rotation, the dynamical phase difference acquired by nearest-neighbor point charges over $T_B$ is $\bG{\cdot} (\bvarpi_1{-}\bvarpi_2){=}2\pi/3$ mod $2\pi$. Only after three fundamental periods do all pairwise phase differences become integer multiples of $2\pi$, hence returning the point charges to their initial state (modulo a global phase) at time zero. The previously-mentioned ultracold-atomic experiment has measured this three-fold multiplication, as we elaborate further in \s{cold}. \\

To what extent is this classical description by point charges valid? This description is classical in the sense that the coordinates of Wannier centers are simultaneous eigenvalues of the projected position operators $PxP$ and $PyP$, with $P$ projecting to the above-mentioned band subspace. It is not a priori obvious that these two operators commute. A famous counter-example where strong quantum fluctuations (with respect to $PxP$ and $PyP$) are topologically quantized is a band subspace with a nonzero Chern number,\cite{Thouless1982} as we further explore in \s{classA}. Since such Chern bands break time-reversal symmetry, one may hope that imposing time-reversal symmetry guarantees the above commutation. This expectation is correct for single-band subspaces, if we \emph{further} impose a spatial point group symmetry -- that of spatial inversion.\cite{Marzari1997} In general, no symmetry enforces $[PxP,PyP]{=}0$ for multi-band subspaces such as that of \fig{fig:helix}(c); we remind the reader that period-multiplied Bloch oscillations only occur in multi-band subspaces. We are therefore motivated to formulate criteria -- going beyond group-theoretic criteria -- that identify multi-band subspaces with commuting projected position operators, i.e., subspaces whose dynamics can be described by classical point charges. Such subspaces are referred to as strong elementary band representations (strong EBRs). Then by imposing symmetry restrictions on the positions of these point charges (just as six-fold symmetry related $\bvarpi_{1}$ and $\bvarpi_{2}$ within one unit cell in the above example), we identify a subclass of strong EBRs which exhibit period multiplication. This program is carried out exhaustively for the 2D space groups (known as wallpaper groups) in \s{perp}. \\

Spatially-extended functions such as Bloch functions represent the opposite extreme to  classical point charges. Yet period multiplication  may also be  understood from the complementary perspective of adiabatic transport of Bloch functions. In the honeycomb example, we consider transport along the bent loop $\calc_3$, as illustrated in \fig{fig:helix}(d); this loop was chosen to exploit the point group symmetry of the honeycomb: half the loop is mapped to the other half by three-fold rotation. The analog of two nearest-neighbor Wannier functions acquiring a dynamical phase difference of $2\pi/3$ is that the Bloch functions  acquire a geometric Zak phase difference of $2\pi/3$ along $\calc_3$. This is elaborated in \s{Zak}, where we also present a general theorem that identifies symmetry-protected Zak phases in space-group-symmetric band subspaces. \\

The culmination of this analogy between real- and $\bk$-space perspectives is our derivation of a one-to-one correspondence between symmetry-protected Zak phases and symmetry-protected Wannier centers. Such multi-band Zak-Wannier relations are developed in \s{sec:zakstrongatomic}-\ref{ZakEBR} for strong EBRs, and  more generally for band representations in \s{composite}.

\subsection{Bloch oscillations with $\mu_{\perp}{>}1$ from the perspective of Wannier functions}\la{perp}

\begin{figure}[!htbp]
\centering
\includegraphics[width=7.0cm]{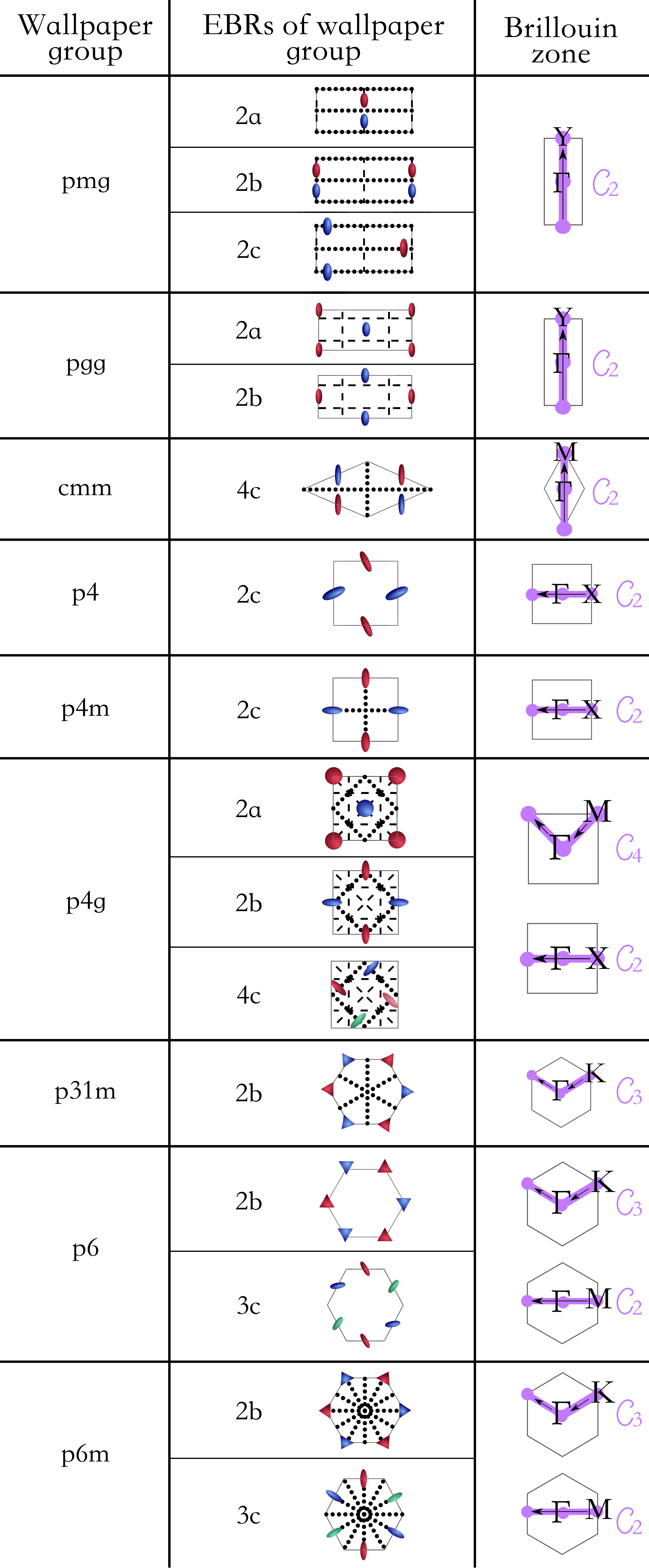}
\caption{Examples of strong elementary band representations (EBRs) which manifest period-multiplied Bloch oscillations; the corresponding period multipliers are listed in \tab{wallpaper}.  Each EBR is specified by a wallpaper group (first column), a Wyckoff position (second column, left), and the symmetry representation of Wannier functions centered on this Wyckoff position. Each Wannier function is  illustrated as a blob over  the real-space Wigner-Seitz unit cell (second column, right). The  shape of the blob indicates the local symmetry of the Wannier function: circles stand for four-fold rotations, triangles for three-fold rotations, and ellipses either for two-fold rotations or reflections; same-colored  blobs are related by lattice translations, different-colored blobs by point-group operations. Dotted (resp.\ long-dashed) lines indicate mirror-invariant (resp.\ glide-invariant) lines.  In the corresponding Brillouin zones (third column), we illustrate the loops ($\calc_n$, violet lines) and the $g_n$-invariant wavevectors ($\Gamma$ and $\bK_n$, violet dots). \la{fig:elem} }
\end{figure}

To simplify the presentation in the remainder of this work, we specialize to a field that is perpendicular to a rotational axis or a reflection plane. Both rotations and reflections involve no fractional lattice translations ($p{=}0$), and are classified as symmorphic symmetries. Hence, we may just as well drop the second subscript in $g_{n,p}$ {(the definitions of $g_{n,p}, p$ and $n$ are summarized in \tab{symbols})}. Essentially the same physics occurs for a field that is perpendicular to a nonsymmorphic screw axis or glide plane.\footnote{As explained in the previous section, the nontrivial effect of nonsymmorphic symmetries originates from the monodromy of symmetry representations as $\bk$ is advanced by a reciprocal vector $\bG_{\parallel}$;  $\bG_{\parallel}$ is parallel to the screw axis or glide plane. This monodromy does not affect adiabatic transport that is induced by a field lying perpendicular to $\bG_{\parallel}$. } \\

Our motivational example on a honeycomb lattice belongs to a broader class of band subspaces which form a representation of a space group $G$ on Wannier functions (in short, a \textit{Wannier representation} of $G$). The Wannier functions are centered on a 2D lattice (illustrated in \fig{fig:elem}), where each lattice site ($\bvarpi$) is invariant under an order-$n$ point group symmetry ($g_n$) that is tabulated in \tab{wallpaper}. For example, the honeycomb lattice has the space group $G{=}p6m$ and the point group $\calp{\simeq}C_{6v}$ ($\simeq$ denotes  a group isomorphism); the latter is generated by a six-fold rotation and a reflection. Each honeycomb vertex is invariant under a three-fold rotation $g_3{=}C_{3,z}$. \\

We shall further particularize to band subspaces whose Wannier functions are locally-symmetric. Such band subspaces are known as band representations (BRs) and have been introduced in \s{sumo}; here we would provide a more precise formulation. Central to the formulation of BRs is the notion of a Wyckoff position.  Given a  space group $G$, a Wyckoff position $\bvarpi$ is simply a point in space with an associated symmetry: the subgroup   of $G$ that preserves $\bvarpi$. This subgroup is defined as the \textit{stabilizer} of $\bvarpi$ and denoted by $\calp_{\varpi} {\subset}G$; each $g{\in}\calp_{\varpi}$ has a trivial action on $\bvarpi$: $g{\circ}\bvarpi{=}\bvarpi$. Any stabilizer  is by construction a point group, however it need not be isomorphic to the point group $\calp{=}G/\calt$ of $G$. For example, if $\bvarpi$ is a generic position, then $\calp_{\varpi}$ is the trivial group. In the motivational example of \s{honey} illustrated in \fig{fig:helix}(c), a honeycomb vertex $\bvarpi_1$ (denoted as $2b$ in the International Tables for Crystallography\cite{Hahn1984}) is invariant under three-fold rotation and reflection (as indicated by the dotted line in the corresponding entry of \fig{fig:elem}) -- these symmetries generate the point group $\calp_{\varpi}{\simeq}C_{3v}$, which is a subgroup of the point group $\calp{\simeq}C_{6v}$ of $G{=}p6m$. In comparison, the center of the honeycomb plaquette has a stabilizer that is isomorphic to $\calp$. To recapitulate,
\begin{definition}\la{def:wyckoff}\normalfont
A \textit{Wyckoff position} $\bvarpi$ of a $d$-dimensional space group $G$ is defined as a point in $\R^d$ with an associated stabilizer $\calp_{\varpi}{\subset} \calp$; for points in $\R^d$ obtained by action of $G$ on $\bvarpi$ (denoted $g{\circ} \bvarpi$ for $g{\in} G$), the corresponding stabilizers are conjugate to $\calp_{\varpi}$.  For symmetries $h {\in} \calp/\calp_{\varpi}$ that do not stabilize $\bvarpi$, $h {\circ} \bvarpi$ is called a \textit{different representative of $\bvarpi$}.
\end{definition}
\noindent A useful characterization of a Wyckoff position is its multiplicity, which counts the number of distinct but symmetry-related positions in the primitive unit cell.\footnote{The multiplicity $M_{\varpi}$ used in the International Table for Crystallography\cite{Hahn1984} differs from our definition for unit cells that are not primitive, e.g. $cm$ and $cmm$. In \app{app:symmex} we are more explicit about this difference and its implication on $\mu_{\perp}$} 
\begin{definition}\la{def:multiplicity}\normalfont
If $g{\circ}\bvarpi{-}\bvarpi$ is a Bravais lattice vector for all $g{\in}G$, then we say that $\bvarpi$ has \textit{unit multiplicity}. Generally, the multiplicity $M_{\varpi}$ of a Wyckoff position $\bvarpi$ (with associated space group $G$, point group $\calp$ of space group, and stabilizer $\calp_{\varpi}$) is equal to $M_{\varpi}{=}|\calp|/|\calp_{\varpi}|$.
\end{definition}
\noindent Here, $|H|$ denotes the order of a group $H$. For the honeycomb example, $M_{\varpi_1}{=}2$ because $\bvarpi_1$ is mapped onto the different representative $\bvarpi_2$ by $C_6 {\in} \calp/\calp_{\varpi}$. We are ready to define a BR:
\begin{definition}\la{def:BR}
\normalfont A \textit{band representation} (BR) of $G$ is a Wannier representation of $G$ that satisfies a local symmetry condition: for each Wyckoff position $\bvarpi$,  all Wannier functions centered at $\bvarpi$ form a unitary, finite-dimensional representation ($V_{\varpi}$) of the stabilizer $\calp_{\varpi}$. 
\end{definition}
\noindent $V_{\varpi}$ shall be referred to as the on-site representation of $\calp_{\varpi}$. 
In the honeycomb example, the single $s$-like Wannier function centered on $\bvarpi_1$ forms a trivial representation ($E$) of $C_{3v}$, hence this example may  be identified as a BR of $G{=}p6m$ with the identifying data $(\bvarpi{=}2b,V_{\varpi}{=}E)$. A BR of $G$ is fully specified by the data $\{ (\bvarpi,V_{\varpi}) \}_{\varpi}$ for all symmetry-unrelated $\bvarpi$. An equivalent\cite{Alexandradinata2018} definition of BRs (by induced representations) is also available in the literature.\cite{Zak1981,Evarestov1984,Bacry1993}  In general, not all Wannier representations of $G$ are BRs of $G$; a counter-example is the Kane-Mele model explored in \s{classAII}. \\

\begin{table}
\scalebox{0.8}{
\begin{tabular}{|c|c|c|c|c|c|}
\hline
 $G$ & $\bvarpi$ & $g_n$ & $\calp_{\varpi}$ & $h {\in} \calp/\calp_{\varpi}$ & $\chi_{\perp}$  \\
 \hline \hline
  $pmg$ & ($2a$, $2b$, $2c$) & ($C_2$, $C_2$, $M_x$) & ($C_{2}$, $C_{2}$, $C_s$) & ($g_y$, $g_y$, $g_y$) & ($2$, $2$, $2$) \\
  \hline
  $pgg$ & ($2a$, $2b$) & ($C_{2}$, $C_{2}$) & ($C_{2}$, $C_{2}$) & ($g_x$, $g_x$) & ($2$, $2$) \\
  \hline
  $cmm$ & $4c$ & $C_2$ & $C_{2}$ & $M_x$ & $2$ \\
  \hline
  $p4$ & $2c$ & $C_{2}$ & $C_{2}$ & $C_4$ & $2$ \\
  \hline
  $p4m$ & $2c$ &$C_{2}$ & $C_{2v}$ & $C_4$ & $2$ \\
  \hline
  $p4g$ & ($2a$, $2b$, $4c$) & ($C_{4}$, $C_{2}$, $M_{x+y}$) & ($C_4$, $C_{2v}$, $C_s$) & ($g_x$, $C_{4}$, $C_{4}$) & ($2$, $2$, $2$) \\
  \hline
  $p31m$ & $2b$ & $C_{3}$ & $C_3$ & $M_x$ & $3$ \\
  \hline
  $p6$ & ($2b$,$3c$) & ($C_{3}$,$C_{2}$) & ($C_3$,$C_2$) & ($C_{6}$,$C_{6}$) & ($3$,$2$) \\
  \hline
  $p6m$ & ($2b$,$3c$) & ($C_{3}$,$C_{2}$) & ($C_{3v}$,$C_{2v}$) & ($C_{6}$,$C_{6}$) & ($3$,$2$) \\
  \hline
\end{tabular}
}
\caption{Data of \fig{fig:elem}; cf. \tab{symbols} for summary of symbols; we follow standard notation to denote point groups;\cite{Tinkham2003} for both $\calp$ and $\calp_{\varpi}$ isomorphic point groups are listed. For the point group elements $g_n, h$, we denote a $2\pi/n$-rotation about an axis perpendicular to the plane by $C_n$, a reflection that inverts the coordinate $a$ by $M_{a}$, a reflection that maps $(x,y) {\rightarrow} (y,x)$ by $M_{x+y}$, and a glide  composed of  $M_a$ with half a lattice translation by $g_a$. $h {\in} \calp/\calp_{\varpi}$ is a symmetry that relates distinct $g_n$-invariant Wyckoff positions. Wyckoff positions ($\bvarpi$) are labeled as $Mq$, with multiplicity $M{>}1$, and $q$ a label for points in the unit cell (following the notation of the International Tables for Crystallography\cite{Hahn1984}). \la{wallpaper}}
\end{table}

All illustrated band subspaces in \fig{fig:elem} are BRs that exhibit period-multipied Bloch oscillations (with a multiplier $\mu_{\perp}{>}1$ that divides $n$) in their \emph{atomic limit}.

\begin{definition} \la{define:atomic} \normalfont
Suppose a BR of a space group $G$ is specified by the data $\{ (\bvarpi,V_{\varpi}) \}_{\varpi}$. The \textit{atomic limit} of this BR (in short, \textit{atomic BR}) is defined as a $G$-symmetric  separation of all its Wannier centers, i.e., the separations $| g\circ \bvarpi {-} g'\circ \bvarpi' |{\rightarrow} \infty$ for all $g,g' {\in} G$ and for any pair of Wannier centers $\bvarpi, \bvarpi'$, while fixing ratios of (distinct) Wannier separations.\footnote{In electronic solids where the Fermi level lies in a spectral gap, atomic BRs may be identified with the colloquially-coined `atomic insulator'. In this work, we do not consider insulators in the context of Bloch oscillations.}
\end{definition}

\noindent There is a less general but  operationally useful definition of an atomic BR. This definition applies only to BRs $\{ (\bvarpi,V_{\varpi}) \}_{\varpi}$ that are energy bands of a tight-binding Hamiltonian; it is further supposed that the tight-binding basis functions are centered on $\{\bvarpi\}_{\varpi}$. Then the atomic limit of a BR, defined with respect to this set of $\bvarpi$, is obtained by symmetrically tuning all tight-binding matrix elements (i.e., hopping amplitudes) between different $\bvarpi$ to zero. Since large Wannier separations imply exponentially-weak hopping amplitudes, we expect that both definitions should coincide within a tight-binding formalism. From the perspective of transport experiments, the atomic BR well approximates dynamics under a field $F$ if the characteristic hopping strength (or characteristic energy splitting within the BR) is much smaller than $Fa$, with $a$ a lattice constant; this is elaborated in \s{sec:finitewidth}. We will demonstrate (in \s{sec:zakstrongatomic}) that the atomic BR is a realistic approximation of cold atoms trapped in optical lattices (with deep, well-separated troughs).\cite{Li2016} In principle, the atomic EBR may also be a good description of some organic solids.\cite{Seo2004,Eo2013}\\






Let us first expand upon the heuristic discussion of continuous-time Bloch oscillations for the honeycomb lattice in \s{honey}. This discussion applies more generally to all BRs illustrated in \fig{fig:elem}. We assume that each of these BRs is a low-energy band subspace of a translation-invariant Hamiltonian $H_0$; this subspace is projected by the operator $P{=}\sum_{\bR}\sum_{j=1}^M P_{j,\bR}$, where ${\bR}$ is summed over the Bravais lattice, and $P_{j,\bR}$ is the projection operator to Wannier functions centered at $\bvarpi_j{+}\bR$, with $j{=}1,{\ldots},M_{\varpi}$ labeling all different Wannier centers within one primitive unit cell. For the honeycomb lattice illustrated in \fig{fig:helix}(c), $P_{j,\textbf{0}}$ projects onto a single $s$-orbital centered at site $\bvarpi_j$ for $j{=}1,2$.\\

In the presence of a field, we assume that dynamics is adiabatic within the low-energy subspace projected by $P$. In the atomic limit, Wannier functions centered on different coordinates are decoupled, hence the adiabatic propagator factorizes multiplicatively as 
\e{ \hat{U}^{\text{atomic}}(T_B) = e^{-i E_0 T_B/\hbar} \sum_{j,\bR} e^{i \bG_n \cdot \bvarpi_j} P_{j,\bR}. \la{atomiclimitpropagator} }  
$E_0$  is the degenerate energy of the spectrally-flattened bands; this flattening originates from the absence of hybridization between spatially-separated, exponentially-localized Wannier functions.\cite{Ashcroft} The reciprocal vector $\bG_n{=}\int_0^{T_B} \bF(t)dt/\hbar$  is chosen for $g_n$-symmetric lattices as
\e{ \bG_n = \check{g}_n \bK_n - \bK_n, \la{Gcond} }
where $\bK_n$ is a $g_n$-invariant wavevector on the boundary of the first BZ. \\


All BRs in \fig{fig:elem} satisfy that nearest-neighbor, $g_n$-invariant Wannier centers $(\bvarpi_{j'},\bvarpi_{j})$ are unrelated by a lattice translation, but are instead related by a point-group transformation $h$ (possibly composed with a lattice translation). $h$ is necessarily distinct from $g_n$, since each of $\bvarpi_{j'}$ and $\bvarpi_{j}$ is $g_n$-invariant; formally, we say $h{\in}\calp/\calp_{\bvarpi_j}$ with $g_n{\in}\calp_{\bvarpi_j}$. In the honeycomb example, we have already noticed that a six-fold rotation ($h$) relates nearest-neighbor Wannier centers. For all lattices in \fig{fig:elem}, the dynamical phase difference acquired over one fundamental period (cf.\ \q{atomiclimitpropagator}), for a $h$-related Wannier pair, equals 
\e{ | \bG_n {\cdot} (\bvarpi_{j'}-\bvarpi_{j}) | = \tf{2\pi}{\chi_{\perp}} \;\text{mod } 2\pi, \; (\chi_{\perp}{>}1) \text{ divides $n$.} \la{specialG}}
More generally, for any pair of Wannier functions separated by $\bvarpi_j{+}\bR{-}\bvarpi_{j'}$, 
\e{ e^{i\chi_{\perp}\bG_n\cdot (\bvarpi_j+\bR-\bvarpi_{j'})}{=}1,\tag{10'} \la{genLaue}}
for all lattice vectors $\bR$. $\chi_{\perp}{=}3$ for the honeycomb lattice; the data $\{g_n,h,\chi_{\perp}\}$ for the other lattices in \fig{fig:elem} is tabulated in \tab{wallpaper}.\\ 

\noindent \textbf{\textit{Example of $\chi_{\perp}{=}2$.}} Consider $G{=}pmg$, $\bvarpi{=}2a$ and $g_n{=}C_{2,z}$ (\tab{wallpaper}): the two $C_{2,z}$-symmetric Wannier functions per unit cell (cf.\ \fig{fig:elem}) are mapped onto each other by a glide ($h{=}g_y$) which inverts the $y$-coordinate and translates by half a lattice constant in the $x$-direction; the two Wannier centers are therefore separated by half a lattice constant in the $x$-direction, thus $\chi_{\perp}{=}2$.\\

 

The combination of \q{atomiclimitpropagator} and \q{genLaue} implies that stroboscopic expectation values $\braket{O}(jT_B)$ of an observable $O$ only have Fourier peaks at  integer multiples of $2\pi/(\chi_{\perp}T_B)$. We have thus established an identity between, $\chi_{\perp}$, a property of the real space Wyckoff positions of bands, and $\mu_{\perp}$, the period multiplier in the stroboscopic Bloch oscillations (cf.\ Definition \ref{define:stroboscopic}):
\e{\text{For atomic BRs,}\as\chi_{\perp}=\mu_{\perp}.\la{equalityWanniermultiplier}}\\


We have claimed that \fig{fig:elem} represents a subclass of BRs which satisfy \qq{atomiclimitpropagator}{genLaue} (and hence realize period-multiplied Bloch oscillations). \qq{atomiclimitpropagator}{genLaue} hold because of three characteristic properties of this subclass. These properties will be stated and exemplified in the subsequent three subsections (\s{buildblock} to \s{Wyckoffcond}), before we prove that they sufficiently lead to \qq{atomiclimitpropagator}{genLaue} (cf.\ \s{threeproof}).

\subsubsection{First property: elementary band representations}\la{buildblock}

The first characteristic is that the BRs are elementary.
\begin{definition}\la{def:EBR}
\normalfont An \textit{elementary band representation} (EBR) of $G$ is a BR of $G$ that cannot be split into smaller BRs of $G$. An EBR is specified by a single Wyckoff position $\bvarpi$, as well as a unitary irreducible representation $V_{\varpi}$ of the stabilizer $\calp_{\varpi}$.\cite{Bacry1988} A BR that is not elementary is defined to be composite.
\end{definition}
\noindent 
For an EBR of $G$ with data $(\bvarpi,V_{\varpi})$ and associated multiplicity $M_{\varpi}$ of $\bvarpi$,  the total number of Wannier functions in one primitive unit cell is equal to $N{=}M_{\varpi} \dim V_{\varpi}$; this is equivalently the number of independent Bloch functions at each wavevector. EBRs may be viewed as the building blocks of composite  BRs (i.e., non-elementary BRs), in analogy with how irreducible representations of finite groups are the building blocks of reducible representation.\footnote{It should be clarified, though, that an EBR is not an irreducible representation of a space group\cite{Bacry1993}} \\

\noindent \textit{\textbf{Example of an EBR in a reflection-asymmetric checkerboard lattice.}}
We consider an EBR of the space group $G{=}p4$ and Wyckoff position $\bvarpi{=}2c$ (\tab{wallpaper}). The real space unit cell contains $M_{\varpi}{=}2$ $s$-like Wannier functions (or simply, $s$-orbitals) centered at $\bvarpi_1$ (red blobs in the corresponding row of \fig{fig:elem}) and $\bvarpi_2$ (blue blobs). These two centers are individually invariant under $g_2{=}C_{2,z}$; $\bvarpi_1$ and $\bvarpi_2$ are mutually related by $h{=}C_{4,z} {\in} \calp/\calp_{\varpi}$. This implies that the $s$-orbitals are individually $C_{2,z}$-invariant and are mapped onto each other (or each others translate) by $C_{4,z}$. 

\subsubsection{Second property: strong band representations}\la{sec:strongBR}

A second characteristic of the subclass is that their Wannier functions are strongly localized. 
\begin{definition}\la{strongd}
\normalfont Let a BR  of space group $G$ be specified by the data $\{ (\bvarpi,V_{\varpi}) \}_{\varpi}$ and $P$ be the projection operator for this BR. This BR is \textit{strong} if $G$ enforces that $PxP$ and $PyP$ commute in the atomic limit (defined with respect to $\{\bvarpi\}_{\varpi}$ in Definition \ref{define:atomic}). Equivalently,\cite{Marzari1997} a BR is \textit{strong}  if $G$ enforces that the non-Abelian Berry curvature $\mathcal{F}^{xy}(\bk)$, defined as $\partial_{k_y} A^x(\bk) {-} \partial_{k_x} A^y(\bk) {+} i[ A^y(\bk), A^x(\bk)]$ (cf. $\bA$ in \q{nonAB}), vanishes in the atomic limit.  
\end{definition}
\noindent The above definition applies to both elementary and composite BRs. A BR that is not strong is defined to be \textit{weak}. In \s{sec:symmetrystrong}, we propose a sufficient symmetry criterion to identify strong BRs.

\subsubsection{Third property: conditions  on the Wyckoff position}\la{Wyckoffcond}

The third and last characteristic of the subclass are a set of conditions on the Wyckoff position $\bvarpi$. 
\begin{enumerate}
\item[(i)] $g_n {\in} \calp_{\varpi}\colon$ $g_n$ is an on-site symmetry of order $n{>}1$. Such $\bvarpi$ is also referred to as nongeneric Wyckoff position.
\item[(ii)] $M_{\varpi}{>}1\colon$ The multiplicity of the Wyckoff position is greater than one. 
\item[(iii)] At least two different representatives ($\bvarpi_j$, $\bvarpi_{j'}$) of $\bvarpi$ lie on spatially-separated $g_n$-invariant points for $g_n$ a rotation (resp.\ $g_2$-invariant lines for $g_2$ a reflection) which are not related by lattice translations.
\end{enumerate}
Condition (ii) rules out 1D space groups. All three conditions are only satisfied for Wyckoff positions in nine out of seventeen wallpaper groups, which are all illustrated in \fig{fig:elem} and summarized in \tab{wallpaper}. If $g_n$ is a mirror, two-fold or four-fold rotational symmetry ($n{=}2,4$), there exists an additional mirror, glide, two-fold or four-fold rotational symmetry which results in  $\chi_{\perp}{=}2$; if $g_n$ is a three-fold rotational symmetry ($n{=}3$), a mirror or six-fold rotational symmetry signifies $\chi_{\perp}{=}3$; no case exists with $g_n$ being $C_{6,z}$ or glide. 
One may verify that in the previous reflection-asymmetric checkerboard example (\s{buildblock}), conditions (i{-}iii) are satisfied, therefore \qq{Gcond}{genLaue} hold for $n{=}2$ and $\chi_{\perp}{=}2$.


\subsubsection{The three properties sufficiently lead to \qq{atomiclimitpropagator}{genLaue} }\la{threeproof}

Composite BRs are less favored to realize period-multiplied Bloch oscillations due to the following reason: a composite BR may split into multiple EBRs which have different on-site energies in the atomic limit -- this would invalidate the factorization of \q{atomiclimitpropagator}. Let us presently focus on the Bloch-oscillatory phenomena of EBRs, and postpone a discussion of composite BRs to \s{composite}.  \\

In the atomic limit of a strong EBR, its Wannier functions  simultaneously diagonalize $PxP$ and $PyP$; in this sense do Wannier functions resemble classical point charges that simultaneously diagonalize $x$ and $y$. The spectrum of $P\br P$ is generated from $\bvarpi$ by action of $G$. Further utilizing the exponential-localization\cite{Cloizeaux1964,Brouder2007} of these Wannier functions, together with the spectral-flattening of bands in the atomic limit, we derive \q{atomiclimitpropagator}.\\

In \app{app:symmex}, we prove that (i-iii) (in \s{Wyckoffcond}) are  necessary and sufficient conditions on the Wyckoff position $\bvarpi$ such that \qq{specialG}{genLaue} are satisfied. This completes the proof.






\subsubsection{Sufficient symmetry criterion for strong BRs}\la{sec:symmetrystrong}

\begin{figure}
\centering
\includegraphics[width=8.5cm]{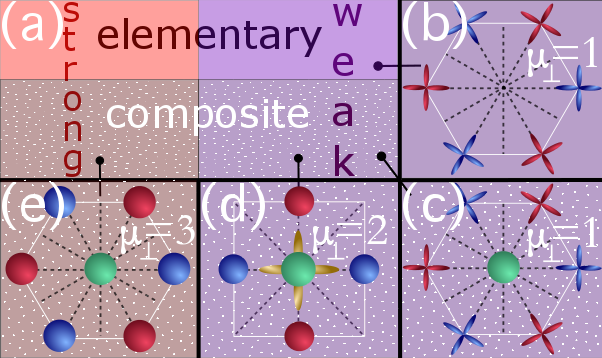}
\caption{(a) Venn diagram of the possible types of band representations (BRs) of wallpaper groups: strong (red background) vs. weak (purple), elementary (plain) vs. composite (dotted). (b-e) illustrate the real-space unit cells of four BRs; these four examples represent three of the four categories in (a). The  Wigner-Seitz unit cell is delineated by a solid white line, and mirror-invariant lines are dashed. Wannier functions are illustrated schematically as circular blobs (for $s$-like orbitals) and cross-like blobs (for $p_x$- and $p_y$-orbitals). The center of (d) illustrates three Wannier functions (s,$p_x,p_y$) localized on the same site. For the BRs illustrated in (d{-}e), their corresponding symmetry eigenvalues (at high-symmetry wavevectors) and Zak phase factors are summarized in \tab{evalEx} with the same labels: (d{-}e). \la{fig:venn} }
\end{figure}


Let us propose a sufficient symmetry criterion to identify strong BRs.  Equivalently stated, it is a sufficient symmetry criterion for the vanishing of the non-abelian curvature $\calf^{xy}$ for band representations in their atomic limit. \\

It is tempting to believe that  EBRs, being in some sense the simplest types of gapped subspaces, should \textit{all} have vanishing curvature in the atomic limit. 
We provide a counter-example in \fig{fig:venn}(b), which illustrates $p_x,p_y$-orbitals on each vertex of a honeycomb lattice. Under application of an in-plane field, the $p$-orbitals on each vertex generically hybridize and split away from the honeycomb vertex; this  reflects the non-commutativity of $PxP$ and $PyP$. This exemplifies an EBR that is weak, as we further elaborate  in the \textbf{\textit{Example}} below.\\

This motivates a precise symmetry criterion that distinguishes between BRs which are elementary vs. composite (not elementary), and strong vs. weak (not strong); a Venn diagram is illustrated in \fig{fig:venn}(a). To clarify, an EBR consists of a single-particle Hilbert space spanned by Bloch functions at each $\bk$; the Hilbert space of a composite BR which consists of two EBRs is the direct sum of the Hilbert spaces of the individual EBRs at each $\bk$. We will further refer to such a composite BR as being obtained by `stacking' of the two EBRs. Examples of stacking are illustrated in \fig{fig:venn}(c{-}e).\\

We propose a sufficient criterion for 2D strong BRs, which utilizes the simplification of
\e{ [PxP,PyP] \xrightarrow{\text{atomic limit}} \sum_{j,\bR} [P_{j,\bR}x P_{j,\bR},P_{j,\bR} y P_{j,\bR}], \la{Projat}}
owing to the exponential-localization of Wannier functions. The right-hand side of \q{Projat} vanishes for $P_{j,\bR}$ of rank one; this is the case for all examples shown in \fig{fig:elem}, including the reflection-asymmetric checkerboard example introduced in \s{buildblock}. However, for $P_{j,\bR}$ of rank larger than one, this need not be the case, as has been exemplified by the $p_x$- and $p_y$-orbitals in a honeycomb lattice. A sufficient condition for commutativity of the locally-projected position operators is that the on-site symmetry ($\calp_{\bvarpi_j+\bR}$) enforces
\e{ P_{j,\bR} \;(\br\cdot \bv_j) \; P_{j,\bR} = (\bvarpi_j+\bR)\cdot \bv_j \; P_{j,\bR}, \la{stabcond} } 
for at least one unit vector $\bv_j$ in the 2D plane; generally, $\bv_j$ is allowed to depend on the Wannier center labeled by $j$. The advantage of formulating \q{stabcond} is that \q{stabcond} follows directly from matrix-element selection rules that are  detailed in \app{app:zerocurv} in full generality. To give a flavor of the more general result, we offer one application of these rules to identify a strong and a weak EBR.  \\

\noindent \textbf{\textit{Example of a strong and weak EBR.}} Let us consider EBRs for which symmetry enforces that 
\e{P_{j,\bR}x  P_{j,\bR}=P_{j,\bR}y  P_{j,\bR} =0,\la{bothPrPvanish}}
for a spatial origin that is fixed to the Wannier center; these EBRs are necessarily strong, since \q{stabcond} is satisfied trivially. A case in point is a two-band EBR (of the wallpaper group $G{=}p4m$) which comprises pairs of $p_x,p_y$-orbitals centered on the vertices of a square lattice. Each pair transforms in a two-dimensional irreducible vector-representation ($X$) of the on-site symmetry group $C_{4v}$, which is generated by a four-fold rotation and a mirror symmetry. Moreover, the two-dimensional position operator $\br{=}(x,y)$ also transforms in the vector representation $X$. Applying a matrix-element selection rule,\cite{Tinkham2003} \q{bothPrPvanish} is satisfied if $X^*{\otimes}X{\otimes}X$, with $X^*$ the complex-conjugate representation of $X$, does not contain the trivial representation. Indeed, from $X^*{=}X$ and inspection of the character table for $C_{4v}$,\cite{Tinkham2003} we derive that   $X^* {\otimes} X {\otimes} X{=}4X$, which proves the claim. On the other hand, for pairs of $p_x,p_y$-orbitals centered on the vertices of a honeycomb lattice (wallpaper group $G{=}p6m$), the on-site symmetry group ($C_{3v}$) is generated by a three-fold rotation and a reflection. For $X$ that is the vector, irreducible representation of $C_{3v}$, $X^* {\otimes} X {\otimes} X$ contains the trivial representation, hence \q{bothPrPvanish} is generically not satisfied. Further arguments in \app{pxpy} demonstrate that $P_{j,\bR}x  P_{j,\bR}$ and $P_{j,\bR}y  P_{j,\bR}$ generically do not commute, hence this is an example of a weak EBR. \\

Other than the selection rules detailed in \app{app:zerocurv}, we are not aware of any other group-theoretic rules that ensure the commutativity of $P_{j,\bR}xP_{j,\bR}$ and $P_{j,\bR}yP_{j,\bR}$. Therefore, it is possible to adopt an operationally more useful definition of a strong BR: a BR of a space group $G$ is strong if the symmetries in $G$ enforce that \q{stabcond} is satisfied for all Wannier centers. We believe this definition is equivalent to that in Definition \ref{strongd}.



\subsection{Zak phases of  strong, atomic EBRs}\la{sec:zakstrongatomic}
 
Period-multiplied Bloch oscillations for strong, atomic EBRs can alternatively be described by quantized differences in Zak phases. This dual perspective exists owing to a multi-band Zak-Wannier relation that we propose here. This relation describes, for the first time, a one-to-one correspondence between Zak phases and the centers of Wannier functions (which are localized in all spatial directions). For further motivation, differences in Zak phases are directly measurable in cold-atomic experiments.\cite{Atala2013a,Li2016}\\

Let us first review how Zak phases appear in adiabatic transport.\cite{Zak1989} We consider the adiabatic transport of a Bloch function $\psi_{j',\bK_n}$ {(the definition of $\bK_n$ may be recalled from \tab{symbols}}) along an arbitrary loop $\calc$ that wraps around the Brillouin torus in the direction of a primitive reciprocal vector $\bG[\calc]$. The geometric component ($\W$) of the adiabatic propagator (cf.\ \q{adevo}) is expressible as a path-ordered exponential (denoted $\overline{\text{exp}}$) of the non-Abelian Berry connection $\bA$  (cf.\ \q{nonAB}):\cite{Berry1984,WilczekFrank1984}
\e{ \W[\mathcal{C}] = \overline{\mathrm{exp}} \big[ {i {\oint}_{\mathcal{C}} \bA (\bk) \cdot \mathrm d\bk} \big]. \la{Wilsonloop} }
This is a simple generalization of \q{Wilsonloopst} to loops in higher-dimensional $\bk$-space. $\W$, also known as the Wilson loop or holonomy matrix, is related to the projected position operator as
\m{ \braket{ \psi_{j,\bk_f}| \overline{\mathrm{exp}} \big[ i \oint_{\mathcal{C}} P \br P \mathrm \cdot d\bk \big] | \psi_{j',\bK_n}} = \\ 
\delta( \bk_f{-}\bK_n{-}\bG) \W[\mathcal{C}]_{j,j'}; \la{WilsonPxP} }
we prove this equality in \app{app:weakfield}. \\

Applying \q{WilsonPxP} and the defining properties of strong, atomic EBRs, we prove in \app{specW} that
\e{ \text{For strong, atomic EBRs with data } (\bvarpi,V_{\varpi}), \lin 
    \phi_j[\calc] = \bG \cdot \bvarpi_j \;\text{mod }2\pi,\; j=1,\ldots,N  \la{zakwannierstrongatomic}}
where $\{ e^{i\phi_{j}} \}_{j=1}^N$ are the eigenvalues of $\W[\calc]$, and $\bvarpi_j$ (the Wannier centers) lie on different  representatives of the Wyckoff position $\bvarpi$. The above Zak-Wannier relation applies to strong, atomic of EBRs of any (magnetic) space group, and in any spatial dimension $d$. The Zak phase $\phi_j[\calc]$ is independent of continuous deformations of the path $\calc$ in $\bk$-space.
Precisely, $\phi_j[\calc]$ depends only on $\bG$ (which specifies the homotopy class of loops in the Brillouin torus) but not on specific local-in-$\bk$ details of the trajectory. This robustness of $\phi_j$ exemplifies a more general statement in bundle theory:  that the holonomy of vector bundles with zero non-Abelian curvature depends only on the homotopy class of the loop.\cite{Eguchi1980} \\


We are ready to rederive stroboscopic Bloch oscillations for the strong, atomic EBRs which satisfy conditions (i{-}iii) in \s{Wyckoffcond}. Owing to the triviality of the dynamical phase in the atomic limit, it is sufficient to consider the geometric Zak phase. In particular, we consider the Zak phases for special loops $\calc$ for which $\bG[\calc]$ satisfies \q{Gcond}. Combining \q{zakwannierstrongatomic} with \qq{specialG}{genLaue}, we derive that all Zak phase factors are $n$'th roots of unity; in particular, a pair of Zak phases differ by $\f{2\pi}{\chi_{\perp}}$ with $\mu_{\perp}{=}\chi_{\perp}{>}1$. Only after $\mu_{\perp}$ fundamental periods do all pairwise Zak-phase differences equal an integer multiple of $2\pi$, which completes the rederivation.\\


\noindent \textbf{\textit{Application of Zak-Wannier relation to strong, atomic EBR on the honeycomb lattice.}}  The example introduced in \s{honey} is a strong, atomic EBR of  wallpaper group $G{=}p6m$, with Wyckoff position $2b$ and a trivial on-site representation (one $s$-orbital per honeycomb vertex). Let us utilize \q{zakwannierstrongatomic} to determine the Zak phases for the bent loop $\calc_3$ (cf.\ \fig{fig:helix}(d)), i.e., we input the Wannier centers ($\bvarpi_{1}$ and $\bvarpi_2$ illustrated in \fig{fig:helix}(c)) and the reciprocal vector $\bG$ (\fig{fig:helix}(d)). For any choice of real-space origin that is three-fold symmetric (e.g. the center of the honeycomb plaquette), $\phi_{1,2}[\calc_3]$ are  integer multiples of $2\pi/3$. For any choice of origin (symmetric or generic),  $|\phi_{1}[\calc_3]{-}\phi_2[\calc_3]|{=}2\pi/3$. This quantized difference in Zak phase is responsible for the three-fold period multiplication that was alternatively derived in \s{motivCBR}. The non-uniqueness of individual Zak phases is discussed from a more general perspective in point (ii) of \s{theoremstate}. The invariance of $\phi_j$ (under continuous deformations of the loop) implies that  $\phi_j[\calc_2]{=}\phi_j[\calc_3]$, where $\calc_2$ is the straight loop drawn in \fig{fig:elem}.
It is remarkable that $|\phi_{1}[\calc_2]{-}\phi_2[\calc_2]|{=}2.06(3)\pi/3$ has been measured\cite{Li2016} for a cold-atomic realization of this EBR, as elaborated in \s{cold} -- we view this as evidence that the atomic EBR, despite being an idealization, can nevertheless be a useful description of experimental systems.


\subsection{Relating symmetry-protected Zak phases to point group representations and Wannier centers}\la{Zak}

What may be said about adiabatic transport of strong EBRs (which satisfy conditions (i{-}iii) in \s{Wyckoffcond}) when we deviate from the idealized atomic limit? Due to tunneling between finitely-separated Wannier functions, energy bands (at zero field) would  no longer be flat and degenerate. A nontrivial energy-momentum dispersion would introduce a dynamical correction to the time-evolution propagator [cf.\ \q{atomiclimitpropagator}], and ultimately disrupts period-multiplied Bloch oscillations. The nature of this disruption is elaborated and quantified in \s{sec:finitewidth}.\\


In spite of this disruption, we will demonstrate that a sharp signature persists in the geometric component (cf.\ \q{Wilsonloop}) of the adiabatic evolution. To recapitulate the main result of \s{sec:zakstrongatomic}, we have derived the spectrum of the Wilson loop $\W[\calc]$ for the above-stated EBRs in their atomic limit: the eigenvalues of $\W[\calc]$ are fixed to $n$'th roots of unity, and are  insensitive to  continuous deformations of the loop $\calc$. For finite Wannier separations, we will demonstrate that the eigenvalues of $\W[\calc]$  \textit{remain} fixed to $n$'th roots of unity, \textit{if} we impose an additional symmetry restriction on $\calc$: namely, $\mathcal{C}$ must intersect two distinct $g_n$-invariant wavevectors (the BZ center $\Gamma$ and $\bK_n$), such that half of the loop is mapped to the other half by $g_n$. Such \textit{symmetry-restricted loops} will be denoted by an additional subscript: $\calc_n$. Note that the reciprocal lattice vector $\bG[\calc_n]$ automatically satisfies \q{Gcond}. Various examples of $\calc_n$ (with $g_n$ an $n$-fold rotation) are illustrated in the last column of \fig{fig:elem}.\\ 

The robustness of the Zak phases (over $\calc_n$) is justified by a theorem which will first be presented  in \s{theoremstate}. In \s{ZakEBR} we will apply the theorem to answer the question that has been posed here.


\subsubsection{Theorem on symmetry-protected Zak phases}\la{theoremstate}

\begin{table}
\scalebox{0.8}{
\begin{tabular}{|c|c|c|c||c|c|c|}
\hline
\textbf{(a)} & $\sigma \big( \rho_4(\Gamma) \big)$ & $\sigma \big( \rho_4(M) \big)$ & $\tilde \sigma \big( \W[\calc_4] \big)$ & $\sigma \big( \rho_2(\Gamma) \big)$ & $\sigma \big( \rho_2(X) \big)$ & $\tilde \sigma \big( \W[\calc_2] \big)$  \\
 \hline \hline
  & $\{1,{-}1\} $ & $\{i,{-}i\}$ & $\emptyset$ & $\{1,1\}$ & $\{1,{-}1\}$& $\{1,{-}1\}$ \\
 \hline \hline \hline
 \textbf{(b)}  & $\sigma \big( \rho_3(\Gamma) \big)$ & $\sigma \big( \rho_3(K) \big)$ & $\tilde \sigma \big( \W[\calc_3] \big)$ & $\sigma \big( \rho_2(\Gamma) \big)$ & $\sigma \big( \rho_2(X) \big)$ & $\tilde \sigma \big( \W[\calc_2] \big)$  \\
 \hline \hline
 & $\{1,1\}$ & $\{\omega,\bar \omega\}$ & $\{\omega,\bar \omega\}$ & $\{1,{-}1\}$ & $\{1,{-}1\}$ & $\emptyset$ \\
 \hline \hline \hline
 \textbf{(c)} & $\sigma \big( \rho_3(\Gamma) \big)$ & $\sigma \big( \rho_3(K) \big)$ & $\tilde \sigma \big( \W[\calc_3] \big)$ & $\sigma \big( \rho_2(\Gamma) \big)$ & $\sigma \big( \rho_2(X) \big)$ & $\tilde \sigma \big( \W[\calc_2] \big)$  \\
 \hline \hline
  & $\{1,\omega,\bar{\omega}\}$ & $\{1,\omega,\bar{\omega}\}$ & $\emptyset$ & $\{1,1,1\}$ & $\{1,{-}1,{-}1\}$ & $\{1,{-}1,{-}1\}$ \\
 \hline \hline \hline
\textbf{(d)} & $\sigma \big( \rho_4(\Gamma) \big)$ & $\sigma \big( \rho_4(M) \big)$ & $\tilde \sigma \big( \W[\calc_4] \big)$  & $\sigma \big( \rho_x(\Gamma) \big)$ & $\sigma \big( \rho_x(X) \big)$ & $\tilde \sigma \big( \W[\calc_2] \big)$  \\
 \hline \hline
 & $\{1,1,{-}1,$ & $\{1,i,i,$ & $\emptyset$ & $\{1,1,1,$ & $\{1,1,1,$ & $\{1,1,{-}1\}$ \\
 & $i,{-}i\}$ & ${-}i,{-}i\}$ & $\emptyset$ & $1,{-}1\}$ & ${-}1,{-}1,{-}1\}$ &  \\
 \hline \hline \hline
 \textbf{(e)} & $\sigma \big( \rho_3(\Gamma) \big)$ & $\sigma \big( \rho_3(K) \big)$ & $\tilde \sigma \big( \W[\calc_3] \big)$ & $\sigma \big( \rho_2(\Gamma) \big)$ & $\sigma \big( \rho_2(X) \big)$ & $\tilde \sigma \big( \W[\calc_2] \big)$  \\
 \hline \hline
 & $\{1,1,1\}$ & $\{1,\omega,\bar \omega\}$ & $\{1,\omega,\bar \omega\}$ & $\{1,1,{-}1\}$ & $\{1,1,{-}1\}$ & $\{1\}$ \\
  \hline \hline \hline
  \textbf{(f)} & $\sigma \big( \rho_3(\Gamma) \big)$ & $\sigma \big( \rho_3(K) \big)$ & $\tilde \sigma \big( \W[\calc_3] \big)$ & $\sigma \big( \rho_2(\Gamma) \big)$ & $\sigma \big( \rho_2(M) \big)$ & $\tilde \sigma \big( \W[\calc_2] \big)$  \\
  \hline \hline
 $S_1^+$ & $\{1\}$ & $\{\bar \omega\}$ & $\{\bar \omega\}$ & $\{1\}$ & $\{{-}1\}$ & $\{{-}1\}$ \\
  \hline \hline
 $S_1^-$ & $\{\omega\}$ & $\{\bar \omega\}$ & $\{\omega\}$ & $\{{-}1\}$ & $\{1\}$ & $\{{-}1\}$ \\
  \hline
\end{tabular}
}
\caption{(a{-}f) label various case studies of gapped band subspaces: (a{-}e) refer to band representations with integer-spin on-site representations. The wallpaper groups and Wyckoff positions for (a{-}e) are respectively ($p4,2c$), ($p6m,2b$), ($p6,3c$) (illustrated in \fig{fig:elem}), and $(p4m,1a){\oplus} (p4m,2c)$, $(p6m,1a){\oplus} (p6m,2b)$ (illustrated in \fig{fig:venn}(d{-}e)), where $\oplus$ refers to the stacking of the corresponding subspaces. (f) describes a two-band subspace in the inversion-symmetric Kane-Mele model; this two-band subspace is splittable into two one-band subspaces $S_1^+$ and $S_1^-$, as explained in \s{classAII}. Second, third, fifth and sixth columns: for each case study, we display the eigenvalues ($\sigma(\rho_n)$) of the matrix representations ($\rho_n$) of $g_n$-symmetry at $g_n$-invariant wavevectors ($\Gamma$ and one of $X,K,M$); $\rho_x$  in (d) represents reflection symmetry which inverts $x$ but not $y$. Fourth and seventh columns: we display the $g_n$-protected Zak phase factors ($\tilde{\sigma}(\W)$) which diagonalize the Wilson loop $\W$ for a $\bk$-space loop $\calc_n$. We use $\omega{=}\mathrm{e}^{2\pi i/3}, \bar{\omega}{=}\mathrm{e}^{-2\pi i/3}$, while $\rho_n$, $\W$ and $\calc_n$ are defined in \tab{symbols}. \la{evalEx}}
\end{table}

Briefly stated, the theorem inputs the $g_n$-symmetry representation of the $N$-dimensional gapped subspace at $g_n$-invariant wavevectors, and outputs eigenvalues of $\W[\calc_n]$ that are robustly fixed to $n$'th roots of unity. Specifically, the input are the eigenvalues of the matrix representation  at the two $g_n$-invariant wavevectors $\bK{=}\Gamma$ or ${=}\bK_n$ which are intersected by $\calc_n$. \\

To elaborate, a symmetry $g_n{=}(\check g_n|\bt)$ acts on Bloch functions $\psi_{\bk}(\br)$ as $\hat g_{n} \psi_{\bk}(\br) {=} \psi_k(\check g_n^{{-}1} \br{-}\bt)$; for half-integer-spin representations, there is an additional action on the spinor component of the Bloch function.\cite{Tinkham2003,MelvinLax1974} For an $N$-band subspace, the $N {\times} N$ matrix representation of $g_n$ on the cell-periodic component of $\psi_{\bk}$ is derived from the above action to be 
\e{[\rho_n(\bK)]_{j,j'}=e^{-iF\pi/n} \langle u_{j,\check{g}_n\bK}| \hat{g}_n| u_{j',{\bK}} \rangle_{\text{cell}}, \la{sewnew} }
where $F{=}0$ (resp.\ ${=}1$) for integer-spin (resp.\ half-integer-spin) representations; we remind the reader of the definitions of $\hat g, \check g$ in \tab{symbols}. All eigenvalues of $\rho_n(\bK)$ are $n$'th roots of unity, owing to the assumed periodicity of $\psi_{j,\bK}$ in reciprocal translations, as well as $\hat{g}^n_n{=}e^{iF\pi}$ (composed with a lattice translation if $g_n$ is nonsymmorphic). The output of the theorem is a (possibly empty) subset of the $\W$-spectrum whose elements are also $n$'th roots of unity. This subset is robust to perturbations of the Hamiltonian $H_0$ that preserve the spectral gap as well as the symmetries of discrete translations and $g_n$ -- in short, we say that the subset is $g_n$-protected.\\

Before we state the theorem, let us define $m_l(\bK) {\in} \{0,...,N\}$ as the number of eigenvalues of $\rho_n(\bK)$ that equal $\mathrm{e}^{2\pi i l/n}$ for $\bK {\in} \{\Gamma, \bK_n\}$. We denote by $\bk_* {\in} \{\Gamma, \bK_n\}$ the wavevector at which $\rho_n(\bk_*)$ has the largest degeneracy ($m_{l_*}(\bk_*)$) among all symmetry eigenvalues. That is, for $l_* {\in} \{0,...,n{-}1\}$, we define $m_{l_*}(\bk_*)$ as the degeneracy that satisfies $m_{l_*}(\bk_*) {\ge} m_l(\bK)$ for all $l$ and for both $\bK {\in} \{\Gamma, \bK_n\}$. The other $g_n$-invariant wavevector that is not $\bk_*$ is labeled as $\bk_s$.

\begin{theorem*}
\normalfont If $r_l[\calc_n] {=} m_l(\bk_s){+}m_{l_*}(\bk_*){-}N{>}0$ for any of $l{=}0,...,n{-}1$, there exist exactly $r_l[\calc_n]$ number of Zak phases that are $g_n$-protected to $\phi[\calc_n]{=}{\pm}2\pi(l{-}l_*)/n$, where ${+}$ and ${-}$ respectively applies for $\bk_*{=}\Gamma$ and $\bK_n$. \la{th1}
\end{theorem*}

\noindent \textbf{\textit{Application of theorem to strong EBR on the honeycomb lattice.}} Let us return to the example introduced in \s{honey}. We will apply our theorem to calculate the Zak phases for the loops $\calc_3$ and  $\calc_2$, as illustrated in the last rows of \fig{fig:elem}. From \tab{evalEx}(b), we obtain the required input for the theorem: the eigenvalues of the matrix representations ($\rho_n; n{=}2,3$) at $g_n$-invariant wavevectors: $\Gamma$, $K$ and $M$. \\

\noindent Let us first evaluate the Zak phases for the bent loop $\calc_3$, which is restricted by three-fold rotational symmetry: $g_3{=}C_{3,z}$. $\rho_3(\Gamma)$ is the identity matrix, while $\rho_3(K)$ has eigenvalues $e^{\pm 2\pi i/3}$. Hence, $m_0(\Gamma){=}2$, $m_1(\Gamma){=}0$, $m_2(\Gamma){=}0$ and $m_0(K){=}0$, $m_1(K){=}1$, $m_2(K){=}1$. The highest degeneracy is attained at $\bk_*{=}\Gamma$ for $l_*{=}0$; $\bk_s{=}K$.  Applying the theorem, we deduce the degeneracies of the $C_{3,z}$-protected Zak phases: $r_0[\calc_3]{=}0{+}2{-}2{=}0$ (for $\phi[\calc_3]{=}0$), $r_1[\calc_3]{=}1{+}2{-}2{=}1$ (for $\phi[\calc_3]{=}2\pi/3$), $r_2[\calc_3]{=}1{+}2{-}2{=}1$ (for $\phi[\calc_3]{=}4\pi/3$); this result is summarized in the column `$\tilde{\sigma}\big( \W[\calc_3]\big)$' of \tab{evalEx}(b). The same Zak phases were obtained consistently through a Zak-Wannier relation that is explained in \s{sec:zakstrongatomic}.\\


\noindent Next we evaluate the Zak phases for the straight loop $\calc_2$, which is restricted by two-fold rotational symmetry: $g_2{=}C_{2,z}$. The eigenvalues of $\rho_2$ (at both $\Gamma$ and $M$) are ${\pm}1$, which implies that $m_0(\Gamma){=}m_1(\Gamma){=}m_0(M){=}m_1(M){=}1$. We may pick $\bk_*{=}\Gamma$ and $l_*{=}0$; in all cases,  $r_l[\calc_2]{\le}0$, i.e., there are no $C_{2,z}$-protected Zak phases. It is instructive to compare this result with  our conclusion [derived in \s{sec:zakstrongatomic}] that $|\phi_1[\calc_2]{-}\phi_2[\calc_2]|{=}2\pi/3$; the latter result is strictly valid only in the atomic limit. For finite Wannier separations, the present analysis shows there is no symmetry-based reason for the robustness of   $|\phi_1[\calc_2]{-}\phi_2[\calc_2]|$.\\  



The proof of the theorem is sketched in \s{sec:intuitionproof} and detailed in \app{app:thquant}. Several remarks are in order:

\begin{enumerate}


\item[(i)]  Let us define $2\pi/\xi_{\perp}$ as the smallest absolute difference between two $g_n$-protected Zak phases; $\xi_{\perp}$ is a positive integer that divides $n$. Where there exists none or only one $g_n$-protected Zak phase, we set $\xi_{\perp}{=}1$, which simply reflects the $2\pi$-ambiguity in the definition of a phase. $\xi_{\perp}{=}3$ in the honeycomb example just above. Since $\xi_{\perp}$ is invariant under symmetry- and gap-preserving deformations of the Hamiltonian, it may be viewed  as a $g_n$-protected topological invariant.  \\

For $N$-band subspaces whose energy function $E(\bk)$ is $N$-fold degenerate at each $\bk$ along $\calc_n$, the adiabatic propagator (cf.\ \q{adevo}) equals $e^{-i E_0 T_B/\hbar}\W[\calc_n]$ (cf.\ \q{Wilsonloop}), where $E_0$ is proportional to the average of $E(\bk)$ over $\calc_n$. For such band subspaces, we may identify $\xi_{\perp}$ with the period multiplier $\mu_{\perp}$ of stroboscopic Bloch oscillations, as defined in Definition \ref{define:stroboscopic}, i.e.,
\e{ \hspace{1cm} \xi_{\perp} = \mu_{\perp} \text{ for energy-degenerate subspaces}. \la{ximu} }

\item[(ii)] In claiming that a $g_n$-protected Zak phase is an $n$'th root of unity, we have presupposed a natural choice for the spatial origin -- that it lies at a $g_n$-invariant point. As is evident from \q{WilsonPxP}, a translation of the origin by $\delta \br$ modifies the Zak phases by a $j$-independent phase: $\phi_j {\rightarrow} \phi_j{+}\bG_n {\cdot} \delta \br$; in particular, a different choice of a $g_n$-invariant origin modifies $\phi_j$ by an integer multiple of $2\pi/n$, but does not change the number of $g_n$-protected Zak phases. Moreover, since differences of Zak phases are origin-independent, so is $\xi_{\perp}$; we explain in \s{exp} how differences in Zak phases can be experimentally measured.

\item[(iii)] For band subspaces for which the largest symmetry degeneracy ($m_{l_*}(\bk_*)$) is not unique, the multiple possible choices for ($l_*$,$\bk_*$) all lead to the same $g_n$-protected Zak phases.
\item[(iv)] This mapping of symmetry to holonomy eigenvalues includes the known case of $n{=}2$,\cite{Alexandradinata2014c} and further extends it to $n{=}3,4$. The inclusion of other point group symmetry may result in \emph{additional} symmetry-protected  Zak phases that are not covered by this theorem.\cite{Alexandradinata2014f} Alternatively stated, if a magnetic space group has $g_n$-symmetry, the above theorem sets a lower bound on the number of symmetry-protected Zak phases. 
\item[(v)] The theorem does not just apply to EBRs, but more generally to any gapped subspace that is invariant under $G$. As exemplified in \s{composite} below, $g_n$-protected Zak phases also occur for composite BRs. The theorem also applies to gapped topologically-nontrivial subspaces, such as subspaces with no representation on Wannier functions (\s{classA}), or BRs which have certain restrictions for the allowed symmetries in the stabilizers ($\calp_{\varpi}$) of the Wannier centers (\s{classAII}). The theorem also applies to energy subspaces that are gapped over $\calc_n$ but not necessarily over the full Brillouin torus, e.g. there may exist point degeneracies (Dirac points) in the energy spectrum away from $\calc_n$. 
\item[(vi)] The noncontractible loop $\calc_n$ does not need to intersect the Brillouin center; the theorem also applies to $\calc_n$ that intersects any two, inequivalent $g_n$-invariant wavevectors, with the constraint that half the loop is mapped to the other half by $g_n$.
\end{enumerate}

We derive two corollaries from the theorem, which we prove in \app{app:cor}.\\

\noindent \textbf{Corollaries.} (I) Since $m_l(\bk_s) {\le} m_{l_*}(\bk_*)$, a necessary but insufficient condition for the existence of $g_n$-protected Zak phases is that $m_{l_*}(\bk_*){>}N/2$. (II) All $N$ Zak phases are $g_n$-protected if and only if $m_{l_*}(\bk_*){=}N$, or equivalently, $\rho_n(\bk_*)$ is proportional to the identity. In particular, for $N{=}1$ (a single band), the lone Zak phase is always $g_n$-protected. \\

\noindent \textbf{\textit{Application of corollaries  to strong EBR on the honeycomb lattice.}} Let us return to the motivational example of \s{honey} and demonstrate the utility of the above corollaries. For the loop $\calc_3$, $m_{l_*}(\bk_*){=}2{=}N$ implies that all Zak phases $\phi[\calc_3]$ are $C_{3,z}$-protected (corollary (II)); while for $\calc_2$, $m_{l_*}(\bk_*){=}1{=}N/2$ implies that no Zak phase $\phi[\calc_2]$ is $C_{2,z}$-protected (corollary (I)).

\subsubsection{Intuition behind the theorem} \la{sec:intuitionproof}

To prove the theorem, we would first relate $\W$ to the matrix representations $\rho_n(\bK)$. Observe that $\calc_n$ is decomposable into the two $g_n$-related paths: $\calc_n^1$ and $\calc_n^2$, such that (a) $\calc_n^1$ has base point $\bK_n$ and end point $\Gamma$, and (b) $\calc_n^2{=}{-}g_n{\circ} \calc_n^1$, with the minus sign indicating a  reversal in orientation. Analogously, we may decompose $\W[\calc_n]$ into two Wilson lines: 
\e{ \W[\calc_n] = \W[\calc^2_n]\W[\calc^1_n]= \rho_n(\bK_n) \W[\calc^1_n]^{\dagger} \rho_n(\Gamma)^{\dagger} \W[\calc^1_n], \la{Wdecomp}}
where in the last equality we employed the symmetry transformation of Wilson lines.\cite{Alexandradinata2014c,Alexandradinata2014f} \q{Wdecomp} manifests the Wilson loop as the product of two generically noncommuting unitaries (each of order $n$). Exploiting the $U(N)$ gauge freedom in the gapped subspace, we may simultaneously diagonalize both $\rho_n$ in \q{Wdecomp}; in this basis, we may view  $\W[\calc^1_n]$ as a basis transformation between the eigenbases of $\rho_n$. The $g_n$-protected eigenvalues of $\W[\calc_n]$ are precisely those eigenvalues which are independent of $\W[\calc^1_n]$. \\

\q{Wdecomp} is a useful intermediate relation to prove the theorem. In the simple case where $\rho_n(\Gamma){=} \lambda_* \mathbb{1}_N$ ($\mathbb{1}_N$ is the $N{\times} N$ unit matrix), we may always choose $\bk_*{=}\Gamma$. Then \q{Wdecomp} implies that $\W[\mathcal{C}_n] {=} \bar{\lambda_*} \rho_n(\bK_n)$ is completely fixed by symmetry, where $\bar{\lambda}_*$ is the complex-conjugate of $\lambda_*$. The same conclusion may be obtained from the theorem: since $m_{l_*}(\Gamma) {=} N$, all Zak phases are $g_n$-protected, i.e., $r_{l}[\calc_n]{=}m_{l}(\bK_n)$ for all $l$.
Similarly, if $\rho_n(\bK_n){=} \lambda_* \mathbb{1}_N$, then $\W[\calc^1_n] \W[\mathcal{C}_n] \W[\calc^1_n]^{\dagger} {=} \lambda_* \rho_n(\Gamma)^{\dagger}$ is again fixed by symmetry and has the same eigenvalues as $\W[\mathcal{C}_n]$ because $\W[\calc^1_n]$ is unitary. That all Zak phases of $\W[\mathcal{C}_n]$ are $g_n$-protected, is another application of the theorem. \\

In the case where neither $\rho_n(\Gamma)$ nor $\rho_n(\bK_n)$ equals to $\lambda_* \mathbb{1}_N$, then not all of the eigenvalues of $\W[\mathcal{C}_n]$ are $g_n$-protected, according to corollary (II). $g_n$-protected eigenvalues exist if and only if the maximally-degenerate eigenspace of $\rho_n(\bk_*)$  (with eigenvalue $\mathrm{e}^{2\pi i l_*/n}$) robustly intersects any of the eigenspaces of $\rho_n(\bk_s)$, as shown in \app{app:thquant}.


\subsubsection{Application of the theorem to strong EBRs: symmetry-based Zak-Wannier relation}\la{ZakEBR}



Let us apply the theorem to the strong EBRs of wallpaper groups which satisfy conditions (i{-}iii) in \s{Wyckoffcond}; these are the EBRs that manifest period multiplication (in their atomic limit). Our goal is to answer the question that motivated  \s{Zak} (cf.\ first paragraph of \s{Zak}). We will show that for this subclass of strong EBRs, a signature of their nontriviality persists away from the atomic limit -- in the symmetry-protection of their Zak phases. \\

This result follows from a one-to-one correspondence between $g_n$-symmetric Wannier centers (defined modulo lattice translations) and $g_n$-protected Zak phases (defined modulo $2\pi$):
\e{ \text{For strong EBRs which satisfy conditions (i{-}iii),}\lin
\phi_j[\calc_n] = \bG_n \cdot \bvarpi_j  \;\; \text{mod } 2\pi; \as j=1,\ldots,M_{\varpi}. \la{ZakWy}}
As a reminder, `$g_n$-protected' quantities are invariant under continuous deformations of the $M_{\varpi}$-band subspace that preserve the energy gaps (above and below) and the symmetries of $g_n$ and discrete translations. Like \q{zakwannierstrongatomic}, \q{ZakWy} is a multi-band Zak-Wannier relation for Wannier functions that are localized in two independent directions. However, \q{ZakWy} extends \q{zakwannierstrongatomic} to the more realistic regime of finite Wannier separations. Moreover, while \q{zakwannierstrongatomic} holds for any BZ loop, \q{ZakWy} only holds for loops $\calc_n$ which satisfy the symmetry restriction described at the end of \s{Zak}.\footnote{There exists a continuous family of loops satisfying the same symmetry restriction.} \\



We will first provide a heuristic argument for \q{ZakWy}, and postpone a proof to the next paragraph. Let us consider a strong EBR (satisfying conditions (i-iii)) in the atomic limit, where \q{zakwannierstrongatomic} was proven to hold. We then perform the thought experiment of contracting all Wannier separations while preserving the wallpaper group; in particular, the symmetries of $g_n$ and discrete translations are preserved. Let us suppose that the Zak phases of all strong EBRs are $g_n$-protected (as will be proven below using our theorem). Since all $g_n$-protected quantities are invariant under contraction, the left-hand-side of \q{zakwannierstrongatomic} is invariant.   The right-hand-side of \q{zakwannierstrongatomic}, given by $\bG_n {\cdot} \bvarpi_j$, is also invariant, since $g_n$-symmetry remains a symmetry of each Wannier center throughout the contraction.\footnote{Under a scaling of the lattice period as $a\rightarrow \lambda a$, $\bG_n \rightarrow \bG_n/\lambda$ and $\bvarpi_j\rightarrow \lambda \bvarpi_j$ } We conclude that, owing to the $g_n$-protection of the Zak phase and the $g_n$-symmetry of the Wannier center, both sides of \q{zakwannierstrongatomic} retain their values regardless of the separation of Wannier centers; this is  exactly the meaning of  \q{ZakWy}. This argument emphasizes the essential role of symmetry in (non-atomic) Zak-Wannier relations; \q{ZakWy} is therefore termed a \emph{symmetry-based Zak-Wannier relation}. Such symmetry-based relations are studied in a broader context in \s{composite}.\\


\noindent \textit{Proof of \q{ZakWy}.} For the strong EBRs of wallpaper groups which satisfy conditions (i{-}iii) in \s{Wyckoffcond} (their Wyckoff positions are listed in \tab{wallpaper}), the on-site symmetry representation ($V_{\varpi}$) of $\calp_{\varpi}$ is one dimensional, i.e., $V_{\varpi}(g_n)$ is just a phase factor. This is because all but one of the stabilizers $\calp_{\varpi}$ listed in \tab{wallpaper} only admit one-dimensional irreducible representations $V_{\varpi}$; the sole exception ($G{=}p6m$, $\bvarpi{=}2b$, $\calp_{\varpi}{=}C_{3v}$) has one two-dimensional irreducible (vector) representation; however the corresponding EBR is weak, as proven in \app{pxpy}. Note that the present discussion excludes half-integer-spin EBRs of \textit{magnetic} wallpaper groups which include time-reversal symmetry. The representation of $g_n$ on cell-periodic functions at $\Gamma$ is simply a direct sum of $V_{\varpi}(g_n)$ -- one direct summand for each of the $M_{\varpi}$ Wannier centers $\bvarpi_j$ in one unit cell. To prove this, we linearly combine the $M_{\varpi}$ Wannier functions of one unit cell, and their Bravais lattice translates, to obtain a set of $M_{\varpi}$ Bloch functions:  
\e{\psi_{j,\bk} =\f1{\sqrt{\calv}} \sum_{\bR} e^{i\bk \cdot \bR}W_{j,\bR}, \la{blochwannier}}
with $\bR$ a Bravais lattice vector and $\calv$ the volume of the Brillouin zone. Note that each Wannier function is weighted by the coefficient $e^{i\bK_n\cdot (\bR+\bvarpi_j)}$, where $\bR{+}\bvarpi_j$ is the coordinate of the Wannier center. Then by applying the symmetry operation $\hat{g}_n$ on the cell-periodic component of these Bloch functions at $\bk=\Gamma$ and $\bK_n$, we derive the $M_{\varpi}{\times}M_{\varpi}$ matrix representations (cf. \q{sewnew}): 
\e{ \left[\rho_n(\Gamma)\right]_{j,j'}&=\delta_{j,j'}V_{\varpi}(g_n),\lin  
\left[\rho_n(\bK_n)\right]_{j,j'} &= \delta_{j,j'}V_{\varpi}(g_n) \mathrm{e}^{i \bG_n \cdot \bvarpi_j}. \la{ebrrep}} 
The additional phase factor $e^{i\bG_n\cdot \bvarpi_j}$ in the second line is nontrivial whenever $g_n$ maps a Wannier center $\bvarpi_j$ to a Wannier center $\check{g}_n \bvarpi_j$ in a distinct unit cell; this induces an additional phase factor $e^{i\check{g}_n\bk \cdot (\bvarpi_j-\check{g}_n \bvarpi_j)}$ due to the plane-wave coefficients in \q{blochwannier}. This phase is trivial for $\bk{=}\Gamma$, but for $\bk{=}\bK_n$,
\e{\check{g}_n \bK_n {\cdot}  (\bvarpi_j-\check{g}_n \bvarpi_j) = (\check{g}_n\bK_n-\bK_n) {\cdot} \bvarpi_j = \bG_n \cdot \bvarpi_j,\notag}
where in the last equality we applied \q{Gcond}. Inserting \q{ebrrep} for $\rho_n$ into \q{Wdecomp}, we derive the desired result. The theorem in \s{theoremstate} further implies that all $M_{\varpi}$ Zak phases are $g_n$-protected. \hfill\(\Box\) 



\section{Symmetry-based Zak-Wannier relations for band representations}\la{composite}

\subsection{Motivational example: composite band representations of 1D space groups}\la{motivCBR}
\begin{figure}
\centering
\includegraphics[width=8.5cm]{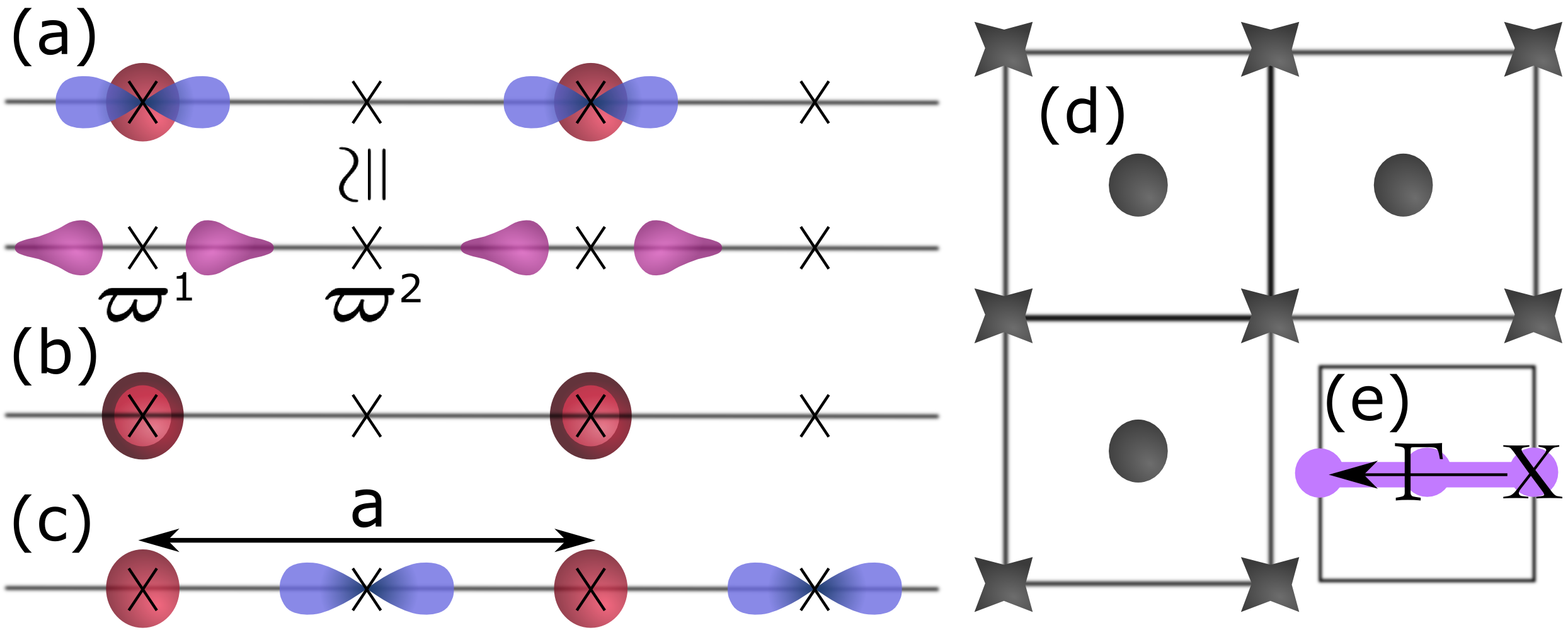}
\caption{ (a-c) Composite band representations of a 1D space group with inversion symmetry. Crosses denote inversion-symmetric points ($\bvarpi^{1,2}$). $s$ ($p$)-orbitals are illustrated by red (blue) blobs; the $s{-}p$ hybrid orbital is colored pink. (d) A square lattice with four-fold rotational symmetry. Wannier functions are centered on the centers and vertices of square plaquettes. (e) Brillouin zone corresponding to the square lattice, with $\calc_2$-loop indicated in violet. \la{fig:1dex} }
\end{figure}

In the previous \s{sec:perpfield}, we have identified a class of strong EBRs which manifest period-multiplied Bloch oscillations; this multiplication can be explained through the adiabatic transport of either Bloch or Wannier functions -- this complementarity is summed up in a Zak-Wannier relation (cf. \q{ZakWy}). In this section we investigate the Bloch oscillations and Zak-Wannier relations of a broader class of BRs which include the composite (i.e., non-elementary) BRs.\\

Precisely, a composite band representation (CBR) is defined as a stack of multiple EBRs, and may therefore be specified by specifying the data that labels all EBRs labeled by $\nu$, namely the Wyckoff position $\bvarpi^{\nu}$ {and} the corresponding irreducible representation $V^{\nu}$ of the on-site symmetry group $\calp_{\varpi^{\nu}}$ (e.g. \tab{wallpaper}). It is possible that $\bvarpi^{\nu}$ for different $\nu$ are equal.    \\

Let us illustrate these notions for the simplest nontrivial space group in 1D; the point group is generated solely by spatial inversion ($g_2$), an order-two symmetry.  There are only four inequivalent EBRs, as exemplified by $s$- (i.e., inversion-even) and $p$- (inversion-odd) orbitals localized on either of the two inequivalent inversion centers $\varpi^1, \varpi^2$; $\varpi^1$ and $\varpi^2$ are separated by half of the lattice period $a$. Fixing the spatial origin to $\varpi^1$, the inversion eigenvalues at high-symmetry wavevectors ($0$ and $\pi/a$) are derived from \q{ebrrep} to be:
\begin{align}
\begin{array}{|c||c|c|}
\hline
   (\varpi,\text{ orbital}) & \rho_2(0) & \rho_2(\pi/a) \\
  \hline \hline
  (\varpi^1,s) & {+}1 & {+}1 \\
  \hline
  (\varpi^1,p) & {-}1 & {-}1 \\
  \hline
  (\varpi^2,s) & {+}1 & {-}1 \\
  \hline
  (\varpi^2,p) & {-}1 & {+}1 \\
  \hline
\end{array}\notag
\end{align}
for the four inequivalent EBRs. The possible two-band CBRs are obtained from stacking any two of the four EBRs. We illustrate in \fig{fig:1dex}: (a) the stacking of $s$ and $p$ on the same $\varpi{=}\varpi^1$ , (b) $s$ and $s$ on the same $\varpi{=}\varpi^1$, and (c) $s$ and $p$ on the different $\varpi^1$, $\varpi^2$. \\ 

These three simple examples already give a good representation of the possible Bloch oscillations and Zak-Wannier relations for CBRs. In \fig{fig:1dex}(a), $s$ (colored red) and $p$ (blue) orbitals may mutually hybridize (pink) and split equidistantly away from $\bvarpi^1$, hence none of the Wannier centers are $g_2$-fixed. Neither are the Zak phases,\cite{Alexandradinata2014c} as one may confirm from application of corollary (I) on the inversion eigenvalues $\rho_2(0), \rho_2(\pi/a){=}{\pm}1$. The Wannier centers and Zak phases in \fig{fig:1dex}(b{-}c) are $g_2$-fixed, but only in case (c) is the Zak phase difference of $\pi$ equal to $G_2(\varpi^2-\varpi^1)$ (hence $\xi_{\perp}{=}\chi_{\perp}{=}2$).\\

In \s{sec:perpfield} we found for certain atomic BRs that symmetry-protected Zak phase differences of $2\pi/\xi_{\perp}$ (with $\xi_{\perp}{>}1$) lead to  Bloch oscillations with multiplier $\xi_{\perp}$. However, the atomic limit of case (c) does not necessarily exhibit period doubling, because the $s$ and $p$ orbitals, being unrelated by symmetry, do not necessarily have the same on-site energies: $E_s$ and $E_p$  (cf.\ the beginning of \s{buildblock}); recall that period multiplication in a $\perp$-field requires that bands are energy-degenerate at all $\bk$ (cf.\ remark (i) in \s{theoremstate}). Nevertheless, since only one parameter need be tuned for $E_s{=}E_p$, it is  plausible that this degeneracy may arise experimentally -- especially in the well-controlled setting of ultracold atoms in optical lattices, as elaborated in \s{cold}. Furthermore, if the deviation from degeneracy $|E_s{-}E_p|$ is small compared to the Wannier-Stark ladder spacing $Fa$, then the Bloch-oscillatory behavior is essentially indistinguishable from that of period doubling (as elaborated in \s{sec:finitewidth}). For these reasons, we are  motivated to identify band subspaces with $\xi_{\perp}{>}1$.\\

In principle, the theorem of \s{theoremstate} should allow us to compute $\xi_{\perp}$ for any band subspace, and $\xi_{\perp}{>}1$ is expressible as a condition on the symmetry eigenvalues at high-symmetry wavevectors. In practice, knowing the symmetry eigenvalues (or constraints thereof) does not directly indicate how one would physically realize this band subspace, e.g., in a tight-binding model or in experiments. On the other hand, if we knew that a  $\xi_{\perp}{>}1$ band subspace is a CBR of a space group, then the Wannier functions of this CBR naturally form a localized basis for a tight-binding model.   It is therefore desirable to first establish a Zak-Wannier relation for CBRs (cf.\ \s{symmbased}), so that one may translate $\xi_{\perp}{>}1$ to a condition on the tight-binding Wannier centers (cf.\ \s{sec:applyzwfind}).

\subsection{Definition and application of the symmetry-based Zak-Wannier relation}\la{symmbased}

In this section we would formulate generalized Zak-Wannier relations for CBRs. Generally, a Zak-Wannier relation is a one-to-one correspondence between Zak phases and Wannier centers. Wannier functions $\{W_{n,\bR}\}_{n=1}^N$ can always be viewed as Fourier transforms of Bloch functions $\{\psi_{n,\bk}\}_{n=1}^N$ which are analytic in $\bk$. Wannier functions only exist for band subspaces with vanishing Chern number,\cite{Brouder2007} and these functions are exponentially-localized in real space due to the analyticity of the Bloch functions.\cite{Cloizeaux1964}  \\


There remains however an ambiguity in the definition of Wannier functions, owing to an arbitrariness in how we choose Bloch functions at each $\bk$; we refer to the freedom in performing unitary transformations 
\e{\psi_{n,\bk} \rightarrow \sum_{m=1}^N \psi_{m,\bk}S_{m,n}(\bk), \ins{with} S(\bk) \in U(N) \la{gaugeambiguity}} 
and analytic in the wavevector $\bk$, as a gauge ambiguity.\cite{Marzari1997} Consequently, individual Wannier centers $\bvarpi$ (defined as the quantum expectation value of the position operator wrt. the Wannier functions) are gauge-dependent; the sum of Wannier centers, however, is uniquely defined modulo a Bravais lattice vector.\cite{King-Smith1993} On the other hand, Zak phases are gauge-invariant modulo $2\pi$.\cite{Alexandradinata2014c} To have any correspondence between Zak phases and Wannier centers in multi-band subspaces, it is therefore necessary that restrictions are imposed on the gauge so that an individual Wannier center is not completely arbitrary. Ideally, these gauge restrictions would be physically motivated. For example, if we impose that Wannier functions are maximally localized in 1D real space (equivalently, that they are eigenstates of the 1D projected position operator),\cite{Marzari1997}  their Wannier centers indeed have a one-to-one correspondence with Zak phases.\cite{Alexandradinata2014c} To formulate a novel Zak-Wannier relation between $g_n$-protected Zak phases and the centers of Wannier functions that are localized over 2D real space, we propose that gauge  restrictions  by \textit{symmetry}  affords us a Zak-Wannier relation. \\

Let us state the Zak-Wannier relations first, then subsequently elaborate on the gauge restrictions, examples and proofs. Not all CBRs satisfy a Zak-Wannier relation. Let us consider a class of  CBRs (each characterized by the data $\{ (\bvarpi^{\nu}, V^{\nu}) \}_{\nu}$ of the constituent EBRs) that satisfy: 

\begin{enumerate}
\item[(a)] All Wyckoff positions $\bvarpi^{\nu}$ of the constituent EBRs are $g_n$-invariant, i.e., $g_n{\in} \calp_{\varpi^{\nu}}$ for all $\nu$. \\
\item[(b)] A symmetry condition on the little groups in $\bk$-space: all the Bloch functions -- at wavevector $\bk_*{=}\Gamma$ or $\bK_n$ (or both) -- transform under the same 1D representation of $g_n$.  \\
\end{enumerate}

\noindent These CBRs satisfy the Zak-Wannier relation:
\e{ \phi^{\nu}_{j_{\nu},\alpha_{\nu}}[\calc_n] = \bG_n {\cdot} \bvarpi^{\nu}_{j_{\nu}}, \la{zakwannier}}
which holds for all $\nu$, all integers $j_{\nu}$ running from $1$ to the multiplicity $M_{\nu}$ of $\bvarpi^{\nu}$, and all $\alpha_{\nu}$ from $1$ to $\dim V^{\nu}$ ($\phi, \calc_n, \bG_n, \bvarpi^{\nu}_{j_{\nu}}, M_{\nu}$ are all defined in \tab{symbols}).  \q{zakwannier} especially says that the Zak phases $\phi^{\nu}_{j_{\nu},\alpha_{\nu}}$ are independent of $\alpha_{\nu}$, i.e., the minimal degeneracy of $\phi^{\nu}_{j_{\nu}}$ is $\dim V^{\nu}$. The degeneracy is greater in cases where $\phi^{\nu}_{j_{\nu}}$ are equal for different $\nu$ or $j_{\nu}$. \\

The decomposition of a CBR into its constituent EBRs is not always unique.\cite{Zeiner2000} This is another manifestation of the gauge ambiguity of Wannier and Bloch functions (cf.\ \q{gaugeambiguity}); choosing a different gauge, we might decompose the same subspace into a different set of EBRs.
However, \q{zakwannier} is agnostic to this arbitrariness, i.e., we will see that the proof is equally valid for any choice of decomposition. On the other hand, we remind the reader that Zak phases are gauge invariant modulo $2\pi$.\cite{Alexandradinata2014c} The combined implication is that the Zak-Wannier relation \q{zakwannier} applies to \emph{any} decomposition into EBRs.\\

The gauge ambiguity is actually larger than was alluded to {in} the previous paragraph. Where previously we only considered decomposing a CBR into multiple EBRs, actually the same CBR can be split into multiple subspaces that are not EBRs. Indeed, if we begin with a decomposition into EBRs and then perform an arbitrary unitary transformation (cf.\ \q{gaugeambiguity}) that mixes together different EBRs -- the resultant set of Wannier functions are generically asymmetric, and do not lie on  $g_n$-invariant Wyckoff positions. From the perspective of numerically constructing Wannier functions by Fourier transformation of Bloch functions (obtained from tight-binding models or an \textit{ab initio} calculations), asymmetric Wannier functions are the norm, rather than the exception; one typically has to do more work to construct symmetric Wannier functions.\cite{Sakuma2013} To recapitulate, the Zak-Wannier correspondence of \q{zakwannier} applies to a $g_n$-symmetric gauge for the Wannier functions, which we now carefully define.

\begin{definition}\la{define:symmbasedzw}
\normalfont An $N$-band subspace ($N{\geq} 1$) satisfies a \emph{symmetry-based Zak-Wannier relation}, if it is a direct sum of EBRs with the following properties: (a) the Wannier functions of each  EBR lie on $g_n$-invariant Wyckoff positions $\bvarpi$, i.e., $g_n {\in}P_{\bvarpi}$. In total there are $N$ Wannier functions which are not related by lattice translations, and correspondingly $N$ translation-inequivalent Wannier centers $\{\bvarpi_j\}_{j=1}^N$. (b) There is a one-to-one correspondence between these Wannier centers (defined modulo lattice translations)  and the Zak phases (defined modulo $2\pi$): $\bG {\cdot} \bvarpi_j{=}\phi_j$ mod $2\pi$ for all $j{=}1,{\ldots},N$. This correspondence holds for all $N$ Zak phases that encode the $N$-band holonomy of a loop that wraps the Brillouin torus in the direction of  $\bG$, a primitive reciprocal vector.
\end{definition}

\noindent \q{zakwannier} is an example of a symmetry-based Zak-Wannier relation. In comparison, a previously-known multi-band Zak-Wannier relation\cite{Zak1991,King-Smith1993,Alexandradinata2014c} is based on the gauge condition of maximal-localization in one spatial direction;\cite{Marzari1997} this condition is agnostic to the symmetry of the band subspace. The multi-band Zak-Wannier relation discussed in \s{parbloch} is of the maximal-localization type, since eigenstates of the projected position operator $PzP$ are maximally localized in the $z$-direction.\cite{Marzari1997}\\

\noindent \textit{Proof of \q{zakwannier}.} Let us consider a CBR which satisfies (a,b). If $\rho_n(\bk_*)$ is proportional to the identity (condition (b)), then \q{Wdecomp} implies that  $\W[\calc_n]$ is unitarily equivalent to $\rho_n(\bK_n)\dg{\rho_n(\Gamma)}$ (cf. $\rho_n, \W$ in \tab{symbols}). The representation $\rho_n$ of $g_n$ for the CBR is a direct sum  of representations $\rho_n^{\nu}$ of $g_n$ for the EBRs labeled by $\nu$. If condition (a) holds, then $\rho_n^{\nu}$ is given by (cf.\ \q{ebrrep})
\e{ \left[\rho_n^{\nu}(\Gamma)\right]_{j_{\nu},l_{\nu}}^{\alpha_{\nu},\beta_{\nu}} &=\delta_{j_{\nu},l_{\nu}} [V^{\nu}(g_n)]_{\alpha_{\nu},\beta_{\nu}},\lin  
\left[\rho_n^{\nu}(\bK_n)\right]_{j_{\nu},l_{\nu}}^{\alpha_{\nu},\beta_{\nu}} &= \delta_{j_{\nu},l_{\nu}} [V^{\nu}(g_n)]_{\alpha_{\nu},\beta_{\nu}} \mathrm{e}^{i \bG_n \cdot \bvarpi^{\nu}_{j_{\nu}}}, \la{ebrrep2} }
where the integer indices $l_{\nu}, j_{\nu}$ run from 1 to the multiplicity of $\bvarpi^{\nu}$ and $\alpha_{\nu}, \beta_{\nu}$ run from 1 to $\dim V^{\nu}$.  Applying the unitarity of $V^{\nu}(g_n)$,
\e{\big[ \rho_n^{\nu}(\bK_n)\dg{\rho_n^{\nu}(\Gamma)} \big]_{j_{\nu},l_{\nu}}^{\alpha_{\nu},\beta_{\nu}} {=} \delta_{j_{\nu},l_{\nu}} \delta_{\alpha_{\nu},\beta_{\nu}} \mathrm{e}^{i \bG_n \cdot \bvarpi^{\nu}_{j_{\nu}}},\la{zakky}} 
which is unitarily equivalent to $\W[\calc_n]$.  Further applying that unitarily-equivalent matrices have the same spectrum, we derive \q{zakwannier}.  \hfill\(\Box\) \\

\noindent \textit{\textbf{Example of symmetry-based Zak-Wannier relation.}} We consider a lattice that is composed of a triangular and a honeycomb sublattice; the $N{=}3$ $s$-orbitals per unit cell are localized at the $C_{3,z}$-invariant Wyckoff positions $\bvarpi^1{=}1a$ (triangular sublattice), $\bvarpi^2{=}2b$ and $\bvarpi^2_2{=}C_{6,z} {\circ} \bvarpi^2$ (honeycomb sublattice), illustrated in \fig{fig:venn}(e) by a green, red and blue blob, respectively. \tab{evalEx}(e) summarizes the representations of $C_{3,z}$ ($\rho_3$) at high-symmetry wavevectors: The maximal degeneracy is $m_{l_*}(\bk_*){=}3$ at $\bk_*{=}\Gamma$ for the eigenvalue $1$ ($l_*{=}0$), i.e., $\rho_3(\Gamma)$ is the identity matrix (cf. $m_l, l_*$ in \tab{symbols}). Therefore, (a,b) apply and \q{zakwannier} provides the following $C_{3,z}$-fixed Zak phases: $\phi^1_{1,1}[\calc_3] {=} \bG {\cdot} \bvarpi^2_{2,1} {=} 0$, $\phi^2_{1,1}[\calc_3] {=} \bG {\cdot} \bvarpi^2 {=} 2\pi/3$ and $\phi^2_{2,1}[\calc_3] {=} \bG {\cdot} \bvarpi^2_2 {=} {-}2\pi/3$, where $\bG$ is a reciprocal vector directed along the horizontal axis, as shown in \fig{fig:bands}(d). Taking differences of these Zak phases, we also find that $\xi_{\perp}{=}3$.

\subsubsection{Symmetry-based Zak-Wannier relations for subspaces with a single Wyckoff position}\la{symmWF}
Let us particularize \s{symmbased} to a subclass of CBRs which satisfy conditions (a,b), for which we can rephrase condition (b) as a condition on the on-site symmetries. The additional conditions are:
\begin{enumerate}
\item[(c)] $\bvarpi^{\nu} {=} \bvarpi$ for all $\nu$ and $\bvarpi$ has unit multiplicity,
\item[(d)] $\bvarpi^{\nu} {=} \bvarpi$ for all $\nu$ and $\bvarpi$ satisfies conditions (i{-}iii) in \s{Wyckoffcond}. Note that condition (a) of \s{symmbased} and (i) of \s{Wyckoffcond} are equivalent. 
\end{enumerate}

\noindent For CBRs satisfying (a,b,c), we find that (b) --  a condition on the symmetry representations in $\bk$-space -- is equivalent to the following condition on symmetry representations in real space:
\begin{enumerate}
\item[(b')] $V^{\nu}(g_n)$ is proportional to the identity with proportionality constant independent of $\nu$, i.e., all Wannier functions transform under the same 1D representation of $g_n$ (which is an on-site symmetry).
\end{enumerate}
\noindent This equivalence implies that the Zak-Wannier relation (\q{zakwannier}) holds, in a special case, for CBRs satisfying (a,b',c).\\

For the class of  CBRs that satisfy (a) and (c), we would now prove that conditions (b) and (b') are equivalent. Given (b') and \q{ebrrep}, we derive that $\rho_n(\Gamma)$ is proportional to the identity, hence (b) is satisfied. Let us further show that (b) implies (b'). First, condition (c) implies that $j_{\nu}$ is an unnecessary index, which we henceforth neglect, and that $\mathrm{e}^{i \bG_n \cdot \bvarpi^{\nu}}{=}\mathrm{e}^{i \bG_n \cdot \bvarpi}$ does not depend on $\nu$. If $\bk_*{=}\Gamma$, combining condition (b) with \q{ebrrep} gives $[V^{\nu}(g_n)]^{\alpha_{\nu},\beta_{\nu}}{=}\lambda_* \delta_{\alpha_{\nu},\beta_{\nu}}$ for some $\lambda_* {\in} U(1)$ that is independent of $\nu$, hence (b') is satisfied. For $\bk_*{=}\bK_n$, combining (b) with \q{ebrrep} imposes $[V^{\nu}(g_n)]^{\alpha_{\nu},\beta_{\nu}}{=} \mathrm{e}^{-i \bG_n \cdot \bvarpi} \lambda_* \delta_{\alpha_{\nu},\beta_{\nu}}$ for all $\nu$, hence (b') holds.\\

For CBRs satisfying (a,b,d), we find that necessarily $\bk_* {=} \Gamma$ in condition (b); (b) with $\bk_* {=} \Gamma$ is also equivalent to (b'). Let us show this. The set $\{ \mathrm{e}^{i \bG_n \cdot \bvarpi_{j_{\nu}}} \}_{j_{\nu}=1}^{M_{\nu}}$ { has at least two distinct elements; this is due to \q{specialG} which holds because of conditions (i{-}iii) of \s{Wyckoffcond}, as proven in \app{app:cond}. Consequently,} $\rho_n^{\nu}(\bK_n)$ in \q{ebrrep} cannot be proportional to the identity, hence (b) cannot be satisfied with $\bk_*{=}\bK_n$. The equivalence of (b') and (b) with $\bk_*{=}\Gamma$  can be derived in close analogy with the previous paragraph.\\

For CBRs which satisfy (a,c) or (a,d), the on-site representation is reducible, i.e., $V{=}\oplus_{\nu} V^{\nu}$, and corollaries (I{-}II) in \s{Zak} simplify to: (I') the necessary condition for the existence of $g_n$-protected Zak phases is that $v_*{>}\dim V/2$, where $v_*{\le} \dim V$ is the dimension of the maximal eigenspace of $V(g_n)$. (II') All Zak phases are symmetry-protected if and only if $V(g_n)$ is proportional to the identity ($v_*{=} \dim V$).

\subsubsection{Symmetry-based Zak-Wannier relations for elementary band representations}\la{ZWebr}

Definition \ref{define:symmbasedzw} also applies to $N$-band subspaces which are themselves EBRs, in which case the `direct sum of EBRs' should be interpreted trivially. \\

The statements in \s{symmbased} and \s{symmWF} also apply to EBRs if we simply substitute `CBR' with `EBR' and remove the unnecessary index $\nu$. The class of EBRs that satisfy such a  modified symmetry-based Zak-Wannier relation include the strong EBRs discussed in \s{ZakEBR} (cf.\ \q{ZakWy}), as well as strong EBRs with unit-multiplicity Wyckoff positions. Note that all single-band EBRs fall into the latter category.

\subsubsection{Band representations without symmetry-based Zak-Wannier relations}
To juxtapose against the CBRs considered in \s{symmbased} and \s{symmWF}, we remind the reader of \fig{fig:1dex}(a), which describes a CBR that is a stack of two single-band EBRs. Single-band EBRs individually have symmetry-protected Zak phases and satisfy a symmetry-based Zak-Wannier relation according to \s{ZWebr}, whereas the two-band CBR does not. \\

The general criterion for a two-band CBR to have symmetry-protected Zak phases (which is necessary for the existence of a Zak-Wannier relation) is as follows: Consider a two-band CBR that is made from stacking two single-band EBRs ($(\bvarpi^1,V^1)$, and $(\bvarpi^2,V^2)$). This two-band subspace does not have $g_n$-protected Zak phases (as defined for parallel transport within two bands), if the subspace can be split into two single-band subspaces ($S_1$, $S_2$), whose single-band Zak phases (defined individually for each of $S_1, S_2$) are distinct from those of $(\bvarpi^1,V^1)$ and $(\bvarpi^2,V^2)$.

\subsection{Application of symmetry-based Zak-Wannier relation to identify subspaces with $\xi_{\perp}{>}1$} \la{sec:applyzwfind}

Combining the condition  $\xi_{\perp}{>}1$ with the Zak-Wannier relation \q{zakwannier}, we derive that 
\e{|\bG_n {\cdot} (\bvarpi^{\nu}_{j_{\nu}}-\bvarpi^{\nu'}_{j'_{\nu}})|=\f{2\pi}{\xi_{\perp}}, \;\; \xi_{\perp}>1}
for at least one set of $\{\nu,\nu',j_{\nu},j'_{\nu'}\}$. $\nu{=}\nu'$ and $j_{\nu}{\neq}j'_{\nu}$ label two $g_n$-invariant Wannier centers which are related by a point-group symmetry other than $g_n$; this case was essentially described in \s{perp} and can be generalized to strong CBRs. \\

 The case $\nu{\neq}\nu'$ labels two $g_n$-invariant Wannier centers belonging to distinct EBRs, as we have exemplified by case (c) in \s{motivCBR} (cf. \fig{fig:1dex}). A 2D generalization of case (c) is a two-band CBR with s-orbitals centered at two inequivalent four-fold-invariant Wyckoff positions, as illustrated in \fig{fig:1dex}(d). The Zak phases associated to the $\calc_2$-loop (illustrated in \fig{fig:1dex}(e)) differ by $\pi$, hence $\xi_{\perp}{=}2$.



\section{Period multiplication in time-reversal-asymmetric Chern bands}\la{classA}


In this section we discuss the possibility of period multiplication for Chern bands (band subspaces with nonzero Chern number), which belong to class A in the Wigner-Dyson symmetry classification.\cite{Kitaev2009a,Schnyder2009} Our discussion is split into two parts:\\

\noi{A} Our previous study of Zak-Wannier relations have demonstrated that symmetry-protected Zak phases can be directly related to symmetry-protected Wannier centers; in essence, we have found that certain band subspaces may be modelled by classical point charges. However, Chern bands admit no such classical description due to unavoidable quantum fluctuations with respect to the noncommuting projected position operators, i.e., there exists no representation of Chern bands on Wannier functions. Nevertheless, the existence of  Bloch functions (which are non-analytic over the Brillouin torus) allows for their characterization by Zak phases -- which can be symmetry-protected. Our study in \s{sec:zakforbidden} demonstrates that Chern bands can realize \textit{classically-forbidden} Zak phases, that is, Zak phases which are impossible to realize in \textit{any} classically-localizable band subspace with zero Chern number. \\

\noi{B} The discussion of (A) suggests the possibility of a classically-forbidden period multiplier for Chern bands. As proof of principle, we construct a $C_{6,z}$-symmetric Chern band  that exhibits stroboscopic oscillations with the classically-forbidden multiplier $\mu_{\perp}{=}2$. Our construction of this model utilizes a novel and generally applicable approach to obtaining Chern bands -- from  splitting a multi-band EBR into multiple fewer-band subspaces.


\subsection{Zak-Chern relations and classically-forbidden Zak phases}\la{sec:zakforbidden}

\begin{figure}
\centering
\includegraphics[width=8.5cm]{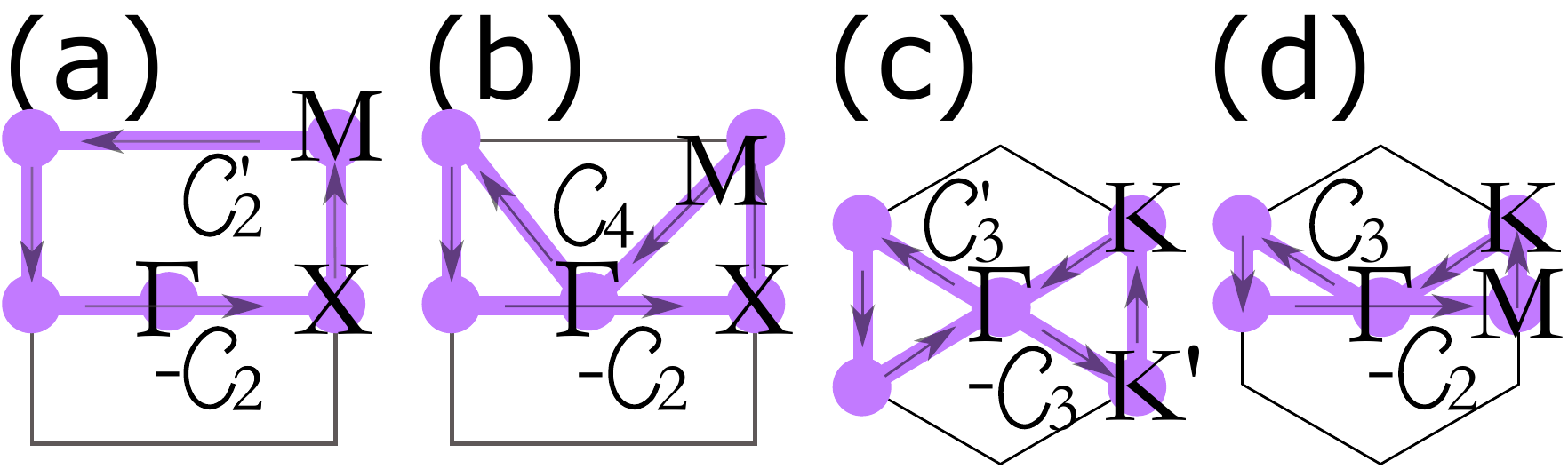}
\caption{Contractible loops (violet lines) in BZs with $C_2$, $C_4$, $C_3$ and $C_6$ symmetry respectively; these loops are used to determine the Chern number mod $n$ ($n{=}2,4,3,6$, respectively). \la{fig:loops} }
\end{figure}

In a $C_{4,z}$-symmetric crystal, the Chern number ($C$) of a single band is determined modulo four by the $C_{n,z}$-fixed Zak phases $\phi[\calc_n]$ for $n{=}2,4$ as
\e{ C_{4,z}:\as C= 2\f{\phi[\calc_2]-\phi[\calc_4]}{\pi} \ins{mod } 4, \la{c4constraint}}
as we prove in \app{app:ZakCh}. Here, $\calc_2$ intersects $\Gamma$ and $X$, and $\calc_4$ intersects $\Gamma$ and $M$, as illustrated in \fig{fig:loops}(b). One consequence of \q{c4constraint} is that $C{=} {-}2\phi[\calc_4]/\pi$ mod $2$, since $\phi[\calc_2] {\in} \{0,\pi\}$ from the theorem (in \s{th1}). In other words, $\phi[\calc_4]=\pm \pi/2$ are two classically-forbidden values that would imply a nonzero Chern number.\\

For $C_{n,z}$-symmetric crystals with $n{\in}\{2,3,6\}$, there exist analogous constraints that relate the Chern number to symmetry-protected Zak phases:
\e{ &C_{2,z}:\as C= \f{\phi[\calc_2]-\phi[\calc_2']}{\pi} &&\ins{mod } 2,\la{c2constraint}  \\
&C_{3,z}:\as C= \f{3}{2}\f{\phi[\calc_3]-\phi[\calc_3']}{\pi} &&\ins{mod } 3,\la{c3constraint} \\
&C_{6,z}:\as C= 3\f{\phi[\calc_2]-\phi[\calc_3]}{\pi} &&\ins{mod } 6, \la{c6constraint}}
where the $\calc_n$ and $\calc_n'$ are loops illustrated in \fig{fig:loops}. {The proofs of \qq{c2constraint}{c6constraint} are closely analogous to the $C_{4,z}$ case in \app{app:ZakCh}.}
\q{c6constraint} implies for a $C_{6,z}$-symmetric band that $\phi[\calc_2]{=}\pi$ is classically forbidden, and also either of $\phi[\calc_3]=\pm 2\pi/3$. \\

All of \qq{c4constraint}{c6constraint} may be generalized to constraints on the Chern number of an $N$-band gapped subspace, if we replace $\phi[\calc_n]$ by the phase of det$\,\W[\calc_n]$, with $\W$ the $N {\times} N$ matrix representation of holonomy (cf.\ \q{Wilsonloop}). These constraints between $g_n$-protected Zak phases and Chern numbers provide a complementary perspective to previously-developed constraints that relate $g_n$-eigenvalues to Chern numbers.\cite{Fang2012,Hourglass2016}

\subsection{Chern bands with classically-forbidden period multipliers } \la{sec:recipe}

The identification of classically-forbidden Zak phases in \s{sec:zakforbidden} implies the existence of classically-forbidden period multipliers in Bloch oscillations. For example, the classically-forbidden $\phi[\calc_2]{=}\pi$ (resp.\ $\phi[\calc_3]{=}{\pm} 2\pi/3$) in $C_{6,z}$-symmetric band subspaces implies that $\mu_{\perp}{=}2$ (resp.\ $\mu_{\perp}{=}3$) is forbidden for the $\calc_2$ (resp.\ $\calc_3$) loop; this follows from the relation between Zak phase differences and $\mu_{\perp}$ that is described in remark (i) of \s{theoremstate}. One motivation of this section is to construct, as proof of principle, a Chern band with a forbidden multiplier. \\ 

Before this specific construction [detailed in \s{sec:proofofprinciple}], let us discuss some generalities about the construction of Chern bands from tight-binding models. 
A band representation (BR), elementary or composite, defines a set of spatially-localized functions which may be used as a basis in a tight-binding model. The net Chern number of any EBR (and hence of any BR) vanishes; this follows from Bloch functions of an EBR (cf.\ \q{blochwannier}) being analytic and periodic over the BZ.\cite{Cloizeaux1964,Brouder2007} There are therefore only two ways to obtain Chern bands: from `band inversion' between distinct EBRs, and from `splitting' a single, multi-band EBR. By `band inversion,' we refer to the well-known process of a gap closing between two EBRs, with an accompanying transfer of topological `charge,' which in this case is the Chern number.  By `splitting' an EBR into two subspaces, we mean to decompose this EBR (having $N{>}1$ bands) into two subspaces $S_1$ and $S_2$ (having dimensions $N_1$ and $N_2$ which sum to $N$ at each $\bk$), such that each of $S_1, S_2$ individually transforms in a representation of $G$. This implies that $S_1$ and $S_2$ can be separated energetically at all $\bk$. \\

The splitting of EBRs into Chern bands is a novel construction that we elaborate upon in \s{sec:split}. In \s{sec:proofofprinciple}, we apply this construction to obtain a Chern band with a classically-forbidden multiplier.

\subsubsection{Obtaining Chern bands from splitting elementary band representations}\la{sec:split}

Let us propose a general recipe to obtain nontrivial Chern bands by splitting EBRs of certain nonmagnetic wallpaper groups ($G$) that we specify below. By `nonmagnetic', we mean that no element of $G$ involves time reversal ($T$). Our recipe may be viewed as sure-fire instructions to cook up models of Chern bands: a splittable EBR defines a tight-binding basis, and  \textit{completely generic} $G$-symmetric matrix elements will split the EBR into nontrivial Chern bands. Our recipe also helps to identify the wallpaper groups for which splittable EBRs may be found, which provides guidance to the concrete materialization of Chern bands. Our EBR-to-Chern recipe is the class-A analog of a recent proposal to obtain nontrivial $\Z_2$ topological insulators in class AII from splitting half-integer-spin EBRs of magnetic wallpaper groups.\cite{Bradlyn2017}\\

To recapitulate, an EBR is a BR that is not splittable into a direct sum of BR's. Most EBRs are BRs  that each satisfy: (i) the Wyckoff position $\bvarpi$ is nongeneric, with an on-site symmetry group $\calp_{\varpi}$ that is maximal, i.e., $\calp_{\varpi}$ is not a subgroup of any other on-site symmetry group $\calp_{\varpi'}$, (ii) the on-site symmetry representation $V$ of $\calp_{\varpi}$ is irreducible.  (i-ii), combined with: (iii) the BR is not an exception listed in Ref.\ \onlinecite{Bacry1988,Cano2017}, might be viewed as an equivalent definition an EBR that is operationally more useful for identification.  \\

When an EBR is split into $m$ subspaces ($S_1$, $S_2$, ..., $S_m$), it must be that at least one of them, say $S_1$, is not a representation of $G$ on Wannier functions; this follows immediately from the first definition of an EBR. The absence of a Wannier representation of $G$ is a topological obstruction;\cite{Soluyanov2011,Winkler2015a,Bradlyn2017} if $S_1$ is a single band, then the topological obstruction must correspond to a nontrivial Chern number.  The orthogonal subspace $\oplus_{i=2}^m S_i$ must then have a nonzero Chern number of opposite sign, so that the net Chern number of the EBR vanishes.  \\

\noindent \textit{\textbf{Example of splitting an EBR into Chern bands.}}
We consider the reflection-asymmetric checkerboard lattice discussed in \s{buildblock} and split the two-band EBR which comprises two $s$-orbitals in each unit cell. There are in principle four possible splittings into single bands ($S_1$, $S_2$) from different combinations of $C_{4,z}$-eigenvalues at the $C_{4,z}$-invariant wavevector $M$ with $C_{2,z}$-eigenvalues at the $C_{2,z}$-invariant wavevector $X$, which are tabulated in \tab{evalEx}(a); in all splittings, neither $S_1$ nor $S_2$ can be time-reversal-invariant, because the Bloch functions at $M$ transform in a 2D complex-conjugate representation (with eigenvalues ${\pm}i$ under $C_{4,z}$) that is irreducible in the presence of $T$ symmetry. That time-reversal symmetry must be broken suggests that $S_1$  and $S_2$ have nonzero and canceling Chern numbers; we confirm this prediction by a Zak phase analysis of $S_1$, combined with the general relation between Zak phases and Chern numbers in \q{c4constraint}. 
From the $C_{4,z}$ eigenvalues at $\Gamma$ and $M$ (cf.\ \tab{evalEx}(a)), we deduce, via the theorem, that $\phi[\calc_4]{=}{\pm}\pi/2$ attains a classically-forbidden value -- \q{c4constraint} then informs us that $C$ must be odd, and hence nonzero.\\

More examples of such splittings are provided in \app{app:BReval}, including: (i) the splitting of a two-band EBR on a honeycomb lattice (with wallpaper group $p6$ and Wyckoff position $2b$, illustrated in \fig{fig:elem}) into two Chern bands, and (ii) the  splitting of a three-band EBR on a Kagome lattice (wallpaper group $p6$ and Wyckoff position $3c$, also illustrated in \fig{fig:elem}) into three one-band subspaces, with at least two of them carrying nonzero Chern numbers. (ii) is elaborated upon in \s{sec:proofofprinciple} using a tight-binding model.\\

Generally, if an $N$-band EBR is split into $N$ single-band subspaces, then at least two of $N$ subspaces must carry nontrivial Chern numbers. This may be proven -- on a case-by-case basis -- by a Zak phase analysis (as we have done above for the checkerboard EBR). In fact the statement is generally true, as we prove in \ocite{Alexandradinata2018}. \\


Not all nonmagnetic wallpaper groups allow for the splitting of $N$-band EBRs into $N$ single-band subspaces. This splitting is forbidden in wallpaper groups ($G$) which constrain the Chern number of any gapped subspace to vanish.  All such $G$ may be identified from the known symmetry transformation of the $U(1)$ Berry curvature under a wallpaper element $g$.\cite{Fang2012} For example, a reflection symmetry $M_x$ that inverts $x$ constrains the curvature as Tr$\calf^{xy}({-}k_x,k_y){=}{-}$Tr$\calf^{xy}(k_x,k_y)$ {(cf. $\calf$ in \tab{symbols})}, which guarantees that the Chern number vanishes. Consequently, the EBRs of all wallpaper groups with reflection symmetry (e.g. $cmm$, $p4m$, $p31m$, $p6m$) are unsplittable into single bands. This argument may be supported by analysis of the little groups at high-symmetry wavevectors, where the reflection symmetry enhances the energy degeneracy. In simple words, robust band touchings that originate (in part) from reflection symmetry prevent an EBR from splitting.\\

\noindent \textit{\textbf{Example of an unsplittable EBR.}}
A case in point is the reflection-\textit{symmetric} checkerboard lattice with $s$-orbitals -- this is an EBR of the wallpaper group $p4m$, which differs from a previously-discussed EBR on the reflection-\textit{asymmetric} checkerboard lattice (group $p4$); cf.\ \fig{fig:elem}. The little group of the $C_{4,z}$-invariant wavevector $M$ is now the point group $C_{4v}$ (in contrast to $C_4$ previously), and the complex-conjugate representation $\rho_4(M)$ {(cf. $\rho_4$ in \tab{symbols})} with eigenvalues ${\pm}i$ is irreducible owing to the additional reflection symmetry. 


\subsubsection{Example of Chern band with classically-forbidden multiplier}\la{sec:proofofprinciple}

We construct a tight-binding model on a Kagome lattice with wallpaper group $p6$. Each of the $M{=}3$ Wannier functions in each unit cell transform as $s$-orbitals, and they are centered on sites which are invariant under $g_2{=}C_{2,z}$, i.e., with Wyckoff position $3c$ (cf.\ \fig{fig:elem}). The corresponding three-band space is a strong EBR (cf. Definition \ref{strongd}); we shall demonstrate that a two-band subspace of it has a classically-forbidden multiplier $\mu_{\perp}{=}2$.\\


For a complex nearest-neighbor hopping, all three bands are generically split in energy. The corresponding symmetry eigenvalues ($\rho_{3}, \rho_{2}$; cf. \q{sewnew})  at high-symmetry wavevectors, the single-band Zak phases ($\phi[\calc_3], \phi[\calc_2]$),  as well as the Chern number ($C$) of each band are listed here:
\begin{align}
\begin{array}{|c||c|c||c|c|c||c|}
\hline
  \rho_3(\Gamma) & \rho_3(K) & \phi[\calc_3] & \rho_2(\Gamma) & \rho_2(M) & \phi[\calc_2] & C \\
  \hline \hline
  \mathrm{e}^{2\pi i /3}  & \mathrm{e}^{-2\pi i /3} & 2\pi/3 & 1 & {-}1 & \pi & 1\\
  \hline
  1 &  1 & 0 & 1 & 1 & 0 & 0\\
  \hline
  \mathrm{e}^{ -2\pi i /3} & \mathrm{e}^{2\pi i /3} & {-}2\pi/3 & 1 & {-}1 & \pi & {-}1 \\
  \hline
\end{array}.\notag
\end{align}
One may verify that this table is consistent with the theorem in \s{theoremstate}, as well as the Zak-Chern relations of \qq{c2constraint}{c6constraint}.\\

Let us consider adiabatic transport along $\calc_2$ in the top two bands with net Chern number $1$. Applying the relation between Zak phases and Bloch oscillations (cf.\ remark (i) of \s{theoremstate}), the Zak phase difference of $\pi$ leads to a classically-forbidden multiplier of $\mu_{\perp}{=}2$, assuming that both energy bands are fine-tuned to degeneracy at each $\bk$.\\

Generally for integer-spin representations of space groups, we expect that no spatial symmetry enforces the energy-degeneracy of multiple bands  for all $\bk$ along a line. Finite-energy splitting leads to a deviation of the Fourier peak (of the stroboscopic time evolution) away  from $2\pi/(\mu_{\perp} T_B)$, as discussed further in \s{sec:finitewidth}. In \s{classAII}, we will demonstrate that period multiplication can manifest in time-reversal-symmetric bands which are topologically nontrivial; this multiplication does not require any fine-tuning, because the energy degeneracy at each $\bk$ can be guaranteed by space-time inversion symmetry.


\section{Period multiplication in topological band subspaces with time-reversal symmetry}\la{classAII}


The Kane-Mele model on a honeycomb lattice is the paradigmatic example of a $\Z_2$ topological band subspace with time-reversal symmetry (Wigner-Dyson symmetry class AII).\cite{Kane2005b} This model is potentially realizable in ultracold atoms in optical lattices, where microwave driving and lattice shaking can artificially induce spin-orbit coupling.\cite{Grusdt2017,Aidelsburger2013} Here, we will demonstrate that this model realizes stroboscopic Bloch oscillations with multiplier  $\mu_{\perp}{=}3$. This period multiplication may be understood from the perspective of symmetry-protected Zak phases (\s{sec:kanemelezak}), but \textit{not} from the perspective of symmetry-protected Wannier functions, as elaborated in \s{sec:kanemelewannier}.

\subsection{Zak phase analysis of Kane-Mele model}\la{sec:kanemelezak}


The Kane-Mele model is obtainable from splitting a half-integer-spin EBR of the magnetic wallpaper group which combines $p6m$ with time-reversal symmetry $T$.\cite{Bradlyn2017} This EBR is characterized by the Wyckoff position $\bvarpi{=}2b$ (in \tab{wallpaper}), with Kramers-degenerate $p_z$-orbitals on each Wannier center. Following \ocite{Bradlyn2017}, we split this four-band EBR into two two-band subspaces ($S_1$ and $S_2$), and for $S_1$ we collect in \tab{evalEx}(f)  the $C_{2,z}$ and $C_{3,z}$ eigenvalues at high-symmetry wavevectors. \\

We would now demonstrate these symmetry eigenvalues  imply a nontrivial $\Z_2$ Kane-Mele invariant. This demonstration is simplified by imposing an additional 3D-spatial-inversion ($\cali$) symmetry which lies outside $p6m$ (a 2D space group); even if $\cali$ symmetry is not a symmetry of the Hamiltonian, $S_1$ may be adiabatically deformed to have this additional symmetry. We may then split $S_1$ into two single-band subspaces ($S_1^{\pm}$) which transform in orthogonal representations of the reflection $\cali C_{2,z}$.  The $C_{n,z}$  eigenvalues and Zak phases of $S_1^{\pm}$ are listed in \tab{evalEx}(f); the Zak phases may be obtained from the symmetry eigenvalues by application of our theorem in \s{th1}. Applying the Zak-Chern relation (\q{c6constraint}), we then determine the Chern number ($C^{\pm}$) in each single-band subspace (also known as a mirror Chern number\cite{Teo2008})  as $C^{\pm}{=}{\pm}1$ mod $6$. The proof is completed by utilizing the well-known equivalence between an odd mirror Chern number and a nontrivial $\Z_2$-invariant.\cite{Teo2008}\\

Note in particular that the Zak phases over $\calc_3$ are $g_3$-fixed to $\phi_{\mp}[\calc_3]{=}{\pm} 2\pi/3$ {(cf. $\phi, \calc_n$ in \tab{symbols})}. This implies that stroboscopic Bloch oscillations (along $\calc_3$) occur with multiplier $\mu_{\perp}{=}3$, assuming that both bands in $S_1$ are energy-degenerate at all $\bk \in \calc_3$. This degeneracy condition is guaranteed  -- without fine-tuning -- if the band subspace is additionally symmetric under spatial inversion $\cali$.\\



From a broader perspective of the Kane-Mele model, we might ask what $C_{n,z}$-fixed Zak phases are possible in gapped two-band subspaces which are both $T$- and $\cali$-invariant, as well as transform in a half-integer-spin representation. To have $C_{n,z}$-fixed Zak phases, our theorem limits the possible representations of $C_{n,z}$ at $C_{n,z}$-invariant wavevectors ($\bK$). At the same time, $T \cali$ commutes with $C_{n,z}$ and is also in the little group of $\bK$ -- this implies that eigenvalues of $e^{iF\pi/n} \rho_n(\bK)$ form complex-conjugate pairs at each $\bK$ (cf. $\rho_n, F$ in \tab{symbols}). If none of the eigenvalues is real, corollary (I) guarantees that there are no $C_{n,z}$-fixed Zak phases. Consequently, a necessary condition for $C_{n,z}$-fixed phases is the existence of real eigenvalues of $e^{iF\pi/n} \rho_n(\bK)$ -- this can only occur for $n{=}3$, which may be deduced from the observation that all eigenvalues of $e^{iF\pi/n} \rho_n(\bK)$ are $n$'th roots of ${-}1$ for half-integer-spin representations ($F{=}1$). Particularizing now to two-band, half-integer-spin subspaces with $C_{6,z}$, $T$ and $\cali$ symmetry, we find that $\xi_{\perp}{=}3$  is equivalent to a nonzero mirror Chern number, as we prove in \app{app:graphene}. A particular case of this is the $\cali$-symmetric Kane-Mele model.

\subsection{Wannier-function analysis of Kane-Mele model}\la{sec:kanemelewannier}

In the real space perspective, we begin with an EBR that is composed of Kramers-degenerate $p_z$-orbitals centered on a honeycomb lattice. The splitting of this EBR into $S_1$ and $S_2$ amounts to separating the Kramers pair on each honeycomb vertex.\cite{Soluyanov2011,Winkler2015a} That is, one may choose the Wannier functions of $S_1$ to lie on $C_{3,z}$-invariant  $\bvarpi$, but each Wannier function -- having no on-site Kramers partner -- cannot individually form a representation of time-reversal symmetry.  This obstruction is intrinsic to to the $\Z_2$-topological phase.\cite{Soluyanov2011,Fu2006b} \\

Despite being centered on a $C_{3,z}$-invariant Wyckoff position, we point out that the Wannier function also does \emph{not} form a representation of $C_{3,z}$.\footnote{This will be proven in a future work; for now, we appeal to the wealth of numerical evidence.} This is supported by numerical constructions of Wannier functions in various works,\cite{Soluyanov2011,Winkler2015a,Bradlyn2017} which agree that the  spin expectation value of the Wannier function cannot be parallel to $\be_z$. These works demonstrate that a Wannier representation of a (magnetic) space group $G$ need not form a band representation of $G$.\\

Let us discuss the implications for a hypothetical Zak-Wannier relation in the Kane-Mele phase. Any multi-band Zak-Wannier relation requires a  prescription to reduce the gauge ambiguity of Wannier functions; our symmetry-based prescription of \s{composite} evidently does not work, because each Wannier function does not locally form a representation of $g_3{=}C_{3,z}$. Alternatively stated, the Kane-Mele phase exhibits symmetry-protected Zak phases and period multiplier, but not a symmetry-based Zak-Wannier relation (cf.\ Definition \ref{define:symmbasedzw}).


\section{Experimental feasibility of period-multiplied Bloch oscillations}\la{exp}

Let us discuss physical parameters which determine if period multiplication is observable -- given a $N$-band subspace, a field $\bF$ and the shortest reciprocal lattice vector $\bG$ in the direction of $\bF$. For simplicity, we shall assume that this subspace is lowest in energy;\footnote{In principle we may consider a scenario where the $N$-band subspace is not lowest in energy. For Bloch oscillations to be observable, we would have to bound the probability that a state initialized in  the $N$-band subspace  is found (at a later time) outside the $N$-band subspace. This would require a simple generalization of the analysis provided in \s{adiabatic}\cite{Nenciu1987}, see also Supplementary material\textsuperscript{43} } every other band is said to belong to the high-energy subspace, and  $E_G$ is defined as the energy gap that separates low- and high-energy subspaces. Precisely, $E_G$ is the difference between the lowest energy of the high-energy bands and the highest energy of the low-energy bands, as illustrated in \fig{fig:engap}. Period multiplication can only occur for a low-energy subspace that comprises multiple bands; for simplicity, we assume there are two bands with corresponding energy functions $E_1(\bk)$ and $E_2(\bk)$, and $E_g$ is defined as the gap that separates these two bands (cf.\ \fig{fig:engap}(a)); $E_g {=} 0$ if both bands are connected as a graph, e.g., due to a nonsymmorphic symmetry (cf.\ \fig{fig:bands}(b), \s{sec:parallelfield}). The energy width $\Delta_{E}$ is defined as the maximum over $\bk$ of $E_2(\bk){-}E_1(\bk)$. \\

Beside these three  energy scales, the other relevant parameters are the magnitude of the field ($F$), the magnitude of the reciprocal lattice vector ($G$), the mass ($m$) of the particle, a relaxation time ($\tau$) induced by many-body or impurity-induced scattering, as well as the maximum over position $\br$ of the energy width of the translation-invariant potential $\Delta V {=} \sup_{\br} |V(\br)| {-} \inf_{\br} |V(\br)| {<} \infty$. \\

In the following subsections, we discuss: A.  a constraint on the parameters ($F,G, m, \Delta V, E_G$)  which  justify the adiabatic approximation for the low-energy subspace, B. a condition on  $(F,G, m, \Delta V,E_g)$ which encourages all bands in the low-energy subspace to participate in transport, and C. a constraint on ($F,G, \tau$) which guarantees that Bloch oscillations are not smeared out by many-body or impurity-induced scattering. A subtlety about transport along bent loops is discussed in \s{sec:sudden}.  In \s{sec:finitewidth}, we discuss and bound deviations from integer period multiplication in $\perp$-fields; these deviations originate from the dynamical component of the adiabatic propagator. We remind the reader that, in contrast, there are no analogous deviations for $\parallel$-fields. Cold atoms in optical lattices allow for parameter ranges that are optimal to observe period multiplication; in \s{cold} we summarize the experimental setup and techniques used in a recent cold-atom experiment\cite{Li2016} that is directly relevant to our theory.

\begin{figure}
\centering
\includegraphics[width=8cm]{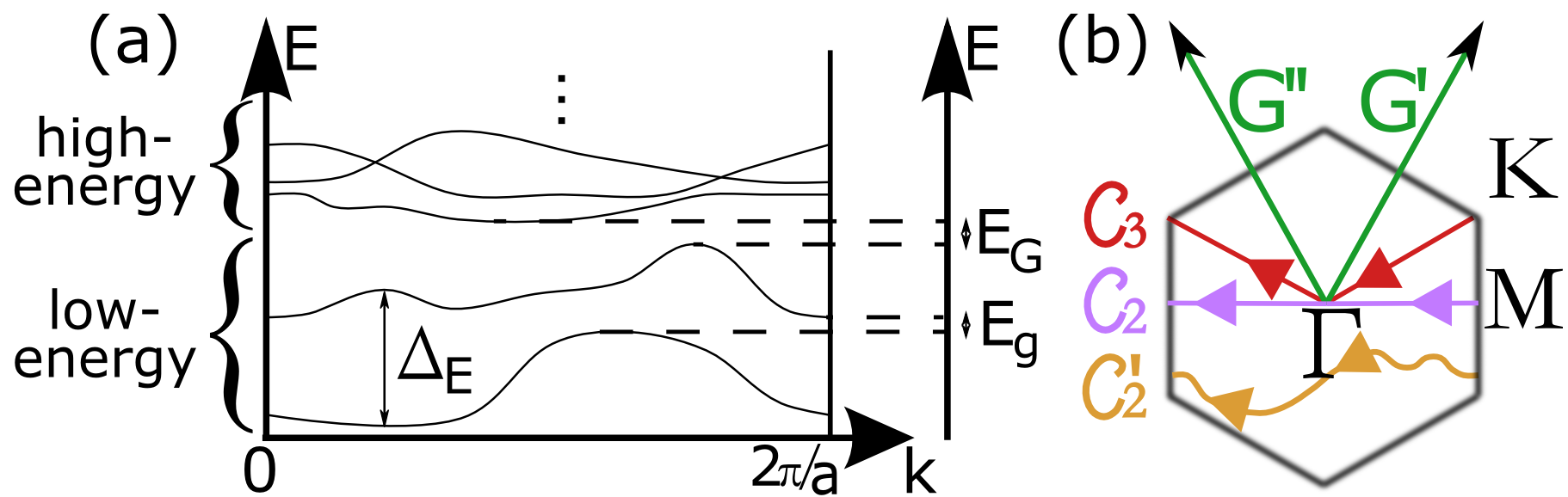}
\caption{(a) Illustration of the energy gap ($E_G$), energy separation ($E_g$) and energy width ($\Delta_{E}$). (b) Possible loops in the Brillouin zone of the honeycomb lattice. \la{fig:engap} }
\end{figure}

\subsection{Bounding leakage by the adiabatic theorem}\la{adiabatic}

To observe Bloch oscillations, it is necessary that an initial state in the low-energy subspace remains (to good approximation) in said subspace on time scales much larger than the oscillatory period ($\mu T_B$; with $\mu$ and $T_B$ defined in \tab{symbols}). Defining the leakage $L(t)$ as  the probability that an initial state is found in the higher-energy subspace at a later time $t$, we would like $L(\mu T_B){\ll}1$. \\

For a tightly-bound particle ($E_G {\ll} \Delta V$), we apply the adiabatic theorems developed by A. and G. Nenciu,\cite{Nenciu1980,Nenciu1987,Nenciu1991} to bound  the leakage as
\e{ L(\mu T_B) \le \f{2\pi F}{G} \f{E_V}{E_G^2} \big( \alpha + \mu \beta \f{E_V}{E_G} \big), \label{leakTBpar} }
to lowest order in $F$, with $\beta {=} (1{+}8/\pi)16/\pi {\approx} 9$, and $\alpha{=}0$ (resp.\ ${=}4/\pi$)  for the $\parallel$-field (resp.\ $\perp$-field).\footnote{The difference between $\perp$- and $\parallel$-fields originates from the distinct symmetry transformation of the operator $\bF \cdot P\br (1-P)$, which is relevant in the adiabatic theorem for field-deformed bands.\cite{Nenciu1991} This distinction is further elaborated in the Supplementary material\textsuperscript{43} } In the simplest case of a one-dimensional crystal, $2\pi F/G{=}F a$ is the potential difference (due to a field $F$) between Wannier functions displaced by the Bravais lattice period $a$ (cf.\ \s{perp}).\footnote{In general, $2\pi F/G{=}\bF \cdot \Delta \bvarpi$ where $\Delta \bvarpi$ is the minimal distance between hybrid Wannier functions\cite{Maryam2014} that are localized along $\bF$}
The leakage depends not only on the obvious energy scales: $2\pi F/G$ and the gap $E_G$; it depends also on a third scale $E_V {=} \big(\hbar G/2 \big) \sqrt{ 2 \Delta V/m }$  formed by $(G,m,\Delta V)$.\footnote{We offer the following interpretation of $E_V=\hbar \omega_V$. Suppose a one-dimensional periodic potential has the form $(\Delta V/2)\cos(G x)$; this potential may be expanded near the minimum ($x{=}0$) as $m\omega_V^2 x^2/2$ (to quadratic order and dropping the constant) where $\omega_V{=}G (\Delta V/2m)^{1/2}$ is the characteristic frequency of the corresponding harmonic oscillator } Combining $L {\ll} 1$ with the upper bound in \q{leakTBpar}, we derive a condition on the force:
\e{ F \ll \f{2\pi}{G} \f{E_G^2}{E_V} \big( \alpha + \mu \beta \f{E_V}{E_G} \big)^{-1}.\la{forcerest}}
For the general form of $L(t)$ which applies beyond the tight-binding regime ($E_G {\ll} \Delta V$), we refer the interested reader to the Supplemental Material.\footnotemark[43]

\subsection{Condition for multi-band transport in the low-energy subspace}


For multi-band transport within the low-energy subspace, we would transport to be non-adiabatic with respect to any of the two bands in this subspace. Equivalently, we would like the leakage from any of the two bands to be large. A conservative estimate for favorable parameters is provided by inverting the inequality of  \q{forcerest} (with $E_G$ replaced by $E_g$):
\e{ F \gg \f{2\pi}{G} \f{E_g^2}{E_V} \big( \alpha + \mu \beta \f{E_V}{E_g} \big)^{-1} }
with constants as discussed after \q{leakTBpar}.

\subsection{Relaxation time}

Thus far, we have not yet discussed the presence of impurities, lattice excitations (phonons) or electron-electron (or particle-particle) interactions. Indeed, scattering is the main limiting process of coherent electronic phenomena like Bloch oscillations; period-multiplied Bloch oscillations can only be observed if 
\e{ F \gg \mu \f{\hbar G}{\tau}, }
with $\mu$ the period multiplier. For solids, realistic parameters are $\f{2\pi}{G}{\approx} 0.1{-}0.5$ nm and $\tau {\approx} 10^{-16}{-}10^{-13}$ s,\cite{Ashcroft} thus the necessary forces to observe at least one Bloch period are large, i.e., $F {\approx} 1{-}4000$ MeV/cm. Lattices with large lattice constants include semiconductor superlattices ($\f{2\pi}{G} {\approx} 10$ nm)\cite{Mendez1993} and cold atoms in optical lattices ($\f{2\pi}{G} {\approx} 1$ $\mu$m);\cite{Li2016} the latter also have longer relaxation times. 

\subsection{Bounding leakage by the sudden approximation for bent loops} \la{sec:sudden}


Let us consider transport along the bent loops $\mathcal{C}_3$ and $\mathcal{C}_4$ (with $\mu_{\perp}{=}3$ and ${=}2$, respectively; cf. $\calc_n$ in \fig{fig:loops}, \tab{symbols}). The kink in either loop corresponds to an instantaneous switch in the direction of $\bF(t)$ (the $\perp$-field) after every half period ($T_B/2$). At this kink, the Hamiltonian is continuous but not first-order differentiable with respect to time.\footnote{This holds in the time-dependent gauge for the vector potential, which is proportional to ${\int}^t_0\bF(t')dt'$.} Since the adiabatic theorem discussed in \s{adiabatic} applies only to first-order differentiable Hamiltonians, the theorem can be used to bound the leakage everywhere on the loop except at the kinks. The leakage at the kink is instead bound by the sudden approximation, i.e., the leakage vanishes in the limit $\bF$ is modified instantaneously. More realistically, if $\delta$ is the time needed to switch the direction of $\bF(t)$, then the leakage at the kink is of order $O(\delta^2)$.\footnote{Sudden changes of a time-dependent Hamiltonian are of two types. In the conventional sudden approximation, an instantaneous change of the Hamiltonian results in it being discontinuous in time. In the context of the kinked loop, an instantaneous change in the Hamiltonian results in it being continuous in time but not first-order differentiable. Further details on this type of `sudden approximations' will be provided in a separate publication.} 

\subsection{Transport signatures for quasi-energy-degenerate bands} \la{sec:finitewidth}

Period multiplication in a $\perp$-field only occurs for band subspaces with exactly degenerate energies (cf.\ remark (i) of \s{theoremstate}). Such degeneracy is either symmetry-imposed (cf.\ \s{classAII}), or originates from an atomic limit (cf.\ \s{perp}), or otherwise requires fine-tuning (cf.\ \s{motivCBR} and \s{sec:proofofprinciple}). In the latter two categories, we are motivated to study transport signatures of band subspaces whose energies are quasi-degenerate at each $\bk$.\\

For brevity of presentation, we would only discuss the case of strong, atomic EBRs with period multiplier $\mu_{\perp}{=}\chi_{\perp}{>}1$ (cf.\ \s{perp}); similar arguments can be applied to other quasi-degenerate subspaces. For finite Wannier center separations, bands generically split in energy (on a scale $\Delta_{E}$) due to tunneling between Wannier centers. A Fourier peak in the stroboscopic expectation value of an observable then deviates from the frequency $2\pi/(\mu_{\perp} T_B)$ by a quantity of order $\Delta_{E}/\hbar$, as derived in \app{app:degen}. Moreover, the ratio of the deviation of the peak over $2\pi/(\mu_{\perp} T_B)$ is of order $\Delta_{E} T_B/(2\pi \hbar){=} \Delta_{E} G/(2\pi F)$. We see that dynamics is well-approximated by an atomic BR if the energy splitting $\Delta_{E}$ (induced by tunneling) is much smaller than the spacing ($2\pi F/G$) between adjacent rungs of the Wannier-Stark ladder.

\subsection{Experimental realization with cold atoms in optical lattices}\la{cold}

A three-fold-periodic Bloch oscillation of a band population was recently measured by T. Li et al.\cite{Li2016} for cold bosonic atoms ($^{87}$Rb) in an optical honeycomb lattice. In this section, we summarize their experimental setup  to stimulate further experimental investigations.\\

Li performed near-adiabatic transport within a two-band subspace that may be identified with the strong EBR discussed in \s{honey}. The lattice (with period $\f{2\pi}{G} {\approx} 500$\ nm) was created by interfering three laser beams of wavelength $\lambda_l{=}755$ nm.\footnote{Private communication with K. Wintersperger} Let us introduce the convenient energy scale $E_r{=}\hbar^2/2m\lambda_l^2{\approx}16.6$ peV, which may be interpreted as the zero-point energy of a $^{87}$Rb atom confined to one wavelength of the laser. The following parameters are obtained from \ocite{Li2016}: the lowest two energy bands are nearly degenerate ($\Delta_{E} {\approx} 0.75 E_r$) and separated from higher-energy bands by a gap $E_G{\approx} 3.7 E_r$; the potential width is $\Delta V {\approx} 62 E_r$. The parameters suggest that the low-energy subspace is well-approximated by an atomic EBR. As explained in \s{honey}, such an atomic EBR manifests Bloch oscillations with multiplier $\mu_{\perp}{=}3$. \\

Trapped bosonic atoms were first condensed into the lowest-energy, single-particle Bloch function at zero wavevector ($\Gamma$).\cite{Zenesini2010,Kling2010} This macroscopically-occupied Bloch state was then driven along the straight path  $\calc_2$ (illustrated in \fig{fig:loops}(d)) by acceleration of the optical lattice. The acceleration is accomplished by linearly sweeping the frequency of the laser beams. Independent control of the sweep rates of two laser beams allows one to vary the magnitude and direction of $\bF(t)$;\cite{Li2016} this allows one to drive a Bloch state along a kinked loop, as exemplified by $\calc_3$ in \fig{fig:engap}(b).\footnote{Private communication with T. Li and I. Bloch.} For a time-independent force $F {\approx} 2$ $\mu$eV/cm, $\mu_{\perp} T_B {\approx} 200$ $\mu$s is just smaller than the exponential decay time ${\approx} 400$ $\mu$s of the measured oscillations.\footnote{The decay time can be obtained from a fit of figure S6A of the Supplementary material for \ocite{Li2016}; the value $400$ $\mu$s was confirmed through private communication with K. Wintersperger }  \\

An example of an oscillatory observable is the band operator, defined by $O_{\bk(t)} {=} \sum_{j{=}1}^N \ket{u_{j,\bk}} j\bra{u_{j,\bk}}$ for energy eigenstates $u_{j,\bk}$; $\bk(t)$ is determined by the acceleration theorem, and $N{=}2$ in the current context. The time-dependent expectation value of the band operator was obtained in Li's experiment by repeated, time-delayed measurements of band populations in the low-energy subspace. Though it was not the motivation of Li's experiment to measure period-multiplied Bloch oscillations, $\mu_{\perp}{=}3$ can be inferred from a recurrence of the lowest-energy band population after three fundamental periods; cf.\ Figure 3B in \ocite{Li2016}. Li's main motivation was to measure Zak-phase differences by generalized Ramsey interferometry,\cite{Zenesini2010,Kling2010}and their measured $\Delta \phi{=}2.06(3)\pi/3$ compares favorably with the theoretical value of $\Delta \phi{=}2\pi/3$ (as calculated in \s{sec:zakstrongatomic}).\\


We now present two physical implications of our theory, that go beyond what has been measured in \ocite{Li2016}. 

\noindent (i) In a band subspace with zero non-Abelian Berry curvature, the Zak phase depends only on how the loop wraps the Brillouin torus, and is insensitive to local-in-$\bk$ details of the loop, as explained in \s{sec:zakstrongatomic}. This may be compared with a well-known phenomenon in electromagnetism: the Aharonov-Bohm phase is insensitive to continuous deformations of the trajectory in regions with zero magnetic field; the field is the \textit{Abelian} $U(1)$ curvature for the electromagnetic vector potential in \textit{real space}. The Aharonov-Bohm phase depends only on the number of times the trajectory winds around magnetic flux tubes. In contrast, the experimental setup in \ocite{Li2016} describes a band subspace with zero \textit{non-Abelian} $U(2)$ Berry curvature  in \textit{$\bk$-space}. One implication is that a continuous deformation of the loop $\calc_2$ (e.g. to $\calc_2'$ or $\calc_3$ in \fig{fig:engap}(b)) would not change $\Delta \phi{=}2\pi/3$. On the other hand, $\Delta \phi{=}0$ for the loop parallel to $\bG'{+}\bG''$ (illustrated in \fig{fig:engap}(b)), which is homotopically inequivalent to $\calc_2$. 

\noindent (ii) The non-Abelian curvature acquires finite value away from the atomic limit. This may be accomplished by reducing the potential barrier that separates honeycomb vertices, so as to  allow for inter-vertex tunneling. One implication would be an increased sensitivity of $\Delta \phi$ to continuous deformations of the loop, e.g. $\Delta \phi$ for $\calc_2$ and $\calc_2'$ generically deviates from $2\pi/3$; however, it is remarkable that $\Delta \phi{=}2\pi/3$ remains quantized for the $\calc_3$ loop, for which half the loop is mapped to the other half by a three-fold rotation. Indeed, the Zak phase difference associated to $\calc_3$ may be viewed as a topological invariant protected by three-fold rotational symmetry, as explained in the example of \s{theoremstate}. A further  implication is that $\Delta \phi{=}2\pi/3$ remains quantized under a perturbation (e.g. a sublattice-dependent potential) that breaks six-fold but preserves three-fold rotational symmetry.

\section{Discussion and outlook}\la{sec:discussin}

The topological classification of band systems continues to be an active field that is enriched by K-theory\cite{Kitaev2009a,Teo2010a,Morimoto2013,Freed2013,Thiang2015,Shiozaki2014,Shiozaki2016,Shiozaki2018,Read2017,bandcombinatorics_kruthoff}, the theory of vector bundles,\cite{DeNittis2014c,DeNittis2014b} and the theory of band representations.\cite{Bradlyn2017,Cano2017,Cano2017a,Alexandradinata2018,Bouhon2018} Some classification schemes can provide model (Dirac) Hamiltonians\cite{Ryu2009,Stone2011,Chiu2013} and topological invariants which are calculable in tight-binding models.\cite{Schnyder2009,Soluyanov2011a,Shiozaki2014,Shiozaki2015,Shiozaki2016,Shiozaki2017,Alexandradinata2014c,Alexandradinata2016,Yu2011c,Maryam2014} Only some of these topological invariants have been associated to experimental signatures -- the majority of these signatures rely on identifying  the energy-momentum dispersions of edge or bulk states,\cite{Fidkowski2011} either through photoemission\cite{Hasan2010} or tunneling spectroscopy.\cite{Drozdov2014} Far fewer invariants have been connected with transport experiments,\cite{Thouless1982,Konig2007,Vayrynen2013,Alexandradinata2018b} and this work represents an attempt to bridge this gap. \\

Let us summarize some of our results from a different perspective than has been presented. We considered the adiabatic transport of Bloch waves, and focused on oscillatory observables which are translation-invariant, as exemplified by the band operator. For all multi-band subspaces that exhibit Bloch oscillations with period multiplier $\mu{>}1$, a unifying property is that their multi-band Zak phases ($\phi$) differ pairwise by integer multiples of $2\pi/\mu$, owing to the point group of the crystal. Generally $\mu$ divides $n$, which is the order of a point group element $g$. The robustness of the Zak phases only occurs for certain orientations of the driving field with respect to a crystallographic axis characteristic of $g$. For example, if $g$ is a screw rotation, two distinct types of Bloch oscillations may arise when we align the field parallel ($\parallel$) or perpendicular ($\perp$) to the screw axis. \\


It is not a priori obvious that a Zak phase difference of $2\pi/\mu$ leads to Bloch oscillations with multiplier $\mu$. More directly, Bloch oscillations originate from pairwise phase differences ($\varphi_j{-}\varphi_{j'}$) in the spectrum of the adiabatic propagator (cf.\ \q{adevo}), where $\varphi_j$ includes both geometric and dynamical contributions. If a $\parallel$-field is applied to a nonsymmorphic crystal, we have found that $\varphi$ \emph{always} has the same ladder structure as $\phi$, hence period multiplication is guaranteed. In the case of the $\perp$-field, pairwise phase differences in $\phi$ only equal that of $\varphi$ if bands are  energy-degenerate at all points on the $\bk$-space loop. This degeneracy is naturally realized by symmetry (as exemplified by the Kane-Mele model with period multiplier $\mu{=}3$), or by taking the atomic limit of a band representation (as exemplified by $\mu{>}1$ strong elementary band representations). \\

While period-multiplied Bloch oscillations can always be understood from the perspective of Zak phases, a complementary, real-space explanation through Wannier functions is not always attainable -- owing to a topological obstruction. A case in point are bands with nontrivial Chern number; these bands are not representable by Wannier functions,\cite{Panati2007,Brouder2007} but may nevertheless be associated to a classically-forbidden multiplier $\mu{>}1$ (cf.\ \s{classA}). We have also studied the Kane-Mele model for $\Z_2$ topological order; though locally three-fold symmetric Wannier functions cannot exist,\cite{Soluyanov2011} this model exhibits a three-fold period multiplication (cf.\ \s{classAII}). \\

A real-space perspective exists for  all our other case studies.  For $d$-dimensional crystals in a $\parallel$-field, this perspective is attained by  Wannier functions that are localized in the direction of  the field and extended as a Bloch function in the other $d{-}1$  direction(s). For $d$-dimensional crystals in a $\perp$-field, we consider 
(hybrid) Wannier functions that are localized in the plane perpendicular to the field, and otherwise extended in $d{-}2$ direction(s).     \\

Underlying the complementary perspectives on Bloch oscillations, is a multi-band Zak-Wannier relation for $N$-band subspaces, i.e.,  a one-to-one correspondence between all $N$ Wannier centers and all $N$ Zak phases. (i) For the $\parallel$-field, the Zak-Wannier relation exists for (hybrid) Wannier functions which are maximally localized\cite{Alexandradinata2014c} in the direction of the field. (ii) For the $\perp$-field, the Zak-Wannier relation exists for  (hybrid) Wannier functions satisfying a {newly-formulated} symmetry condition -- namely, that they correspond to $g$-symmetric elementary band representations. Such \emph{symmetry-based Zak-Wannier relations} are introduced by us in this work.\\

More generally, a one-to-one correspondence may hold between a subset of the N Zak phases and a subset of the N Wannier centers; such correspondences would be referred to as incomplete. Band subspaces with an incomplete Zak-Wannier correspondence may still manifest Bloch oscillations with multiplier $\mu{>}1$ -- if at least two $g$-protected Zak phases differ by $2\pi/\mu$; this is left to future investigations. Band subspaces with $g$-protected Zak phases (having either a complete or incomplete Zak-Wannier correspondence)  should be identifiable by application of our theorem (cf.\ \s{theoremstate}). This theorem inputs, for any band subspace, the representation of $g$ at high-symmetry wavevectors, and outputs the subset of Zak phases which are fixed to integer multiples of $2\pi/n$ with $n$ the order of $g$. \\

Though $\mu{\geq}1$ Bloch oscillations do not  occur for band insulators,\cite{Ashcroft} they may in principle occur for band metals, and more realistically for bosonic cold atoms in optical lattices (cf.\ \s{cold}). Underlying this broad range of applications is that $\mu$, when formulated as a quantized difference in Zak phases, is a topological invariant that characterizes band wavefunctions -- independent of the filling or particle statistics. From this perspective, we may compare $\mu$ with other space-group-protected topological invariants that characterize the filled bands of insulators.\cite{Fu2011,Liu2014,Alexandradinata2014g,Alexandradinata2014c,Shiozaki2014,Fang2015,Shiozaki2015,Hourglass2016,Shiozaki2016,Shiozaki2017,kane2005A,Kane2005b,QSHE_bernevig,HgTe_bernevig,QSHE_Rahul,fu2006,fukanemele_3DTI,Inversion_Fu,moore2007,Rahul_3DTI,Classification_Chiu,AZ_mirror,spacegroupclass_slager,bandcombinatorics_kruthoff,Bradlyn2017,Cano2017} In stable\cite{Kitaev2009a,Ryu2009,Stone2011,Chiu2013,Morimoto2013,Freed2013,Shiozaki2014,Shiozaki2016,Shiozaki2017,bandcombinatorics_kruthoff} classifications of topological insulators, the corresponding topological invariants are invariant against symmetric deformations of the Hamiltonian that preserve a \textit{single} gap -- the gap that separates filled and empty bands. Such stable topological invariants do not change upon addition of trivial (filled or empty)  bands.\footnote{By `trivial bands', we mean band representations of the space group\cite{Alexandradinata2018}}  In contrast, $\mu$ \textit{may} change under the addition of trivial bands. That is to say, if we evaluate $\mu$ for two band subspaces (that differ only by an addition of trivial bands), we may arrive at distinct values for $\mu$. In this sense, $\mu$ is more closely analogous to invariants of fragile topological insulators.\footnote{A fragile topological insulator\cite{Po,Alexandradinata2018,Cano2017a,Bouhon2018} is defined by two properties: (a) Wannier functions exist but cannot be locally symmetric. (b) By addition of trivial bands (corresponding to locally-symmetric Wannier functions, or equivalently band representations), a fragile topological insulator can be trivialized. It should be clarified that a band subspace with $\mu{>}1$ is not necessarily a fragile topological insulator. }

\begin{acknowledgments}
We are especially grateful to Gheorghe Nenciu for clarifying his adiabatic theorem, and to Tracy Li and Immanuel Bloch for informative discussions that linked this work to their $^{87}$Rb experiment. We thank Nicholas Read, Chen Fang, Ken Shiozaki and Wang Chong for discussions of holonomies in point group symmetric crystals.
We acknowledge support by NSF grant No. DMR-1408916 (JH) and the Yale Postdoctoral Prize Fellowship (AA).
\end{acknowledgments}

\vspace{2cm}

\newpage

\appendix

\begin{widetext}

\begin{table}
\begin{tabular}{|c|c|}
\hline
 Symbol(s) & Description  \\
  \hline \hline
  $\bk, \bk_0$ & wavevector, with subscript: initial wavevector \\
  \hline
  $\br$ & position operator or position vector  \\
  \hline
  $a, \ba$ & lattice constant, primitive lattice vector\\
  \hline
  $F, \bF$ & magnitude of force, force vector\\
  \hline
  $T_B$ & Bloch period\\
  \hline
  $\mu, \mu_{\parallel}, \mu_{\perp} {\in} \Z_{>0}$ & integer in period multiplication (of $T_B$), subscript specifies direction of $\bF$ relative to some symmetry axis\\
  \hline
  $N {\in} \Z_{>0}$ & number of bands, number of atoms per unit cell\\
  \hline
  $O$ & arbitrary operator (in Bloch representation: assumed to be translation-invariant) \\
  \hline
  $P$ & projection operator onto $N$-dimensional subspace \\
  \hline
  $\psi_{\bk}$ & Bloch function in the subspace projected by $P$ \\
  \hline
  $u_{\bk}$ & cell-periodic component of $\psi_{\bk}$ \\
  \hline
  $\bA(\bk)$ & Berry connection \\
  \hline
  $\calf^{xy}(\bk)$ & Berry curvature in 2D or for a plane in 3D \\
  \hline
  $H_0$ & translation-invariant, free-particle Hamiltonian \\
  \hline
  $\mathcal{H}$ & effective Hamiltonian of $N$-dimensional subspace \\
  \hline
  $U$ & propagator generated by $\mathcal{H}$, expressed in Bloch basis at $\bk_0$ in $P$ \\
  \hline
  $\overline{\mathrm{exp}}$ & time- or path-ordered exponential \\
  \hline
  $J {\in} \Z_{>0}$ & number of Wannier-Stark ladders \\
  \hline
  $G$ & space group \\
  \hline
  $\calp$ & point group \\
  \hline
  $g{=}(\check g|\bt) {\in} G$ & symmetry element, $\check g$ leaves the spatial origin invariant, $\bt$ is a translation \\
  \hline
  $g_{n,p}, g_n {\in} G$ & symmetry of order $n$: $(\check g_{n,p})^n, (\check g_{n})^n {=} e$ (identity), $(g_{n,p})^n$ is a translation by $p$ primitive lattice vectors\\
  \hline
  $n {\in} \Z_{\ge 0}$ & order of $g_{n,p}, g_n$ \\
  \hline
  $p {\in} \Z_{\ge 0}$ & determines fractional translation of $g_{n,p}$ \\
  \hline
  $C_{n,j}, C_n {\in} G$ & $n$-fold rotation about axis with unit vector $\be_j$ or $\be_z$ (if unspecified) \\
  \hline
  $M_j {\in} G$ & reflection that inverts the $j$th coordinate \\
  \hline
  $\bvarpi, \bvarpi^{\nu}$ & Wyckoff position; superscript labels an EBR \\
  \hline
  $\bvarpi_j, \bvarpi^{\nu}_{j_{\nu}}$ & symmetry-related Wyckoff positions (to $\bvarpi, \bvarpi^{\nu}$, respectively) \\ 
  \hline
  $P_{j,\bR}$ & projection operator to Wannier functions localized at $\bvarpi_j{+}\bR$\\
  \hline
  $M_{\varpi}, M_{\nu} {\in} \Z_{>0}$ & multiplicity of a Wyckoff position $\bvarpi, \bvarpi^{\nu}$, respectively: number of symmetry-related Wannier centers per unit cell \\
  \hline
  $\calp_{\varpi}, \calp_{\varpi^{\nu}}$ & on-site symmetry group (stabilizer) of $\calp_{\varpi}, \calp_{\varpi^{\nu}}$, respectively \\
  \hline
  $V_{\varpi}, V^{\nu}$ & on-site symmetry representation of $\calp_{\varpi}, \calp_{\varpi^{\nu}}$, respectively \\
  \hline
  $\bK, \bK_n, \Gamma$ & high-symmetry, or $g_n$-invariant, wavevector; $\Gamma$ denotes the origin of the BZ \\
  \hline
  $\bG, \bG_n$ & reciprocal vector, with subscript: satisfies \q{Gcond} \\
  \hline
  $\chi_{\perp}{\in}\Z_{>0}$ & $2\pi/\chi_{\perp}$ is the smallest difference in Wannier centers times $\bG_n$ \\
  \hline
  $\calc, \calc_n$ & noncontractible loop in the BZ, with subscript: goes through $\Gamma$, maps half of the loop to other half by $g_n$ \\
  \hline
  $\W$ & Wilson loop or matrix representation of holonomy \\
  \hline
  $\phi$ & Zak phase: $\mathrm{e}^{i \phi}$ is an eigenvalue of $\W$ \\
  \hline
  $\xi_{\perp}{\in} \Z_{>0}$ & $2\pi/\xi_{\perp}$ is the smallest difference in Zak phases \\
  \hline
  $\hat g$ & regular representation of $g {\in} G$ on Bloch or Wannier functions \\
  \hline
  $F {\in} \Z_2$ & $F {=} 0$ for spinless or integer-spin representations, $F{=}1$ for half-integer spin representations \\
  \hline
  $\rho_n(\bk)$ & $N$-dimensional matrix representation on Bloch functions at $\bk$ of $\hat g_n$ \\
  \hline
  $m_l(\bk) {\in} \Z_N$ & degeneracy of eigenvalue $\mathrm{e}^{2\pi i l/n}$ in $\rho_n(\bk)$, $l {\in} \Z_n$ \\
  \hline
  $\bk_*$ & wavevector for which $m_l(\bK)$ is maximal among $\bK {\in} \{\Gamma, \bK_n \}$ \\
  \hline
  $l_* {\in} \Z_n$ & eigenvalue $\mathrm{e}^{2\pi i l_*/n}$ for which $m_{l_*}(\bk_*)$ is maximal \\
  \hline    
  $C \in \Z$ & (first) Chern number of $N$-dimensional subspace over the BZ \\
  \hline
\end{tabular}
\caption{Summary and description of symbols used throughout the main text. \la{symbols}}
\end{table}
\end{widetext}

\section{Organization of the appendix} 
We start the appendix with a table to summarize the symbols that were used throughout the main text; for a symbol that appears in the main text but not in this table, its definition ought to be nearby.  In \app{app:perm} we derive the spectrum of the adiabatic propagator within a subspace with nonsymmorphic symmetry, under the application of a $\parallel$-field; we also describe a symmetry criterion for orbital splitting in the Wannier-Stark effect. 
A symmetry criterion to obtain strong band representations is presented in \app{app:zerocurv}. Then, the sufficiency and necessity of conditions (i{-}iii) in \s{Wyckoffcond}, to satisfy \qq{specialG}{genLaue} for at least one $\bG_n$ of the form of \q{Gcond}, is proven in \app{app:exhaust}. We relate the Wilson loop to the projected position operator in \app{app:weakfield}, i.e., we derive \q{WilsonPxP}, and derive the Zak-Wannier relation for strong, atomic EBRs (\q{zakwannierstrongatomic}). 
We then prove the theorem and the corollaries of \s{th1} in \app{app:quant}. Next, we detail on several case studies: In \app{app:ZakCh}, we prove the Zak-Chern relation for the reflection-asymmetric checkerboard lattice, stated in \s{sec:split}; in \app{app:BReval}, we present a general analysis of split subspaces in the hexagonal and Kagome lattices from a more general perspective than presented in \s{sec:proofofprinciple}; in \app{app:graphene}, we prove the claim that $\mu_{\perp}{=}3$ if and only if the mirror Chern number is nonzero for systems such as the Kane-Mele model studied in \s{sec:kanemelezak}.
At last, \app{app:degen} derives corrections to the frequency $2\pi/(\mu_{\perp} T_B)$ occurring in the Bloch oscillation for a $\perp$-field applied to a nearly-degenerate band subspace characterized by $\mu_{\perp}$.

\section{Conventions}
The chapters are self-contained if no explicit reference to other chapters is made.
We use the following conventions: 
\begin{itemize}
\item We sum over repeated indices (Einstein's summing convention); if the sum is only over some of the repeated indices, we explicitly write the sum.
\item For an index $j=0,...,N-1$ we write $j \in \Z_N$.
\item `Translation-invariant' is a shorthand for invariance under discrete translations in real space.
\item The identity operator in a Hilbert space is denoted by $\mathbb{1}$; if the Hilbert space is of finite dimension $N$, we denote the identity operator by $\mathbb{1}_N$.
\item Unitarily conjugate operators or matrices $A, B$ are denoted by $A \cong B$, whereas for groups $G, H$, $G \cong H$ means that there exists a group isomorphism between $G$ and $H$.
\item Complex conjugation of a complex number $z$ is denoted by $\bar{z}$ while the adjoint of an operator $A$ is $A^{\dagger}$.
\end{itemize}

\section{Adiabatic evolution under a $\parallel$-field}\la{app:perm}

In crystals with a nonsymmorphic symmetry $g_{n,p}$ (cf. \tab{symbols} for the definition of $g_{n,p}$), energy bands divide into subspaces with minimal dimension of $\mu_{\parallel}$ (given by \q{munon}) per wavevector, such that within each subspace all energy bands are connected (as a graph)\cite{Michel1999} over the BZ (cf. \fig{fig:helix}(b), \fig{fig:bands}(e,h)). This is because the Bloch wavefunctions which describe energy bands along $g_{n,p}$-invariant lines, can be chosen as eigenstates of $\hat g_{n,p}$ (cf. \tab{symbols}).
We will show that, related to the division of energy bands into subspaces of minimal dimension $\mu_{\parallel}$, is that the adiabatic propagator $U(T_B)$ \big(defined in \q{adevo}\big) describes a permutation of order $\mu_{\parallel}$; from this we will then derive the ladder structure of the spectrum of $U(T_B)$. \\

The ladder structure of $U(T_B)$  is valid for any energy matrix, including the zero energy matrix. This gives us an alternate proof for the ladder structure of the Zak phases ($\phi$). The proof is presented separately for $\mu_{\parallel}{=}n$ and for $\mu_{\parallel}{<}n$. For $\mu_{\parallel}{<}n$, we further distinguish between two cases: $U(T_B)$ either describes a single $\mu_{\parallel}$-cycle or multiple $\mu_{\parallel}$-cycles; for $\mu_{\parallel}{=}n$, only single $\mu_{\parallel}$-cycles exist. A $\mu_{\parallel}$-cycle is a cyclic permutation of order $\mu_{\parallel}$ where no element is mapped to itself when permuted less than $\mu_{\parallel}$ times. $U(T_B)$ describes multiple $\mu_{\parallel}$-cycles if there exist symmetry-eigenvalues which are not mapped onto each other by any multiple of $U(T_B)$. For $u$ labeling different $\mu_{\parallel}$-cycles, the degeneracy $J_u$ of the $u$'th cycle is the dimension of a symmetry-eigenspace in the $u$'th cycle. The total degeneracy $J {=} \sum_u J_u$ equals the number of orbitals in $1/\mu_{\parallel}$ of the primitive unit cell; $J$ also equals the number of Wannier-Stark ladders. \\

For $J$ Wannier-Stark ladders, we may ask if the ladders are all degenerate. To every Wannier-Stark ladder (indexed by $\iota_u{=}1,{\ldots},J_u$) we associate a phase $\gamma^{(\iota_u)}$; then the offset between the $\iota_u$'th and the $\iota'_{u'}$'th Wannier-Stark ladder is
\e{ \Delta \gamma_{\iota_u,\iota'_{u'}} = \gamma^{(\iota_u)}- \gamma^{(\iota'_{u'})}. \la{gammadiff}} 
For a single cycle of degeneracy $J$, we will show that
\e{ \{ \mathrm{e}^{i \gamma^{(\iota)}} \}_{\iota=1}^J = \sigma \big( U(\mu_{\parallel} T_B)_{0,0} \big), \la{gammas}}
where $\sigma$ denotes the spectrum of a matrix and $U(\mu_{\parallel} T_B)_{0,0}$ is the $J{\times}J$ matrix that describes adiabatic transport for $\mu_{\parallel} T_B$ of a symmetry-degenerate eigenspace; the spectrum is independent of the choice of eigenspace. For multiple cycles, the $u$'th cycle is associated with phases $\{ \gamma^{(\iota_u)} \}_{\iota_u=1}^{J_u}$ which are determined by an equation analogous to \q{gammas} but with $U(\mu_{\parallel} T_B)_{0,0}$ restricted to the $u$'th cycle. This restriction is possible due to a block-diagonal form of $U(T_B)$ wrt. the different cycles,as we elaborate in \s{app:lln}. \\

An example of a single 4-cycle is shown in \fig{fig:helix}(a{-}b) for $g_{4,1}$ a four-fold screw ($\mu_{\parallel} {=} n {=} 4$); \fig{fig:bands}(d{-}e) shows a single 2-cycle of symmetry degeneracy $J{=}2$ for $g_{4,2}$ a four-fold screw ($\mu_{\parallel} {=} 2$ while $n{=}4$). 

\subsection{$\mu_{\parallel}{=}n$}\la{app:singlecycle}
If $\mu_{\parallel} {=} n$, then the spectrum of the matrix representation $\rho_{n,p}(\bk)$ of $g_{n,p}$ \big($\rho_{n,p}(\bk)$ is defined as in \q{sewnew}\big) has to comprise all the eigenvalues $\{ \omega_{j,n}{=}\mathrm{e}^{2\pi i j p/n} \}_{j {\in} \Z_{\mu_{\parallel}}}$,\footnotemark[43] where each eigenvalue has equal degeneracy $J$. 
Assuming first that $U(T_B)$ cyclically permutes all $n$ eigenspaces of $\hat g_{n,p}$, or more precisely, $U(T_B)$ describes a $\mu_{\parallel}$-cycle wrt. the eigenspaces of $\hat g_{n,p}$, we now outline how this permutation results in a ladder structure of the spectrum of $U(T_B)$ (details below): For a $\mu_{\parallel}$-cycle, concatenating $\mu_{\parallel}$ permutations gives no permutation at all. Therefore, $U(\mu_{\parallel} T_B)$ is block-diagonal, where each block has dimension $J{\times}J$. Each of these $\mu_{\parallel}$ blocks is unitarily equivalent to every other block, and may be viewed as the unitary representing the adiabatic evolution of an initial state in one ($J$-dimensional) eigenspace of $\hat g_{n,p}$ over a $\mu_{\parallel}$-fold-expanded BZ, as shown \fig{fig:helix}(b). Let us denote the eigenvalues of each block as $\{ \mathrm{e}^{i \gamma^{(\iota)}} \}_{\iota=1}^J$, with $\gamma^{(\iota)} \in [0,2\pi)$. The spectrum of $U(T_B)$ is then obtained from $U(\mu_{\parallel}T_B)$ by, loosely speaking, taking the $\mu_{\parallel}$'th root of $U(\mu_{\parallel} T_B)$. The spectrum of $U(T_B)$ can thus be organized into $J$ sets of order $\mu_{\parallel}$. Each set, labeled by $\iota {=} 1,\ldots,J$, has the ladder structure: $\{ \mathrm{e}^{i \gamma^{(\iota)}/\mu_{\parallel}} \mathrm{e}^{2\pi i l/\mu_{\parallel}} \}_{l=0}^{\mu_{\parallel}{-}1}$ (illustrated in \fig{fig:bands}(c,f,i)), with $\mathrm{e}^{2\pi i l/\mu_{\parallel}}$ originating from the diagonalization of the order-$\mu_{\parallel}$ permutation matrix. \\

\noindent \textit{Proof.} Let us first explain what it means for $U(T_B)$ to describe a $\mu_{\parallel}$-cycle wrt. the eigenspaces of $\hat g_{n,p}$. We denote the eigenstates of $\rho_{n,p}(\bk)$ by $\{ \ket{ u^{\alpha}_{j,\bk} }_{\text{cell}} \}_{j {\in} \Z_{\mu_{\parallel}},\alpha {\in} \Z_J}$ with $j$ labeling the symmetry representation and $\alpha$ the basis states which span a degenerate eigenspace. We will further assume that these states are energy eigenstates. We will show that in the basis $\mathcal{B} {=} \{ \ket{ u^{\alpha}_{j,\bk_0} }_{\text{cell}} \}_{\alpha {\in} \Z_J,j {\in} \Z_{\mu_{\parallel}}}$ chosen at the base point $\bk_0$, $U(T_B)$ takes the form
\e{ U(T_B)^{\beta,\alpha}_{j,j'} =_{\mathcal{B}} \, \delta_{j,j'-1} D_{(j')}^{\beta,\alpha}. \la{adiabaticNonSymm} }
Here, the subscript $\mathcal{B}$ reminds us of the special basis choice, and $\delta_{j,j'-1}$ is a unitary and faithful matrix representation of the $\mu_{\parallel}$-cycle that maps $\{ \ket{ u_{j,\bk_0}^{\alpha} }_{\text{cell}} \}_{\alpha} {\to} \{ \ket{ u_{j-1,\bk_0}^{\alpha} }_{\text{cell}} \}_{\alpha}$ for all $j {\in} \Z_{\mu_{\parallel}}$ with $\ket{ u_{0,\bk_0}^{\alpha} }_{\text{cell}} {\equiv} \ket{ u_{\mu_{\parallel},\bk_0}^{\alpha} }_{\text{cell}}$. Explicitly, the matrix reads as
\[
U(T_B) =_{\mathcal{B}}
\left[
\begin{array}{c|c|c|c|c}
0 & D_{(1)} & 0 & ... & 0  \\
\hline
0 & 0 & D_{(2)} & ... & 0 \\
\hline
0 & 0 & 0 & \ddots & 0 \\
\hline
0 & 0 & 0 & ... & D_{(\mu_{\parallel}-1)} \\
\hline
D_{(0)} & 0 & 0 & ... & 0 
\end{array}
\right],
\]
where $D_{(j')}$ are unitary $J {\times} J$ matrices indexed by $j' {\in} \Z_{\mu_{\parallel}}$ and with matrix elements labeled by $\alpha, \beta {\in} \Z_J$ (no sum over $j'$):
\m{ D^{\beta,\alpha}_{(j')} = \braket{ u^{\beta}_{j'-1,\bk_0}| \mathrm{e}^{i \bG \cdot \br}|u^{\nu}_{j',\bk_0+\bG} }_{\text{cell}} \\
\bigg( \overline{\mathrm{e}}^{i \int_{0}^{T_B} \big[\;\bA (\bk(t')) \cdot \bF(t')- \mathfrak{E}(\bk(t'))\;\big] dt'/\hbar}\bigg)_{j',j'}^{\nu,\alpha}. \la{diagonaladiabatic}}
The non-Abelian Berry connection $\bA$ is defined as in \q{nonAB} wrt.\ the symmetric basis $\{ \ket{ u^{\alpha}_{j,\bk} }_{\text{cell}} \}_{\alpha {\in} \Z_J,j {\in} \Z_{\mu_{\parallel}}}$, $F$ is the driving force, and $\mathfrak{E}$ the energy matrix \big(defined in \q{ematrix}\big). The matrix $D_{(j')}$ describes adiabatic evolution over the loop $\calc_n\colon \bk_0 {\to} \bk_0 {+} \bG$ within the eigenspace $\{ \ket{ u_{j',\bk}^{\alpha} }_{\text{cell}} \}_{\alpha {\in} \Z_J}$. \\

To prove \q{adiabaticNonSymm}, we use that $\bra{u_{j,\bk_f}} = \mathrm{e}^{i \bG \cdot \br} \bra{u_{j,\bk_0}}$, and that $\mathrm{e}^{i \bG \cdot \br} \ket{ u^{\nu}_{j+1,\bk_0+\bG} }_{\text{cell}}$ and $\ket{ u^{\nu}_{j,\bk_0} }_{\text{cell}}$ belong to the same representation of $\hat g$, because $\mathrm{e}^{-i p \bG \cdot \ba/n} \omega_{j+1,n}{=}\omega_{j,n}$. Then \q{adevo} reads as (no sum over $j,j'$) 
\m{ U(T_B)^{\beta,\alpha}_{j,j'} =_{\mathcal{B}} \, \braket{ u^{\beta}_{j,\bk_0}| \mathrm{e}^{i \bG \cdot \br}|u^{\nu}_{j+1,\bk_0+\bG} }_{\text{cell}} \\
{\times} \bigg( \overline{\mathrm{e}}^{i \int_{0}^{T_B} \big[\;\bA (\bk(t')) \cdot \bF(t')- \mathfrak{E}(\bk(t'))\;\big] dt'/\hbar}\bigg)_{j+1,j'}^{\nu,\alpha}  \\
= \delta_{j+1,j'} \braket{ u^{\beta}_{j'-1,\bk_0}| \mathrm{e}^{i \bG \cdot \br}|u^{\nu}_{j',\bk_0+\bG} }_{\text{cell}}  \\
{\times} \bigg( \overline{\mathrm{e}}^{i \int_{0}^{T_B} \big[\;\bA (\bk(t')) \cdot \bF(t')- \mathfrak{E}(\bk(t'))\;\big] dt'/\hbar}\bigg)_{j',j'}^{\nu,\alpha}. \la{Udiag} }
The $\delta_{j{+}1,j'}$-factor comes from the following two observations: first, \q{ematrix} reads as
\e{ \braket{ u^{\beta}_{j,\bk}| \mathrm{e}^{-i \bk \cdot \br} H_0 \mathrm{e}^{i \bk \cdot \br} |u^{\alpha}_{j',\bk}}_{\text{cell}} = \delta_{j,j'} \big[ \mathfrak{E}(\bk) \big]^{\beta,\alpha}_{j,j} \la{Energydiag} } 
by our assumption that the states $\ket{ u^{\alpha}_{j,\bk} }_{\text{cell}}$ are simultaneously energy and symmetry eigenstates. Secondly, the non-Abelian Berry connection satisfies
\m{ \bA(\bk)^{\beta,\alpha}_{j,j'} = \braket{u^{\beta}_{j,\bk}|i \nabla_k| u^{\alpha}_{j',\bk}}_{\text{cell}}  \\
= \omega_{j} \bar \omega_{j'} \braket{ u^{\beta}_{j,\bk}| \hat g^{\dagger} i \nabla_k \hat g | u^{\alpha}_{j',\bk}}_{\text{cell}}  \\
= \omega_{j} \bar \omega_{j'} \bA(\bk)^{\beta,\alpha}_{j,j'}, \label{Ansymm}}
using $\hat g^{\dagger} i \nabla_k \hat g {=} i \nabla_k$. This implies that the non-Abelian Berry connection is also block-diagonal wrt.\ $j$, i.e., for all $j' {\neq} j$ and all $\alpha,\beta {\in} \Z_J\colon \bA(\bk)^{\beta,\alpha}_{j,j'} {=} 0$. Therefore, \q{Udiag} proves \qq{adiabaticNonSymm}{diagonaladiabatic}.\\

We now calculate the eigenstates $\{ \bff^{l,\iota} \}_{l {\in} \Z_{\mu_{\parallel}}, \iota {\in} \Z_J}$ and -values $\{ \mu^{l,\iota} \}_{l {\in} \Z_{\mu_{\parallel}}, \iota {\in} \Z_J}$ of $U(T_B)$, which satisfy
\e{ \sum_{j' \in \Z_{\mu_{\parallel}}, \beta \in \Z_J} \big( U(T_B)_{j,j'}^{\alpha,\beta} - \mu^{l,\iota} \delta_{j,j'} \delta_{\alpha,\beta} \big) f^{l,\iota}_{j',\beta} =_{\mathcal{B}} \, 0 \la{evaleq}}
for $j, l {\in} \Z_{\mu_{\parallel}}$, $\alpha, \iota {\in} \Z_J$. We will show that 
\e{ \mu^{l,\iota} = \mathrm{e}^{i(\gamma^{(\iota)}+2\pi l)/\mu_{\parallel}}, \as \bff^{l,\iota} = \f{1}{\sqrt{\mu_{\parallel}}} \mathrm{e}^{i(\gamma^{(\iota)}+2\pi l) j/\mu_{\parallel}} \bar \bv^{j,\iota}, \la{eveceval}}
for $\gamma^{(\iota)} \in [0,2\pi)$ generically independent of each other and $\bar \bv^{j,\iota} {\in} \C^{\mu_{\parallel}} \otimes \C^J \cong \C^N$ an orthonormal basis. For a fixed $\iota {\in} \Z_J$, each set $\{ \mu^{l,\iota} \}_{l {\in} \Z_{\mu_{\parallel}}}$ corresponds to one Stark-Wannier ladder. \\

In the simplest case of a spinless system with $J{=}1$, we can neglect the indices $\alpha, \beta, \iota$. Then $D_{(j)} {=} \mathrm{e}^{i \theta_j}$ for some $\theta_j {\in} [0,2\pi)$. Let 
\e{ f^l_j = \f{1}{\sqrt{\mu_{\parallel}}} \mathrm{e}^{i(\gamma+2\pi l) j/\mu_{\parallel}} \bar \bv^j, \as \bar v^j_{j'} = \delta_{j,j'} \mathrm{e}^{-i \sum_{0\le j''\le j} \theta_{j''}} \la{evec}}
with $\gamma {=} \sum_{j \in \Z_{\mu_{\parallel}}} \theta_{j}$, then 
\m{ \sum_{j' \in \Z_{\mu_{\parallel}}} U(T_B)_{j,j'} f_{j'}^l  \\
=_{\mathcal{B}} \, \f{1}{\sqrt{\mu_{\parallel}}} \sum_{j' \in \Z_{\mu_{\parallel}}} \delta_{j+1,j'} \mathrm{e}^{i \theta_{j'}} \mathrm{e}^{i(\gamma+2\pi l) j'/\mu_{\parallel} - i \sum_{0\le j''\le j'} \theta_{j''}}  \\
= \f{1}{\sqrt{\mu_{\parallel}}} \mathrm{e}^{i(\gamma+2\pi l) (j+1)/\mu_{\parallel} - i \sum_{0\le j''\le j} \theta_{j''}} = \mu^l f_{j}^l }
where $\mu^l {=} \mathrm{e}^{i (\gamma {+}2\pi l)/\mu_{\parallel}}$ satisfies the eigenvalue \q{evaleq}.

\begin{widetext}
We now prove that in the general case of $J{>}1$
\[
\sbox0{$\begin{matrix}0&1&0&...&0\\ 0&0&1&...&0\\  & & & \ddots &\\   0&0&0&...&1\\  \mathrm{e}^{i \theta^{(0)}} & 0 & 0 & ... & 0\end{matrix}$}
\sbox2{$\begin{matrix}0&1&0&...&0\\ 0&0&1&...&0\\  & & & \ddots & \\  0&0&0&...&1\\  \mathrm{e}^{i \theta^{(J-1)}} & 0 & 0 & ... & 0\end{matrix}$}
U(T_B) =_{\bar{\mathcal{B}}} \mathrm{diag} \left[
\begin{array}{c c c}
\left( \usebox{0} \right), & \makebox[\wd0]{\large $...$}, & \left( \usebox{2} \right)
\end{array}
\right]
\]
where each block has dimension $\mu_{\parallel}$. $U(T_B)$ is expressed in the orthonormal basis
\e{ \bar{\mathcal{B}} = \{ \bar v^{j,\iota}_{j',\alpha} \ket{ u_{j',\bk_0}^{\alpha} }_{\text{cell}} \}_{j {\in} \Z_{\mu_{\parallel}}, \iota {\in} \Z_J}, }
 of $\C^N$, where the vectors $\bar \bv^{j,\iota}$ are as in \q{eveceval}.
We note that the reducible representation of the $\mu_{\parallel}$-cycle in the basis $\mathcal{B}$ \big(see above \q{diagonaladiabatic}\big) has been decomposed into $J$ irreducible $\mu_{\parallel}$-cycles in $\bar{\mathcal{B}}$. The eigenvalues of $U(T_B)$ consist of the eigenvalues of each of its $\mu_{\parallel} {\times} \mu_{\parallel}$ blocks. Using the result from $J{=}1$ studied above (particularized to $\theta_1{=}\theta_2{=}...{=}\theta_{\mu_{\parallel}-1}{=}0$ and $\theta_0 {=} \theta^{(\iota)}$), then $\gamma^{(\iota)} {=} \theta^{(\iota)}$ and we obtain the eigenvalues and {-}vectors stated in \q{eveceval}. \\

We first notice that
\[
\bigg( U(T_B) \bigg)^{\mu_{\parallel}} =_{\mathcal{B}}
\left[
\begin{array}{c|c|c|c}
D_{(1)} D_{(2)} ... D_{(\mu_{\parallel}-1)} D_{(0)} & & &  \\
\hline
 & D_{(2)} D_{(3)} ... D_{(0)} D_{(1)} & &  \\
\hline
 & & \ddots &  \\
\hline
 & & & D_{(0)} D_{(1)} ... D_{(\mu_{\parallel}-2)} D_{(\mu_{\parallel}-1)}
\end{array}
\right],
\]
is block-diagonal, with all blocks on the diagonal being unitarily equivalent. The $j$'th block ($j {\in} \Z_{\mu_{\parallel}}$) corresponds to adiabatic transport with initial symmetry-degenerate states $\{ \ket{ u_{j,\bk_0}^{\alpha} }_{\text{cell}} \}_{\alpha {\in} \Z_J}$ for $\mu_{\parallel}$ BZ's. Especially, the eigenvalues of different blocks, corresponding to different initial symmetry eigenvalues, are the same.
Let $\{ \bv^{\iota}, \mathrm{e}^{i \theta^{(\iota)}} \}_{\iota {\in} \Z_J}$ be the normalized eigenvectors and -values of the last block $D_{(0)} D_{(1)} ... D_{(\mu_{\parallel}{-}2)} D_{(\mu_{\parallel}{-}1)}$, i.e., (in the basis $\mathcal{B}$)
\e{ \sum_{\alpha {\in} \Z_J} \bigg( \big( U(T_B) \big)^{\mu_{\parallel}} \bigg)^{\beta,\alpha}_{j',\mu_{\parallel}-1} v^{\iota}_{\alpha} &=_{\mathcal{B}} \, \delta_{j',\mu_{\parallel}-1} \mathrm{e}^{i \theta^{(\iota)}} v^{\iota}_{\beta}. }
Absent any other symmetries, the different phases $\theta^{(\iota)} {\in} [0,2\pi)$ are generically independent. We now define the ordered basis 
\e{ &\{\bar{\bv}^{0,0}, \bar{\bv}^{1,0}, ..., \bar{\bv}^{\Lambda-1,0}, \bar{\bv}^{0,1}, ..., \bar{\bv}^{\Lambda-1,1}, ..., ..., \bar{\bv}^{0,J-1}, ..., \bar{\bv}^{\Lambda-1,J-1}\}\colon \lin 
&\ass \bar v^{\Lambda-1,\iota}_{j',\alpha}=_{\mathcal{B}} \, \delta_{j',\Lambda-1} v^{\iota}_{\alpha}, \ass \bar \bv^{j,\iota} =_{\mathcal{B}} \, \big( U(T_B) \big)^{\Lambda-j-1} \bar \bv^{\Lambda-1,\iota} }
for $j, j' {\in} \Z_{\mu_{\parallel}}$, $\alpha, \iota {\in} \Z_J$. The left-right ordering of this basis corresponds to the top-down and left-right ordering of the row and column indices of $U(T_B)$ in $\bar{\mathcal{B}}$, respectively. The basis $\bar{\mathcal{B}}$ is normalized because it is constructed from the iterative application of a unitary $U(T_B)$ to a normalized vector $\bar \bv^{\mu_{\parallel}{-}1,\iota}$. States in $\bar{\mathcal{B}}$ are orthogonal wrt. $\iota$ because of the orthogonality of $\{ \bv^{\iota} \}_{\iota {\in} \Z_J}$ in $\C^J$. States in $\bar{\mathcal{B}}$ are orthogonal wrt. $j$ because $\big(U(T_B)\big)^l$ changes a state with symmetry eigenvalue $\omega_{j,n}(\bk_0)$ to a state with symmetry eigenvalue $\omega_{j{-}l,n}(\bk_0) {\neq} \omega_{j,n}(\bk_0)$ for all $l{=}1,...,\mu_{\parallel}{-}1$.
By construction, application of $U(T_B)$ on each basis element of $\bar{\mathcal{B}}$ gives
\e{ \bar \bv^{l-1,\iota} =_{\mathcal{B}} \, U(T_B) \bar \bv^{l,\iota},\ass U(T_B) \bar \bv^{0,\iota} =_{\mathcal{B}} \, \big( U(T_B) \big)^{\mu_{\parallel}} \bar \bv^{\mu_{\parallel}-1,\iota} = \mathrm{e}^{i \theta^{(\iota)}} \bar \bv^{\mu_{\parallel}-1,\iota}; }
hence $U(T_B)$ expressed in the basis $\bar{\mathcal{B}}$ is indeed as stated above. We have found that the phases $\{ \gamma^{(\iota)} \}_{\iota=1}^J$ in \q{eveceval} equal the eigen-phases of the last, or equivalently, first, block of $\bigg( U(T_B) \bigg)^{\mu_{\parallel}}$; this proves \q{gammas}. \hfill\(\Box\)
\end{widetext}

\subsection{$\mu_{\parallel}{<}n$}\la{app:lln}

For $\mu_{\parallel}{<}n$, the spectrum of $\rho_{n,p}$ has two possible forms: either there exists a single (possibly symmetry-degenerate) $\mu_{\parallel}$-cycle or there exist multiple (possibly symmetry-degenerate) $\mu_{\parallel}$-cycles. In either case, the spectrum does not have to contain all the eigenvalues $\{ \omega_{j_0,n} \}_{j_0 {\in} \Z_n}$. The case of a single $\mu_{\parallel}$-cycle is analogous to the discussion in \app{app:singlecycle}. All that is left to study is adiabatic transport in a subspace with multiple $\mu_{\parallel}$-cycles. \\

We begin the study with an example of monodromy of the energy bands for $\mu_{\parallel} {=} 2 {<} n {=} 6$, $p {=}3$. If the symmetry eigenvalues of $\rho_{n,p}$ are $\{ 1, {-}1 \}$ for one cycle with no symmetry degeneracy ($J_1{=}1$), and for the other cycle $\{ \omega, {-}\omega \}$ ($\omega {=} \mathrm{e}^{2\pi i/3}$) with degeneracy two ($J_2{=}2$); the total number of bands is $(J_1{+}J_2) \mu_{\parallel} {=} 3 {\times} 2 {=} 6$. In this example, the propagator can be block-diagonalized to two $2$-cycles with $J_1{=}1$ and $J_2{=}2$:
\[
\sbox0{$\begin{matrix}
0 & D_{(0,1)} \\
D_{(0,0)} & 0
\end{matrix}$}
\sbox1{$\begin{matrix}
0 & D_{(-1,1)}  \\
D_{(-1,0)} & 0
\end{matrix}$}
U(T_B)=\left[
\begin{array}{c|c}
\usebox{0}&\makebox[\wd0]{\large $0$} \\
\hline
\makebox[\wd0]{\large $0$}&\usebox{1}
\end{array}
\right].
\]
Here, $D_{(0,0)}, D_{(0,1)} {\in} U(1)$ and $D_{(-1,0)}, D_{(-1,1)} {\in} U(2)$. We see that $U(T_B)$ is block-diagonal wrt. to different $\mu_{\parallel}$-cycles; this reduces the problem back to single $\mu_{\parallel}$-cycles.\\

\noindent \textit{Proof.} Let $U(T_B)$ consist of multiple $\mu_{\parallel}$-cycles (labeled by index $u {\in} \Z_{\mathrm{gcd}(p,n)}$), then we show that $U(T_B)$ is block-diagonal wrt. $u$, i.e., $U(T_B)_{u,j; u',j'}^{\alpha_{j_0},\beta_{u'}} {=} 0$ for all $u {\neq} u'$. We denote the simultaneous eigenstates of the Bloch Hamiltonian and the symmetry by $\{ \ket{ u^{\alpha_{u}}_{u,j,\bk} }_{\text{cell}} \}_{u, \alpha_{u}, j}$ where $\alpha_{u}$ labels different eigenstates with the same eigenvalue $\mathrm{e}^{2\pi i u/n} \omega_{j,n}$, similar to before, and $j {\in} \Z_{\mu_{\parallel}}$. We proved in \qq{Energydiag}{Ansymm} that the Bloch Hamiltonian and the non-Abelian Berry connection are diagonal in such a basis. Therefore, \q{Udiag} still holds with the replacements $j' \to (u',j'), j \to (u,j), \beta \to \beta_{u'}, \alpha \to \alpha_{u}$. Because of this block-diagonal form of $U(T_B)$ in different cycles, its eigenvalues and {-}vectors consist of the eigenvalues and {-}vectors of $U(T_B)$ restricted to each single (possibly degenerate) cycle, i.e., they are all of the form of \q{eveceval}, but with possibly different degeneracies $J_u$. \hfill\(\Box\) \\

\subsection{Orbital splitting by the Wannier-Stark effect}\la{deg}

For $J{>}1$ orbitals localized on the same real space center ($\bvarpi$), we may ask in what situations are the Wannier-Stark ladders nondegenerate. We will see that a driving force $\bF$ cannot split a Kramers degeneracy in a Wannier-Stark ladder, though it might split other degeneracies. We will formulate symmetry-based criteria to determine when degeneracies remain in the presence of a $\parallel$-field.\\

The orbitals form a representation of a 2D (magnetic) on-site symmetry group $\calq$, which is the subgroup of $G$ that remains a symmetry in the presence of the field; for a 1D chain embedded in 3D space, as exemplified in \fig{fig:helix}(a) and \fig{fig:bands}(a,d,g), $G$ is generally a line group.\cite{Damnjanovic2010} In the simplest case of the force directed parallel to the chain (along $\be_z$), $\calq$ is the subgroup of $G$ that preserves the $z$-coordinate. \\

Since time-reversal symmetry ($T$) acts locally in space, $T$ is not spoilt upon application of a force. We may therefore expect for half-integer-spin representations of $T$ that (i) $J{\in} 2\Z$ due to spin-doubling, and (ii) the spectrum of $\calh$ is Kramers-degenerate.\footnote{Consequences (i{-}ii) are verified in the Supplementary material\textsuperscript{43}} As an example, we apply (i{-}ii) to a glide-symmetric ($g_{2,1}$) crystal with two atoms per unit cell, and a Kramers-degenerate orbital ($J{=}2$) localized on each atom (cf.\ \fig{fig:bands}(g)); its band structure and degenerate ladder spectrum are illustrated in \fig{fig:bands}(h{-}i). The Kramers degeneracy of the spectrum of $\calh$, combined with its ladder structure, results in period doubling ($\mu_{\parallel}{=}2$). \\

Though a driving force $\bF$ cannot split a Kramers degeneracy, it may yet split a degeneracy originating from the orbital degree of freedom -- this orbital splitting due to the Wannier-Stark effect is an analog of spin splitting due to the Zeeman effect. Owing to a symmetry reduction in the presence of an electric field, the degeneracy in the Wannier-Stark spectrum may be less than the number of orbitals $J$ -- we refer to this as orbital splitting. \\

For a band representation with Wyckoff position $\bvarpi$ and on-site representation $V$, application of a field $\bF$ reduces the on-site symmetry group $\calp_{\varpi}$ to a subgroup $\calq$. The symmetry-based criterion for orbital splitting is that orbital splitting generically occurs if the unitary representation $V$ is reducible under $\calq$. \\

Let us exemplify such a splitting for a 1D chain embedded in 3D. For a band representation having a single Wannier center $\bvarpi$ per unit cell, we consider an irreducible vector representation $V$ (e.g. $p_x{\pm} ip_y$ orbitals) of $\calp_{\varpi}{=}C_{4v}$, which is generated by four-fold rotation about $\be_z$ (parallel to the 1D chain) and a reflection ($M_y$) that inverts $y$. If the field is oriented in the $xz$-plane and has a nonzero component in $\be_x$, then $\calq{=}\{e,M_y\}$; $\calq$ only has 1D irreducible representations. This implies that the two Wannier-Stark ladders are generically nondegenerate, i.e., orbital splitting occurs.\\

Similarly, we might consider a one-dimensional lattice with point group generated by $M_y$ and $g_{6,3}$ (a sixfold-screw with half a lattice translation along $\be_z$); then, energy bands are ($\mu_{\parallel}{=}2$)-fold connected according to \q{munon} and the discussion in \s{parbloch}. A $p_x{\pm} ip_y$-orbital forms a two-dimensional irreducible representation of the on-site symmetry group $\calp_{\varpi}{=}\{e,C_{3,z}, M_y, C_{3,z} M_y\}$. For a field as described above, the reduced symmetry group is $\calq{=}\{e,M_y\}$, which results in orbital splitting. If we consider an additional $s$-orbital localized at another Wyckoff position, the corresponding Wannier-Stark ladders with ladder spacing $\f{2\pi}{\mu_{\parallel}}{=}\pi$ are generically nondegenerate: One offset $\Delta \gamma^{(1)} {\neq} 0$ exists between the Wannier-Stark ladders corresponding to the $p$-orbitals, and another one, $\Delta \gamma^{(2)} {\neq} 0$, between a Wannier-Stark ladder from a $p$-orbital and one from the $s$-orbital.

\section{Sufficient symmetry criteria for strong band representations}\la{app:zerocurv}

As described in \s{sec:symmetrystrong}, a sufficient  symmetry criteria for strong band representations (BRs) is that  \q{stabcond} is satisfied for each Wannier center $\bvarpi_j$. The goal of this section is to formulate sufficient symmetry criteria for \q{stabcond}.\\

These criteria are formulated  for each Wannier center  individually. By choosing  the spatial origin to lie at $\bvarpi_j{+}\bR$, \q{stabcond2} simplifies to
\e{ P_{j,\bR} \;(\br\cdot \bv_j) \; P_{j,\bR} = 0 \la{stabcond2} }
for some unit-magnitude two-vector $\bv_j$. Two possibilities emerge: either (a') symmetry constrains both $P_{j,\bR} x P_{j,\bR}{=}P_{j,\bR} y P_{j,\bR}{=}0$, or (b') symmetry constrains only one component (corresponding to $\bv_j$) of the projected position operator to vanish. We expect that the isotropic condition (a') occurs only for nongeneric Wyckoff positions at isolated high-symmetry points, while the anisotropic condition (b') occurs for nongeneric Wyckoff positions that is movable along a high-symmetry line. Let us utilize known matrix-element selection rules\cite{Tinkham2003} to formulate sufficient criteria for (a') and (b'), as detailed in \s{sec:aprime} and \s{sec:bprime} respectively. Examples of weak BRs which do not satisfy these criteria are provided in \s{examplesweakcompositeBR} and \s{pxpy}.

\subsection{Sufficient criterion for isotropic, strong EBRs}\la{sec:aprime}

We denote the on-site symmetry group at $\bvarpi_j{+}\bR$ by $\calp_j$, as a short-hand for $\calp_{\varpi_j+R}$. $X_j$ is its 2D vector representation, and $V_j$ its (possibly reducible) on-site representation on Wannier functions; $X_j^*$ and $V_j^*$ are their respective conjugate representations. Applying a matrix-element selection rule,\cite{Tinkham2003}a sufficient condition for (a') is that (a) $(V_j)^* \otimes X_j \otimes V_j$, when decomposed into unitary irreducible representations of $\calp_j$, does not contain the trivial representation.
An example of an isotropic strong EBR has been described at the end of \s{sec:symmetrystrong}; here we exemplify an isotropic strong CBR:\\

\noindent \textbf{\textit{Example of isotropic strong CBRs.}} A case in point are BRs of wallpaper groups  which are characterized by: (i) a single Wyckoff position ($\bvarpi$) with multiplicity $M_{\varpi}$ being unity or greater, (ii) an on-site symmetry group $\calp_{\varpi}$ that is neither trivial nor $C_s$ ($C_s$ is generated only by reflection\cite{Tinkham2003}), and (iii) the on-site representation $V_{\varpi}$ of $\calp_{\varpi}$ is a direct sum of the same one-dimensional irreducible representations. First note that the tensor product  $V_{\varpi}^* {\otimes} V_{\varpi}$, with $V_{\varpi}^*$ the complex conjugate representation of $V_{\varpi}$,  is just the trivial representation. This implies that $V_{\varpi}^{*} {\otimes} X {\otimes} V_{\varpi}{=}(\dim V_{\varpi})^2 X$, with $X$ the two-dimensional vector representation of $\calp_{\varpi}$. Condition (ii) implies that $\calp_{\varpi}$ is one of the $C_n$ or $C_{nv}$ groups, for which the vector representation $X$  does \textit{not} contain the trivial representation.

\subsection{Sufficient criterion for anisotropic, strong EBRs}\la{sec:bprime}

For a 2D unit vector $\bv_j$, we define the 1D on-site symmetry group $\calp_j^{\bv_j}$ as consisting of symmetries in $\calp_j$ that map $\bv_j$ to ${\pm} \bv_j$. In wallpaper groups, $\calp_j^{\bv_j}$ must therefore be a subgroup of $C_{2v}$, i.e., equal to the trivial group, $C_2$, $C_s$ or $C_{2v}$; $\calp_j^{\bv_j}$ is necessarily Abelian for integer-spin representations, which we assume henceforth. All the unitary irreducible representations of an Abelian group are 1D.
For example, if (i) $\calp_j{=}\{e,C_{2,z}\}{=}C_2$ then $\calp_j^{\bv_j}{=}C_2$; similarly if (ii) $\calp_j$ has a reflection that inverts $\bv_j$, then $\calp_j^{\bv_j}$ also inherits it. For $\calp_j^{\bv_j}$ we denote the corresponding 1D vector representation by $X_j^{\bv_j}$; for a given representation $V_j$ of $\calp_j$, we define $V_j^{\bv_j}$ as the restricted representation of $\calp_j^{\bv_j}$. In example (i) above and $\bv_j{=}\be_x$, $X_j^{\be_x}$ is the (unique) nontrivial 1D unitary irreducible representation of $\calp_j^{\be_x}{=}C_2$; if $V_j$ represents a $p_y$-orbital, then $V_j^{\be_x}$ is the trivial representation of $C_2$; on the other hand, if $V_j$ represents a $p_x$-orbital, then $V_j^{\be_x}{\equiv}X_j^{\be_x}$.\\ 

Then, a sufficient condition for (b') is that (b) $(V_j^{\bv_j})^* \otimes X_j^{\bv_j} \otimes V_j^{\bv_j}$, when decomposed into into unitary irreducible representations of $\calp_j^{\bv_j}$, does not contain the trivial representation.  \\


\noindent \textbf{\textit{Example of anisotropic strong EBRs.}} Let us consider the wallpaper group $G{=}pmg$ and the Wyckoff position $\bvarpi_j{=}2c$ (cf. \, \fig{fig:elem}), which is invariant under $\calp_j{=}C_s{=}\{e,M_x\}$ where $e$ is the trivial element and $M_x$ is the reflection that inverts $x$ and leaves $y$ invariant. For a single orbital with representation $V_j$ localized at $\bvarpi_j$, $(V_j)^*{\otimes}V_j{=}A_1$ is the trivial representation of $\calp_j$. The vector representation $X_j$ of $\calp_j$ is $A_1{\oplus}A_2$ where $A_2$ is the nontrivial 1D irreducible representation of $C_s$. Hence, $(V_j)^* {\otimes} X_j {\otimes} V_j {=} A_1{\oplus}A_2$ contains the trivial representation, i.e., criterion (a) is not satisfied. 
We now show that criterion (b) holds for $\bv_j {=} \be_x$ but not for $\bv_j {=} \be_y$: The reduced on-site symmetry group remains $\calp_j^{\be_x}{=}\calp_j$ with on-site symmetry representation $V_j^{\be_x}{=}V_j$; however, the reduced position operator ($x$) corresponds to the one-dimensional representation $X_j^{\be_x}{=}A_2$. Then $(V_j^{\bv_j})^* {\otimes} X_j^{\bv_j} {\otimes} V_j^{\bv_j}{=}A_2$ does not contain the trivial representation, i.e., criterion (b) holds. In the $y$-direction, $\calp_j^{\be_y}{=}\{e\}$ implies that all representations are trivial, thus (b) cannot be satisfied for $\bv_j{=}\be_y$.


\subsection{Examples of weak and composite BRs}\la{examplesweakcompositeBR}

Consider composite BRs of the space group with a \textit{trivial} point group. A subset of these composite BRs have  reducible on-site representations on at least one Wannier center (indexed by $j$). Since the stabilizer $\calp_j$ is trivial, a reducible, integer-spin representations is necessarily at least two-dimensional. In the absence of any on-site symmetry, the direct product of any representation is always the trivial representation; physically stated, in the absence of any on-site symmetry, Wannier functions can hybridize and split away from $\bvarpi_j$. Such BRs are weak.\\

Similarly, we might consider composite BRs of a space group with a \textit{nontrivial} point group. A subset of these composite BRs have a reducible on-site representation on at least one \textit{generic} Wyckoff position, i.e., a Wannier center with a trivial stabilizer. The same argument as in the previous paragraph leads to the conclusion that such BRs are weak.

\subsection{Examples of weak and elementary BRs}\la{pxpy}


By exhausting all EBRs for integer-spin representations of wallpaper groups, we have found a single example of a \textit{weak and elementary} BR. For future investigation, we believe that more examples may be uncovered in  half-integer-spin EBRs of wallpaper groups, and EBRs of 3D space groups. \\

Let us consider $p_x {\pm} i p_y$-orbitals on the vertices of a honeycomb lattice (with Wyckoff position $\bvarpi$); they form the two-dimensional irreducible (vector) representation $V_{\varpi}{=}X$ of the on-site symmetry group $\calp_{\varpi}{=}C_{3v}$, which is generated by $M_y$ and $C_3$. For convenience, we place the origin in $\bvarpi$.\\


Let us argue, in two steps, that generically $PxP$ and $PyP$ do not commute. In the first step, we observe that $PxP$ and $PyP$ are generically nonzero because $X^* {\otimes} X {\otimes} X$ contains the trivial representation, i.e., condition (a) is not satisfied. To elaborate, let us denote the trivial representation of $C_{3v}$ by $A_1$, the nontrival 1D irreducible representation by $A_2$ and the 2D irreducible representation by $E$; the vector representation falls into the latter category: $X{=}E$. Then $X{\otimes} X{=}A_1 {\oplus}A_2 {\oplus}E$. Since $X^*{=}X$, it follows that $X^* {\otimes} X {\otimes} X$ contains $X{\otimes}X$, which contains $A_1$, the trivial representation. \\

In the second step, we would argue that $PxP$ and $PyP$ do not commute. Indeed, the only nonzero matrix elements in both operators can be derived from knowing the only $C_{3v}$-invariant function that is cubic in powers of $x$ and $y$: $x^3{-}3xy^2$;\cite{Tinkham2003} we have assumed here that $C_{3v}$ includes a mirror that inverts $y$. Consequently, 
\e{PxP\eq\ket{p_x}\braket{ p_x | x | p_x }\bra{p_x}+\ket{p_y}\braket{ p_y | x | p_y }\bra{p_y}, \lin 
PyP\eq\ket{p_x}\braket{ p_x | y | p_y }\bra{p_y}+ \ket{p_y}\braket{ p_y | y | p_x }\bra{p_x}. }
In the basis of $\ket{p_x}$ and $\ket{p_y}$, we observe that $PxP$ is a diagonal matrix with generically different diagonal elements, and $PyP$ is off-diagonal -- therefore they do not commute, for the same reason that the Pauli matrices $\sigma_3$ and $\sigma_1$ do not commute.



\section{$\chi{>}1$ for EBRs}\la{app:exhaust}
We show first that criteria (i{-}iii) in \s{Wyckoffcond} are sufficient to prove \qq{specialG}{genLaue} ($\chi{>}1$) for at least one $\bG_n$ of the form of \q{Gcond}.
Next, we exhaust all EBRs of wallpaper groups, to show that criteria (i{-}iii) in \s{Wyckoffcond} are actually equivalent to \q{genLaue}.
We notice that \q{Gcond} rules out $\check g_n{=}C_{6,z}$, because there exists no $C_{6,z}$-invariant wavevector on the first-BZ boundary. Moreover, by condition (i) in \s{Wyckoffcond}, $g_n$ cannot be a glide reflection.  

\subsection{Conditions (i-iii) in \s{Wyckoffcond} $\Rightarrow$ \q{genLaue}}\la{app:cond}
Since $\bvarpi_j{+}\bR$ is generated from $\bvarpi_j$ by $\bR {\in} \calt$, $e^{i \bG_n{\cdot} \bvarpi_j}$ is independent of the representative Wannier center of $\bvarpi_j{+}\bR$. Moreover, we may infer $e^{i n\bG_n{\cdot} \bvarpi_j}{=}1$ from the following demonstration: consider a plane wave with 2D wavevector $\bK_n$ and with coordinates restricted to $\bvarpi_j{+}\bR$; by condition (ii), such a discrete plane wave forms a scalar representation of $g_n$, i.e., $g_n$ maps $e^{i\bG_n {\cdot} \bvarpi_j}$ as
\e{ g_n:\; e^{i\bK_n \cdot \bvarpi_j}\; \rightarrow \;e^{i\bK_n\cdot \check{g}_n^{-1}\bvarpi_j}= e^{i\bG_n\cdot \bvarpi_j} e^{i\bK_n\cdot \bvarpi_j},\la{app:phase}}
with $e^{i\bG_n{\cdot} \bvarpi_j}$ an $n$'th root of unity, owing to the triviality of $g_n^n$. By condition (iii), $\bvarpi_{j'}$ and $\bvarpi_j$ are related by an element in $\calp$ that is not in $\calp_{\varpi}$, and therefore $\Delta \bvarpi{=}\bvarpi_{j'}{-}\bvarpi_j$ cannot be a Bravais lattice vector. It follows that if $\Delta \bvarpi$ and $\bG_n$ are not orthogonal, $e^{i\bG_n{\cdot} \Delta \bvarpi}$ must equal a nontrivial root of unity.\\  

We address the possibility that $\bG_n{\cdot} \Delta \bvarpi{=}0$ for $g_n$ a reflection or rotation. In the former case, this is impossible owing to (a) $\bvarpi_{j'}$ and $\bvarpi_j$ lying on distinct, parallel mirror lines, due to condition (iii), and (b) $\bG_n$ being orthogonal to both mirror lines, due to \q{Gcond}. For $g_n$ a rotation of order $2,3$ or $4$, and supposing $\bG_n {\cdot} \Delta \bvarpi{=}0$, there exists a linearly-independent $\bG_n'$ which also satisfies \q{Gcond} for a distinct wavevector $\bK_n'$ on the BZ boundary -- this is a well-known property of rotationally-invariant points.\cite{Tinkham2003} It follows that $\bG_n'{\cdot}\Delta \bvarpi$ is neither zero, nor an integer multiple of $2\pi$; $e^{i\bG_n'{\cdot} \Delta \bvarpi}$ must therefore be a nontrivial root of unity. This completes the proof that \q{specialG} is satisfied. \\

At last, we show that all other pairs of $g_n$-invariant Wannier separations $\Delta \bvarpi'$ are (possibly trivial) multiples of $2\pi/\mu_{\perp}$, i.e. \q{genLaue} is satisfied. We first notice that $e^{i\bG_n{\cdot} \Delta \bvarpi'}$ is an $n$'th root of unity, as shown above; thus $\bG_n{\cdot} \Delta \bvarpi'$ is a multiple of $2\pi/n$. For $n{=}2,3,4$, all divisors that are not one, are either equal to $n$, or divide another divisor (in which case $\mu_{\perp}$ is the largest divisor for which \q{specialG} is satisfied). More explicitly, for $n{=}2,3$, $\mu_{\perp}{=}n$, thus if $\bG_n{\cdot} \Delta \bvarpi'$ is not zero, it is trivially a multiple of $2\pi/\mu_{\perp}$; for $n{=}4$ and $\bG_n{\cdot} \Delta \bvarpi'$ nonzero, either $\mu_{\perp}{=}4$, or $\mu_{\perp}{=}2$, such that $\bG_n{\cdot} \Delta \bvarpi'$ also being an $n$'th root of unity, must be a multiple of $2\pi/\mu_{\perp}$. This completes the proof that under conditions (i{-}iii), \qq{specialG}{genLaue} are satisfied. \hfill\(\Box\) 

\subsection{Wallpaper groups: Conditions (i{-}iii) in \s{Wyckoffcond} $\Leftrightarrow$ \q{genLaue}}\la{app:symmex}
In 2D, we went through all Wyckoff positions of the $17$ wallpaper groups to study when conditions (i{-}iii) in \s{Wyckoffcond} and \q{genLaue} are independently satisfied. We found that they are simultaneously satisfied for $16$ Wyckoff positions in $9$ wallpaper groups, while for all other Wyckoff position, neither of them hold. We therefore conclude that in 2D, conditions (i{-}iii) and \q{genLaue} are equivalent.\\

There are several ways in which a Wyckoff position of a wallpaper group fails to satisfy conditions (i{-}iii) in \s{Wyckoffcond}, as well as \q{genLaue}:
\begin{itemize}
\item The Wyckoff position is nongeneric (condition (i) is not satisfied); equivalently, $\bG {\cdot} \bvarpi_j$ are not fixed by symmetry to any special value, i.e., \q{genLaue} is generically not satisfied.
\item If $M_{\varpi}{=}1$, condition (ii), as well as \q{specialG}, are not satisfied.
\item $2a$ in $cm$, ($2a$, $2b$) in $cmm$: Here, $2a$ and $2b$ in the International Table of Crystallographie\cite{Hahn1984} refer to multiplicity two with a centered unit cell, whereas we define $M_{\varpi}$ wrt.\ primitive unit cells; $M_{\varpi}{=}1$ and condition (ii), as well as \q{specialG}, are not satisfied.
\item ($4d$, $4e$) in $cmm$, ($2e$, $2f$, $2g$, $2h$) in $pmm$, ($4d$, $4e$, $4f$) in $p4m$, $3d$ in $p3m1$, $3c$ in $p31m$, ($6d$, $6e$) in $p6m$: the non-trivial symmetry in $\calp_{\varpi}$ is a reflection; these Wyckoff positions lie on a reflection-invariant line (rather than a point, as for rotational symmetries), hence $\bG {\cdot} \bvarpi_j$ is not fixed to symmetry by any special value, i.e., \q{genLaue} is generically not satisfied. Though $M_{\varpi}{>}1$, all symmetry-related Wannier centers lie on non-parallel lines, and condition (iii) is not satisfied.
\end{itemize}

\section{Relations between the projected position operator and the Wilson loop}\la{app:weakfield}

The first part of this appendix proves \q{WilsonPxP}. The second part proves the atomic Zak-Wannier relation (\q{zakwannierstrongatomic}) for strong, atomic EBRs  of a space group $G$. For notational simplicity, let us drop the subscript $n$ of $\bG$ in \q{WilsonPxP}. 

\subsection{Proof of \q{WilsonPxP}}

The integral on the left-hand side of \q{WilsonPxP} may be reparametrized by a time variable $\mathrm dt$ as
\e{\bra{ \psi_{j,\bk_f} } \overline{\mathrm{e}}^{i \int_0^{T_B}  P \bF(t) \cdot \br P \mathrm dt } \ket{ \psi_{j',\bk_0} },\la{repara} }
with a time-dependent force satisfying $\bF(t){=}\hbar d\bk/dt$, and $T_B$ is the time taken to complete the loop $\calc$. \q{repara} is equivalent to the problem of a Bloch function ($\psi_{j',\bk_0}$ being the initial state)  evolving under a time-dependent field $\bF(t)$. The role of $P$ in \q{repara} is to restrict the field-induced dynamics to a low-energy subspace spanned by $N$ Bloch functions at each wavevector in the Brillouin zone; this restriction is physically justified by the adiabatic approximation, which holds for sufficiently large $T_B$.\footnote{The adiabatic theorem holds for $\calc$ that is smooth. If $\calc$ is kinked, the restriction is further justified by the sudden approximation discussed in \s{sec:sudden}.}  By the acceleration theorem,\cite{Nenciu1980} the wavevector of the initial Bloch function evolves in accordance with $d\bk/dt{=}\bF(t)/\hbar$ and $\bk_0{=}\bk(0)$ as the initial condition. We may therefore replace $P$ in \q{repara} by $P(\mathcal{C})\subset P$, where $P(\mathcal{C})$ is the line integral of $P(\bk){=}\sum_{n=1}^N \ket{\psi_{n,\bk}} \bra{ \psi_{n,\bk} }$ over the loop $\bk {\in} \mathcal{C}$.
 \q{WilsonPxP} is then equivalent to the identity:
\m{\delta(\bk_f-\bk_0-\bG) \W[\calc]_{j,j'} \\ 
= \bra{ \psi_{j,\bk_f} } \overline{\mathrm{e}}^{i \oint_{\mathcal{C}} P(\mathcal{C}) \br P(\mathcal{C}) \mathrm \cdot d\bk} \ket{ \psi_{j',\bk_0} }. \la{app:WilsonPxP} }
For a reminder of the definitions of various symbols, we refer the reader to  \tab{symbols}.\\

We begin by deriving an analog of \q{app:WilsonPxP} for an infinitesimal path in $\bk$-space, with the eventual goal of concatenating infinitesimal paths into a finite loop $\calc$. The infinitesimal analog is (writing out sums explicitly)
\m{ \limit{|\delta \bk|\rightarrow 0} \int_{\mathcal{C}} \sum_{j,j'=1}^N \ket{\psi_{j,\bk+\delta \bk}}\big( \;e^{i\bA(\bk)\cdot \delta \bk}\;\big)_{j,j'}\bra{\psi_{j',\bk}} \mathrm d\bk \\ 
=  \limit{|\delta \bk|\rightarrow 0} e^{iP(\calc)\br P(\calc) \cdot \delta \bk}P(\calc). \la{analdiscrete}}
The left-hand-side of \q{analdiscrete} is the operator that induces parallel transport from $\bk \rightarrow \bk+\delta \bk$ for any $\bk$ and $\bk+\delta \bk \in \calc$. \\

We start the proof of \q{analdiscrete} by applying the idempotency of $P(\calc)$, such that the right-hand-side of \q{analdiscrete} becomes
\begin{multline*}
 e^{iP(\calc)\br P(\calc) \cdot \delta \bk}P(\calc) \\
=\bigg[I+iP(\calc)\br P(\calc) \cdot \delta \bk+O(|\delta \bk|^2)\bigg]P(\calc) \\
= P(\calc) \bigg[I+i\br \cdot \delta \bk+O(|\delta \bk|^2)\bigg]P(\calc) \\
=P(\calc) \,e^{i\br \cdot \delta \bk}\,P(\calc) +O(|\delta \bk|^2).
\end{multline*}
By definition of $P(\calc)$
\begin{multline*}
 P(\calc) \,e^{i\br \cdot \delta \bk}\,P(\calc) \\ 
= \sum_{j,j'=1}^N \int_{\mathcal{C}} \int_{\mathcal{C}} \ket{\psi_{j,\bk}} \braket{\psi_{j,\bk}| e^{i\br \cdot \delta \bk}| \psi_{j',\bk'}}\bra{\psi_{j',\bk'}} \mathrm d\bk \mathrm d\bk'
\end{multline*}
and using
 \es{ \braopket{\psi_{j,\bk}}{e^{i\br \cdot \delta \bk}}{\psi_{j',\bk'}} &= \delta(\bk-\bk'-\delta \bk)\braketaa{ u_{j,\bk}}{j_{n',\bk'}}_{\text{cell}} \lin 
 &= \delta(\bk-\bk'-\delta \bk)\big( e^{i\bA(\bk')\cdot \delta \bk} \big)_{j,j'}, }
the infinitesimal version is proven.\\

To relate \q{analdiscrete} to \q{app:WilsonPxP}, it is useful to decompose $\calc$ into the path-ordered product $\calc_{K+1}\ldots\calc_2\calc_1$, where $K$ is number of kinks (possibly zero) in $\calc$, and each of the $\calc_i$ is an oriented straight path beginning at wavevector $\bk_{i-1}$ and ending at $\bk_i$. Let $\delta \bk_i$ by an infinitesimal vector parallel to $\calc_i$. Then by path-ordered concatenation of \q{analdiscrete} over the path $\calc_i$, we derive
\begin{multline*}
\delta(\bk_i-\bk_{i-1}-\delta \bk_i) \W[\calc_i]_{j,j'} \\ 
= \delta(\bk_i-\bk_{i-1}-\delta \bk_i) \big( \overline{\mathrm{e}}^{i \int_{\calc_i} \bA (\bk) \cdot \mathrm d\bk} \big)_{j,j'} \\
= \braopket{ \psi_{j,\bk_{i}}}{ \overline{\mathrm{e}}^{i \int_{\mathcal{C}_i} P(\mathcal{C}) \br P(\mathcal{C}) \mathrm \cdot d\bk} }{\psi_{j',\bk_{i-1}}},
\end{multline*}
and finally
\begin{multline*}
\delta(\bk_{K+1}-\bk_0-\bG) \W[\calc]_{j,j'} \\ 
=\delta(\bk_{K+1}-\bk_0-\bG) \big( \W[\calc_{K+1}]\,\ldots \,\W[\calc_2]\,\W[\calc_1] \big)_{j,j'} \\
=\braopket{ \psi_{j,\bk_{K+1}}}{ \overline{\mathrm{e}}^{i \oint_{\mathcal{C}} P(\mathcal{C}) \br P(\mathcal{C}) \mathrm \cdot d\bk} }{\psi_{j',\bk_0}}. 
\end{multline*}
\hfill\(\Box\)\\

\subsection{Zak-Wannier relation for strong, atomic BRs}\la{specW}
The atomic Zak-Wannier relation (\q{zakwannierstrongatomic}) says that the eigenvalues of $\W$  are equal to $\{ \mathrm{e}^{i \bG \cdot \bvarpi_j} \}_{j=1}^N$, where $\{ \bvarpi_j \}_{j=1}^N$ are different representatives of the Wyckoff position $\bvarpi_1$ (cf.\ Definition \ref{def:wyckoff}) of the strong, atomic EBR.\\

Let us begin from the just-proven identity (\q{WilsonPxP}), which applies generally to any $N$-band subspace projected by $P$. We then particularize to strong BRs, which satisfy the defining property that $PxP$ and $PyP$ commute in the atomic limit. That is to say, there exists a set of Wannier functions $W_{j,\bR}$ (projected by $P_{j,\bR}$) that are eigenstates of $P\br P$  with eigenvalues $\bvarpi_j{+}\bR$. Consequently, the operator on the left-hand-side of \q{WilsonPxP} may be expressed as 
\e{\overline{\mathrm{e}}^{i \oint_{\mathcal{C}} P \br P  \cdot d\bk}=\sum_{j,\bR} e^{i \bG \cdot \bvarpi_j} P_{j,\bR}, \la{pxpstrong}}
where $\bG$ is the primitive reciprocal vector that connects $\calc$ across the Brillouin zone. In deriving the above expression, we have used that $[PxP,PyP]{=}0$ for every $\bk{\in}\calc$, hence the path-ordering may be ignored, and $P\br P$ may be replaced by directly by its spectral decomposition $\sum_{j,\bR}(\bvarpi_j{+}\bR)P_{j,\bR}$.\\

The Fourier transform of $W_{j,\bR}$ (cf.\ \q{blochwannier}) defines a set of Bloch functions  $\psi_{j,\bk}$ which satisfy
\e{ \sum_{\bR} \braket{ \psi_{j,\bk_f}| P_{l,\bR}| \psi_{j',\bk_0} } = \delta(\bk_f-\bk_0-\bG) \delta_{j,j'} \delta_{j,l}. \la{strongbloch}}
It follows from \q{pxpstrong} and \q{strongbloch} that
\m{ \braket{ \psi_{j,\bk_f}| \overline{\mathrm{e}}^{i \oint_{\mathcal{C}} P \br P  \cdot d\bk}| \psi_{j',\bk_0} } \\
= \delta(\bk_f-\bk_0-\bG)  \delta_{j,j'} \mathrm{e}^{i \bG \cdot \bvarpi_j}. \la{ad1}}
Combining the above equation with \q{WilsonPxP}, we derive
\e{ \W[\calc]_{j,j'} = \delta_{j,j'} \mathrm{e}^{i \bG \cdot \bvarpi_j}. \la{holidx} }
This equality is valid in the basis of Bloch functions that are the Fourier transforms of projected-position eigenstates. In a more general basis given by $\psi_{\alpha,\bk} {=}\sum_{j=1}^N \psi_{j,\bk} S_{j,\alpha}(\bk)$ with unitary $S$, the Wilson loop $\W$ need not be diagonal. Independent of basis, we may identify the eigenphase of $\W[\calc]$ (i.e., the Zak phase) as $\phi_j[\calc]{=}\bG{\cdot} \bvarpi_j$ mod $2\pi$, which proves our claim.  \hfill\(\Box\)\\



\section{Theorem for symmetry-protected Zak phases}\label{app:quant}
Here we provide the proof for the theorem and its corollaries stated in \s{theoremstate}.

\subsection{Proof of the theorem}\la{app:thquant}
We first notice that we can write the loop $\mathcal{C}_n {=} \check g_n \bK_n {-} \bK_n$ as the concatenation of the two lines: $\mathcal{C}_n^1 {=} \Gamma-\bK_n$ and $\mathcal{C}_n^2 {=} \check g_n \bK_n-\Gamma$. Wilson lines at symmetry-related wavevectors satisfy\cite{Alexandradinata2014c,Alexandradinata2014f}
\e{ \W[\check g_n \bk' \leftarrow \check g_n \bk] = \rho_n(\bk') \W[\bk' \leftarrow \bk] \rho_n(\bk)^{\dagger} }
(this was shown in terms of the sewing matrix but can equivalently be shown for $\rho_n$).\footnotemark[43] \\

Let $\W[\mathcal{C}_n^1]$ be the Wilson line with base point $\bK_n$ and end point $\Gamma{=}\check g_n \Gamma$, then using the above relation, we find
\e{ \W[\mathcal{C}_n] &= \W[\check g_n \bK_n \leftarrow \bK_n] = \W[\check g_n \bK_n \leftarrow \check g_n \Gamma] \W[\Gamma \leftarrow \bK_n]\lin
&= \breve{g}_n(\bK_n) W[\bK_n \leftarrow \Gamma] \breve{g}_n(\Gamma)^{\dagger} \W[\Gamma \leftarrow \bK_n]\lin
&= \breve{g}_n(\bK_n) Z^{\dagger} \breve{g}_n(\Gamma)^{\dagger} Z = \rho_n(\bK_n) Z^{\dagger} \rho_n(\Gamma)^{\dagger} Z. \la{app:Wdecomp}}
The unitary Wilson line $Z {=} \W[\mathcal{C}_n^1]$ conjugates the unitary matrix $\rho_n(\Gamma)^{\dagger}$, but does not change its eigenvalues. Varying parameters in the Hamiltonian that preserve the symmetry $g_n$, $Z$ can be any element of $U(N)$ (assuming no other symmetries); the $g_n$-protected Zak phase factors exactly correspond to the eigenvalues of $\W[\mathcal{C}_n]$ that are independent of $Z$.\\

We use the spectral decomposition of $\rho_n(\bK_n)$ and $Z^{\dagger} \rho_n(\Gamma)^{\dagger} Z$ to obtain a spectral decomposition of $\W[\mathcal{C}_n]$. Let $\mathfrak{P}_*$ be the projection operator onto the eigenspace of $Z^{\dagger} \rho_n(\Gamma)^{\dagger} Z$ with eigenvalue $\bar{\lambda}_*{=} \mathrm{e}^{-2\pi i l_*/n}$; $\mathfrak{Q}_* {=} \mathbb{1}_N {-} \mathfrak{P}_*$ projects to its orthogonal complement. Let $\mathfrak{P}_l$ be the projection operator onto the eigenspace of $\rho_n(\bK_n)$ with eigenvalue $\mathrm{e}^{2\pi il/n}$. Decomposing the Wilson loop as
\m{ \W[\mathcal{C}_n] = \sum_{l} \mathrm{e}^{2\pi i(l-l_*)/n} \mathfrak{P}_l \mathfrak{P}_*  \\
+ \sum_{l} \mathrm{e}^{2\pi il/n} \mathfrak{P}_l \mathfrak{Q}_* Z^{\dagger} \rho_n(\Gamma)^{\dagger} Z \mathfrak{Q}_*, }
we find that the rank of the projection $\mathfrak{P}_l \mathfrak{P}_*$ varies with $Z$. If the rank $m_l(\bK_n)$ of $\mathfrak{P}_l$ is larger than the rank $N{-}m_{l_*}(\bk_*)$ of $\frakQ_*$, then $\mathfrak{P}_l \mathfrak{P}_*$ must have positive rank. This means that there exists a subspace of minimal rank $r_l[\calc_n]{=}m_l(\bK_n){-}(N{-}m_{l_*}(\bk_*))$, independent of $Z$, which projects to the Wilson loop eigenvalue $\mathrm{e}^{2\pi i(l{-}l_*)/n}$. If we try to apply the same reasoning to $\mathfrak{P}_l \mathfrak{Q}_*$, we find that $m_l(\bk_s)$ cannot be larger than $m_{l_*}(\bk_*)$, by assumption. Therefore $\mathfrak{P}_l \mathfrak{Q}_*$ has no $Z$-independent subspace.\\  

In the case where $\bk_*{=}\bK_n$, we first notice that $Z \W[\mathcal{C}_n] Z^{\dagger} {=} Z \rho_n(\bK_n) Z^{\dagger} \rho_n(\Gamma)^{\dagger}$ and $\W[\mathcal{C}_n]$ are spectrally equivalent. Then we can apply a similar reasoning as for $\bk_*{=}\Gamma$. This proves the theorem. \hfill\(\Box\)  

\subsection{Proof of corollaries (I-II)}\la{app:cor}
To prove corollary (I), we notice that $m_l(\bk_s) {\le} m_{l_*}(\bk_*)$, therefore $r_l[\calc_n] {\le} 2m_{l_*}(\bk_*){-}N$ is bigger than zero if and only if $m_{l_*}(\bk_*) {>} N/2$. \\

For corollary (II), we observe that all Zak phases are $g_n$-protected if and only if $r_l[\calc_n]{=}m_l(\bk_s)$, which is equivalent to $N{=}m_{l_*}(\bk_*)$. \\

At last, we prove that even for non-unique $(l_*,\bk_*)$, $(l'_*,\bk'_*)$, i.e., $m_{l_*}(\bk_*){=}m_{l'_*}(\bk'_*)$, the output of the theorem is unique. We distinguish two scenarios: (i) If $\bk_* {=} \bk'_*$ but $l_* {\neq} l'_*$ then $m_{l_*}(\bk_*){=}m_{l'_*}(\bk'_*) {\le} N/2$. By corollary (I), there can be no symmetry-protected Zak phases. (ii) If $l_* {=} l'_*$ but $\bk_* {\neq} \bk'_*$, then for all $l {\neq} l_*$ holds that $m_l(\bk_*) {\le} N{-}m_{l_*}(\bk_*)$ and thus $r_l[\calc_n] {\le} 0$. Therefore, the only symmetry-protected Zak phases arise for $l{=}l_*$, for which $r_l[\calc_n]{=}2m_{l_*}(\bk_*){-}N$ Zak phases are $g_n$-protected to ${\pm} 2\pi(l_*{-}l'_*)/n$, where ${+}$ (resp.\ ${-}$) applies for $\bk_*{=}\Gamma, \bk'_*{=}\bK_n$ (resp.\ $\bk_*{=}\bK_n, \bk'_*{=}\Gamma$). Interchanging the roles of $(l_*,\bk_*)$ and $(l'_*,\bk'_*)$ leads to the same $g_n$-protected Zak phases. \hfill\(\Box\)

\section{Case studies with topologically nontrivial subspaces in class A/AII}

\subsection{Zak-Chern relation in $C_{4,z}$-symmetric lattices}\la{app:ZakCh}

To prove \q{c4constraint}, we concatenate $\calc_4$ with ${-}\calc_2$ (the minus sign indicates a reversal in orientation) to form the contractible loop $\calc'$ which bounds a quarter of the the Brillouin torus. Due to the periodicity of the Bloch functions on a torus, the Berry phase of $\calc'$ satisfies $\phi[\calc']{=}\phi[\calc_4]{-}\phi[\calc_2]$. Moreover, via Stoke's theorem, $\phi[\calc']$ equals a quarter of the total integrated curvature, modulo $2\pi$; the latter equals ${-}2\pi C/4$, hence we obtain \q{c4constraint}. The mod-four arbitrariness in \q{c4constraint} originates from the mod-$2\pi$ ambiguity in the Berry phase.
\hfill\(\Box\)  

\subsection{Class A: honeycomb and Kagome lattices}\la{app:BReval}
An example of splitting an EBR into Chern insulators is given by the honeycomb lattice with $s$-orbitals at the Wyckoff positions $2b$ (cf. \tab{wallpaper}); here we consider the wallpaper group $p6$ in contrast to $p6m$ used in the main text. The corresponding eigenvalues of $C_{3,z}$ ($\rho_3$) and $C_{2,z}$ ($\rho_2$) are also listed in \tab{evalEx}(b) ($p6$ is a subgroup of $p6m$). There are four ways to split the EBR into two single bands, depending on the combination of $C_{3,z}$ and $C_{2,z}$ eigenvalues at the high-symmetry wavevectors. One single band has Zak phase $2\pi/3$ for the $\calc_3$-loop, while the other band has Zak phase ${-}2\pi/3$. Using \q{c3constraint}, we also find that the Chern numbers of the two single bands must be equal to ${\pm} 1$ modulo $3$ for all four splittings. \\

We now consider splitting the EBR of the Kagome lattice with $s$-orbitals at the Wyckoff position $3c$ (cf. \tab{wallpaper}). We have seen an explicit tight-binding model in \s{sec:proofofprinciple}, while here we discuss what are all the possible splittings in general.
From \tab{evalEx}(c) we infer that the three-band $\calc_2$-Wilson loop results in $\mu_{\perp}{=}2$, while the $\calc_3$-loop has no symmetry-protected Zak phases ($\mu_{\perp}{=}1$). There are in principle $72$ possible splittings of the three-band subspace into three single-band subspaces. The $\calc_2$-Wilson loop after the splitting results in three $C_{2,z}$-fixed Zak phases, with two of them differing by $\pi$; there are many different possibilities for the three $C_{3,z}$-fixed Zak phases and therefore for the Zak phases of the $\calc_3$-Wilson loop. For each splitting, we calculate the three single-band Chern number modulo $6$ using \q{c6constraint} and combine them into an unordered triple: each component corresponds to the Chern number of a single band. These triples are of the form $(0,1,{-}1), (0,3,{-}3), (1,1,{-}2), ({-}1,{-}1,2), (1,2,{-}3)$. The three Chern numbers must sum to zero, because the three-band subspace is an EBR (which has zero three-band Chern number). We notice that the triples $(0,2,-2), (2,2,2)$ do not appear in the above combinations, which is equivalent to say that not all Chern numbers can be even (indeed, two single-band Chern numbers are odd and one is even).

\subsection{Proof of equivalence of $\mu_{\perp}{=}3$ and nonzero mirror Chern number in class AII}\la{app:graphene}
We prove that for the set of two-band subspaces with half-integer spin representation, as well as time-reversal $T$, six-fold rotation $C_{6,z}$, and inversion $\cali$ symmetry, holds: $\mu_{\perp}{=}3$ if and only if the mirror Chern number $C^{\pm}$ is nonzero. \\

Due to $T \cali$ symmetry, all $C_{3,z}$ eigenvalues (of $\rho_3$) in the two-band subspace come in pairs of only two kinds: $\{\mathrm{e}^{2\pi i/3}, \mathrm{e}^{2\pi i/3}\}$ and $\{1,\mathrm{e}^{{-}2\pi i/3}\}$. This is because the spectrum of $\mathrm{e}^{iF\pi/n} \rho_n$ must be invariant under complex-conjugation. Let us exhaust all possible combinations of $C_{3,z}$ eigenvalues that would give rise to $\mu_{\perp}{=}3$. If only $C_{3,z}$ symmetry existed, the theorem (in \s{th1}) states that either
\e{ (a) &\sigma\big( \rho_{3}(\Gamma) \big)=\{\mathrm{e}^{2\pi i/3}, \mathrm{e}^{2\pi i/3}\},\, \sigma\big( \rho_{3}(K) \big)=\{1,\mathrm{e}^{{-}2\pi i/3}\},\lin
(b) &\sigma\big( \rho_{3}(\Gamma) \big)=\{1,\mathrm{e}^{{-}2\pi i/3}\},\, \sigma\big( \rho_{3}(K) \big)=\{\mathrm{e}^{2\pi i/3}, \mathrm{e}^{2\pi i/3}\}. }
Since $M_z{=} C_{2,z} \cali$ is a symmetry by assumption, and commutes with $C_{6,z}$, for any single-band subspace of $M_z$, the $C_{3,z}$ eigenvalues at $K$ and $K'$ are identical. Let us apply Chen-Gilbert-Bernevig's criterion\cite{Fang2012} which relates the Chern number modulo three to the $C_{3,z}$ eigenvalues; applying the criterion within each mirror subspace, we determine the mirror Chern number modulo three, but not its parity. When we restrict to a single band that transforms in one representation of $M_z$,
\e{ e^{-2\pi i C/3} = \rho_3(\Gamma) \rho_3(K) \rho_3(K')=\rho_3(\Gamma) \big(\rho_3(K)\big)^2. }
In case (a), the product of eigenvalues within one $M_z$ subspace equals either of $\rho_3(\Gamma) \big(\rho_3(K) \big)^2{=}\mathrm{e}^{2\pi i/3}$ or ${=}\mathrm{e}^{-2\pi i/3}$, which implies that the mirror Chern number is nonzero. This is also the case for (b): $\rho_3(\Gamma) \big(\rho_3(K)\big)^2{=}\mathrm{e}^{-2\pi i/3}$ or ${=}\mathrm{e}^{2\pi i/3}$. \\

We have not yet exhausted all cases with $\mu_{\perp}{=}3$. Since $M_z$ symmetry allows us to split the two-band subspace into two single-band subspaces, we should apply the theorem within each single band. To obtain a Berry phase of ${\pm} 2\pi/3$ (where the ${+}$ occurs for one $M_z$ subspace and ${-}$ for the other), we could also pair up $1$ (at one invariant wavevector) with $\mathrm{e}^{{-}2\pi i/3}$ (at the other invariant wavevector). Then $\rho_3(\Gamma) \big( \rho_3(K) \big)^2{=}\mathrm{e}^{{+}2\pi i/3}$ or ${=}\mathrm{e}^{{-}2\pi i/3}$ implies that the mirror Chern number is nonzero. \hfill\(\Box\) 

\section{Applying a $\perp$-field to a nearly-degenerate band subspace}\la{app:degen}
The atomic limit of crystals describes a scenario where bands are not dispersive. We have identified band subspaces which Bloch oscillate with frequency $2\pi/(\mu_{\perp} T_B)$ in the atomic limit. 
In the vicinity of the atomic limit, bands have a small but finite energy dispersion, i.e., the band width at each $\bk$ is
\e{ \Delta_{E}(\bk) = \max_{i, j=1,...,N} &| \braket{ u_{i,\bk}| H_0(\bk) | u_{i,\bk} }_{\text{cell}}\lin
&-\braket{ u_{j,\bk}| H_0(\bk) | u_{j,\bk} }_{\text{cell}}| > 0, }
where $\ket{ u_{j,\bk} }_{\text{cell}}$ span the $N$-dimensional subspace at each $\bk {\in} \mathcal{C}_n$, and $H_0(\bk){=} \mathrm{e}^{-i \bk \cdot \br} H_0 \mathrm{e}^{i \bk \cdot \br}$ is the Bloch Hamiltonian. The overall band width is defined by $\Delta_{E} {=} \max_{\bk \in \mathcal{C}_n} \Delta_{E}(\bk) {>} 0$. We will show that the shift in the frequency ($\delta \omega$) from $2\pi/(\mu_{\perp} T_B)$ is bounded by 
\e{ |\delta \omega | \le \f{1}{\hbar T_B} \int_0^{T_B} \| \Delta_{E} \big( \bk(t) \big) \| \mathrm dt \le \f{\Delta_{E}}{\hbar} \la{shift}}
where $\bk(t)$ describes the parametrization along the loop $\mathcal{C}_n$ with base point $\bk_0 {\in}\calc_n$ and which wraps around the BZ in the direction of the reciprocal vector $\bG_n$ (defined in \q{Gcond}). \\

The time-ordered integral in $U(T_B)$ ($U(T_B)$ is defined in \q{adevo}) can be expressed explicitly by discretizing time $T_B {=} (L{+}1) \delta t$ for $L {\in} \mathbb{N}$ large and $\delta t {>} 0$ small, and therefore also discretizing the wavevector $\bk_l {=} \bk(l\delta t)$ for $l{=}0,..,L$ such that $\bk_L {=} \bk_0 {+} \bG_n$ with increment $\boldsymbol{\delta k}_l {=} \bk_l {-} \bk_{l{-}1}$ for all $l{=}1,...,L$. Eventually, we will take the continuum limit $\delta t {\to} 0, L {\to} \infty$ with $(L{+}1) \delta t {\equiv} T_B$ constant. Let us first express the infinitesimal propagator $U(\delta t)$ to first order in $\delta t$ as
\m{ U(\delta t)_{i,j} = \big( \mathbb{1}_N-\f{i}{\hbar} \mathfrak{E} (\bk_l) \delta t + i \bA (\bk_l) \cdot \boldsymbol{\delta k}_l \big)_{i,j}  \\
= \delta_{i,j}-\f{i}{\hbar} \mathfrak{E}_{i,j}(\bk_l) \delta t - \braket{ u_{i,\bk_l}| \nabla_k | u_{j,\bk_l} }_{\text{cell}} \cdot \boldsymbol{\delta k}_l \\
= \braket{ u_{i,\bk_l} | u_{j,\bk_l} }_{\text{cell}} -\f{i}{\hbar} \mathfrak{E}_{i,j}(\bk_l) \delta t + \braket{\nabla_k u_{i,\bk_l} | u_{j,\bk_l} }_{\text{cell}} \cdot \boldsymbol{\delta k}_l  \\
= \braket{ u_{i,\bk_{l+1}}| \mathrm{e}^{-i H_0(\bk_l) \delta t/\hbar} | u_{j,\bk_l} }_{\text{cell}}, \la{Udiscinf} }
where $\mathfrak{E}$ is the energy matrix, defined in \q{ematrix}, and $\bA$ the non-Abelian Berry connection (\q{nonAB}).
We write the projection operator on cell-periodic functions as
\e{ \frakP(\bk) = \sum_{j \in \Z_N} \ket{ u_{j,\bk} } \bra{ u_{j,\bk} }_{\text{cell}}. \la{Proju}}
The adiabatic evolution operator $U(T_B)$ in discrete time, $U(T_B) {=} \prod_{l=0}^{L-1} U (\delta t)$,  is then\cite{Alexandradinata2014c}
\es{ U(T_B)_{i,j} = \braket{ u_{i,\bk_0+\bG}| \big( \prod_{\mathcal{C}_n: l=0}^{L-1} \frakP(\bk_{l+1}) \mathrm{e}^{-i H_0(\bk_l) \delta t/\hbar} \big) | u_{j,\bk_0} }_{\text{cell}} }
where the notation $\prod_{\mathcal{C}_n: l{=}0}^{L-1}$ clarifies that the product is path-ordered along the loop $\mathcal{C}_n$. Note that the discretization error is now of order $\mathcal{O}( L \delta t^2 ) {=} \mathcal{O}( \delta t )$, while expanding the exponential to first order in $\delta t$ gives a contribution of order $\mathcal{O}( L \delta t ) {=} \mathcal{O}( 1 )$.  \\

In the energy-degenerate case where $\mathfrak{E}_{i,j}(\bk_l) {=} \delta_{i,j} E_0(\bk_l)$, the discretized propagator $U(T_B)$ simplifies to
\m{ U^{\mathrm{atomic}}(T_B)_{i,j} = \mathrm{e}^{-i \sum_{l=1}^L E_0(\bk_l) \delta t/\hbar}  \\
{\times} \braket{ u_{i,\bk_0+\bG}| \big( \prod_{\mathcal{C}_n: l=1}^L \frakP(\bk_l) \big) | u_{j,\bk_0} }_{\text{cell}} + \mathcal{O}( \delta t ). \la{Deltaz}} \\

Away from that limit, the energy matrix is not proportional to the identity, so let
\es{ \mathfrak{E}_{i,j}(\bk) = E_0(\bk) \delta_{i,j} + \delta E_{i,j}(\bk) }
with $E_0(\bk) {=} \f{1}{N} \mathrm{tr} \mathfrak{E}(\bk)$ and $\delta E_{j,i}(\bk) {=} \overline{\delta E}_{i,j}(\bk)$ a Hermitian, traceless $N{\times} N$ matrix (for a complex number $z$, we denote by $\bar z$ its complex conjugate), which is bounded by the band width $\Delta_{E} (\bk)$ at each $\bk$.
Then \q{Udiscinf} together with the equalities $H_0(\bk) {=} E_0(\bk) \frakP(\bk) {+} \big( H_0(\bk){-}E_0(\bk) \frakP(\bk) \big)$ and $\frakP^2(\bk) {=} \frakP(\bk)$, gives to zeroth order in $\delta t$:
 \begin{multline*}
 U(T_B)_{i,j} = \mathrm{e}^{-i \sum_{l=1}^L E_0(\bk_l) \delta t/\hbar} \bra{ u_{i,\bk_0+\bG} } \prod_{\mathcal{C}_n: l=0}^{L-1} \bigg( 1 - \\
 \f{i \delta t}{\hbar} \big( H_0(\bk_l) - E_0(\bk_l) \big) + \frakP(\bk_{l+1}) \bigg) \frakP(\bk_{l}) \ket{ u_{j,\bk_0} }_{\text{cell}} \\
= U^{\mathrm{atomic}}(T_B)_{i,j} - \f{i \delta t}{\hbar} \mathrm{e}^{-i \sum_{l=1}^L E_0(\bk_l) \delta t/\hbar} \bra{ u_{i,\bk_0+\bG}} \\ 
\sum_{l'=0}^{L-1} \bigg( \prod_{l>l'} \frakP(\bk_l) \big( H_0(\bk_{l'}) - E_0(\bk_{l'}) \frakP(\bk_{l'}) \big) \prod_{l''<l'} \frakP(\bk_{l''}) \bigg) \\ 
\ket{ u_{j,\bk_0} }_{\text{cell}}
\end{multline*}
where in the last equation $\f{\delta t}{\hbar} \delta E(\bk)$ appears exactly once at each position $\bk_l$ in the path-ordered product. This is the discretized version (to first order in $\lambda$) of the well-known identity
\es{ \bar{\mathrm{e}}^{\int_0^t \big( A + \lambda B \big) \mathrm dt'} = \bar{\mathrm{e}}^{\int_0^t A \mathrm dt'} \, \bar{\mathrm{e}}^{\int_0^t \big( \bar{\mathrm{e}}^{-\int_0^{t'} A \mathrm ds} \lambda B \, \bar{\mathrm{e}}^{\int_0^{t'} A \mathrm ds} \big) \mathrm dt'}   . }
Using the triangle and Cauchy-Schwartz inequalities several times, as well as $\| \frakP(\bk) \| {=} 1$ for all $\bk {\in} \mathcal{C}_n$, we obtain for the operator norm
\m{ \| U(T_B)-U^{\mathrm{atomic}}(T_B) \| \le \f{\delta t}{\hbar} \sum_{l'=1}^L \| \delta E(\bk_{l'}) \| + \mathcal{O}( \delta t ) \\
\le \f{\delta t}{\hbar} \sum_{l'=1}^L \Delta_{E}(\bk_{l'}) + \mathcal{O}( \delta t ). \la{shift2}}\\

To relate $\| U(T_B)-U^{\mathrm{atomic}}(T_B) \|$ to the shift in oscillation frequency, we consider the eigenvalues and (normalized) -vectors $\{ \mathrm{e}^{i \varphi_m}, \ket{ \psi_{m,\bk_0} } \}_{m {\in} \Z_N}$ of $U^{\mathrm{atomic}}(T_B)$ (\q{Deltaz}); in this basis, we denote matrix elements by $\tilde{U}^{\mathrm{atomic}}(T_B)_{m,l} {=} \delta_{m,l} \mathrm{e}^{i \varphi_m}$. Since the right-hand-side of \q{shift2} is small, we can use a first order perturbation to find $\tilde{U}(T_B)_{m,l}{=}\delta_{m,l} \mathrm{e}^{i (\varphi_m {+} \epsilon_m)}+\mathcal{O}(\Delta_E^2)$ where $\mathcal{O}(\epsilon_m)=\mathcal{O}(\Delta_E)$ for all $m {\in} \Z_N$. For simplicity, we assumed that $\varphi_m {\neq} \varphi_l$ for $m {\neq} l$, but degeneracies can be treated analogously using standard methods of degenerate perturbation theory; the result, \q{shift}, is the same.
The operator norm $\| \, . \, \|$ can then be bounded from below as follows: for all $m {\in} \Z_N$
\m{ \| U(T_B)-U^{\mathrm{atomic}}(T_B) \|^2 = \| \tilde{U}(T_B)-\tilde{U}^{\mathrm{atomic}}(T_B) \|^2 \\ 
\ge \sum_{l \in \Z_N} \big( \tilde{U}(T_B)- \tilde{U}^{\mathrm{atomic}}(T_B) \big)^{\dagger}_{m,l} \big( \tilde{U}(T_B)- \tilde{U}^{\mathrm{atomic}}(T_B) \big)_{l,m} \\
= | 1-\mathrm{e}^{i \epsilon_m} + \mathcal{O}(\Delta_E^2) |^2 = |\epsilon_m|^2 + \mathcal{O}(\Delta_E^3); \la{interm} }
especially, $\| U(T_B){-}U^{\mathrm{atomic}}(T_B) \| {\ge} \max_{m} |\epsilon_m|$. \\

In the atomic limit, the Bloch oscillation frequency is obtained from the phases $\{ \varphi_m \}_{m\in \Z_N}$ as $\f{1}{T_B} \min_{m {\neq} 0} |\varphi_m{-} \varphi_0|$; because of the dependency of the phases $\{ \varphi_m \}_{m\in \Z_N}$ on the choice of origin, as elaborated in \s{theoremstate}, we can assume without loss of generality that $\varphi_0 {=} 0$. The shift in oscillation frequency is therefore 
\es{ |\delta \omega| &= \f{1}{T_B} \min_{m {\neq} 0} |\epsilon_m| \le \f{1}{T_B} \max_{m \neq 0} |\epsilon_m| \lin 
&\le \f{1}{T_B} \| U(T_B){-}U^{\mathrm{atomic}}(T_B) \| \lin 
&\le \f{1}{\hbar T_B} \int_0^{T_B} \| \Delta_{E} \big( \bk(t) \big) \| \mathrm dt, }
where we took the continuum limit of \q{shift2} for the last equality. This proves \q{shift}. \hfill\(\Box\)

\newpage
\begin{widetext}
\section{Organization of the supplementary material}

This supplementary material provides helpful tools for the beginner to understand the main text. Much of the material is available from various sources in the literature, and collected here for pedagogy.
In the preliminaries, we first review Bloch theory \big(\s{app:Bloch}\big), different electromagnetic gauge choices \big(\s{app:gauge}\big), basics about (magnetic) space groups \big(\s{app:Blochsymm}\big) and its representation on Bloch functions. A detailed introduction to band representations is given in \s{app:band}, followed by basics on Wilson loops and its gauge-invariant formulation -- the Kato Hamiltonian \big(\s{app:gapped}\big). In \s{app:weakfield} we prove a relation between the adiabatic propagator (which includes $P\br P$ in its generator) and the Wilson loop.
In \s{altern} we provide two alternative derivations of the Zak-Wannier relation (Eq. (21) in the main text) for strong EBRs in the atomic limit. At last, in \s{app:leakage}, we give an extended derivation of the adiabatic theorem and its leakage, following closely A. Nenciu's thesis.\cite{Nenciu1987}

\section{Preliminaries}

\subsection{Review of the Bloch Hamiltonian}\la{app:Bloch}
We review the theory of Bloch states: They are eigenstates under translation and reduce the study of translation-invariant systems to a unit cell.\\

To describe (i) electronic materials with light elements and consequently weak spin-orbit coupling, or (ii) bosons in a crystalline lattice, we apply the field-free Schr\"odinger Hamiltonian, defined as
\e{H_0 =\f{ \bp^2}{2m}+V(\br); \la{fieldfreeschrodinger}}
otherwise, we apply the Pauli Hamiltonian
\e{H_0 \eq \f{1}{2m}\left( \bp + \f{\hbar}{4mc^2}\bsigma \times \nabla V \right)^2+V(\br). \la{fieldfreepauli}}
We define the lattice potential $V(\br) = V(\br + \bR)$ that is invariant under Bravais-lattice translations $\bR$; $\bsigma$ is the vector of Pauli matrices. We use the same symbol $H_0$ for both Hamiltonians and we assume that expressions with $H_0$ apply to both types of Hamiltonians.
Each eigenstate of $H_0$ can be labeled by an integer $j$ and has the form of a Bloch wave
\e{\psi_{j,\bk} =\eikr u_{j,\bk},\la{Bloch}}
where $u_{j,\bk} = u_{j,\bk}(\br)$ in the Schrodinger case, and $=u_{j,\bk}(\br,s)$ with additional spin index $s$ in the Pauli case.
In both cases, $u_{j,\bk}$ is periodic with respect to Bravais-lattice translations $\br \rightarrow \br +\bR$, and shall henceforth be referred to as cell-periodic functions; they form an orthonormal set for each $\bk$ which is complete with respect to periodic functions. The Hamiltonian acts on cell-periodic functions as
\e{ H_0(\bk) = e^{-i\bk \cdot \br}H_0e^{i\bk \cdot \br}; \la{BlochHam}}
we will refer to $H_0$ as the Hamiltonian, and $H_0(\bk)$ as the Bloch Hamiltonian.\\

A cell-periodic function can be expanded as
\e{ \braket{\alpha| u_{j,\bk}}_{\text{cell}} \eq u_{j,\bk}(\alpha), \as \ket{u_{j,\bk}}_{\text{cell}} = \sum_{\alpha}u_{j,\bk}(\alpha)\ket{\alpha}_{\text{cell}},\lin
  \braket{u_{j,\bk}| \alpha}_{\text{cell}} \eq \bar{u}_{n\bk}(\alpha), \as \bra{u_{j,\bk}}_{\text{cell}} = \sum_{\alpha} \bar{u}_{n\bk}(\alpha) \bra{\alpha}_{\text{cell}},\la{expandcellperiodic}}
where $\alpha$ is a shorthand for either $\bt$ or $(\bt,s)$ in the Pauli, with $s$ a spin index, and $\bt$ the cell-periodic position coordinate that is defined with the equivalence $\bt \sim \bt +\bR$ ($\bR$ being a Bravais-lattice vector). We define $\sum_{\alpha}{=}\calv \int_{\text{unit cell}} \mathrm d\bt$, where $\calv$ is the volume of the BZ,\cite{Fiorenza2015} and there is an additional sum over the spin index $s$ for spinor wavefunctions. The inner product over one unit cell is then defined as
\e{ \braket{u| u'}_{\text{cell}} = \sum_{\alpha} \bar{u}(\alpha)u'(\alpha), \as \braket{\alpha| \beta}_{\text{cell}} = \delta_{\ab},}
where $\delta_{\ab}$ is a shorthand for the product of a Dirac delta function in real space and a Kronecker delta function in spin space, if present. We remark that we can also interpret $\alpha$ as a discrete label for a basis of \low orbitals in tight-binding models.\cite{Slater1954,Lodwin1950}\\

\subsection{Electromagnetic gauge freedom}\la{app:gauge}
A particle in a crystalline potential $V$ and in a spatially homogeneous electric field $\bE(t)$ satisfies the Schr\"odinger equation with Hamiltonian
\e{ H(t) = \frac{1}{2m} \big( \bp- e \bA(\br,t) \big)^2 + V(\br). \la{temporal}}
This Hamiltonian is in the temporal gauge, so called because the vector potential $\bA(t) = - \int_t \bE(t') \mathrm dt'$ is time-dependent and the scalar potential is zero. The Hamiltonian in the scalar gauge reads as 
\e{ H'(t) &= \frac{\bp^2}{2m} + V(\br) - e \bE(t) \cdot \br, \la{scalar}}
which can be obtained from \q{temporal} by a gauge transformation $G(\br,t) = \mathrm{e}^{-i e \bA(t) \cdot \br}$.
When we discuss systems with spin-orbit coupling, $\bp$ is replaced by $\bp + \f{\hbar}{4 m c^2} \bsigma \times \nabla V$ \big(see \q{fieldfreepauli}\big). We assumed the term $\f{|\bF|}{c^2}$ is negligible,\cite{Winkler2003} then spin-orbit coupling does not break spatial symmetries.\\

\subsection{Symmetries in the Bloch Hamiltonian}\la{app:Blochsymm}
In addition to translational invariance, a solid can have point-group symmetries as well as symmetries that reverse time. Let us discuss how they act on Bloch functions and then specialize to nonsymmorphic symmetries.\\

\subsubsection{Generalities about space groups}\la{app:symm}
The set of all symmetries of a solid is defined as the (magnetic) space group $G$. Any spatial symmetry $g \in G$ has the form $g = (\check g|\bt)$ where $\check g$ is the point group component of $g$ and $\bt$ is a translation. For application to half-integer spin representations, we define $\bar e$ as a $2\pi$-rotation that is not equal to the identity, $\bar e \neq e$, but $\bar e^2 = e$. \\

Lattice translations in $d=1,2,3$ dimensions are denoted by $\mathcal{T} \cong \mathbb{Z}^d$ and all other symmetries are elements of the point group $\calp=G/\mathcal{T} = \{ [g_1], ..., [g_M] \}$ with $[g_n] = g_n \mathcal{T}$ a coset. $\calp$ is isomorphic to another point group, which consists of symmetries that leave a point in space invariant. In our definition, point group symmetries are more generally representatives $g_n$ in $\calp$, which can contain fractional, or even full lattice translations. A space group $G$ is symmorphic if there exists an origin that is preserved by all point group elements, otherwise it is called nonsymmorphic. The noncommutative multiplication rule for two elements $g_1=(\check g_1|\bt_1), g_2=(\check g_2|\bt_2) \in G$ and the inverse of $g_1$ are given by
\e{ g_1 g_2 &= (\check g_1|\bt_1) (\check g_2|\bt_2) = (\check g_1 \check g_2| \check g_1 \bt_2 + \bt_1),\la{groupmult} \\ 
g_1^{-1} &= (\check g_1^{-1}| -\check g_1^{-1} \bt_1).\la{inverse}}\\

For magnetic space groups we also consider $g$ that reverses time. The time reversal operation $g=T$ acts trivially on space ($\check{g}=e, \bt=0$), and is represented by $\hat{T}=U_T K$, with $U_T$ a unitary transformation and $K$ the complex conjugation operation; $\hat{T}^2=(-1)^{\sma{F}}$, where $F=0$ for integer-spin representations ($U_T=e$), and $F=1$ for half-integer-spin representations ($U_T=-i\sigma_y$ in spinor space). It is useful to introduce a $\Z_2$-index $s$ that distinguishes between transformations which are purely spatial (and therefore have a unitary representation $\hat{g}$), and transformations which involve time reversal, possibly composed with a spatial operation (in which case $\hat{g}$ is antiunitary): 
\begin{align}
s(g)=\begin{cases}0, \as \hat{g} \; \ins{unitary,} \\ 1, \as \hat{g} \; \ins{antiunitary}\end{cases}. \la{definesg}
\end{align}\\
 
For a space group element $g=(\check g|\bt) \in G$ its representation in real space is\cite{MelvinLax1974}
\e{ (g \circ \br)_i = \check{g}_{ij} r_j+t_i, \ass \check{g}^{\mo}=\check{g}^T \in\R, \la{gactsonposition}}
for all $i=1,...,d$, which we shorten notationally as $g \circ \br = \check{g} \br+\bt$.
The linear representation of $G$ on scalar functions $f(\br)$ of $\br$ is defined by
\e{ \hat g f(\br) = f(g^{-1} \circ \br), \as \hat g = \widehat{ (\check g| \bt) }, \as \hat g_1 \hat g_2 = \widehat{g_1 g_2} \la{linrep} }
for $g, g_1, g_2 \in G$. We denote the group element obtained by the group multiplication of $g_1$ with $g_2$ by $g_1 g_2$, as defined in \q{groupmult}. For example, the representation of $g=(\check g|\bt) \in G$ on a plane waves is
\e{ \hat{g} e^{i\bk \cdot \br} = e^{[(-1)^{s(g)}i]\bk \cdot [\check{g}^{-1} (\br-\bt)]}=e^{i[(-1)^{s(g)} \check{g} \bk]\cdot( \br-\bt)}.\la{gactsoneikr2}}
Consequently, a Bloch wave at wavevector $\bk$, when operated upon by $g$, transforms with a possibly distinct wavevector 
\e{ \bk' \equiv (-1)^{s(g)}\check{g} \bk, \as \p{k'_{\alpha}}{k_{\beta}}=(-1)^{s(g)}\check{g}_{\ab}, \la{app:gactsonk2}}
as can be ascertained from \big(\q{Bloch}\big)
\e{\hat{g} e^{i\bk\cdot \br} u_{\bk}(\br) = e^{i\bk' \cdot \br} \hat{g}(\bk) u_{\bk}(\br). \la{gactsonBlochwave}}
Here, we have combined $\hat{g}$ and the phase factor in \q{gactsoneikr2} into
\e{\hat{g}(\bk) \equiv e^{-i(-1)^{s(g)}[\check{g} \bk]\cdot \bt}\hat{g}. \la{sewingg}}
If we are interested in how $\hat g$ acts on $\ket{ u_{\bk} }$ (without the $\bk$-dependent phase factor), we set $\bt = 0$.
\\

If $g$ is a symmetry of the Hamiltonian, meaning that $\hat{g} H_0 \hat{g}^{-1} = H_0$, then, applying \q{BlochHam}, \q{gactsoneikr2} and \q{gactsonBlochwave} gives the equalities
\e{ \hat{g}(\bk) H_0(\bk)\hat{g}(\bk)^{-1} &= \big( e^{-i\bk' \cdot \br} \hat{g} e^{i\bk\cdot \br} \big) \big( e^{-i\bk \cdot \br}H_0 e^{i\bk \cdot \br} \big) \big( e^{-i\bk\cdot \br} \hat{g}^{-1} e^{-i\bk' \cdot \br} \big) = e^{-i\bk' \cdot \br} \hat{g} H_0 \hat{g}^{-1} e^{-i\bk' \cdot \br}\lin
&= e^{-i\bk' \cdot \br} H_0 e^{i\bk' \cdot \br} = H_0( \bk' ) = H_0\big( (-1)^{s(g)}\check{g} \bk \big). \la{Hamsymm}}
This implies that if $\ket{u_{\bk}}$ is an eigenstate of $H_0(\bk)$ with eigenvalue $E(\bk)$, then ${\hat{g}(\bk)}\ket{u_{\bk}}K^{s(g)}$ is an eigenstate of $H_0\big( (-1)^{s(g)}\check{g} \bk \big)$ with the same eigenvalue $E(\bk)$.\\

\subsubsection{Sewing matrix}\la{app:sew}
For a subspace spanned by a quasi-periodic basis $\{ \ket{ u_{j,\bk} }_{\text{cell}} \}_{j=1}^N$, for which $u_{j,\bk+\bG}(\br) = \mathrm{e}^{-i \bG \cdot \br} u_{j,\bk}(\br)$ for all reciprocal lattice vectors $\bG$, and that is closed under the action of $g= (\check g| \bt) \in G$, its unitary matrix representation is
\e{ \breve{g}(\bk)_{i,j} = \braket{ u_{i,(-1)^{s(g)} \check{g} \bk}| \hat{g}(\bk) | u_{j,\bk} }_{\text{cell}} K^{s(g)}. \la{app:gactsonu2}}
We refer to this unitary matrix colloquially as the sewing matrix,\cite{Fang2013a} owing to its function in `sewing' together the cell-periodic functions by symmetry. The corresponding Bloch basis \big(\q{Bloch}\big) and the sewing matrix are both periodic in $\bk$. At $\bk$ which is $\check g$-invariant, defined as $\check g \bk = \bk + \bG$ for a reciprocal lattice vector $\bG$, the eigenvalues of $\breve g(\bk)$ are invariant under basis transformation of quasi-$\bk$-periodic bases. In the main text we defined $\rho_n(\bk):= \mathrm{e}^{-i F \pi /n} \breve{g}_n(0)$ for $g_n$ with $\nu(g_n)=1$; its advantage compared to $\breve{g}$ is that the set of possible eigenvalues of $\rho_n$ does not depend on $F$ (discussed in the paragraph after next). \\

For each $g_n \in P$ that is not in $[(e|0)]=\mathcal{T}$, we define the order of $g_n$ as the smallest integer $n$ such that
\e{ g_n^n = \big( \bar e^{\nu(g_n)}| \bR(g_n) \big), \ass \bR(g_n) = p \ba \la{nonsymm} }
with $\ba$ the shortest primitive Bravais lattice vector such that the translational part of $g_n$ is parallel to $\ba$, $p \in \Z_n$ and $\nu(g_n)$ a $\Z_2$-index. For example, $\nu(C_n) = 1$ for an $n$-fold rotation $C_n$, $\nu(M) = 1$ for a reflection $M$ and $\nu(\cali)=0$ for inversion $\cali$. For symmorphic transformations there exists an origin such that $p=0$ and $\check g_n \ba = \ba$. For nonsymmorphic transformations $p\neq 0$ for any choice of origin; there exists an origin such that $\bR(g_n)$ is parallel to the axis of rotation if $g_n$ is a screw, or lies in the plane of reflection if $g_n$ is a glide (\cite{Evarestov}, Chapter 3.1.5). Our convention is that `screw' is short-hand for screw rotation and `glide' for glide reflection; we use `rotation' to mean a screwless rotation, and `reflection' a glideless reflection.\\

The eigenvalues of $\breve g_n(\bk)$ at a $\check g_n$-invariant $\bk$ have the form 
\e{ \omega_{j,n}(\bk) = \mathrm{e}^{-i p \bk \cdot \ba/n} \mathrm{e}^{i \pi \nu(g_n) F/n} \omega_{j,n} = \mathrm{e}^{-i p \bk \cdot \ba/n} \mathrm{e}^{i \pi(\nu(g_n) F+ 2 j p)/n} \la{sewingeval} }
for some $j=0,...,n-1$ and $\omega_{j,n}=\mathrm{e}^{2\pi i j p/n}$. This follows from $\breve g_n^n(\bk) = (-1)^{\nu(g_n) F} \mathrm{e}^{-i p \bk \cdot \ba}$. If we do not specify the order $n$ of $g$, we write $\omega_j(\bk)$; if $F=0$, $\omega_{j,n}(0) = \omega_{j,n}$ are the possible eigenvalues of $\rho_n$. \\

\subsubsection{Nonsymmorphic symmetries}\la{app:nonsymm}
For all nonsymmorphic symmetries $g$ there exists an origin, such that $\ba$ in \q{nonsymm} is one of the primitive lattice vectors. There always exists a $g$-invariant line (or plane), wherein each wavevector is invariant under the screw (or glide) $g$: let $\ba_1:=\ba$ and $\bb_1$ be the primitive dual vector to $\ba_1$, which we believe is either parallel to $\ba_1$ (for $g$ a screw) or lies in the glide plane (for $g$ a glide). Then the line $\bk(t)=\bk_0+t\bb_1$ with $t \in [0,1]$ is $g$-invariant for any $g$-invariant $\bk_0$. 
Our belief originates from the following observation: given a primitive basis $\{ \ba_i \}_i$, the primitive dual basis $\{ \bb_j \}_j$ is uniquely defined through $\bb_j \cdot \ba_i = 2\pi \delta_{i,j}$. Since $\bb_1$ is primitive, also $\check g^{-1} \bb_1$ is primitive, thus $2\pi = \bb_1 \cdot \ba_1 = [\check g \bb_1] \cdot [\check g \ba_1]=[\check g \bb_1] \cdot \ba_1$ (the second equality uses that $\check g$ does not change the scalar product); this implies that $\check g \bb_1 = \bb_1+n_2 \bb_2 + n_3 \bb_3$ where $n_2, n_3 \in \{0,\pm 1\}$. In all cases with screw symmetry that we studied, one can choose $\ba_2, \ba_3$ orthogonal to $\ba_1$, thus $n_2=n_3=0$; for glide, we could always choose $\ba_3$ orthogonal to $\ba_1$ and $\ba_2$, thus $n_3=0$ but in general $n_2 \neq 0$, which can be seen from the following example. For the glide reflection ($n=2$) with $\ba = \ba_1 + \ba_2$ and $\ba_1, \ba_2$ primitive, we note that also $\ba$ and $\ba_1$ are primitive. The dual vector to $\ba$ is $\bb = \bb_2$, which is invariant under reflections, but not parallel to $\ba$. \\

For the nonsymmorphic symmetry $g$ characterized by $n$ and $p$ (as in \q{nonsymm}), the possible eigenvalues $\omega_{j}(\bk)$ \big(defined in \q{sewingeval}\big) of $\breve g(\bk)$ are continuous in $\bk$ and cannot be made periodic/single-valued over one BZ:
\e{ \omega_{j}(\bk+\bb) = \mathrm{e}^{-i \bb \cdot \ba p/n} \omega_{j}(\bk) = \mathrm{e}^{-2\pi i p/n} \omega_{j}(\bk) = \omega_{j-1}(\bk) \neq \omega_{j}(\bk). \la{monodromy}}
Consequently, every eigenvalue in $\{ \mathrm{e}^{2\pi i j_0/n} \omega_{0}(\bk-j\bb)= \mathrm{e}^{2\pi i j_0/n} \omega_{j}(\bk)\}_{j=0}^{N-1}$ must have the same degeneracy, and its eigenfunctions $\{ \ket{u^{\alpha_{j_0}}_{j_0,j,\bk}}_{\text{cell}} \}_{\alpha_{j_0}}$ cannot be periodic over one BZ. This is captured by a $J\times J$ unitary transformation $c_{(j)}^{j'_0,\beta_{j'_0};j_0,\alpha_{j_0}} \in \mathbb{C}$ that relates $\{ u^{\alpha_{j_0}}_{j_0,j,\bk+\bb}(\br) \}_{\alpha_{j_0}}$ to $\{ \mathrm{e}^{-i \bb \cdot \br} u^{\alpha_{j_0}}_{j_0,j-1,\bk}(\br) \}_{\alpha_{j_0}}$ and satisfies $\sum_{j_0, j'_0, \alpha_{j_0}, \beta_{j'_0}} | c_{(j)}^{j'_0,\beta_{j'_0};j_0,\alpha_{j_0}}|^2 = 1$ for all $j \in \Z_{n}$; equivalently stated, there exists a transition matrix
\e{ \braket{ u^{\beta_{j'_0}}_{j'_0,j',\bk_0}| \mathrm{e}^{i \bG \cdot \br}| u^{\alpha_{j_0}}_{j_0,j,\bk_0+\bb} }_{\text{cell}} = \delta_{j',j-1} c_{(j)}^{j'_0,\beta_{j'_0};j_0,\alpha_{j_0}}, \ass c_{(j)}^{\dagger} = c_{(j)}^{-1}. \la{trans}}
Along the $g$-invariant line, there exist simultaneous eigenstates $\ket{ u^{\alpha_{j_0}}_{j_0,j,\bk} }_{\text{cell}}$ of the sewing matrix $\breve g(\bk)$ and the Bloch Hamiltonian $H_0(\bk)$. Consequently, \q{monodromy} also implies permutations of energy bands along the $g$-invariant line. This example of monodromy has been formulated for elementary band representations,\cite{Michel1999} but we point out that it applies more generally to $g$-invariant subspaces which are gapped over the entire $g$-invariant line.\\

For a $g$-invariant subspace that is gapped and of dimension $N$ for each $\bk$ over the entire $g$-invariant line, each eigenvalue $\omega_{j_0-j}(\bk)$, and therefore also the corresponding functions $\ket{ \psi^{\alpha_{j_0}}_{j_0,j,\bk} }$ (related to $\ket{ u^{\alpha_{j_0}}_{j_0,j,\bk} }_{\text{cell}}$ by \q{Bloch}), are periodic over $\Lambda$ BZ's where
\e{ \Lambda = \f{n}{\mathrm{gcd}(p,n)}. \la{order} }
If $\Lambda=n$ then all symmetry eigenvalues of a subspace at $\bk$ are given by $\{ \omega_{-j}(\bk) \}_{j\in \Z_{\Lambda}}$, thus all its $\Lambda$ energy bands are necessarily connected. If $\Lambda<n$, the subspace may contain another symmetry eigenvalue $ \mathrm{e}^{2\pi i j_0/n} \omega_{0}(\bk)$ that is not contained in this set, but then the subspace must also contain the eigenvalues $\{  \mathrm{e}^{2\pi i j_0/n} \omega_{-j}(\bk) \}_{j\in \Z_{\Lambda}}$. This shows that for a subspace with $N$ bands,
\e{ \exists J \in \mathbb{N} \colon \as N=J \Lambda. \la{nonsymmdeg} }
Its $N$ symmetry eigenvalues are of the form $\{  \mathrm{e}^{2\pi i j_0/n} \omega_{-j}(\bk) \}_{j_0,j}$ for some $j_0 \in \Z_p$ and all $j \in \Z_{\Lambda}$.\\

We prove that $\omega_{j}(\bk)$ is periodic over $\Lambda$ BZ's by finding that the smallest integer $m$ such that $\f{p}{n} m \in \mathbb{N}$ is given by $m = \Lambda$. 

\noindent \textit{Proof.} The greatest common divisor of $p$ and $n$, $\mathrm{gcd}(p,n)$, satisfies $p=\mathrm{gcd}(p,n) m_1, n = \mathrm{gcd}(p,n) m_2$ with $m_1, m_2 \in \mathbb{N}$ and $\mathrm{gcd}(m_1,m_2)=1$. Then we rewrite $\f{p}{n} m = \f{m_1}{m_2} m$; this can only be an integer if $m$ divides $m_2$. The smallest such integer is indeed $m = \f{n}{\mathrm{gcd}(p,n)} = \Lambda$.\hfill\(\Box\)\\

Next, we prove that for every symmetry-eigenbasis $\{ \ket{ u^{\alpha_{j_0}}_{j_0,j,\bk_0} }_{\text{cell}} \}_{j_0, \alpha_{j_0}, j}$ over the base point $\bk_0$, their parallel-transported states $\ket{ \hat u^{\alpha_{j_0}}_{j_0,j,\bk} }_{\text{cell}}$ are orthogonal at each $\bk$ and are eigenstates of the sewing matrix with eigenvalue $ \mathrm{e}^{2\pi i j_0/n} \omega_{j}(\bk)$. 
The parallel-transported states are defined by\cite{Alexandradinata2014c}
\e{ \ket{ \hat u^{\alpha_{j_0}}_{j_0,j,\bk} }_{\text{cell}} = \hat \W(\bk \leftarrow \bk_0) \ket{ u^{\alpha_{j_0}}_{j_0,j,\bk_0} }_{\text{cell}} \la{PTstate}}
with unitary $\hat{\W}(\bk \leftarrow \bk_0)$ that is the path-ordered product of projection operators $\frakP(\bk)$ \big($\frakP(\bk)$ is given explicitly in \q{Proju}\big)
\e{ \hat{\W}(\bk \leftarrow \bk_0) = \prod_{\bq\colon \bk \leftarrow \bk_0} \frakP(\bq). }

\noindent \textit{Proof.} $\ket{ \hat u^{\alpha_{j_0}}_{j_0,j,\bk} }_{\text{cell}}$ is continuously differentiable in $\bk$, hence its eigenvalue $ \mathrm{e}^{2\pi i j_0/n} \omega_{j}(\bk)$ cannot discontinuously change to a different $ \mathrm{e}^{2\pi i j'_0/n} \omega_{j'}(\bk)$ by varying $\bk$. Parallel-transported states with different eigenvalues are automatically orthogonal, because states in different eigenspaces are orthogonal. Within the same eigenspace, the parallel-transported states $\ket{ \hat u^{\alpha_{j_0}}_{j_0,j,\bk} }_{\text{cell}}, \ket{ \hat u^{\beta_{j'_0}}_{j'_0,j,\bk} }_{\text{cell}}$ are also orthogonal:
\e{ \braket{ \hat u^{\beta_{j'_0}}_{j'_0,j,\bk}| \hat u^{\alpha_{j_0}}_{j_0,j,\bk} }_{\text{cell}} = \braket{ u^{\beta_{j'_0}}_{j'_0,j,\bk_0}| \hat \W^{\dagger}(\bk \leftarrow \bk_0) \hat \W(\bk \leftarrow \bk_0)| u^{\alpha_{j_0}}_{j_0,j,\bk_0} }_{\text{cell}} = \braket{ u^{\beta_{j'_0}}_{j'_0,j,\bk_0}| u^{\alpha_{j_0}}_{j_0,j,\bk_0} }_{\text{cell}} = 0, \la{PTortho}}
where we used the unitarity of $\hat \W(\bk \leftarrow \bk_0)$. \hfill\(\Box\)\\

For a subspace with $N=J\Lambda$ bands, additional symmetries might increase the number of connected bands from $\Lambda$ to a multiple of $\Lambda$.
We now prove that for half-integer spin representations with time-reversal symmetry the connectivity is always $2\Lambda$ instead of $\Lambda$; we will refer to this as $J$ even in \q{nonsymmdeg}.

\noindent \textit{Proof.} Irreducible representations at time reversal-invariant points, i.e., $-\bk_* = \bk_*+\bG$ for a reciprocal lattice vector $\bG$, are real and even-dimensional (Kramers pairs), hence the total number of bands must be even, $N = 2 N'$. Along the $\check g$-invariant line there exists at least one time reversal-invariant point $\bk_*$, which we set to be the base point $\bk_0 = \bk_*$ for parallel transport. The statement follows from the claim that \q{nonsymmdeg} holds for $N'$, i.e., $N' = J' \Lambda$ for a $J' \in \mathbb{N}$. To show this, we first apply \q{nonsymmdeg} to $N$, i.e., $N=J \Lambda$ for a $J \in \mathbb{N}$. For all $\bk$ on the $g$-invariant line, let $\{ \ket{ \hat u^{\alpha_{j_0}}_{j_0,j,\bk} }_{\text{cell}} \}_{j_0, \alpha_{j_0}, j}$ \big(defined in \q{PTstate}\big) be the symmetry-eigenbasis obtained from parallel-transport. Then we can iteratively construct a new orthonormal symmetry-eigenbasis $\{ \ket{ \hat u'^{\alpha_{j_0}}_{j_0,j,\bk} }_{\text{cell}}, \ket{ \hat u'^{\alpha_{j_0}}_{\bar j_0,j,\bk} }_{\text{cell}} \}_{j_0, \alpha_{j_0}, j}$, where the number of different $j_0$'s and $\alpha_{j_0}$'s is $J' = J/2$, as follows: 
let $\ket{ \hat u'^{\alpha_1}_{1,j,\bk} }_{\text{cell}} = \ket{ \hat u^{\alpha_1}_{1,j,\bk} }_{\text{cell}}$ and define $\ket{ u^{\alpha_1}_{\bar 1, j,\bk_*} }_{\text{cell}} = \hat T \ket{ u^{\alpha_1}_{1,j,\bk_*} }_{\text{cell}}$, such that $\braket{ u^{\alpha_1}_{\bar 1, j,\bk_*}| u^{\alpha_1}_{1,j,\bk_*} }_{\text{cell}} = 0$. Its parallel-transported state $\ket{ \hat u'^{\alpha_1}_{\bar 1,j,\bk} }_{\text{cell}}$ is orthogonal to $\ket{ \hat u'^{\alpha_1}_{1,j,\bk} }_{\text{cell}}$ because of a similar calculation to \q{PTortho}. Secondly, if there exists a $\alpha_2$, using $\braket{ u^{\alpha_2}_{1,j,\bk_*} | u^{\alpha_1}_{1,j,\bk_*} }_{\text{cell}} = 0$, and Gram-Schmidt orthogonalization for $\ket{ u^{\alpha_2}_{1,j,\bk_*} }_{\text{cell}}$, result in a $\ket{ u'^{\alpha_2}_{1,j,\bk_*} }_{\text{cell}}$ that is also orthogonal to $\ket{ u^{\alpha_1}_{\bar 1,j,\bk_*} }_{\text{cell}}$. Similarly, we define the parallel-transported states $\ket{ \hat u'^{\alpha_2}_{1,j,\bk} }_{\text{cell}}, \ket{ \hat u'^{\alpha_2}_{\bar 1,j,\bk} }_{\text{cell}}$ by first Gram-Schmidt orthogonalizing them, and then using \q{PTstate}. The construction of $\ket{ \hat u'^{\alpha_{j_0}}_{j_0,j,\bk} }_{\text{cell}}$ for $j_0 > 1$ works exactly the same way. This results in the desired basis after a finite number of steps and finishes the proof.\hfill\(\Box\)\\

\subsection{Review of Wannier functions, Bloch functions and band representations}\label{app:band}
Wannier functions are commonly used as a spatially-localized basis for (i) tight-binding models, as well as for (ii) trivial energy subspaces that are gapped over the entire BZ, and are trivial in the sense of having vanishing Chern number. Both (i) and (ii) correspond to a (complex) vector space spanned by Wannier functions $\{ w_{n,\bR}^{\alpha_n} \}_{n,\alpha_n,\bR}$ that are localized at the Wannier centers
\e{ \bvarpi_n + \bR = \braket{ w_{n,\bR}^{\alpha_n}| \br | w_{n,\bR}^{\alpha_n} } = \int_{\mathbb{R}^d} \br |w_{n,\bR}^{\alpha_n}|^2 \mathrm d\br, \la{center}}
where $n$ labels the different Wannier centers in one unit cell, $\alpha_n$ labels different Wannier functions localized at $\bvarpi_n$ (in general, the number of Wannier functions localized at $\bvarpi_n$ varies with $n$), $\bR$ is a Bravais lattice vector and $d$ is the spatial dimension. Equivalently, (i) and (ii) is spanned by Bloch functions, defined by
\e{ \tilde \psi_{n,\bk}^{\alpha_n} (\br) = \f{1}{\sqrt{\calv}} \sum_{\bR} \mathrm{e}^{i \bk \cdot \bR} w_{n,\bR}^{\alpha_n}(\br), \la{Blochdef}}
which are periodic and smooth in $\bk \in BZ$, and where $\calv$ is the volume of the BZ.
The tight-binding basis or the trivial gapped energy subspace is a band representation, or colloquially an atomic band insulator, if for all $n,\bR$ the Wannier functions $\{ w_{n,\bR}^{\alpha_n} \}_{\alpha_n}$ form a representation of the little group of $\bvarpi_n + \bR$ (defined in the next paragraph); this little group may include time-reversal symmetry. \\

Band representations describe concisely how Wannier and Bloch functions transform under symmetries of a space group $G$: at a single position $\bvarpi$ within one unit cell, we specify the symmetry representation $V$ of the little group $\calp_{\varpi}$ on Wannier functions $\{ w_{1,0}^{\alpha} \}_{\alpha=1}^{\mathrm{dim}(V)}$, and then deduce how symmetries act on Bloch functions at all $\bk$ in the BZ.
The little group (or stabilizer subgroup) $\calp_{\varpi}$ consists of all $h\in G$ that leave $\bvarpi$ invariant, $h \circ \bvarpi = \bvarpi$ \big(this action was defined in \q{gactsonposition}\big). The little group $\calp_{\varpi}$ can have less symmetries than the point group $\calp = G/\mathcal{T}$ (where $\mathcal{T} = \{ (e|\bR); \bR \text{ is a Bravais lattice vector}\}$ is the group of translations), in which case the multiplicity, defined as the index of $\calp_{\varpi}$ in $\calp$
\e{ M = \f{ |\calp| }{ |\calp_{\varpi}|, } \label{app:mult}}
is bigger than one. 
The construction of the band representation is summarized as follows: utilizing representatives $g_n \in \calp$ of the set of left cosets, 
\e{ \calp/\calp_{\bvarpi=0} = \{ [g_1=(e|0)],...,[g_M] \}, \ass \calp = \cup_{n=1}^M g_n \calp_{\varpi=0} \la{Pcoset} }
with $[g_n] = g_n \calp_{\varpi=0}$, and Bravais lattice translations $(e|\bR) \in \mathcal{T}$, define symmetry-related Wannier functions 
\e{ w_{n,0}^{\alpha}(\br) = \hat{g}_n w_{1,0}^{\alpha}(\br) = w_{1,0}^{\alpha}( g_n^{-1} \circ \br), \as w_{n,\bR}^{\alpha}(\br) = \widehat{ (e|\bR) } w_{n,0}^{\alpha}(\br) = w_{n,0}^{\alpha}(\br-\bR) \la{symmBR} }
that are centered at $\bvarpi_n = g_n \circ \bvarpi$ and $\bvarpi_n + \bR$, respectively, for all $\alpha=1,...,\mathrm{dim}(V)$. We use the convention $g_1 = (e|0)$ such that $\bvarpi_1 = \bvarpi$. \q{center} follows from this construction, but $\braket{ w_{n',\bR'}^{\beta}| \br | w_{n,\bR}^{\alpha} }$ is generally nonzero for $n \neq n'$ or $\alpha \neq \beta$. A symmetry $g = (\check{g}| \bt) \in G$ \big(using the notation of \q{gactsonposition}\big) maps a Wannier function $w_{n,\bR}^{\alpha}$ localized at $\bvarpi_n + \bR$ onto a Wannier function $w_{n',\check g \bR + \bR_{n',n}(g)}^{\alpha}$ localized at $\bvarpi_{n'} + \check g \bR + \bR_{n',n}(g)$ with a Bravais lattice vector $\bR_{n',n}(g)$ defined by
\e{ \bR_{n',n}(g) = g \circ \bvarpi_n - \bvarpi_{n'} = \check g \bvarpi_n + \bt - \bvarpi_{n'} = \bR_{n',n}(\check g) + \bt. \la{app:translationvec}}
The Wannier functions transform under a space group symmetry $g\in G$ as
\e{ \hat g \ket{ w_{n,\bR}^{\alpha} } = \ket{ w_{n',\check{g} \bR + \bR_{n',n}(g) }^{\beta} } V( \check g_{n'}^{-1} \check g \check g_n )_{\beta,\alpha}, \la{WannierBR}}
as derived in \app{BRconstruction}. Equivalently,
\e{ V( \check g_{n'}^{-1} \check g \check g_n )_{\beta,\alpha} = \braket{ w_{n',\check{g} \bR + \bR_{n',n}(g) }^{\beta}| \hat g | w_{n,\bR}^{\alpha} }. \la{unirrep} }
The symmetry $g$ acts on Bloch functions by
\e{ \hat g \ket{ \tilde \psi_{n,\bk}^{\alpha} } = \mathrm{e}^{- i [\check{g} \bk] \cdot \bR_{n',n}(g)} \ket{ \tilde \psi_{n', \check{g} \bk}^{\beta} } V(\check g_{n'}^{-1} \check g \check g_n)_{\beta,\alpha}, \la{BlochBR}}
therefore the sewing matrix \big(\q{app:gactsonu2}\big) reads as
\e{ \breve g(\bk)^{\beta,\alpha}_{n',n} = \mathrm{e}^{- i [\check{g} \bk] \cdot \bR_{n',n}(g)} V(\check g_{n'}^{-1} \check g \check g_n)_{\beta,\alpha}. \la{app:sewingBR} }\\

\subsubsection{Equivalent band representations}\la{app:equiv}
Two band representations with Wannier functions $\{ w_{n,\bR}^{\alpha} \}_{n,\alpha,\bR}, \{ W_{m,\bR'}^{\beta} \}_{m,\beta,\bR'}$ are equivalent (or isomorphic representations) if there exists a unitary transformation $T$, a Bravais lattice vector $\bS$ and $n,m$ with $N_n = N_m$ such that for all Bravais lattice vectors $\bR$
\e{ \ket{ w_{n,\bR}^{\alpha_n} } = \sum_{\beta_m} T_{n,m; \bS}^{\alpha_n,\beta_m} \ket{ W_{m,\bR-\bS}^{\beta_m} } \la{app:repeqw} }
(we write the sum over $\beta_m$ explicitly to clarify that there is no sum over $m$ and $\bS$). The transformation $T$ describes combinations of the following three operations: (i) on-site basis transformations, (ii) a different choice of `equivalent' initial Wannier center $\bvarpi$ (defined shortly as a Wyckoff position) resulting in permuted labels $n,m$, and (iii) a different choice of unit cell such that the Wannier functions are all shifted by a Bravais lattice vector $\bS$.
It is important to notice that our definition of equivalence is for infinite crystals and different from the one used for finite crystals with Born-von Karman boundary conditions.\cite{Bacry1988} For the space group $F222$ (No. 22) it was observed that if two band representations are inequivalent for a finite crystal, then they are inequivalent for an infinite crystal, but not necessarily the other way around.\cite{Bacry1988a}\\

In terms of the corresponding Bloch functions $\{ \tilde \psi_{n,\bk}^{\alpha_n} \}_{n,\alpha_n,\bk}, \{ \tilde{\Psi}_{m,\bk'}^{\beta_m} \}_{m,\beta_m,\bk'}$, equivalence means that there exists a unitary transformation $T(\bk)$ and some $n,m$ with $N_n = N_m$ such that for all $\bk \in BZ$
\e{ \ket{ \tilde \psi_{n,\bk}^{\alpha_n} } = \sum_{\beta_m} T_{n,m}^{\alpha_n,\beta_m}(\bk) \ket{ \tilde{\Psi}_{m,\bk}^{\beta_m} },\ass T_{n,m}^{\alpha_n,\beta_m}(\bk) = \mathrm{e}^{i \bk \cdot \bS} T_{n,m; \bS}^{\alpha_n,\beta_m}, \la{app:repeq}}
hence by definition $T(\bk)$ is periodic and analytic in $\bk \in BZ$. Therefore, equivalent band representations describe systems in the same symmetry-protected topological phase: there exists an analytic interpolation $T(\bk)$ that preserves all spatial symmetries. On the other hand, for two inequivalent band representations, no such $T(\bk)$ exists, thus the two systems are in different symmetry-protected topological phases. We conclude that a symmetry-protected topological phase of an atomic band insulator is equal to an equivalence class of a band representation.
We note that there are other symmetry-protected topological phases that are not band representations, for example Chern insulators.\\

\subsubsection{Wyckoff positions}\la{app:Wyckoff}
An important concept to find equivalence classes of band representations are Wyckoff positions. A Wyckoff position $[\bvarpi]$ is a set (more precisely, an equivalence class) of points in real space that all have conjugate (in $G$; not only isomorphic) little groups $\calp_{\varpi}$. Every Wyckoff position consists of infinitely many points because of translational symmetry, hence we can restrict (representatives of) Wyckoff positions to the Wigner-Seitz torus. The simplest example of a Wyckoff position is the generic one, which consists of all points that have the identity element in their little group- that is almost every point, or more topologically, a dense and open set in the Wigner-Seitz torus (with the standard topology). A less trivial example can be found in the wallpaper group $p4$ that has two Wyckoff positions with little groups isomorphic to the point group $C_4$, but they are conjugate by a fractional translation that does not lie in $G$, so they are inequivalent Wyckoff positions.\\

All elements of a crystallographic orbit of a point $\bvarpi$, defined by the set $G \circ \bvarpi = \{ g \circ \bvarpi; g \in G \}$, have conjugate little groups and therefore they all belong to the same Wyckoff position $[\bvarpi]$. If a Wyckoff position varies with a parameter, it consists of infinitely many crystallographic orbits. Two representatives $\bvarpi, \bvarpi'$ of the same Wyckoff position $[\bvarpi]$ induce equivalent sets of band representations because conjugate little groups have equivalent sets of unitary irreducible representations. Non-generic Wyckoff positions $[\bvarpi]$ are closed as sets; they can be zero-dimensional, i.e., inversion-symmetric or rotationally-invariant points, one-dimensional, i.e., reflection-invariant or rotationally-invariant lines, or two-dimensional, i.e., mirror planes. 
From now on, we will lighten and abuse notation by referring to $\bvarpi$, a representative of a Wyckoff position $[\bvarpi]$ in the Wigner-Seitz torus, as Wyckoff position itself.\\

\subsubsection{Elementary band representations}
An elementary band representation is a band representation that cannot be decomposed into band representation, and all band representations are obtained from elementary ones (by direct summation). 
It was shown in\cite{Bacry1988} that band representations induced from nongeneric Wyckoff positions and unitary irreducible representations are elementary, except for $40$ special cases that are listed in a table for integer-spin representations (they are real two-dimensional irreducible of four different point groups); this list has recently been extended to half-integer spin representations.\cite{Bradlyn2017,Cano2017} \\

\subsubsection{Construction of elementary band representations}\la{BRconstruction}
\begin{figure}
\centering
\includegraphics[width=15cm]{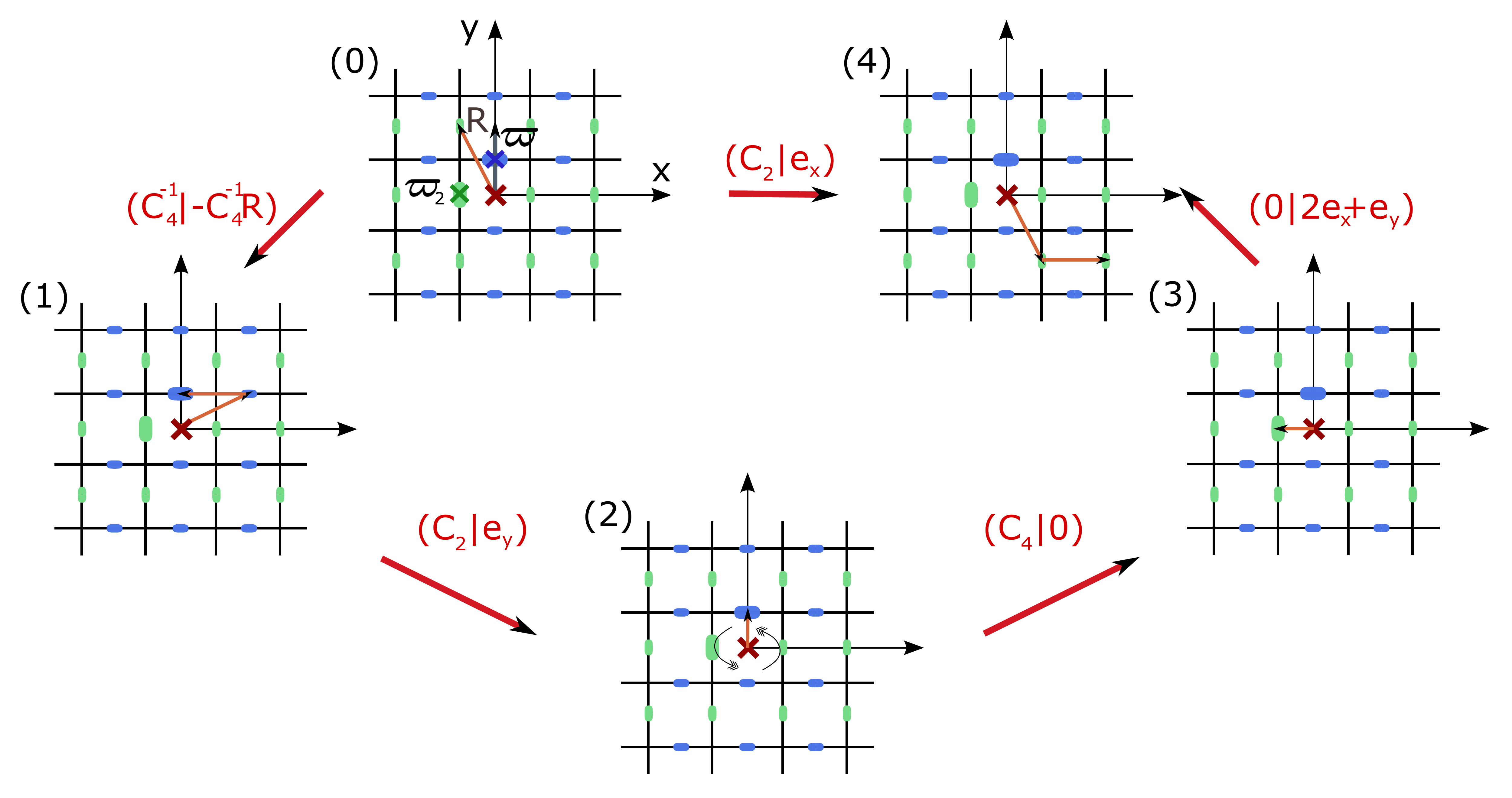}
\caption{Demonstration of the decomposition of $\hat g$ acting on $w_{n,\bR}$, i.e., $\hat g w_{n,\bR} = \widehat{ \big( e|\check g \bR + \bR_{n',n}(g) \big) } \hat g_{n'} \hat h \hat g_n^{-1} \widehat{ (e|-\bR) } w_{n,\bR}$, on the checkerboard lattice with $g = (C_2,\be_x), g_n = g_{n'} = g_2=(C_4|0), h=(C_2|\be_y)$ and $\bR=(0|\be_y)$ \big(grey arrow in (0)\big). The space group is $G=\calt \rtimes C_{4v}$, with $N=1$ orbital per Wannier center. The blobs indicate $C_2$-symmetric Wannier functions.
(0) The origin (red cross) has the full $\calp=C_{4v}$ point group symmetry. The Wyckoff position $\bvarpi$ lies at the blue cross in the enlarged blue blob and has little group $\calp_{\varpi} = \{e,(C_{2}|\be_y)\}$; there are $M=2$ \big(defined in \q{app:mult}\big) symmetry-related Wannier functions per unit cell, indicated by the blue and green blobs. We decompose the transformation of the Wannier function $w_{2,\bR}$ under $g = (C_2|\be_x)$, given by the arrow from (0) to (4), in four steps (in each step the orange arrow points to the transformed Wannier function):
(1) rotate and translate $w_{2,\bR}$ to $w_{1,0}$ ,
(2) apply the little group element $h$ to $w_{1,0}$,
(3) rotate the resulting Wannier function back,
(4) translate the Wannier function to the final position. \la{banconstr} }
\end{figure}

Here we give more details in the construction of an elementary band representation and derive \qq{WannierBR}{BlochBR} for any choice of origin.\cite{Evarestov1984} 

\noindent \textit{Proof.} First, pick a Wyckoff position $\bvarpi$ in the Wigner-Seitz unit cell and note that elements of its little group are generally of the form $h=(\check h|\bs_h) \in \calp_{\varpi}$ with $\check h$ centered at the origin, and the translation 
\e{ \bs_h = \bvarpi-\check h \bvarpi \la{wtransl} } 
is a Bravais lattice vector if the origin is a $h$-invariant point. A $h$-invariant point $\bx$ is invariant under $h$ modulo Bravais lattice translations. Then for a $h$-invariant origin and $h \in \calp_{\varpi}$: $h \circ 0 = \bs_h$ must be a Bravais lattice vector. In the particular case that $\varpi = 0$ then $\bs_h = 0$, and $\calp_{\varpi}$ assumes the canonical form $\calp_{\varpi=0} = \{ (\check h|0); (\check h|\bs_h) \in \calp_{\varpi} \}$. From a unitary irreducible representation $V$ of $\calp_{\varpi=0}$ with orthonormal basis $\{ w^{\alpha} \}_{\alpha=1}^N$, we define a unitary irreducible representation of $\calp_{\varpi}$ for a general choice of origin such that $\bvarpi$ need not be 0: for the orthonormal basis $w_1^{\alpha}(\br) = w^{\alpha}(\br-\bvarpi)$ we define the representation of $h \in \calp_{\varpi}$ by
\e{ \hat{h} w_1^{\alpha}(\br) &= w_1^{\alpha}(\br)(h^{-1} \circ \br) = w_1^{\alpha}(\br)\big(\check h^{-1} (\br - \bs_h)\big) = w_1^{\alpha}(\br)\big(\check h^{-1} (\br - \bvarpi + \check h \bvarpi)\big)\lin 
&= w_1^{\alpha}(\br)\big(\check h^{-1} (\br - \bvarpi) + \bvarpi)\big) = w^{\alpha}\big( \check h^{-1}(\br-\bvarpi) \big) = w^{\beta}(\br-\bvarpi) V(\check h)_{\beta,\alpha}\lin 
&= w_1^{\beta}(\br) V(\check h)_{\beta,\alpha}. \la{pointgr}}
Equivalently,
\e{ V(\check h)_{\beta,\alpha} = \int \bar w_1^{\beta}(\br) \hat{h} w_1^{\alpha}(\br) \mathrm d\br \la{unirrep2}} 
for all $\alpha,\beta =1,...,M$. Note that $V$ only depends on the point-group component $\check h$ of $h$ and not on $\bs_h$. By construction, the Wannier functions $w_1^{\alpha}$ are centered at the Wyckoff position $\bvarpi$,
\e{ \braket{ w^{\alpha}| \br | w^{\alpha} } = 0, \ass \braket{ w_1^{\alpha}| \br | w_1^{\alpha} } = \bvarpi. }\\

Next, generate more Wannier functions utilizing representatives $g_n \in \calp$ defined in \q{Pcoset}, which we choose such that all $\bvarpi_n = g_n \cdot \bvarpi$ lie in the Wigner-Seitz unit cell. With the definition of \q{symmBR}: 
\e{ w_n^{\alpha} (\br) = \hat{g}_n w_1^{\alpha}(\br) = w_1^{\alpha}( g_n^{-1} \circ \br ) = w^{\alpha}( g_n^{-1} \circ \br - \bvarpi ) = w^{\alpha} \big( g_n^{-1} \circ (\br - g_n \circ \bvarpi) \big) }
for all $\alpha=1,...,N$, we find that these Wannier functions are indeed centered at $\bvarpi_n$ (we made the choice $g_1 = (e|0)$ such that $\bvarpi_1 = \bvarpi$ is consistent with our notation). Translation by Bravais-lattice vectors gives additional Wannier functions $w_{n,\bR}^{\alpha}(\br) = w_n^{\alpha}(\br-\bR)$ centered at $\bvarpi_n+\bR$ and finishes the construction of the elementary band representation.\\

The transformation of a Wannier function $w_{n,\bR}^{\alpha}$ for an element $g = (\check{g}| \bt) \in G$ given in \q{WannierBR} is obtained as follows \big(demonstrated in \fig{banconstr}\big): for a Bravais lattice vector $\bR'$ we can use the coset-decomposition \q{Pcoset} for $( e| -\bR' ) g g_n \in \cup_{n'=1}^M g_{n'} \calp_{\bvarpi}$, hence there exists a $g_{n'}$ and a $h=(\check h| \bs) \in \calp_{\varpi}$ such that
\e{ ( e| -\bR' ) g g_n = g_{n'} h, \ass \check{g} \check g_n = \check g_{n'} \check h. \la{decomp}}
Using $g_n = (\check g_n| \bt_n), g_{n'} = (\check g_{n'}| \bt_{n'})$, \q{groupmult} and \q{wtransl}, we identify $\bR' = \bR_{n',n}(g)$ as defined in \q{app:translationvec}. We obtain the representation of $g$ on $w_{n,\bR}^{\alpha}$, that is \q{WannierBR} for $\check h = \check g_{n'}^{-1} \check{g} \check g_n$, through
\e{ \hat g w_n^{\alpha}(\br-\bR) & \stackrel{\eqref{symmBR}}= \hat g \widehat{ (e|\bR) } w_n^{\alpha}(\br) \stackrel{\eqref{symmBR}}= \widehat{ g (e|\bR) } \hat g_n w_1^{\alpha}(\br) \stackrel{\eqref{groupmult}, \eqref{linrep}}= \widehat{(e|\check g \bR)} \hat g \hat g_n w_1^{\alpha}(\br)\lin 
& \stackrel{\eqref{decomp}}= \widehat{(e|\check g \bR)} \widehat{ \big(e|\bR_{n',n}(g)\big) g_{n'} h } w_1^{\alpha}(\br) \stackrel{\eqref{pointgr}}= \widehat{ \big(e|\check{g} \bR +\bR_{n',n}(g)\big) } \hat g_{n'} w_{1}^{\beta}( \br ) V(\check h)_{\beta,\alpha}\lin 
& \stackrel{\eqref{symmBR}}= \widehat{ \big(e|\check{g} \bR +\bR_{n',n}(g)\big) } w_{n'}^{\beta}( \br ) V(\check h)_{\beta,\alpha} \stackrel{\eqref{pointgr}}= w_{n'}^{\beta}\big( \br- \check{g} \bR - \bR_{n',n}(g) \big) V(\check h)_{\beta,\alpha}. }
From \q{decomp} follows $\check h = \check g_{n'}^{-1} \check{g} \check g_n$, such that the last equation is indeed equal to \q{WannierBR}. Note that multiplication with $\bar w_{n'}^{\beta}\big( \br- \check{g} \bR - \bR_{n',n}(g) \big)$ and integration over $\br$ gives \q{unirrep} \big(in comparison to \q{unirrep2}\big). We emphasize that the representation $\hat g$ only maps Wannier functions $w_{n,\bR}^{\alpha}$ onto $w_{n',\bR'}^{\alpha}$ if their symmetries $g_n$ and $g_{n'}$ are related through \q{decomp}.
The representation on Bloch functions, given by \q{BlochBR}, is derived as follows
\e{ \hat g \tilde \psi_{n,\bk}^{\alpha} (\br) &= \f{1}{\sqrt{\calv}} \sum_{\bR} \mathrm{e}^{i \bk \cdot \bR} \hat g w_{n,\bR}^{\alpha}(\br) = \f{1}{\sqrt{\calv}} \sum_{\bR} \mathrm{e}^{i \bk \cdot \bR} w_{n'}^{\beta}\big( \br-\check g \bR - \bR_{n',n}(g) \big) V(\check h)_{\beta,\alpha}\lin
&= \f{1}{\sqrt{\calv}} \mathrm{e}^{- i [\check g \bk] \cdot \bR_{n',n}(g)} \sum_{\bR'} \mathrm{e}^{i [\check g \bk] \cdot \bR'} w_{n'}^{\beta}( \br-\bR' ) V(\check h)_{\beta,\alpha} = \mathrm{e}^{- i [\check g \bk] \cdot \bR_{n',n}(g)} \tilde \psi_{n', \check g \bk}^{\beta} (\br) V(\check h)_{\beta,\alpha}. } \hfill\(\Box\)\\

\subsection{Projection operators and Wilson loops}\la{app:gapped}
An energy subspace is gapped in a region $D \subset BZ$ if: (i) there exist projection operators $P(\bk), Q(\bk)$ of constant rank for all $\bk \in D$ such that $P(\bk) \oplus Q(\bk) = \mathbb{1}$, and (ii) the minimal gap, defined by
\e{ E_G = \inf \{ | \braket{ \psi_{n,\bk}| H_0 | \psi_{n,\bk} }-\braket{ \psi_{\bar n,\bk}| H_0 | \psi_{\bar n,\bk} }|; \bk \in D, \ket{ \psi_{n,\bk} } \text{ in } P(\bk), \ket{ \psi_{\bar n,\bk} } \text{ in } Q(\bk) \}, \la{gap}}
is a positive real number.
An $N$-dimensional gapped energy subspace, is projected by
\e{ P(D) = \int_{D} P(\bk) \mathrm d\bk = \int_{D} \sum_{n=1}^N \ket{ \psi_{n,\bk} }\bra{ \psi_{n,\bk} } \mathrm d\bk; \la{ProjBZ}}
where $\{ \ket{ \psi_{n,\bk} } \}_{n=1}^N$ are Bloch functions.. In the case $D=BZ$ we write $P(BZ) \equiv P$. The projection onto the cell-periodic functions $\{ \ket{ u_{n,\bk} }_{\text{cell}} \}_{n=1}^N$ in the gapped subspace \big(\q{Bloch}\big) is denoted by
\e{ \frakP(\bk) = \sum_{n=1}^N \ket{ u_{n,\bk} }\bra{ u_{n,\bk} }_{\text{cell}}, \ass \frakP(D) = \int_{D} \frakP(\bk) \mathrm d\bk, \ass \frakP(BZ)\equiv \frakP. \la{Proju} }
The inner product $\braket{ u_{m,\bk}| u_{n,\bk} }_{\text{cell}} = \f{1}{\calv} \delta_{n,m}$ should be understood as $\sum_{\alpha} \bar u_{m,\bk}(\alpha) u_{n,\bk}(\alpha)$ with $\alpha$ summing/integrating over one unit cell \big(as already discussed below \q{expandcellperiodic}\big), while $\braket{ \psi_{m,\bk}| \psi_{n,\bk'} } = \delta_{m,n} \delta(\bk-\bk')$ is a distinct inner product over $\mathbb{R}^d$ ($d$ is the spatial dimension), tensored with spin space.\\

We often deal with $U(N)$ basis transformations in the cell-periodic functions in $\frakP(\bk)$:
\e{\ket{u_{n,\bk}}_{\text{cell}}\rightarrow \sum_{m=1}^N \ket{u_{m,\bk}}_{\text{cell}}T_{m,n}(\bk), \ass T^{\mo}=\dg{T}.\la{basistransf}}
We will refer to this as a gauge transformation within $\frakP$. With respect to this transformation, certain quantities are invariant (such as the projection $\frakP$ itself); other quantities transform covariantly, i.e., they change only in being conjugated by the unitary $T$ (e.g. the Wilson loop); even others have a more complicated transformation rule, e.g. the Berry connection transforms as\cite{WilczekFrank1984}
\e{\bA \rightarrow T^{-1}\bA T+iT^{-1}\nabk T. \la{gaugetransform}}\\

Wilson loops for loops $\mathcal{C}$ in the BZ are a generalization of the Berry and Zak phases\cite{Berry1984,Zak1989} for parallel transport in gapped energy subspaces with dimension $N>1$ at each wavevector $\bk$. In the presence of an electric field, the wavevector $\bk$ depends on time $t$ through\cite{Nenciu1980}
\e{ \bk(t) = \bk_0 + \int_{0}^t \bF(t') \mathrm dt'/\hbar \la{kevo}. }
In its gauge-invariant formulation, parallel transport within $\frakP(\bk)$ is generated by the Kato Hamiltonian\cite{Kato1950,Budich2013}
\e{ \mathcal{A}(t) = i [\frakP\big( \bk(t) \big), \mathrm d_t \frakP\big( \bk(t) \big)], \ass \W_{n',n}(t) = \braket{ u_{n',\bk(t)}| \overline{\mathrm{e}}^{i \int_0^{t} \mathcal{A}(t') \mathrm dt'} | u_{n,\bk(0)} }_{\text{cell}} \la{Kato} }
for all $n=1,...,N$ where $\overline{\mathrm{e}}$ denotes the time-ordered exponential. An equivalent definition of the Wilson line can be given with the non-Abelian Berry connection
\e{ \bA(\bk)_{n',n} = \braket{ u_{n',\bk}| i \nabla_k| u_{n,\bk} }_{\text{cell}},\ass \W_{n',n}(t) = \big( \overline{\mathrm{e}}^{i \int_0^{t} \bA\big(\bk(t')\big) \cdot \dot \bk(t') \mathrm dt'} \big)_{n',n}. \la{app:nonAB} }
It is important to note that the non-Abelian Berry connection is defined wrt. to a differentiable (in $\bk$) basis $\{ \ket{u_{n,\bk}}_{\text{cell}} \}_n$. On the other hand, the Kato Hamiltonian does not depend on the choice of basis and assumes only that the projection operator $\frakP(\bk)$ is differentiable in $\bk$. \\

The parametrization $\{ \bk(t) \}_{t \in [0,T_B]}$ describes a loop $\mathcal{C} \in BZ$ if there exists a time $t=T_B$ such that $\bk(T_B) = \bk(0)+\bG$ with $\bG$ being a reciprocal lattice vector (possibly zero). Non-contractible loops ($\bG\neq 0$) exist because the fundamental group of the Brillouin-torus is non-trivial. Quasi-$\bk$-periodicity of the basis $\{ \ket{u_{n,\bk}}_{\text{cell}} \}_n$ ensures that the Wilson loop $\W(T_B)$ transforms covariantly under gauge transformations, hence we assume $u_{n,\bk(T_B)}(\br) = \mathrm{e}^{-i \bG \cdot \br} u_{n,\bk(0)}(\br)$. The remaining gauge ambiguity in the choice of basis $\{ \ket{u_{n,\bk(0)}}_{\text{cell}} \}_n$ at the base point $\bk(0)$, as expressed in \q{basistransf} for $\bk \to \bk(0)$, does not affect the eigenvalues of $\W(T_B)$. Consequently, the Wilson loop may be expanded as
\e{ \W[\mathcal{C}]_{n',n} \equiv \W(T_B)_{n',n} = \braket{ u_{n',\bk(0)}| \mathrm{e}^{i \bG \cdot \br} \overline{\mathrm{e}}^{i \int_{\mathcal{C}} \mathcal{A}^a(\bk) \mathrm dk_a} | u_{n,\bk(0)} }_{\text{cell}} = \big( \overline{\mathrm{e}}^{i \int_{\bk(0)}^{\bk(0)+\bG} A^a(\bk) \mathrm dk_a} \big)_{n',n} \la{app:Wilsonloop}}
where $\overline{\mathrm{e}}$ is now the path-ordered exponential. In the last step, we also used $\mathcal{A}(t) = \mathcal{A}^a(\bk) \dot k_a= i [\frakP\big( \bk(t) \big), \partial_{k_a} \frakP\big( \bk(t) \big)] \dot k_a$ and $\dot k_a \mathrm dt = \mathrm dk_a$. The Wilson loop is `geometric' in the sense that it only depends on the loop $\mathcal{C}$ and not on the Bloch period $T_B$. \\

\section{Weak-field dynamics}\la{app:weakfield}
We consider dynamics induced by a class of time-dependent fields $\bF(t)$ satisfying the following properties: (i) the direction of $\bF(t)$ may switch instantaneously at isolated times, but is otherwise constant. (ii) The magnitude of $\bF$ may slowly depend on time within the time intervals where the direction of $\bF$ is constant. To reiterate (i-ii), we may express $\bF(t)=|\bF_0|\textbf{n}(t)g(\Omega t)$ such that: $\textbf{n}(t)$ is a unit vector whose direction possibly switches at isolated times; $g$ is a smooth, dimensionless, scalar function; $\Omega$ is a characteristic frequency of the field. This class of fields includes the simplest case of a time-independent field, as well as the time-dependent fields that produce bent loops that we consider in the main text. \\ 

Let us consider an initial state 
\e{\Psi(\br,t=0) =\sum_{n=1}^N f_{n}(0)\psi_{n,\bk_0}(\br)}
in the $N$-dimensional subspace projected by $P(\bk_0)$ for a base point $\bk_0$; the acceleration theorem informs us that at later times the state will have wavevector $\bk(t)$ given in \q{kevo}. We define the curve traced out by $\bk(t)$ on the Brillouin torus as $\calc(t)$; by assumption, $\calc(t)$ is straight except possibly at isolated wavevectors-- we shall refer to these wavevectors (if they exist) as the kinks of $\calc$. In our examples discussed in the main text, we define the Bloch period $T_B$ as the time when $\calc(T_B)$ noncontractibly wraps around the torus exactly once; in general, we need not assume that $\calc(t)$ forms a loop. We claim that in the limits of $|\bF_0|$ and $\Omega \rightarrow 0$, $\Psi(\br,t)$ belongs in the subspace projected by $P(\bk(t))$:
\e{ \Psi(\br,t) \rightarrow \sum_{n = 1}^N f_{n}(t) \psi_{n,\bk(t)}(\br),\as f_n(t)=\sum_{n'=1}^N U(t)_{n,n'}f_{n'}(0), \la{Ansatz}}
with $\{f_n(t)\}$ related to $\{f_n(0)\}$ by the unitary propagator:
\e{ U(t)_{n,n'} \eq \delta\big( \bk-\bk(t) \big) \bigg( \overline{\mathrm{exp}}[i \int_{0}^{t} \big[\;\bA (\bk(t')) \cdot \bF(t')- \mathfrak{E}(\bk(t'))\;\big] dt'/\hbar] \bigg)_{n,n'} \la{BerrySol} \\
\eq \braket{\psi_{n,\bk}| \overline{\mathrm{exp}}[ i\int^t_0\,\bigg( \bF(t')\cdot P(\calc)\br P(\calc)-P(\bk(t'))H_0P(\bk(t')) \bigg) \,dt'/\hbar] | \psi_{n',\bk_0}}, \la{adiabaticevo}}
$\overline{\mathrm{exp}}$ denotes the time-ordered exponential, $\bA$ the non-Abelian Berry connection \big(\q{app:nonAB}\big), $P(\mathcal{C})$ the integral of $P(\bk)$ over $\bk \in \mathcal{C}$ \big(defined in \q{ProjBZ}\big), and $E(\bk)$ an $N\times N$ matrix defined by
\e{ \mathfrak{E}(\bk)_{n,n'} = \braket{ u_{n,\bk}| H_0(\bk) | u_{n',\bk} }_{\text{cell}}. \la{energy}}
Here, the inner product is defined over the unit cell, and the Bloch Hamiltonian is defined in \q{BlochHam}. The propagator in \q{BerrySol} differs by $U(T_B)$ defined in the main text by $\delta\big( \bk-\bk(t) \big)$, which can be seen as a consequence of the acceleration theorem.\cite{Nenciu1980} The propagator of \q{BerrySol} is defined with a basis of Bloch functions which are first-order-differentiable with respect to $\bk \in \calc$, except possibly at isolated wavevectors where $\calc$ is kinked; at these isolated wavevectors, we only impose that $\psi_{n,\bk}$ is continuous across the kink. 

As in \q{app:Wilsonloop}, $U(T_B)$ can be written such that it only depends on the choice of gauge at the base point $\bk_0$:
\e{ U(T_B)_{n,n'} \eq \delta(\bk_f-\bk_0-\bG) \braket{ u_{n,\bk_0}| \mathrm{e}^{i \bG \cdot \br} \overline{\mathrm{exp}}[i \int^{T_B}_0 \Big( \mathcal{A}(t')- \frakP\big(\bk(t')\big) H_0\big(\bk(t')\big) \frakP\big(\bk(t')\big) \Big) \mathrm dt'/\hbar] | u_{n',\bk_0} }_{\text{cell}} \la{BerrySol2} \\
\eq \delta(\bk_f-\bk_0-\bG) \braket{ u_{n,\bk_0}| \mathrm{e}^{i \bG \cdot \br} \overline{\mathrm{exp}}[i\int^{T_B}_0\,\big( \bF(t')\cdot \frakP(\calc)\br \frakP(\calc)-\frakP(\bk(t')) H_0(\bk(t')) \frakP(\bk(t'))\big) \,dt'/\hbar] | u_{n',\bk_0} }_{\text{cell}}. \la{adiabaticevo2} }\\

\q{adiabaticevo} may be particularized to the case where $\calc$ is a noncontractible loop, and $P(\calc)$ projects to bands with the $N$-fold-degenerate dispersion $E_0(\bk)$; the propagator over the Bloch period $T_B$ is then 
\e{ U^{\text{atomic}}(T_B) = \delta\big( \bk-\bk(T_B) \big) e^{-i\int^{T_B}_0 E_0(\bk(t'))dt'/\hbar} \W(\mathcal{C}), \la{app:adiabaticevoDeg} }
where the Wilson-loop matrix \big(defined in \q{app:Wilsonloop}\big) can be written as
\e{\W[\mathcal{C}]_{n,n'} = \braket{ u_{n,\bk_0}| \mathrm{e}^{i \bG \cdot \br} \overline{\mathrm{exp}}[i \oint_{\mathcal{C}} \frakP(\calc)\br \frakP(\calc) \cdot \mathrm d\bk] | u_{n',\bk_0} }_{\text{cell}}, \la{app:WilsonPxP} } 
as shown in appendix F of the main text. \\

We prove the equality of \qq{BerrySol}{adiabaticevo} by employing the time-dependent Schr\"odinger equation and the adiabatic theorem. However, we should emphasize that \qq{BerrySol}{adiabaticevo} are equal as matrix operators; we provide in \s{app:independent2} a more direct proof of their equality that de-emphasizes their application to adiabatic dynamics. Our assumptions on $\bF(t)$ and the energy dispersion merely relate $\W$ to the propagator of electric transport. \\


\subsection{Using the adiabatic theorem}
That $\Psi(\br,t)$ remains in $P(\bk(t))$ (thus justifying the ansatz of \q{Ansatz}) follows from two observations: (i) on parts of $\calc$ where the force direction is constant, we apply the well-known adiabatic approximation in the limit $|\bF_0| \rightarrow 0$\cite{Messiah,Nenciu1980}; we remark that a generalization of the adiabatic theorem exists for fields whose magnitudes are slowly-varying in time $(\Omega \rightarrow 0)$\cite{Nenciu2008}. Transitions between $P$ and $Q$ lead to corrections to the adiabatic approximation; these corrections are bound in \app{app:leakage} in the case of time-independent fields. (ii) The adiabatic approximation is not valid at the kink(s); here, the sudden approximation guarantees that the state remains in $P(\calc)$. In a future publication, we estimate corrections to the sudden approximation (corresponding to transitions to $Q$) as 
\e{ || Q U_{\text{switch}}(\delta) P || \le \big( \f{\delta^2}{2m} \Delta_0 (\bp \cdot \bar{\bV}) \big)^2 + \mathcal{O}(\delta^5) }
where $U_{\text{switch}}(\delta)$ describes the evolution during switching the field, $\delta$ is the switching time, $\bar{\bV} = \f{1}{\delta} \int_0^{\delta} \bF_{\delta}(t') \mathrm dt'/\hbar$ is the average velocity in $\bk$-space and $\Delta_0 (\bp \cdot \bar{\bV})$ is the standard deviation of the momentum operator $\bp$ projected along $\bar{\bV}$. \\

Given that dynamics occurs within $P(\calc)$ (owing to the above argument), \q{BerrySol} is derived by inserting \q{Ansatz} into the time-dependent Schr\"odinger equation 
\e{i \hbar \mathrm d_t \Psi(\br,t) = (H_0-\bF(t)\cdot \br) \Psi(\br,t).\la{timedepse}}
The left-hand-side is calculated as
\es{ i \hbar \mathrm d_t \Psi(\br,t) = \mathrm{e}^{i \bk(t) \cdot \br} \sum_{n'=1}^N \Big( &\big( i \hbar \mathrm d_t f_{n'}(t) \big) u_{n',\bk(t)}(\br) - f_{n'}(t) \bF(t) \cdot \br u_{n',\bk(t)}(\br) + i f_{n'}(t) \big( \nabla_k u_{n',\bk(t)}(\br) \cdot \bF(t) \big) \Big) \la{DDDD}}
where we used that $\hbar d_t \bk(t) = \bF(t)$. The second term on the right-hand-side of \q{DDDD} cancels with the term $\bF(t)\cdot \br \Psi$ on the right-hand-side of \q{timedepse}. Next, we apply $\bra{\psi_{n,\bk(t)}}$ to this equation and use the orthonormality of the Bloch basis, to obtain the equation of motion for the wavefunction $f_{n}(t)$
\e{ i \hbar \mathrm d_t f_{n}(t) = \sum_{n'=1}^N \Big( \mathfrak{E}_{n,n'}\big( \bk(t) \big) - \bF(t) \cdot \bA_{n,n'}\big(\bk(t) \big) \Big) f_{n'}(t), }
from which follows \q{BerrySol}.\\

\q{adiabaticevo} simply describes Hamiltonian dynamics within the restricted subspace $P(\calc)$. It is derived by considering 
the formal, exact solution to \q{timedepse}: 
\e{ \Psi(\br,t)= \overline{\mathrm{e}}^{-i\int^t_0(H_0-\bF(t')\cdot \br)dt'/\hbar}\Psi(\br,0),}
and restricting the dynamics as
\e{ \Psi(\br,t) \approx \overline{\mathrm{e}}^{-i\int^t_0\,P(\calc)(H_0-\bF(t')\cdot \br)P(\calc)\,dt'}\Psi(\br,0).}
The translational invariance of $H_0$ implies that we can replace $P(\calc)H_0P(\calc) \rightarrow P(\bk(t'))H_0P(\bk(t'))$ \emph{under} the time-ordering sign. Finally, applying $\bra{\psi_{n,\bk(t)}}$ to the above equation, we derive \q{adiabaticevo}. \hfill\(\Box\)\\

\subsection{Direct proof}\la{app:independent2}
We begin by deriving an analog of \qq{BerrySol}{adiabaticevo} for an infinitesimal path in $\bk$-space:
\e{ \limit{|\delta \bk|\rightarrow 0} \delta(\bk'-\bk-\delta \bk) \big( \;e^{i\bA(\bk)\cdot \delta \bk-i \mathfrak{E}(\bk)\delta t}\;\big)_{n,n'}= \limit{|\delta \bk|\rightarrow 0} \bra{\psi_{n,\bk'}}e^{iP(\calc)\br P(\calc) \cdot \delta \bk -iP(\bk)H_0P(\bk)\delta t}\ket{\psi_{n',\bk}}, \la{analdiscrete3}}
with $\delta \bk$ defined by $\bF \delta t$. To prove the above equation, we express the right-hand-side as
\e{ &\bra{\psi_{n,\bk'}}e^{iP(\calc)\br P(\calc) \cdot \delta \bk -iP(\bk)H_0P(\bk)\delta t}\ket{\psi_{n',\bk}}\lin
\eq \bra{\psi_{n,\bk'}}e^{iP(\calc)\br P(\calc) \cdot \delta \bk}e^{-iP(\bk)H_0P(\bk)\delta t}\ket{\psi_{n',\bk}}\lin
\eq \sum_{m=1}^N \bra{\psi_{n,\bk'}}e^{iP(\calc)\br P(\calc) \cdot \delta \bk}\ketbra{\psi_{m\bk}}{\psi_{m,\bk}} e^{-iP(\bk)H_0P(\bk)\delta t}\ket{\psi_{n',\bk}} \lin
\eq \delta(\bk'-\bk-\delta \bk) \sum_{m=1}^N \big(e^{i\bA(\bk)\cdot \delta \bk}\big)_{n,m}\big(e^{-i \mathfrak{E}(\bk)\delta t}\big)_{m,n'} = \delta(\bk'-\bk-\delta \bk) \big(e^{i\bA(\bk)\cdot \delta \bk-i \mathfrak{E}(\bk)\delta t}\big)_{n,n'}.}
In the above expressions, all quantities that are second order in $\delta \bk$ are ignored; in the third equality we employed the identity (F2) in the appendix of the main text. We may derive the equality of \qq{BerrySol}{adiabaticevo} by a path-ordered concatenation of \q{analdiscrete3} over the finite loop $\calc$, in close analogy to the proof of \q{app:WilsonPxP} in appendix F of the main text.\hfill\(\Box\)\\

\section{Alternative derivation of Zak-Wannier relation for strong, atomic elementary band representations}\la{altern}
Here we rederive the Zak-Wannier relation \big(Eq. (17) in the main text\big) for strong EBRs in the atomic limit. A proof using a real-space perspective, was provided in Appendix F.2 of the main text; here, we use a $\bk$-space perspective instead.

\noindent \textit{Proof.} The cell-periodic functions are related to Wannier functions as
\e{ u_{j,\bk}^{\alpha_j}(\br) = \f{1}{\sqrt{\calv}} \sum_{\bR} \mathrm{e}^{i \bk \cdot (\bR - \br)} W_{j,\bR}^{\alpha_j}(\br), \la{TBBloch}}
where $\calv$ is the volume of the BZ.\cite{Fiorenza2015} 
For Wannier functions $\{ \ket{W_{j,\bR}^{\alpha_j}} \}_{j,\alpha_j,\bR}$ that are eigenfunctions of $P \br P$ with eigenvalues $\{ \bvarpi_j{+}\bR \}_{j,\bR}$ in the atomic limit (which exist for strong EBRs only), the non-Abelian Berry connection \big(defined in \q{app:nonAB}\big) reads as
\e{ \bA(\bk)_{j',j}^{\alpha'_{j'},\alpha_j} &= \braket{ u_{j',\bk}^{\alpha'_{j'}}| i \nabla_k| u_{j,\bk}^{\alpha_j} }_{\text{cell}}  
= \f{1}{\calv} \sum_{\bR,\bR'} \mathrm{e}^{i \bk \cdot (\bR-\bR')} \braket{ W_{j',\bR'}^{\alpha'_{j'}}| \br-\bR |W_{j,\bR}^{\alpha_j} } 
= \sum_{\bR''} \mathrm{e}^{-i \bk \cdot \bR''} \braket{ W_{j',\bR''}^{\alpha'_{j'}}| \br |W_{j,0}^{\alpha_j} } \lin
&= \sum_{\bR''} \mathrm{e}^{-i \bk \cdot \bR''} \delta_{j,j'} \delta_{\alpha'_{j'},\alpha_j} \delta_{\bR'', 0} \bvarpi_{j} 
= \delta_{j,j'} \delta_{\alpha'_{j'},\alpha_j} \bvarpi_{j}, \la{torsion} }
thus the $N$-band Wilson loop \big(which is defined in \q{app:Wilsonloop}\big) simplifies to
\e{ \mathcal{W}[\mathcal{C}_n]_{j',j}^{\alpha'_{j'},\alpha_j} &= \big( \overline{\mathrm{e}}^{i \int_{\mathcal{C}_n} \bA(\bk) \cdot \mathrm d\bk} \big)_{j',j}^{\alpha'_{j'},\alpha_j} = \delta_{j',j} \delta_{\alpha'_{j'},\alpha_j} \mathrm{e}^{i \bG_n \cdot \bvarpi_j}. \la{WilsonIso2}}
We remark that, while \q{WilsonIso2} has been derived in a particular basis, the eigenvalues of $\mathcal{W}[\mathcal{C}_n]$ are basis independent.\cite{Alexandradinata2014c}  \hfill\(\Box\)\\

\section{Estimate of leakage for adiabatic approximation}\la{app:leakage}
In the adiabatic approximation, a state remains in a gapped energy subspace \big(projected by $P$, as defined in \q{ProjBZ}\big) upon application of a spatially homogeneous time-independent electric force $\bF$. The discussion for time-dependent $\perp$-fields (as defined in the main text), can be reduced to time-independent forces, as we discuss at the beginning of \app{app:weakfield}.
In the scalar gauge \big(reviewed in \q{scalar}\big) the potential $\bF \cdot \br$ is an unbounded operator that is neither translation-invariant nor bounded; for solids the inclusion of this term to the translation-invariant Hamiltonian $H_0$ \big(defined in \q{fieldfreeschrodinger}\big) makes the spectrum of $H'=H_0-\bF \cdot \br$ absolutely continuous\cite{Avron1977}. Therefore it is questionable whether dynamics occurs within the energy subspace projected by $P$ only. The aim of this section is to quantify how much overlap an initial state in $P$ has with states in $Q = \mathbb{1}-P$, the orthogonal subspace to $P$, after time $t$; we call this the leakage 
\e{ L(t) = \| Q \mathrm{exp}(-i H' t/\hbar) P \|_{op} \le 1. \la{leakdef}}
The operator norm for operators acting on the Hilbert space $L^2(\mathbb{R}^d, \mathrm d^d \br)$ ($d=1,2,3$) is defined by $\|A\|_{op} = \sup_{\psi \in L^2; \| \psi \|_2=1} \|A \psi \|_2$ where the $L^2$ norm is $\| \psi \|_2 = \big( \int_{\mathbb{R}^d} |\psi(\br)|^2 \mathrm d\br \big)^{1/2}$; we henceforth omit the subscripts on $\|.\|_{op}$, $\|.\|_2$ and ask the reader to infer the right norm contextually.\\

A simple expression for the leakage can be given in the limit where the potential width $\Delta V = \sup_{\br} |V(\br)| - \inf_{\br} |V(\br)| < \infty$ is much bigger than the energy gap $E_G$:
\e{ \Delta V \gg E_G. \la{simpleCond}} 
In contrast, in the nearly-free limit, the gap is of order of the potential width, therefore we interpret \q{simpleCond} loosely as corresponding to the opposite, tight-binding limit. If the magnitude of the force $F$ is small enough, the leakage out of $P$ under evolution is bounded as
\e{ L(t) &\le \alpha \f{E_V 2\pi F/G}{E_G^2} + \f{\beta}{2 \pi} \big( \frac{2\pi F/G}{E_G} \big)^2 \f{E_V^2 |t|/\hbar}{E_G} + \xi(t), \label{app:leak}\\ 
\xi(t) &= \mathcal{O}\bigg( (\f{E_V 2\pi F/G}{E_G^2})^2, \f{E_V |t|}{\hbar} \f{2\pi F/G}{E_G} (\f{E_V 2\pi F/G}{E_G^2})^2 \bigg) \nonumber }
with reciprocal lattice vector $\bG = F T_B/\hbar$ of magnitude $G$, characteristic energy $E_V = \f{\hbar G}{2} \sqrt{ \f{2 \Delta V}{m} }$, particle mass $m$ and the dimensionless quantities $\alpha=\f{4}{\pi}$, $\beta = \f{128}{\pi^2} + \f{16}{\pi} \approx 9$. Note that in one-dimension $G = 2\pi/a$ and therefore $2\pi F/G = Fa$. \q{app:leak} holds in three-dimensions for all Bravais lattices, using the estimate $\bG \cdot \ba \le G a$. This will be explicitly used in \q{Kinint}.\\

Since our main interest is to estimate the leakage at times $t = \mu T_B = \mu \f{\hbar G}{F}$, we keep terms of order $F, F^2 t$ and drop terms of order $F^2, F^3 t$. 
Bloch oscillations are only observable if the leakage is small over $\mu$ Bloch periods, $L(\mu T_B) \ll 1$. This gives a condition on the force to be small:
\e{ 2\pi \f{F}{G} \ll \f{E_G^2}{\alpha E_V} \f{1}{1 + \f{\mu \beta}{\alpha} \f{E_V}{E_G}}.}
The condition simplifies in the following two limits
\begin{align}
2\pi \f{F}{G} \ll \begin{cases}
 \f{E_G}{\mu \beta} (\f{E_G}{E_V})^2 & \text{,if } 14 \mu \f{E_V}{E_G} \gg 1,\\
 \f{E_G^2}{\alpha E_V} & \text{,if } 14 \mu \f{E_V}{E_G} \ll 1.
\end{cases}
\end{align}
General expressions, not assuming \q{simpleCond}, are long and therefore not explicitly spelled out here. But the interested reader can collect them as follows: for the general condition on the force use \q{forcecond}, \q{B0} and \q{I11}, and for the general expression for the leakage $L$ use \q{P0Int} and \qq{I20}{I12}.\\

The projection operator $P$ is defined at zero field and therefore has all point-group symmetries of the solid. We review in \app{app:deformed} a refined adiabatic approximation which bounds the leakage from field-deformed band subspaces (explicitly, we will study the first-order deformed band projection operator $P_1$) which approach $P$ in the limit of zero field. $P_1$ is the projection to the (assumed) gapped subspace of $H_1:=H_0-(\bF \cdot P\br Q +h.c.)$, where $Q=1-P$. We point out that $H_1$ is translational-invariant (just like $H_0$), respects the nonsymmorphic $g_{n,p}$ symmetry for the $\parallel$-field, but does not respect the point-group symmetry $g_n$ for the $\perp$-field. This difference originates from the symmetry transformation of $\bF \cdot P\br Q$. Therefore, to observe Bloch oscillations in the $\perp$-field, we would like the leakage from $P$ to be small, while for the $\parallel$-field, we impose a weaker condition that the leakage from $P_1$ is small. A bound on the leakage from $P$ is given in \q{app:leak}, and that from $P_1$ is given by
\e{ L_1(t) \le \beta \big( \frac{2\pi F/G}{E_G} \big)^2 \f{E_V^2 |t|/\hbar}{E_G} + \xi_1(t), \as \xi_1(t) = \mathcal{O}\bigg( \f{E_V |t|}{\hbar} \f{2\pi F/G}{E_G} (\f{E_V 2\pi F/G}{E_G^2})^2 \bigg); }
we will see in the derivation why this bound on $L_1(t)$ exactly equals the second term in \q{app:leak}.\\

For the proof, we follow closely A. Nenciu's thesis\cite{Nenciu1987} and do not claim any originality; one advantage we offer is a more general and explicit derivation in English, with the clarification of some technical steps.\\

\begin{figure}
\centering
\includegraphics[width=15cm]{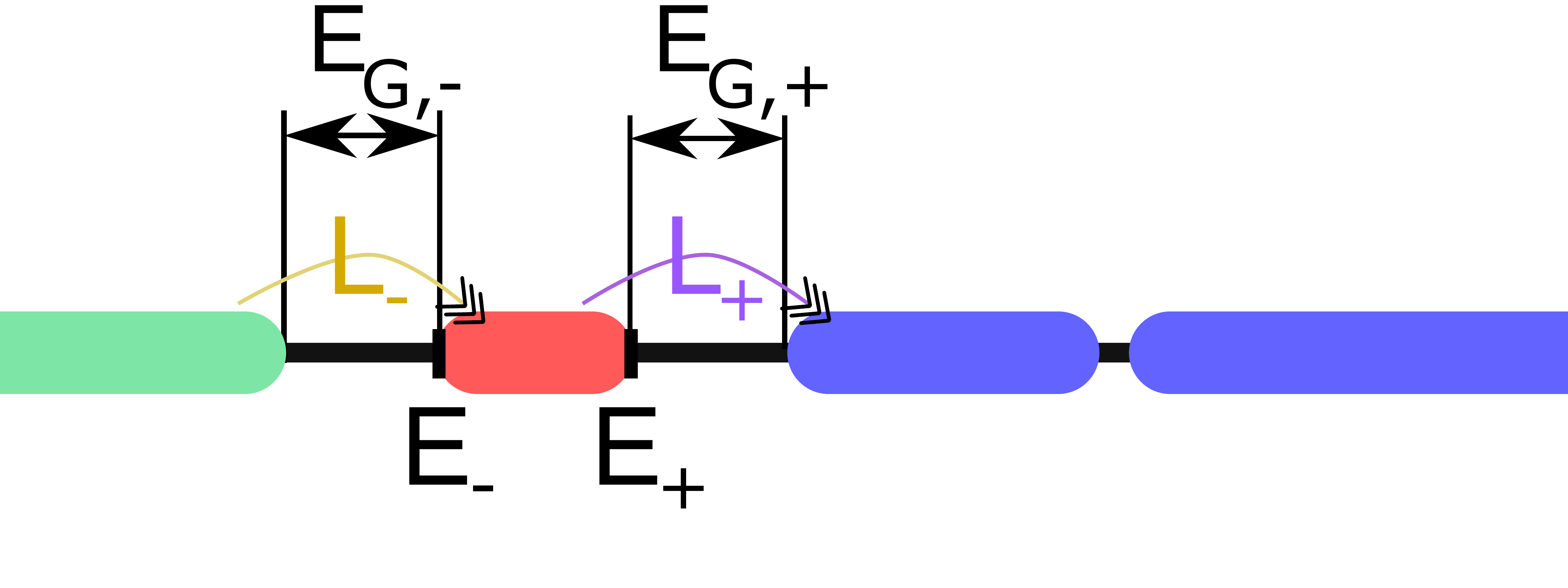}
\caption{Gapped energy subspace (red) in the energy range $[E_-,E_+]$ is separated from high energies (blue) and low energies (green) by energy gaps $d_+, d_->0$, respectively. The yellow arrow $L_-$ symbolizes leakage from the lowest energy subspace to higher energy subspaces, while $L_+$ stands for the leakage from the subspace of energies lower than $E_+$ to energies higher than $E_++d_+$. \la{twogap} }
\end{figure}

The estimate \q{app:leak} applies to the case where $P$ projects to bands having the lowest energy in the spectrum of $H_0$. If we consider a gapped energy subspace that is sandwiched between other energy bands, as shown in Fig. \ref{twogap}, we can obtain the corresponding leakage as follows: The estimate \q{app:leak} applies to (i) $L_+$: the leakage from energies $E \le E_+$ to higher energies $E \ge E_+ +d_+$ (shown by the purple arrow), and (ii) $L_-$: the leakage from energies $E\le E_-$ to higher energies $E\ge E_- + d_-$ (shown by the yellow arrow). We show next that the estimate on $L_-$ is equal to the estimate on the leakage from higher energies $E \ge E_- + d_-$ to lower energies $E \le E_- $ because of the following equalities
\es{ \| P_- U(t) (1-P_-) \| = \| (1-P_-) U(t)^{\dagger} P_- \| = \| (1-P_-) U(-t) P_- \| = L_-(-t). }
Since \q{app:leak} does not depend on the sign of time, the estimates are indeed equal. Consequently, the sum of (i) and (ii), $L_+ + L_-$, gives an upper bound on the total leakage of the gapped energy subspace into higher and lower energies.\\

\subsection{Review of deformed-band construction}\la{app:deformed}
The usual perturbation theory can only handle Hamiltonians that have finite-dimensional degenerate energy subspaces before and after the perturbation. An electric field (in the scalar gauge) is a singular perturbation in that it changes the spectrum of the Bloch Hamiltonian from continuous and gapped to absolutely continuous (no gaps). To handle such a singular perturbation, a more advanced perturbation theory, called deformed-band construction, was introduced\cite{Nenciu1991}. The perturbation parameter is the magnitude of the force $F$. The idea is that we approximate the projection operator of the low-energy subspace $P$ by a series of deformed-band projection operators $P_n$ that (i) converge to $P$ in the limit $F \to 0$, (ii) commute with $H_0 -F(x-B_n)$ where $B_n$ is a bounded operator with decreasing norm for increasing order in deformation, i.e., $\| B_n\| = \mathcal{O}(F^n)$, and (iii) form an asymptotic series in $F$. \\

The projection operators $P_n$ are themselves obtained from corresponding deformed translation-invariant Hamiltonians $H_n$ with continuous and gapped spectra. The technical assumptions for the $n$th-order construction are that the resolvent \big(defined in \q{spectral}\big) of the Hamiltonian $\mathrm{e}^{-i \bF \cdot \br t/\hbar} H' \mathrm{e}^{i \bF \cdot \br t/\hbar}$ \big($H'$ is defined in \q{temporal}\big) is $(n+1)$-times differentiable in $t$, and the deformed Hamiltonian $H_n$ is gapped. 
We also assume that (i) the gap does not close throughout the interpolation from $H_n$ and $H_{n+1}$, and (ii) there exists a single, fixed energy $-E_c$ that lies in the gap throughout this same interpolation. Condition (ii) is needed to employ the analytic techniques that are used in this section.\\

\begin{figure}
\centering
\includegraphics[width=15cm]{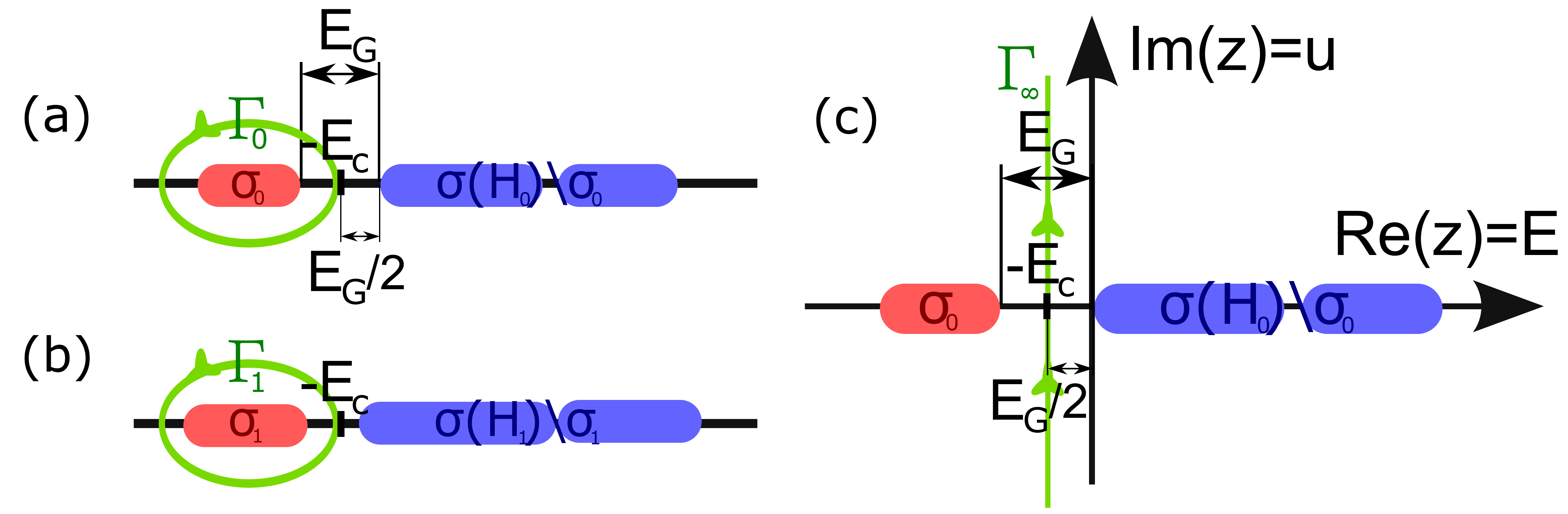}
\caption{Spectra of $H_0$ (red and blue) and loops (green) around them in complex-energy space $z=E+iu \in \mathbb{C}$. (a) The zeroth order gapped energy spectrum $\sigma_0$ (red) is separated from its complement $\sigma(H_0)\backslash \sigma_0$ (blue) by a gap of size $E_G$, the center of the gap is denoted by $-E_c$. We consider a loop $\Gamma_0$ (green) that encircles $\sigma_0$ and goes through $-E_c$. (b) First order gapped energy spectrum $\sigma_1$ (red) can be thought as deformation of $\sigma_0$ in (a). The gap is smaller but does not close. The loop $\Gamma_1$ (green) can be chosen to be equal to $\Gamma_0$. (c) We show that all line integrals can be reformulated as line integrals along the vertical loop $\Gamma_{\infty}$ (green) that crosses the real axis at $-E_c$. \la{spectrum} }
\end{figure}

Let us be more explicit about the first-order deformed-band construction. 
Without loss of generality, we take the force to be along the $x$-direction, $\bF = F \be_x$ and $F>0$. The $x$-component of the off-diagonal position operator is defined by
\e{B_0 = P x Q + Q x P = (Q-P)[x,P]. \label{B0def}} 
$B_0$ is used to define the first-order deformed band Hamiltonian $H_1 = H_0 - F B_0$. We call an operator $A$ off-diagonal wrt. $P_n$ if $P_n A P_n = Q_n A Q_n = 0$, and diagonal if $P_n A Q_n = Q_n A P_n = 0$. 
We can analogously define a reduced off-diagonal (wrt. $P_1$) position operator
\e{B_1 = P_1 x_1 Q_1 + Q_1 x_1 P_1=(Q_1-P_1)[x_1,P_1] \label{B1def} }
where $x_1 = x-B_0$ is diagonal wrt. $P$. \\

Condition (ii), mentioned above, is satisfied for the energy $-E_c$ that lies in the center of the gap of $H_0$ \big(Fig. \ref{spectrum} (a)\big), if we impose the following conservative condition on the force $F$ and the energy gap $E_G>0$ of $H_0$:
\e{F < \frac{\pi E_G \sqrt{m \Delta V/2}}{4 \hbar} \big(1+\f{1}{2} \sqrt{ 1+ (\f{4 \Delta V}{E_G})^2 } \mathrm{atan}(\f{4 \Delta V}{E_G}) \big)^{-1}. \la{forcecond}} 
This condition is derived from the observation that the energies in $\sigma_1$ are perturbations of the energies in $\sigma_0$, and the perturbation can shift an energy by at most $F \| B_0 \|$. To ensure that the gap does not close in the interpolation from $H_0$ to $H_1$, we impose the condition $F \| B_0 \| < E_G/2$. The bound on the norm of $B_0$ \big(that we will find later in \q{B0}, \q{I11}\big) is exactly
\e{ \|B_0 \|\le \f{2 \hbar}{\pi \sqrt{m \Delta V/2}} \big(1+\f{1}{2} \sqrt{ 1+ (\f{4\Delta V}{E_G})^2 } \mathrm{atan}(\f{4\Delta V}{E_G}) \big). \la{B0est}} 
In the limit given by \q{simpleCond}, the condition in \q{forcecond} reads $F < G \f{E_G^2}{2\pi \alpha E_V}$; we did not explicitly state it before because it is implied in the condition given by \q{app:leak}.\\

\subsection{Leakage from deformed bands}
In the construction of deformed bands, the operators $B_n$ play the important role of removing off-diagonal elements of the position operator from the Hamiltonian, order by order in the electric force $F$. As a consequence, higher order deformed-band projection operators $P_n$ lead to less leakage under exact evolution $U(t)= \mathrm{e}^{-i H't/\hbar}$,
\e{ L_n(t) = \| (\mathbb{1}-P_n) U(t) P_n \| = \mathcal{O}(F^n t), }
than the projection operator $P$ at zeroth-order. A bound on $L_n$ and $\| P_n - P\|$ can then be used to obtain a better bound on $L$, utilizing the following inequality:
\e{ L(t) &= \|(\mathbb{1}-P) U(t) P\| = \|(P_n-P) U(t) P + (\mathbb{1}-P_n) U(t) (P-P_n) + (\mathbb{1}-P_n) U(t) P_n\| \nonumber \\
&\le 2\| P_n-P\| + \|(\mathbb{1}-P_n) U(t) P_n \| = 2\| P_n-P\| + L_n(t). \label{firstsecond}}
Here and elsewhere, we frequently utilize the Cauchy-Schwarz inequality, as well as $\|P_n\|=\| \mathbb{1}-P_n\|=\|U(t)\| = 1$. It gets increasingly complicated to calculate the leakage at higher orders and we will only calculate the first order. Some relations hold at all orders $n \in \mathbb{N}_0$, for example \q{firstsecond}, but we will only use them for $n=0,1$ in this appendix. \\

We show next that the leakage out of the $n$th order deformed subspace $P_n$ after evolution for time $t$ is
\e{ L_n(t) = \| (\mathbb{1}-P_n) \mathrm{e}^{-i H' t/\hbar} P_n\| &= \| Q_n \mathrm{e}^{-i (H'+FB_n) t/\hbar} P_n + Q_n \f{i F}{\hbar} \int_0^t \mathrm{e}^{-i (H'+FB_n) (t-s)/\hbar} B_n \mathrm{e}^{-i H' s/\hbar} \mathrm ds P_n\| \nonumber \\
& \le \| B_n\| \f{F |t|}{\hbar} \label{leakbn}}
where we used that $H'+F B_n$ is diagonal wrt. $P_n$. 

\noindent \textit{Proof.} The first equality uses
\e{ \mathrm{e}^{-i H' t/\hbar} = \mathrm{e}^{-i (H'+FB_n) t/\hbar} + \f{i F}{\hbar} \int_0^t \mathrm{e}^{-i (H'+FB_n) (t-s)/\hbar} B_n \mathrm{e}^{-i H' s/\hbar} \mathrm ds. \la{opeqn} }
We prove this by showing that
\es{ A(t) = \mathrm{e}^{-i (H'+FB_n) t/\hbar}-\mathrm{e}^{-i H' t/\hbar} + \f{i F}{\hbar} \int_0^t \mathrm{e}^{-i (H'+FB_n) (t-s)/\hbar} B_n \mathrm{e}^{-i H' s/\hbar} \mathrm ds }
is zero for all times $t$. Note that $A(0)=0$ and therefore the first-order differential equation
\es{ \f{d}{dt} A(t) &= \f{i}{\hbar} H' \mathrm{e}^{-i H' t/\hbar} -\f{i}{\hbar}(H'+F B_n) \mathrm{e}^{-i (H'+FB_n) t/\hbar} \big( 1+ \f{iF}{\hbar} \int_0^t \mathrm{e}^{i (H'+FB_n) s/\hbar} B_n \mathrm{e}^{-i H' s/\hbar} \mathrm ds \big)\lin 
&+\f{i}{\hbar}F B_n \mathrm{e}^{-i H' t/\hbar} = -\f{i}{\hbar}(H'+F B_n) A(t) }
has the unique solution $A(t) = A(0) \mathrm{e}^{-i (H'+FB_n) t/\hbar} = 0$. \hfill\(\Box\) \\

Using \q{leakbn} for $n=0$ we could directly find a bound on the leakage for zeroth order $L(t) \le \|B_0\| \f{Ft}{\hbar}$. The bound on $\|B_0\|$ \big(\q{B0est}\big) is independent of $F$, consequently, leakage at zeroth order over one Bloch period is bounded from above by a field-independent constant, $L(T_B) \le G \|B_0\|$. Such an estimate is not ideal because no improvement on the leakage can be predicted for smaller forces, as one would physically expect. This expectation is however met when we use \q{firstsecond} to get \q{app:leak}, therefore it is necessary to find the estimate through the first-order-deformed band quantities $\|P_1-P\|$ and $\|B_1\|$.\\

\subsection{Resolvent technique and integrals $I_{l,q}$}
\begin{figure}
\centering
\includegraphics[width=15cm]{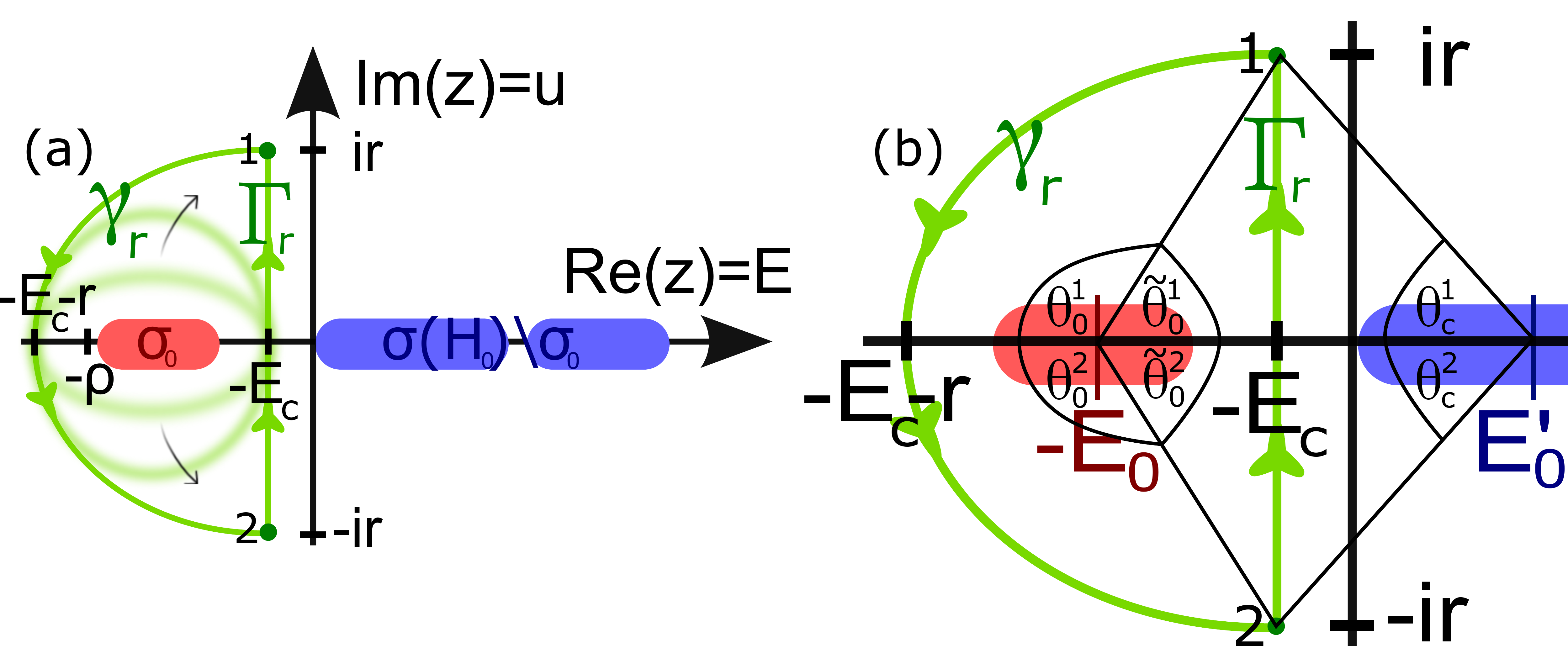}
\caption{(a) Deformation of the loop $\Gamma_0$ from Fig. \ref{spectrum} (a) to the loop formed by $\Gamma_r$ and $\gamma_r$. The vertical line $\Gamma_r$ crosses the real axis at the center of the gap $-E_d$ between $\sigma_0$ (red) and its complement $\sigma(H_0)\backslash \sigma_0$ (blue). (b) For an energy $-E_0<0$ in $\sigma_0$ (red) the angles $\theta_0^r, \theta_0^{-r}$ are used to show that depending on if they are the endpoints of the line $\gamma_r$ or $\Gamma_r$ we have to choose two different branches of the complex logarithm. For an energy $E'_0>0$ in $\sigma(H_0)\backslash \sigma_0$ (blue) the angles $\theta_c^r, \theta_c^r$ to the endpoints of the line $\gamma_r$ or $\Gamma_r$ do not require two different choices of the branch of the complex logarithm. \la{arc} }
\end{figure}

We employ the resolvent technique to find bounds on the norm of operators. The $n$th order projection operator can be rewritten as a Cauchy integral
\e{P_n = -\frac{1}{2\pi i} \oint_{\Gamma_n} R_n(z) \mathrm dz, \as R_n(z)=\big( H_n-z \big)^{-1} \label{spectral}}
where $\Gamma_n$ is a counter-clockwise loop in the complex-energy plane $z=E+iu \in \mathbb{C}$ that encircles the energy interval $\sigma_n$ and $R_n(z)$ is the resolvent. Note that the spectrum of self-adjoint operators is confined to the real line. Fig. \ref{spectrum} (a) is a sketch of the energy spectrum of $H_0$; the low energies (red) are energetically separated from the high energies (blue) by a gap $E_G>0$ and $-E_c<0$ denotes the center of the energy gap. In the sketch of the energy spectrum of $H_1$, Fig. \ref{spectrum} (b), the electric force $F$ is assumed to be small enough \big(\q{forcecond}\big) such that the energy $-E_c$ still lies in the energy gap. By assumption, both $P$ and $P_1$ project onto gapped energy subspaces and the energy interval $\sigma_1$ is obtained from a continuous deformation of $\sigma_0$, therefore we can choose the same loop $\Gamma = \Gamma_0 = \Gamma_1$ to encircle $\sigma_0$ and $\sigma_1$ \big(green loop in Fig. \ref{spectrum} (a), (b)\big). \\

The projection operator $P$ can also be written as a line integral over the infinite vertical loop $\Gamma_{\infty}$ \big(Fig. \ref{spectrum} (c)\big)
\e{ P &= -\f{1}{i \pi} \int_{\Gamma_{\infty}} P R_0(z) \mathrm dz.\la{Proj} }
 For powers of the resolvent $R_0$ times an operator $A$ times $R_0$, the line integral over $\Gamma$ equals the line integral over $\Gamma_{\infty}$ under the following conditions: there exists a constant $M>0$ such that either (a) $\|A\|$ is bounded by $M$ or (b) $\|A R_0(z)\|$ is bounded by $M$ for all $z \in \mathbb{C}$. Then
\e{ \oint_{\Gamma} \big( R_0(z) \big)^l A R_0(z) \mathrm dz &= \int_{\Gamma_{\infty}} \big( R_0(z) \big)^l A R_0(z) \mathrm dz \la{GGinf} }
holds in case (a) for all $l>0$, in case (b) for all $l>1$.
We postpone the proof of \qq{Proj}{GGinf} to \app{app:proofG} and only comment on them here.
\q{Proj} goes beyond the expectation that one can simply replace $\Gamma$ by $\Gamma_{\infty}$ in \q{spectral}. The origin of the additional factor of $2$ can be understood from a simpler example: The standard line integral $\f{1}{2\pi i} \oint \f{\mathrm dz}{z}=1$ can be split into two integrals over semicircles that both give $\f{1}{2}$ and sum up to $1$. In this analogy, $\Gamma_{\infty}$ is one of the semicircles and is only equal to half of the full loop integral \big(\q{spectral}\big). The additional projection operator $P$ appears because $R_0$ has further singularities on the real line.
\q{GGinf} will be used extensively to calculate the estimate on the leakage. One advantage of using $\Gamma_{\infty}$ over $\Gamma$ is that we can bound $\|T^{1/2} R_0(z) \|$ for all $z \in \Gamma_{\infty}$ as
\e{ \|T^{1/2} R_0(z) \| \le \sqrt{ \f{ 1+(4\Delta V/E_G)^2 }{\Delta V} }. \la{T12}} 
This bound will be found later in \qq{Test1}{Test}, \qq{Test2}{Test4}, whereas we do not know how to find such an estimate for $z \in \Gamma$.\\

Using \q{GGinf}, we can reduce the bounds on $\|P_1-P\|$ and $\|B_1\|$ to integrals of the form
\e{ I_{l,q} &= \frac{1}{2\pi} \int_{\Gamma_{\infty}} \| R_0(z)\|^l \| T^{1/2} R_0(z)\| ^q |\mathrm dz| \label{Kinint}}
for $l,q \in \mathbb{N}$ with $l+q/2>1$ \big(this condition is to ensure convergence, as will be explicit in \q{Int1}\big) where $T = \f{\bp^2}{2m}$ is the kinetic energy operator. We will express the leakage with these integrals $I_{l,q}$ as
\e{ L(t) \le 2 \hbar \sqrt{ \f{2}{m} } F I_{1,1} I_{2,0} + \f{\hbar}{m} F^2 |t| \big( 4 I_{2,1} I_{1,1} + I_{2,0} I_{1,2} + \f{1}{4} (I_{2,0})^2 + I_{2,0} (I_{1,1})^2\big) + \xi(t) \label{P0Int}}
where $\xi(t)>0$ for all $t$ and $\xi(T_B) = \mathcal{O}(F^2)$.\\

\noindent \textit{Proof.} We prove \q{P0Int} by using \qq{firstsecond}{leakbn} for $n=1$, i.e., by finding bounds on $\|P_1-P\|$ and $\|B_1\|$.

\subsection{Reduction of estimates of leakage to $I_{l,q}$}\la{app:reduction}
We start by showing that 
\e{ \|P_1-P\| \le F \|B_0\| I_{2,0} + \xi', \as \xi' = F^2 \|B_0\|^2 I_{3,0} + \mathcal{O}(F^3)>0. \la{P1P0}}
Let us first notice that 
\e{ &R_1(z) = \big( 1+F R_1(z) B_0 \big) R_0(z) \la{R1} \\
\Leftrightarrow &\big( 1+F R_1(z) B_0 \big)^{-1} R_1(z) = R_0(z) \nonumber\\
\Leftrightarrow &R_1^{-1}(z) \big( 1+F R_1(z) B_0 \big) = R^{-1}_0(z)\nonumber\\
\Leftrightarrow &(H_1-z) + F B_0 = (H_0-z) \nonumber}
where the last equality holds by definition of $H_1$. Then we can rewrite the first-order deformed projection operator as
\e{ P_1 &= -\frac{1}{2\pi i} \oint_{\Gamma} R_1(z) \mathrm dz = -\frac{1}{2\pi i} \oint_{\Gamma} R_0(z) \mathrm dz - \frac{F}{2\pi i} \oint_{\Gamma} R_1(z) B_0 R_0(z) \mathrm dz\lin 
&= -\frac{1}{2\pi i} \oint_{\Gamma} R_0(z) \mathrm dz - \frac{F}{2\pi i} \oint_{\Gamma} R_0(z) B_0 R_0(z) \mathrm dz + \Xi' = P - \frac{F}{2\pi i} \int_{\Gamma_{\infty}} R_0(z) B_0 R_0(z) \mathrm dz + \Xi' \label{man3}}
where
\e{ \Xi' = -\f{F^2}{2\pi i} \int_{\Gamma_{\infty}} R_0(z) \big( B_0 R_0(z) \big)^2 \mathrm dz + \mathcal{O}(F^3) \la{Ap} }
and we used \q{GGinf}, case (a), to replace the integral over $\Gamma$ with an integral over $\Gamma_{\infty}$. The justification for this replacement is postponed to \app{app:proofG}. Then we find that
\es{ \|P_1-P\| \le \frac{F}{2\pi} \| B_0 \| \int_{\Gamma_{\infty}} \|R_0(z)\|^2 | \mathrm dz| + \xi' \le F \|B_0\| I_{2,0} + \xi'.}
with $\xi' = \| \Xi' \| \le F^2 \|B_0\|^2 I_{3,0} + \mathcal{O}(F^3)$.\\

The reduction of $\|B_1\|$ to the integrals $I_{l,q}$ is more complicated; we will show that it is given by
\e{ \|B_1\| \le F \f{4 \hbar^2}{m} \big( I_{2,1} I_{1,1} + I_{2,0} I_{1,2} + \f{1}{4} (I_{2,0})^2 + I_{2,0} (I_{1,1})^2\big) + \xi'' \label{B1}}
where
\e{ \xi'' = \mathcal{O}\Big( \f{F^2 \hbar^3}{m^{3/2}} (I_{1,1})^2 I_{3,1}, \f{F^2 \hbar^3}{m^{3/2}} I_{1,1} I_{3,0}\big(I_{1,2} + \f{1}{4} I_{2,0} + (I_{1,1})^2\big) \Big). \la{r22}}
The proof of this inequality will be aided by three relations: 
\begin{enumerate}
\item For normal, bounded operators $A$ the norm is equal to the supremum of the absolute value of its eigenvalues, i.e., $\| A \| = \sup \{ |a|; a \in \sigma(A) \}$. For example, $Q_1-P_1$ is a normal, bounded operator with $\|Q_1-P_1\| = 1$, 
\item the reduced position operator $x_1=P x P + Q x Q$ is diagonal wrt. $P$, 
\item from 2., $P + Q = \mathbb{1}$, the triangle inequality and the Cauchy-Schwarz inequality, we find\cite{Nenciu}
\e{ \|[x_1,R_0(z)]\| \le \| [x,R_0(z)] \|. \la{ineq} }
\end{enumerate} 
With the definition of $B_1$ \big(given in \q{B1def}\big), \q{man3}, and relation 2., we get
\e{ B_1 &= -(Q_1 - P_1) [x_1,P_1]\lin 
& = -\frac{F}{2\pi i} (Q_1 - P_1) \oint_{\Gamma} \Big( [x_1,R_1(z)] B_0 R_0(z) + R_1(z) [x_1,B_0] R_0(z) + R_1(z) B_0 [x_1,R_0(z)] \Big) \mathrm dz \lin
&= -\frac{F}{2\pi i} (Q_1 - P_1) \oint_{\Gamma} \Big( [x_1,R_0(z)] B_0 R_0(z) + R_0(z) [x,B_0] R_0(z) + R_0(z) B_0 [x_1,R_0(z)] \Big) \mathrm dz + \Xi''\lin
&= -\frac{F}{2\pi i} (Q_1 - P_1) \int_{\Gamma_{\infty}} \Big( [x_1,R_0(z)] B_0 R_0(z) + R_0(z) [x,B_0] R_0(z) + R_0(z) B_0 [x_1,R_0(z)] \Big) \mathrm dz + \Xi'' \la{B1b}}
with $\Xi''$ an operator of order $F^2$. For the second line we also employed that $x_1=x-B_0$ therefore $[x_1,B_0] = [x,B_0]$; we justify later in \app{app:proofG} why we can replace $\oint_{\Gamma} \mathrm dz$ by $\int_{\Gamma_{\infty}} \mathrm dz$. Using \q{ineq} we find
\e{ \|B_1\| \le \frac{F}{2 \pi} \int_{\Gamma_{\infty}} \big( 2 \| [x,R_0(z)] \| \|B_0 \| \|R_0(z) \| + \| [x,B_0] \| \|R_0(z)\|^2 \big) | \mathrm dz| + \xi'' \la{B1est}}
with $\xi'' = \| \Xi'' \|$.
What remains is to relate (i) $ \| [x,R_0(z)] \|$, (ii) $\|B_0\|$ and (iii) $\| [x,B_0] \|$ to $I_{l,q}$ \big(\q{Kinint}\big). Relation (i) is aided by the following identity
\e{ R_0^{-1}(z) [x,R_0(z)] R_0^{-1}(z) &= [R_0^{-1}(z),x] = [H_0,x] \lin
\Leftrightarrow [x,R_0(z)] &= R_0(z) [H_0,x] R_0(z) = -\f{i \hbar}{m} R_0(z) p_x R_0(z). }
For the last equality we used the Schr\"odinger Hamiltonian \big(\q{fieldfreeschrodinger}\big) with velocity $\f{p_x}{m}$ proportional to the canonical momentum. Further employing 
\e{ \|p_x R_0(z) \| \le \| |\bp| R_0(z)\| = \sqrt{2m} \| T^{1/2} R_0(z)\| \la{px}} 
we are lead to 
\e{ \| [x,R_0(z)] \| \le \hbar \sqrt{ \f{2}{m} } \|R_0(z)\| \|T^{1/2} R_0(z)\|. \label{xRest} }
For (ii) we use \q{B0def}, \q{spectral} and \q{xRest} and find
\e{ \|B_0\| &= \| [P,x] \| \le \hbar \sqrt{ \frac{2}{m} } \f{1}{2\pi} \int_{\Gamma} \|R_0(z)\| \|T^{1/2} R_0(z)\| | \mathrm dz| = \hbar \sqrt{ \frac{2}{m} } \f{1}{2\pi} \int_{\Gamma_{\infty}} \|R_0(z)\| \|T^{1/2} R_0(z)\| | \mathrm dz|\lin 
&= \hbar \sqrt{ \frac{2}{m} } I_{1,1}. \label{B0} }
The proof for the replacement of $\Gamma$ with $\Gamma_{\infty}$ is shown in \app{app:proofG}.
The first step to obtain relation (iii) is to apply the definition of $B_0$ \big(\q{B0def}\big) to derive 
\e{ [x,B_0] = (Q-P) [ [P,x],x ] - 2( [P,x] )^2 \la{xB0} }
with the estimate $\| [P,x] \| = \hbar \sqrt{ \frac{2}{m} } I_{1,1}$ \big(as in \q{B0}\big). Writing $P$ again in terms of $R_0$ leads to
\es{ [[P,x], x] &= -\f{1}{2\pi i} \oint_{\Gamma} \Big( 2 R_0(z) \big( [x,H_0] R_0(z) \big)^2 + R_0(z) [[x,H_0],x] R_0(z) \Big) \mathrm dz\\
&= -\f{1}{2\pi i} \int_{\Gamma_{\infty}} \Big( 2 R_0(z) \big( [x,H_0] R_0(z) \big)^2 + R_0(z) [[x,H_0],x] R_0(z) \Big) \mathrm dz, }
where we defer to \app{app:proofG} for the argument to replace $\Gamma$ with $\Gamma_{\infty}$. Employing \q{px} results in
\es{ \|[[P,x], x]\| \le \f{1}{2\pi} \int_{\Gamma_{\infty}} \Big( \f{4 \hbar^2}{m} \| R_0(z)\| \| T^{1/2} R_0(z) \|^2 + \f{\hbar^2}{m} \| R_0(z)\|^2 \Big) | \mathrm dz| = \f{4 \hbar^2}{m} I_{1,2} + \f{\hbar^2}{m} I_{2,0}. \la{P0xx}}
From \q{xB0} and this last equation follows
\e{ \| [x,B_0] \| &\le \| [ [P,x],x ] \| + 2 \| [P,x] \|^2 \le \f{4 \hbar^2}{m} I_{1,2} + \f{\hbar^2}{m} I_{2,0} + \f{4 \hbar^2}{m} I_{1,1}^2. }
Inserting this equation with \q{xRest} and \q{B0} into \q{B1est} gives the claimed bound on $||B_1||$ \big(\q{B1}\big). Using the derived estimates and explicitly writing out $\Xi''$, one can now find a bound for $\xi'' =\|\Xi''\| $ which will lead to \q{r22}. Finally, collecting equations \q{firstsecond}, \q{leakbn}, \q{B0} and \q{B1} leads to the claimed bound on the leakage $L$ \big(\q{P0Int}\big) with $\xi(t) = 2\xi' + \f{F|t|}{\hbar} \xi''$. \hfill\(\Box\)\\

\subsection{Calculation of $I_{l,q}$}\la{app:calc}
All that is left to finish the proof of \q{app:leak} is to calculate the integrals $I_{2,0}, I_{1,1}, I_{2,1}, I_{1,2}$ \big(\q{Kinint}\big) in \q{P0Int} explicitly, and for the estimates also $I_{3,0}, I_{3,1}$. We will find
\e{ I_{2,0} &= \frac{1}{2 \pi \Delta V} \big( 1+\frac{4\Delta V}{E_G} \mathrm{atan}(\frac{4\Delta V}{E_G}) \big),\la{I20}\\
I_{1,1} &= \frac{2}{\pi \sqrt{\Delta V}} \big( 1 + \frac{1}{2} \sqrt{1 + (\frac{4\Delta V}{E_G})^2} \mathrm{atan}( \f{4\Delta V}{E_G} ) \big) \la{I11}\\
I_{2,1} &= \frac{1}{2 \pi (\Delta V)^{3/2}} \big( \f{2}{3} + (\frac{4\Delta V}{E_G})^2 \big),\la{I21}\\
I_{1,2} &= \frac{1}{\pi \Delta V} \big( 1+\frac{4\Delta V}{E_G} \sqrt{ 1+ (\frac{4\Delta V}{E_G})^2 } \big), \la{I12}\\
I_{3,0} &= \f{1}{8 \pi (\Delta V)^2}+\f{16\Delta V}{\pi E_G^3 \sqrt{1+(\f{4\Delta V}{E_G})^2}}, \la{I30}\\
I_{3,1} &= \f{1}{2\sqrt{\Delta V}} \Big( \f{1}{5\pi (\Delta V)^2} + \f{1}{\pi E_G \Delta V} \f{1}{\sqrt{ 1+(\f{4\Delta V}{E_G})^2 }} + \f{4}{\pi E_G^2} \sqrt{ 1+(\f{4\Delta V}{E_G})^2 } \mathrm{atan}(\f{4\Delta V}{E_G}) \Big) \la{I31} }
which simplify under the assumption $\Delta V \gg E_G$ to
\e{ I_{2,0} &= \frac{1}{E_G},\la{simpleest}\\
I_{1,1} &= \frac{2 \sqrt{\Delta V}}{E_G},\lin
I_{2,1} &= \frac{8\sqrt{\Delta V}}{\pi E_G^2},\lin
I_{1,2} &= \frac{16 \Delta V}{\pi E_G^2},\lin
I_{3,0} &= \f{4}{\pi E_G^2},\lin
I_{3,1} &= \f{4 \sqrt{\Delta V}}{E_G^3}.\nonumber }\\

\noindent \textit{Proof.} To calculate the integrals $I_{l,q}$, we find the norm of the residuum by using a standard formula from functional analysis
\e{ \| R_n(z)\| = \bigg( \inf_{E \in \sigma(H_n)} |E-z| \bigg)^{-1} = \bigg( \inf_{E \in \sigma(H_n)} \sqrt{ (E_c+E)^2 + u^2 } \bigg)^{-1} = \f{1}{ \sqrt{ (E_G/2)^2 + u^2 } } \le \f{1}{u} \la{Resbound}}
for $z=-E_c+iu \in \Gamma_{\infty}$ and $\sigma(H_n)$ is the spectrum of $H_n$. To find an estimate on the norm of $T^{1/2} R_0(z)$ is more problematic because the kinetic energy operator $T$ is unbounded. This norm can only be bounded if the lattice potential is well-behaved. 
For example, if the lattice potential satisfies $\|V\| \le \Delta V < \infty$ \big(setting the zero of energy to be equal to the infimum of $\sigma(H_0) \backslash \sigma_0$\big) then it follows from the semi-positive definiteness of $T$ that 
\e{ \| V(T-z)^{-1}\| \le \|V\| \|(T-z)^{-1}\| \le \frac{\Delta V}{u} \le \f{1}{2} \label{Vest}} 
for all $u=\mathrm{Im}(z) > 2\Delta V$. A more general condition on $V$ to satisfy \q{Vest} can be found in a paper by Avron\cite{Avron1979}.\\

Since the integrands are positive for all $I_{l,q}$, the integrals from $u=-i\infty$ to $u=0$ and from $u=0$ to $u=i \infty$ are equal. Moreover, we split the integrals as
\e{ I_{l,q} &= \frac{1}{2\pi} \int_{\Gamma_{\infty}} \|R_0(z)\|^l \|T^{1/2} R_0(z) \|^q |\mathrm dz| = I_{l,q}^1 + I_{l,q}^2 }
where 
\e{ I_{l,q}^1 &= \frac{1}{\pi} \big( \int_{2\Delta V}^{\infty} \|R_0(z)\|^l \|T^{1/2} R_0(z) \|^q \mathrm du,\\
I_{l,q}^2 &= \int_0^{2\Delta V} \|R_0(z)\|^l \|T^{1/2} R_0(z) \|^q \mathrm du }
for $z=-E_c+iu \in \Gamma$, such that \q{Vest} is satisfied for all $z$ in the integration range of the first integral, $I_{l,q}^1$.\\

For the first integral $I_{l,q}^1$ we use
\e{ (T+V-z) R_0(z) = \mathbb{1} \Leftrightarrow (T-z) R_0(z) + V R_0(z) = \mathbb{1} \Leftrightarrow R_0(z) = (T-z)^{-1} \big( \mathbb{1}- V R_0(z) \big) }
together with \q{Vest} to find
\e{ \|T^{1/2} R_0(z) \| &\le \|T^{1/2} (T-z)^{-1} \| + \| T^{1/2} (T-z)^{-1} \| \| V R_0(z) \| \lin
& \le \|T^{1/2} (T-z)^{-1} \| \Big( 1+ \|V (T-z)^{-1}\| + \|V(T-z)^{-1} V R_0(z) \| \Big) \lin
& \le \|T^{1/2} (T-z)^{-1} \| \sum_{n=0}^{\infty} \f{1}{2^n} = 2 \|T^{1/2} (T-z)^{-1} \| \la{Test1}}
for all $u = \mathrm{Im}(z) > 2\Delta V$. The last expression can now be calculated directly
\e{ \|T^{1/2} (T-z)^{-1} \|^2 &= 2m \sup_{ p \ge 0 } | \frac{p}{p^2-2mz} |^2 = 2m \sup_{p\ge 0} \frac{p^2}{ (p^2+2mE_c)^2+(2mu)^2 }\lin 
&= \frac{ \sqrt{E_c^2 + u^2} }{ \big( \sqrt{E_c^2+u^2}+E_c \big)^2+u^2 } \le \frac{1}{2u} \la{Test}}
where we used that the maximum of the norm is attained at $p^2 = 2m \sqrt{ E_c^2 + u^2 }$.
This inequality together with $\|R_0(z)\| \le \f{1}{u}$ \big(\q{Resbound}\big) gives a bound on the first integral
\e{ I_{l,q}^1 \le \frac{1}{\pi} \int_{2\Delta V}^{\infty} u^{-l} (\f{2}{u})^{q/2} \mathrm du = \frac{2^{q/2}}{\pi (l+q/2-1)} (2\Delta V)^{-(l+q/2-1)}. \la{Int1}}
Convergence of this integral is assured by the assumption that $l + q/2 > 1$. Explicit calculation gives the first term in each of \q{I20}-\eqref{I12}.\\

For the second integral we have to be a bit more careful to prevent it from blowing up at $u= \mathrm{Im}(z) = 0$ and we cannot use \q{Vest} anymore. The idea is to find an intermediate complex number $z_0$ to bound $T^{1/2} R_0(z)$ using the Cauchy-Schwarz inequality
\e{ \|T^{1/2} R_0(z) \| = \|T^{1/2} R_0(z_0) (H_0-z_0) R_0(z) \| \le \|T^{1/2} R_0(z_0) \| \| (H_0-z_0) R_0(z) \|. \la{Test2}}
We choose $z_0=E_c+2i \Delta V$ with $E_b \ge -E_c$ in order to obtain an integrable estimate:
\e{\|(H_0-z_0) R_0(z)\|^2 &= \sup_{E \in \sigma(H_0)} \frac{ (E-E_b)^2 + (2\Delta V)^2 }{ (E+E_c)^2 + u^2 }\lin 
&= \frac{ (E_G/2-E_c-E_b)^2 + (2\Delta V)^2 }{ (E_G/2)^2 + u^2 } \le \frac{(E_G/2)^2 + (2\Delta V)^2}{(E_G/2)^2 + u^2} \la{Test4}}
where the maximum of the norm is attained at $E = \f{E_G}{2}-E_c$. From $\| (T-z_0)^{-1} \| = \big( \inf_{z \in [0,\infty} |z-z_0| \big)^{-1} \le \f{1}{2 \Delta V}$ it follows that $\| V(T-z)^{-1}\| \le \f{ \Delta V }{2 \Delta V} = \f{1}{2}$, hence \q{Vest} is still satisfied. Following the same steps as before, $\|T^{1/2} R_0(z_0) \| \le \f{1}{\sqrt{ \Delta V }}$. Combining the above equations with \q{Resbound}, we get (extracting also an extra factor $4 \Delta V$)
\e{ I_{l,q}^2 \le \frac{(4\Delta V)^{q/2}}{\pi} \big( 1+(\frac{E_G}{4\Delta V})^2 \big)^{q/2} \int_0^{2\Delta V} ( (E_G/2)^2 + u^2 )^{-(l+q)/2} \mathrm du. }
This gives the second term in \q{I20}-\eqref{I12}.\\

Plugging \q{simpleest} back into \q{P0Int} we finally come to the main result, \q{app:leak}, of this section
\e{ L(t) &\le 4 \hbar \sqrt{ \f{2\Delta V}{m} } \f{F}{E_G^2} + \f{4\hbar}{m} F^2 |t| ( \f{16 \Delta V}{\pi E_G^3} + \f{16 \Delta V}{\pi E_G^3} + \f{1}{4 E_G^2} + \f{4 \Delta V}{E_G^3} ) + \xi(t)\lin 
&= 4 \hbar \sqrt{ \f{2\Delta V}{m} } \f{F}{E_G^2} + \f{2\hbar}{m} F^2 |t| (\f{64}{\pi}+8) \f{\Delta V}{E_G^3} + \f{\hbar}{m E_G^2} F^2 |t| + \xi(t)\lin
& = \alpha \f{E_V 2\pi F/G}{E_G^2} + \f{\beta}{2\pi} \big( \frac{2\pi F/G}{E_G} \big)^2 \f{E_V^2 |t|/\hbar}{E_G} + \f{\hbar}{m E_G^2} F^2 |t| + \xi(t) \la{app:leakproof}}
with $\alpha=\f{4}{\pi}$, $\beta = \f{128}{\pi^2}+\f{16}{\pi}$, $E_V = \f{\hbar G}{2} \sqrt{ \f{2 \Delta V}{m} }$ and $\xi(t) = 2\xi'+\f{F |t|}{\hbar} \xi''$ where
\es{ \xi' \le F^2 \|B_0\|^2 I_{3,0} + \mathcal{O}(F^3) = \mathcal{O}\big( (\f{2\pi F/G \, E_V}{E_G^2})^2 \big) } 
and 
\es{ \xi'' = \mathcal{O}\big( \f{F^2 \hbar^3}{m^{3/2}} I_{1,1}^2 I_{3,1}, \f{F^2 \hbar^3}{m^{3/2}} I_{1,1} I_{3,0}(I_{1,2} + \f{1}{4} I_{2,0} + I_{1,1}^2 ) \big) = \mathcal{O}\big( \f{E_V}{E_G} (\f{2\pi F/G \, E_V}{E_G^2})^2 \big). } 
When neglecting the last term in \q{app:leakproof}, we arrive at \q{app:leak}. \hfill\(\Box\)\\

\subsection{Proof for using the loop $\Gamma_{\infty}$}\la{app:proofG}
We start by proving \qq{Proj}{GGinf}, and then confirm the applicability of \q{GGinf} to a number of examples that we have encountered when finding the estimate of the leakage.\\

For the first part of the proof, \q{Proj}, we note first that the resolvent $R_0$ is holomorphic away from the spectrum of $H_0$, therefore we can deform $\Gamma$ without changing the line integral \q{spectral}, so long as the deformed $\Gamma$ does not cross the spectrum. In particular, we can deform $\Gamma$ to the concatenation of the vertical line $\Gamma_{r}$ and the semicircle $\gamma_r$; the radius ($r>0$) of the semicircle is chosen such that it does not cross the spectrum \big(Fig. \ref{arc} (a)\big). For analytic convenience, we define the zero in energy to lie at the lower bound of $\sigma(H_0) \backslash \sigma_0$, i.e., the u-axis in Fig. \ref{arc} (a), and $-E_c <0$ to be the point on the real line at distance $E_G/2$ from both $\sigma_0$ and $\sigma(H_0) \backslash \sigma_0$. 
We prove \q{Proj} by showing that
\e{ \lim_{r \to \infty} \f{1}{2\pi i} \oint_{\gamma_{r}} R_0(z) \mathrm dz = -\f{1}{2} \mathbb{1}, \as \lim_{r \to \infty} \f{1}{2\pi i} \oint_{\Gamma_{r}} R_0(z) \mathrm dz = \f{1}{2} (Q-P). \la{P0rel}}
We start by using the completeness of Bloch functions $\{ \psi_{n,\bk} \}_{n \ge 1, \bk}$, i.e., $\mathbb{1} = \sum_{n\ge 1} \int \ket{\psi_{n,\bk}} \bra{ \psi_{n,\bk} } \mathrm d\bk$, therefore
\e{ R_0(z) = \sum_{n\ge 1} \int (E_{n,\bk}-z)^{-1} \ket{ \psi_{n,\bk} } \bra{ \psi_{n,\bk} } \mathrm d\bk, }
and then calculate the line integral for fixed $E_{n,\bk}$, i.e.
\e{ \oint_{\Gamma} (E_{n,\bk}-z)^{-1} \mathrm dz = \int_{\Gamma_r} (E_{n,\bk}-z)^{-1} \mathrm dz + \int_{\gamma_r} (E_{n,\bk}-z)^{-1} \mathrm dz = - \left. \ln(z-E_{n,\bk}) \right|_{2}^1 - \left. \ln(z-E_{n,\bk}) \right|_{1}^2 }
where the endpoints $1$ and $2$ of the lines $\Gamma_{r}$ and $\gamma_r$ are indicated in Fig. \ref{arc} (a), (b) in dark green.
For each of the line integrals, the logarithm of the radius $|z-E_{n,\bk}|$ cancels, and we are left with the difference in the phase of $z-E_{n,\bk}$ at the endpoints $1$ and $2$. These angles are demonstrated in Fig. \ref{arc} (b).
We note that depending on whether the energy $E_{n,\bk}$ is in $\sigma_0$ or in $\sigma(H_0)\backslash \sigma_0$, we choose a corresponding branch of the logarithm that avoids crossing the branch cut. We denote the principal branch of arctangent by $\mathrm{atan}$. Concretely, let $E_{n,\bk} = -E_0 \in \sigma_0$ with $E_0 > 0$, then for the integral over $\Gamma_r$ we choose the branch $\tilde \theta \in [-\pi,\pi)$ such that $\tilde \theta_0^{2} = - \tilde \theta_0^1 = - \mathrm{atan}(\f{r}{E_0})$
\e{\int_{\Gamma_r} (E_0-z)^{-1} \mathrm dz = i\tilde \theta_0^{2} - i\tilde \theta_0^1 = -2i \mathrm{atan}(\f{r}{E_0}). \la{Gamma0} }
For $\gamma_r$ we choose the branch $\theta \in [0,2\pi)$ such that $\theta_0^{2} = 2\pi - \theta_0^1$ gives
\e{\int_{\gamma_r} (E_0-z)^{-1} \mathrm dz = i\theta_0^{1} - i\theta_0^{2} = 2i \mathrm{atan}(\f{r}{E_0})- 2\pi i. \la{gamma0}}
Therefore, for $E_{n,\bk} \in \sigma_0$
\e{ \oint_{\Gamma} (E_{n,\bk}-z)^{-1} \mathrm dz = - 2\pi i, }
which justifies the resolvent expression for the projection operator $P$ \big(\q{spectral}\big).
On the other hand, let $E_{n,\bk} = E'_0 \in \sigma(H_0) \backslash \sigma_0$. Now for both $\Gamma_r$ and $\gamma_r$ we choose the branch $\theta \in [0,2\pi)$ such that $\theta_c^{1} = \pi - \mathrm{atan}(\f{r}{E'_0})$, $\theta_c^{2} = \pi + \mathrm{atan}(\f{r}{E'_0})$ leads to
\e{\int_{\Gamma_r} (E'_0-z)^{-1} \mathrm dz &= i\theta_c^{2} - i\theta_c^1 = 2i \mathrm{atan}(\f{r}{E'_0}), \la{Gammac}\\
\int_{\gamma_r} (E'_0-z)^{-1} \mathrm dz &= i\theta_c^{1} - i\theta_c^{2} = -2i \mathrm{atan}(\f{r}{E'_0}). \la{gammac}}
So the line integral for $E_{n,\bk} \in \sigma(H_0) \backslash \sigma_0$ is
\e{ \oint_{\Gamma_r} (E_{n,\bk}-z)^{-1} \mathrm dz = 0. }
Now, we employ \q{Gamma0}, \q{Gammac} and the completeness of Bloch functions to evaluate
\e{ -\f{1}{2\pi i} \int_{\Gamma_r} R_0(z) \mathrm dz &= \f{1}{\pi} \sum_{n= 1}^N \int \mathrm{atan}(\f{r}{E_{n,\bk}}) \ket{ \psi_{n,\bk} } \bra{ \psi_{n,\bk} } \mathrm d\bk - \f{1}{\pi} \sum_{n\ge N} \int \mathrm{atan}(\f{r}{E_{n,\bk}}) \ket{ \psi_{n,\bk} } \bra{ \psi_{n,\bk} } \mathrm d\bk\lin 
&\xrightarrow{r\to \infty} \f{1}{2}(P - Q) }
where $N$ is the dimension of the gapped energy subspace at each $\bk$. The limit $r \to \infty$ can be exchanged with the integral $\int \mathrm d\bk$ because of a slight generalization to parametrized functions of Lebesgue's Dominated Convergence Theorem: For all $n,\bk$ the energy $E_{n,\bk} < \infty$ is a possibly large but finite number and the $r$-parametrized functions $\mathrm{atan}(\f{r}{E_{n,\bk}})$ are bounded by $\f{\pi}{2}$ for all $r$. Since the integral $\int \f{\pi}{2} \ket{ \psi_{n,\bk} } \bra{ \psi_{n,\bk} } \mathrm d\bk$ is finite, the theorem allows us to move the limit $r \to \infty$ across the integral, i.e.
\e{ \lim_{r \to \infty} \int \mathrm{atan}(\f{r}{E_{n,\bk}}) \ket{ \psi_{n,\bk} } \bra{ \psi_{n,\bk} } \mathrm d\bk = \int \lim_{r \to \infty} \mathrm{atan}(\f{r}{E_{n,\bk}}) \ket{ \psi_{n,\bk} } \bra{ \psi_{n,\bk} } \mathrm d\bk = \f{\pi}{2} \int \ket{ \psi_{n,\bk} } \bra{ \psi_{n,\bk} } \mathrm d\bk. }
Employing \q{gamma0} and \q{gammac}, we find
\es{ -\f{1}{2\pi i} \int_{\gamma_r} R_0(z) \mathrm dz &= -\f{1}{\pi} \sum_{n= 1}^N \int \big(\mathrm{atan}(\f{r}{E_{n,\bk}})-\pi \big) \ket{ \psi_{n,\bk} } \bra{ \psi_{n,\bk} } \mathrm d\bk \lin 
&+ \f{1}{\pi} \sum_{n\ge N} \int \mathrm{atan}(\f{r}{E_{n,\bk}}) \ket{ \psi_{n,\bk} } \bra{ \psi_{n,\bk} } \mathrm d\bk \xrightarrow{r\to \infty} \f{1}{2} \mathbb{1}. }\\

To prove \q{GGinf}, we use \q{Resbound} and that the distance between the curve $\gamma_r$ and $\sigma(H_0)$ is bounded from below by $r-E_c-\rho$, where $\rho$ is a lower bound on the spectrum of $H_0$ \big(Fig. \ref{arc} (a)\big). This gives $\| R_0(z) \| \le \f{1}{r-E_c -\rho}$ which we then use in
\e{ \| \int_{\gamma_r} \big( R_0(z) \big)^l A R_0(z) \mathrm dz \| \le \int_{\pi/2}^{3\pi/2} \| R_0(z) \|^l \| A R_0(z)\| r \mathrm d \theta \le M (r-E_c) \pi (r-E_c-\rho)^{-l} \xrightarrow{r \to \infty} 0. }
This proves case (b) for all $l>1$. To prove case (a) we use $\|A R_0(z)\| \le \|A \| \|R_0(z)\| \le M \|R_0(z)\|$ such that now
\e{ \| \int_{\gamma_r} \big( R_0(z) \big)^l A R_0(z) \mathrm dz \| \le \int_{\pi/2}^{3\pi/2} \| R_0(z) \|^{l+1} \| A \| r \mathrm d \theta \le M (r-E_c) \pi (r-E_c-\rho)^{-l-1} \xrightarrow{r \to \infty} 0 \la{boundest}}
for all $l>0$. If the norm of a quantity converges to zero in a certain limit, the quantity itself converges to zero which in our case leaves us with \q{GGinf} where $\Gamma_{\infty} = \lim_{r \to \infty} \Gamma_r$.\hfill\(\Box\) \\

Now, let us check how \q{GGinf} applies to certain cases that we have encountered in \app{app:reduction}, for all of which $l=1$.
\begin{itemize}
\item \q{man3}: $\oint_{\Gamma} R_0(z) B_0 R_0(z) \mathrm dz$, $A = B_0$, $M$ is given by \q{B0est}; therefore case (a) applies,
\item \q{Ap}: $\oint_{\Gamma} R_0(z) \big( B_0 R_0(z) \big)^2 \mathrm dz + \mathcal{O}(F^3)$, $A = B_0 R_0(z) B_0$, $M = \f{2 \hbar^2}{m} I_{1,1} I_{2,1}$ from an estimate similar to \q{B0}; therefore case (a) applies,
\item \q{B1b}: We split the expression into four integrals
\begin{itemize}
\item $\oint_{\Gamma} R_0(z) T^{1/2} R_0(z) B_0 R_0(z) \mathrm dz$, $A = T^{1/2} R_0(z) B_0$, $M$ is given by \q{T12} times \q{B0est}; therefore case (a) applies,
\item $\oint_{\Gamma} R_0(z) B_0 R_0(z) T^{1/2} R_0(z) \mathrm dz$, $A = B_0 R_0(z) T^{1/2}$, $M$ is given by \q{T12} times \q{B0est}; therefore case (a) applies,
\item $\oint_{\Gamma} R_0(z) [P,x]^2 R_0(z) \mathrm dz$, $A = [P,x]^2$, $M$ is given by \q{B0}; therefore case (a) applies,
\item $\oint_{\Gamma} R_0(z) [[P,x],x] R_0(z) \mathrm dz$, $A = [[P,x],x]$, $M$ is given by \q{P0xx}; therefore case (a) applies,
\end{itemize}
\item \q{B0}: $\oint_{\Gamma} R_0(z) T^{1/2} R_0(z) \mathrm dz$, $A=T^{1/2} R_0(z)$, $M$ is given by \q{T12}; this time we would like to apply case (b) but $l=1$ is not included in the statement; it cannot be applied. However,\cite{Nenciu} for all $z \in \gamma_r\colon \|(T-z)^{-1}\| \le \f{1}{r}$ because of the non-negativity of $T$. Thus \q{Vest} is still satisfied and similarly \q{Test1}. Then the integrand is of order $r^{-3/2}$, and the integral vanishes for $r \to \infty$.
\end{itemize}

\end{widetext}

\bibliography{BO_total}

\end{document}